\documentclass[11pt]{article}
\usepackage[top=2cm,left=2cm,right=2cm]{geometry}
\usepackage[numbers,square]{natbib}
\newcommand{\rra}{\rangle }
\newcommand{\lla}{\langle}
\usepackage{epsfig}
\usepackage{cancel}
\usepackage{caption}
\usepackage{feynmf} 
\usepackage{amssymb}
\usepackage{amsfonts}
\usepackage{epsf}
\usepackage{rotating}
\usepackage{graphicx}
\usepackage{amsmath}
\usepackage{fancyhdr}
\usepackage{subfigure}
\usepackage{graphics}
\usepackage{pstricks}
\usepackage{color}
\usepackage{frontespizio}
\usepackage{hyperref}
\hypersetup{
    colorlinks,
    citecolor=green,
    filecolor=black,
    linkcolor=blue,
    urlcolor=black
}
\usepackage{type1ec}
\usepackage[T1]{fontenc}
\usepackage{lettrine}
\usepackage{bbold}
\usepackage{calligra}
\usepackage{tikz}
\usepackage{subfigure}
\usepackage{mathrsfs}
\usepackage{curve2e}
\usepackage{setspace}
\usepackage{indentfirst}
\usepackage{emptypage}
\usepackage[babel]{csquotes} %bibliografia
\usepackage[font=small,labelfont=bf,labelsep=quad]{caption} %tabelle e figure
\usepackage{graphicx} %figure
\usepackage{listings} %codici
\usetikzlibrary{patterns}
\usepackage{relsize}
\usetikzlibrary{intersections,positioning}
\usetikzlibrary{decorations.pathmorphing,decorations.markings,arrows,positioning}
\usepackage{braket}
\usepackage{mathrsfs}
\usepackage{stackengine}
\usepackage{calc}
\newlength\shlength
\newcommand\xshlongvec[2][0]{\setlength\shlength{#1pt}%
  \stackengine{-5.6pt}{$#2$}{\smash{$\kern\shlength%
    \stackengine{7.55pt}{$\mathchar"017E$}%
      {\rule{\widthof{$#2$}}{.57pt}\kern.4pt}{O}{r}{F}{F}{L}\kern-\shlength$}}%
      {O}{c}{F}{T}{S}}

\newcommand{\sdfrac}[2]{\mbox{\small$\displaystyle\frac{#1}{#2}$}}

\IfFileExists{dsfont.sty}
{\usepackage{dsfont}
	\let\mathbb=\mathds
	\newcommand{\id}{\mathds{1}}}
{\typeout{Package dsfont.sty was not found, using alternative macros.}
	\let\mathds=\mathbb
	\newcommand{\id}{\mbox{1 \kern-.59em {\rm l}}}}

\usepackage{slashed}
\usepackage{units}
\usepackage{setspace}
\newcommand{\nn}{\nonumber}
\let\a=\alpha   \let\b=\beta   \let\g=\gamma   \let\d=\delta
\let\e=\epsilon         
    \let\k=\kappa  \let\l=\lambda  \let\m=\mu
\let\n=\nu           \let\p=\pi      \let\r=\rho
\let\s=\sigma        
\let\c=\chi         
\let\D=\Delta
\let\d=\delta
\let\s=\sigma
\let\P=\Pi
\newcommand{\figref}[1]{Fig.~\ref{#1}}			% for figures
\newcommand{\secref}[1]{Section~\ref{#1}}		% for sections
\newcommand{\appref}[1]{Appendix~\ref{#1}}		% for appendix references
\renewcommand{\a}{\alpha}

\renewcommand{\r}{\rho}
\newcommand{\G}{\Gamma}

\newcommand{\Tr}{\text{Tr}}

\def\nbox#1#2{\vcenter{\hrule \hbox{\vrule height#2in
			\kern#1in \vrule} \hrule}}
\def\sq{\,\raise.5pt\hbox{$\nbox{.09}{.09}$}\,}
\def\sqb{\,\raise.5pt\hbox{$\overline{\nbox{.09}{.09}}$}\,}

\newcommand{\bea}{\begin{eqnarray}}
\newcommand{\eea}{\end{eqnarray}}
\newcommand{\be}{\begin{equation}}
\newcommand{\ee}{\end{equation}}

\newcommand{\bes}{\begin{subequations}}
	\newcommand{\ees}{\end{subequations}}

\def\nn{\nonumber\\}

\def\Box{\sq}

\numberwithin{equation}{section}

\usepackage{accents}

\begin{document}
\begin{center}
\vspace{1.5cm}
{\Large\bf  The General 3-Graviton Vertex ($TTT$) of Conformal Field Theories \\
 in Momentum Space in $d=4$\\ }

\vspace{0.3cm}

\vspace{0.3cm}

\vspace{0.2cm}

\vspace{0.3cm}

\vspace{2 cm}

{\bf  Claudio Corian\`o and Matteo Maria Maglio \\}

\vspace{0.5cm}

{Dipartimento di Matematica e Fisica {\it "Ennio De Giorgi"}\\
Universit\`a del Salento \\ and \\
 INFN Lecce, \\
Via Arnesano, 73100 Lecce, Italy}

\vspace{0.5cm}

\end{center}
%\date{\version}
\begin{abstract}
We present a study of the correlation function of three stress-energy tensors in $d$ dimensions using free field theory realizations, and compare them to the exact solutions of their conformal Ward identities (CWI's) obtained by a general approach in momentum space. The identification of the corresponding form factors is performed within a reconstruction method, based on the identification of the transverse traceless components $(A_i)$ of the same correlator. The solutions of the primary CWI' s are found by exploiting the universality of the Fuchsian indices of the conformal operators and a re-arrangement of the corresponding inhomogenous hypergeometric systems. We confirm the number of constants in the solution of the primary CWI's of previous analysis.  
In our comparison with perturbation theory, we discuss scalar, fermion and spin 1 exchanges at 1-loop in dimensional regularization. 
Explicit checks in $d=3,4,5$ prove the consistency of this correspondence. By matching the 3 constants of the CFT solution with the 3 free field theory sectors available in $d=4$, the general solutions of the conformal constraints is expressed just in terms of ordinary scalar 2- and 3-point functions $(B_0,C_0)$. We show how the renormalized $d=4$ $TTT$ vertex separates naturally into the sum of a traceless and an anomaly part, the latter determined by the anomaly functional and generated by the renormalization of the 
correlator in dimensional regularization. The result confirms the emergence of anomaly poles and effective massless exchanges as a specific signature of conformal anomalies in momentum space, directly connected to the renormalization of the corresponding gravitational vertices, generalizing the behaviour found for the $TJJ$ vertex in previous works.

\end{abstract}

\newpage
\section{Introduction} 
Exact results in four dimensional conformal field theories (CFT's) have gathered a lot of attention along the years, 
mostly because the enlarged $SO(2,4)$ symmetry of such theories has been essential for determining the structure of the correlators, especially for 2- and 3-point functions. These are derived by imposing on them the corresponding conformal Ward identities (CWI's), which  in even spacetime dimensions are broken by the conformal anomaly  \cite{Duff:1993wm}. \\
Such analysis have traditionally been performed in coordinate space, where the conformal constraints are readily implemented. We recall that the equations for a certain correlator are first solved for separate external spacetime coordinates - giving the homogeneous, conformally invariant solutions - and, at a second stage, the contribution due to the conformal anomaly is taken into account by adding an ultralocal term \cite{Osborn:1993cr, Erdmenger:1996yc}, which is generated when all the external coordinate points coalesce. In dimensional regularization (DR) in $d=4$ this is achieved by the standard $E$ and $C^2$ counterterms, corresponding to the Euler-Poincar\`e density and to the Weyl tensor squared respectively. The anomaly is generated by the $d$-dimensional trace of such counterterms, which is responsible for the appearance of a finite inhomogeneous term as $d\to 4$. One of the goal of our analysis will be to describe step by step the perturbative renormalization for the $TTT$, illustrating the separation between the anomaly (or trace) and the traceless parts in the free field theory realization of such correlator, which may turn useful for future studies of the conformal anomaly action. As we are going to discuss, we will work in full generality, and in $d=3$ and $4$ our analysis matches the most general CFT solution, providing the simplest realization of such correlator. We anticipate that our results are in complete agreement with previous analysis of similar correlators such as the $TJJ$ \cite{Giannotti:2008cv, Armillis:2009pq,Coriano:2018zdo}. In particular, this works extends the perturbative results contained in \cite{Coriano:2018bbe}, where some of the methods used for the transition to momentum space have been extensively discussed and to which we refer for further details.

\subsection{The hierarchy of the CWI's and the BMS reconstruction}
The conformal constraints are hierarchical, with 3-point functions defined in terms of 2-point functions, and so on for higher orders. For 3-point functions they are strong enough to fix the solutions modulo few constants. For 4-point functions the solutions are only partially fixed, due to the presence of an arbitrary function of two conformally invariant ratios. Completely traced correlators of $4-T$'s $6-T$'s and higher are special, since they are fixed by the first 4 correlators of the same type, due to the presence of conformal trace relations \cite{Coriano:2013xua}, given the quartic nature of the Wess-Zumino action, as derived by Weyl gauging \cite{Coriano:2013nja} or by other methods.\\
Undoubtedly, the theoretical interest in this family of correlators is remarkable because of the appearance, beside of the conformal, also of mixed anomalies with axial vector currents \cite{Bastianelli:2012bz,Bastianelli:2012es,Bastianelli:2000hi,Bastianelli:1999ab,Bastianelli:2016nuf, Bastianelli:2018osv} \cite{Bonora:2017gzz,Bonora:2018obr,Bonora:2014qla}. \\
In coordinate space, the solution of the conformal constraints \cite{Osborn:1993cr} has been obtained for several correlation functions, among which the most demanding one was the $TTT$, i.e. the correlator of three stress-energy tensors (with free indices).
Semi-local contributions, which are generated by a partial pinching of the external coordinates, appear on the right hand side of the conformal WI's, multiplied by 2-point functions, showing that a complete reconstruction of tensor correlators is indeed possible in a hierarchical way. In this process, conservation and trace WI's are determined by correlators of lower orders (e.g 2-point functions in the $TTT$ case).

As shown by Osborn and Petkou in coordinate space \cite{Osborn:1993cr} and by Bzowski, McFadden and Skenderis (BMS approach) \cite{2014JHEP...03..111B} in momentum space, the solutions of such equations are determined up to few constants, which characterize a specific CFT. BMS have shown that it is possible to formulate a complete reconstruction procedure for tensor correlators which extends the case of scalar correlators \cite{Coriano:2013jba}.
In the case of a Lagrangian realization of a CFT, such constants are determined by the field content, i.e. the number of scalars, fermions etc. appearing in the theory, and for a sufficient number of independent family sectors and particle multiplicities they are expected to saturate the exact (non Lagrangian) solution and provide its simplest realization.\\
Such perturbative solutions, obviously, remain valid beyond perturbation theory and match the general CFT prediction for 3-point functions in specific cases, as we are going to show. The $d=3$ and $d=4$ cases are such, since the presence of two free field theories sectors, scalars and fermions in $d=3$, and of scalars, fermions and vectors in $d=4$, for instance, allow to perform a matching between general  and Lagrangian solutions in the most general way. \\ The simplifications obtained from the perturbative analysis, respect to the general solution - no matter if expressed in terms of 3K integrals (i.e. integrals of 3 Bessel functions)\cite{2014JHEP...03..111B}, or, as we are going to show, directly in terms of the Appell's function $F_4$ \cite{Coriano:2013jba, Coriano:2018bbe}, discussed in \cite{Coriano:2018bbe} and here in Section \ref{fuchs} -  are remarkable. In fact, in the latter case, we are allowed to use only $B_0$ and $C_0$, the scalar 2- and 3-point functions of ordinary perturbation theory at one loop, to express the full result. Obtaining such expressions for the $TTT$  or for other vertices and using a specific reconstruction method which keeps the number of form factors minimal, allows to achieve a large simplification of the entire approach. At the end of our analysis we will summarize the explicit way to obtain the reconstruction of such a vertex, which may turn useful also for further studies in quantum gravity.

\subsection{The role of perturbation theory in the equivalence} 

While the equivalence between the CFT and the free field theory solutions is obviously expected at some level,  the search for an exact match between the two approaches not only provides a simplification of the results but also offers some physical intuition about the origin of the conformal anomaly, once we move to momentum space. In coordinate space this dynamical interpretation is simply absent.
In fact, conformal and chiral anomalies show a striking similarity, in momentum space, due to the emergence of anomaly poles \cite{Giannotti:2008cv, Armillis:2009pq, Armillis:2009im, Armillis:2010qk}. The general nature of this mechanism is easily understood in momentum space by dispersion theory, and unifies chiral and conformal anomalies, as evident from the anomaly supermultiplet  \cite{Coriano:2014gja}.  They provide a possible unifying tract of both, which otherwise appear to be completely unrelated. A simple physical description of this phenomenon is indeed currently possible in perturbation theory (see the discussion in our companion paper \cite{Coriano:2018bbe}), at least in the $TJJ$ and in its supersymmetric version.\\
 It is associated, in the integration over the loop momentum ($k$), to the region of $k$ describing the decay of a massive external graviton line into two collinear massless particles, and then turning into two photons ($t$-channel cut) \cite{Coriano:2018bbe}. The result of this interaction is in the generation of a term $(\sim \beta(g)/k^2)$, proportional to the $\beta$-function of a theory, which vanishes if the theory is conformal at quantum 
level $(\beta=0)$. We have recently shown that this phenomenon, in the $TJJ$ case, holds beyond perturbation theory  
\cite{Coriano:2018bbe} having compared exact solutions with their Lagrangian realization in QED and it is associated to renormalization.

The pole structure describes in the light-cone behaviour of a theory and of its anomaly in terms of its intrinsic degrees of freedom, without the introduction of a Goldstone mode, such as in the case of the Wess-Zumino action. 
 It is more than an educated guess to predict that in the infrared such an action is expected to take a second form. This would be a spontaneously broken version of the same theory, with an asymptotic degree of freedom in the form of a dilaton, the Goldstone mode of broken conformal simmetry, and with the emergence of 3- and 4- dilaton interactions in the infrared, as described by its Wess-Zumino form \cite{Coriano:2012dg,Coriano:2013xua,Coriano:2012nm}.
For such a reason, the possibility of matching the CFT and the perturbative solutions of the CWI's in momentum 
space for the more complex case of the $TTT$ can help to clarify these issues, at least up to cubic level in the fluctuations of the external gravitational metric. At the same time such analysis is a mandatory initial step for further studies of the exact form of the conformal anomaly action.

 \subsection{Perturbative matchings}
The perturbative matching between the general and the Lagrangian solutions of the $TTT$ can be obtained by investigating a certain number of independent sectors for such correlator in general ($d$) spacetime dimensions. Given the fact that the general solution of the CWI's depends on 2 independent constants in $d=3$, and on 3 for $d \geq 4$  \cite{Osborn:1993cr}, the study of this correspondence is performed, in perturbation theory, by the inclusion of  2 sectors (fermion and scalar) for odd values of $d$, with the addition of a gauge sector in $d=4$, which is conformal invariant only in $d=4$. Compared with the $d=3$ and $d=5$ cases, where the $TTT$ is finite, in $d=4$ the Feynman diagrams need to be renormalized, with the generation of an anomalous contribution, and the matching with the general CFT solution, in this case, is complete. \\
We should also mention that in higher even dimensions the use of antisymmetric forms running in the loops, which take the same role of the spin 1 sector of $d=4$, may allow to extend our analysis, providing a third independent sector. 
In odd spacetime dimensions only the case $d=3$ is entirely matched, since in this case it has been shown that only two constants are necessary in order to characterize the general solution of the CWI's. The analysis presented in \cite{Osborn:1993cr} in coordinate space and in \cite{Bzowski:2013sza,Bzowski:2018fql} in momentum space agree on the presence of 3 independent constants in the general solution for $d > 3$, which clearly cannot be completely 
matched by the two conformally invariant free field sectors (scalars and fermions) which are available in odd dimensions. 
The results for the $TTT$ presented in \cite{Bzowski:2013sza} for $d=3$ and $d=5$ are rather simple, and are in agreement with the result obtained by us by combining the two free field theory sectors (scalars and fermions) which are available in the same dimensions. In $d=5$, for instance, the solution given in \cite{Bzowski:2013sza} is accurately matched, in our case, by such sectors, but it is expected to correspond to a particular solution of the conformal constraints. It is therefore possible that the additional CFT's described by the general solution $d=2 k+1$ correspond    to interacting CFT's which do not find a perturbative free-field theory realization. 

\subsection{Reconstruction}
The explicit expression of the $TTT$ vertex in momentum space is expected to be very involved, unless one is able to identify a specific procedure in order to reduce the number of independent form factors and bring them into correspondence with the available solution in coordinate space. Attempts in this direction have been made in the past \cite{Cappelli:2001pz, Coriano:2012wp}, but a general method that introduces a minimal set of form factors which does the job and allows to reconstruct the entire correlator has been proposed only more recently for the $TTT, TTO, TJJ$ correlators \cite{Bzowski:2013sza}. The method is based on a reconstruction program for such correlators which starts from their transverse traceless sectors and builds up the entire correlator exploiting the endomorphic action of the special conformal transformation, with the longitudinal components determined with the use of the conservation and trace Ward identities. The method is completely autonomous compared to coordinate space \cite{Bzowski:2017poo} and allows to derive scalar equations for the transverse traceless form factors which are then solved in terms of 3K integrals. An alternative approach, which bypasses such integrals has been presented by us for the $TJJ$ in our companion work \cite{Coriano:2018bbe}, which is based on the observation that the Fuchsian indices of the conformal Ward identities are universal for such systems of equations \cite{Coriano:2018bbe}. We will show how to re-arrange the hypergeometric differential equations of the CWI's in such a way to generate their non-homogenous solutions starting from the homogenous ones, extending our method from the $TJJ$ to the $TTT$. This parts of our analysis is quite independent of the rest of the work but it confirms that the set of the primary CWI's is indeed determined by a set of 5 constants, in agreement with \cite{Bzowski:2013sza}. This provides a second independent check on the number of constants present in such solutions before the imposition of the constraints of momentum conservation (secondary Ward identities).  

As we are going to show in a forthcoming work, our approach can be extended in order to look for special solutions for more general correlators, of 4- and higher points, which are, obviously, not completely fixed by the underlying conformal symmetry. We hope to come back to this point in the future.
\subsection{Perturbative solutions}
One of the main issues with the general solution is that it gets very involved in the presence of divergences, and requires 
an entirely new regularization procedure for such 3K integrals, which, however, does not make transparent the fact that the result has to be clearly equivalent to the perturbative one. We have not attempted to compare our results with those of  \cite{Bzowski:2017poo} for $d=4$, but we have verified their complete agreement in $d=3$ and $5$ using our $d$-dimensional computation. By the same token, the anomalous CWI's will be derived using our Lagrangian framework and are as general as those derived in \cite{Bzowski:2017poo}, but in 
a far more direct and simplified form.\\
The study of the matching between the general and the perturbative solutions, and the check of their equivalence, will be done by working in $d=3$ and $5$ dimensions, in order to prove the consistency of our results with the general solution obtained from CFT \cite{Bzowski:2013sza}.  The matching to perturbation theory brings in significant simplification of the general solution in terms of 3K integrals, or  the very same solution in terms of Appell's hypergeometrics that we will present below.\\
 Notice that due to the need or regularizing the solution,
3K integrals \cite{Bzowski:2013sza,Bzowski:2015yxv, Bzowski:2017poo,Bzowski:2018fql} are not the master integrals of perturbation theory, since the propagators appearing in the loop - after a suitable conversion - do not carry integers exponents. They cannot be handled by the ordinary reduction procedures which are typical of the multiloop analysis in QCD, due to the need of shifting the exponents in the Feynman propagators of the integrands by a (real) regulator. This has motivated us to reconsider independently all the BMS reconstruction  \cite{Bzowski:2017poo} from a simple perturbative perspective. While this follows overall the original proposal, some of the relations concerning the projected special CWI's have been reobtained using an independent strategy. For instance, we have made an extensive use of Lorentz Ward identities, not mentioned in the original work, in order to come to a final agreement with the expressions quoted 
in \cite{Bzowski:2017poo}.

\section{The TTT and TTO correlators }
We start by stating our definitions and conventions. 
We introduce the ordinary definition of the energy-momentum tensor in terms of the generating functional of the theory $\mathcal{W}$ in the Euclidean case
 \begin{equation}
\braket{T^{\m\n}(x)}=\sdfrac{2}{\sqrt{g(x)}}\sdfrac{\d \mathcal{W}}{\d g_{\m\n}(x)}
 \end{equation}
 where
 \begin{equation}
\mathcal{W}=\sdfrac{1}{\mathcal N}\int\,\mathcal D\,\Phi\ e^{-S}
 \end{equation}
 with $\mathcal N$ a normalization factor. $\Phi$ denotes all the quantum fields of the theory and $S$ is the quantum action. For the multi-graviton vertices, it is convenient to define the corresponding correlation function as the $n$-th functional variation with respect to the metric of the generating functional $\mathcal W$ evaluated in the flat-space limit
 \begin{equation}
 \begin{split}
\braket{T^{\m_1\n_1}(x_1)\dots T^{\m_n\n_n}(x_n)}\equiv&\left[\sdfrac{2}{\sqrt{g(x_1)}}\dots\sdfrac{2}{\sqrt{-g(x_n)}}\sdfrac{\d^n\mathcal{W}}{\d g_{\m_1\n_1}(x_1)\dots\d g_{\m_n\n_n}(x_n)}\right]_{flat}\\
=&\left.2^n\sdfrac{\d^n\mathcal{W}}{\d g_{\m_1\n_1}(x_1)\dots\d g_{\m_n\n_n}(x_n)}\right|_{flat}
\end{split}\label{defTn}
 \end{equation}
so that it is explicitly symmetric with respect to the exchange of the metric tensors. The 3-point function we are interested in studying is found through \eqref{defTn} for $n=3$
\begin{equation}
\begin{split}
&\braket{T^{\m_1\n_1}(x_1)T^{\m_2\n_2}(x_2)T^{\m_3\n_3}(x_3)}=8\,\bigg\{-\Braket{\sdfrac{\d S}{\d g_{\m_1\n_1}(x_1)}\,\sdfrac{\d S}{\d g_{\m_2\n_2}(x_2)}\,\sdfrac{\d S}{\d g_{\m_3\n_3}(x_3)}}\\
&\hspace{1.5cm}+\Braket{\sdfrac{\d^2 S}{\d g_{\m_1\n_1}(x_1)\d g_{\m_2\n_2}(x_2)}\,\sdfrac{\d S}{\d g_{\m_3\n_3}(x_3)}}+\Braket{\sdfrac{\d^2 S}{\d g_{\m_1\n_1}(x_1)\d g_{\m_3\n_3}(x_3)}\,\sdfrac{\d S}{\d g_{\m_2\n_2}(x_2)}}\\
&\hspace{1.5cm}+\Braket{\sdfrac{\d^2 S}{\d g_{\m_2\n_2}(x_2)\d g_{\m_3\n_3}(x_3)}\,\sdfrac{\d S}{\d g_{\m_1\n_1}(x_1)}}-\Braket{\sdfrac{\d^3 S}{\d g_{\m_1\n_1}(x_1)\d g_{\m_2\n_2}(x_2)\d g_{\m_3\n_3}(x_3)}}\bigg\}
\end{split}\label{expansion}
\end{equation}
where the angle brackets denote the vacuum expectation value. The last term is identically zero in DR, being proportional to a massless tadpole. The first term on the rhs of \eqref{expansion} has the diagrammatic representation of a triangle topology, while the contribution of a second functional derivative times a single derivative of the action is interpreted in the perturbative analysis as a bubble diagram. We will keep in mind such decompositon which will be relevant in the last paper of our work once we come to the perturbative analysis
.
\section{Canonical and trace Ward Identities}
The conformal constraints for the $TTT$ correspond to dilatation and special conformal transformations, beside the 
usual Lorentz symmetries. Generically
\begin{equation}
\sum_{j=1}^3G_g(x_j)\braket{T^{\m_1\n_1}(x_1)\,T^{\mu_2\nu_2}(x_2)\,T^{\mu_3\nu_3}(x_3)}=0,
\end{equation}
where $G_g$ are the generators of the infinitesimal symmetry transformations. Among these, the conservation WI in flat space of the stress-energy tensor can be obtained by requiring the invariance of $\mathcal{W}[g]$ under diffeomorphisms of the background metric 
\begin{equation}
\mathcal{W}[g]=\mathcal{W}[g']
\end{equation}
where $g'$ is the transformed metric under the general infinitesimal coordinate transformation $x^\m\to x'^\mu=x^\mu -\e^\mu$
\begin{equation}
\d g_{\m \n }=\nabla_\m \epsilon_\n +\nabla_\n \e_\m.
\end{equation} It generates the relation
\begin{equation}
\nabla_\n \braket{T^{\m \n}}=0
\label{transverse}
\end{equation}
while naive scale invariance gives the traceless condition
\begin{equation}
g_{\mu\nu}\braket{T^{\m\n}}=0.\label{trace}
\end{equation}
These have been the only constraints taken into account in previous perturbative studies of the $TJJ$  
\cite{Armillis:2009pq,Giannotti:2008cv,Armillis:2010qk} and $TTT$ \cite{Coriano:2012wp}. The functional differentiation of \eqref{transverse} and \eqref{trace} allows to derive ordinary Ward identities for the various correlators. 
For the three point function case these take the form 
\begin{align} \label{WI3PFcoordinate}
\partial_\nu\langle T^{\mu\nu}(x_1)T^{\rho\sigma}(x_2)T^{\alpha\beta}(x_3) \rangle
&=
\bigg[\langle T^{\rho\sigma}(x_1)T^{\alpha\beta}(x_3)\rangle\partial^\mu\d(x_1,x_2) + 
\langle T^{\alpha\beta}(x_1)T^{\rho\sigma}(x_2)\rangle\partial^\mu\d(x_1,x_3) \bigg]\nn 
&\quad-
\bigg[\delta^{\mu\rho}\langle T^{\nu\sigma}(x_1)T^{\alpha\beta}(x_3)\rangle
+     \delta^{\mu\sigma}\langle T^{\nu\rho}(x_1)T^{\alpha\beta}(x_3)\rangle\bigg]\partial_\nu\d(x_1,x_2)\nn
&\quad-
\bigg[\delta^{\mu\alpha}\langle T^{\nu\beta}(x_1)T^{\rho\sigma}(x_2)\rangle
+ \delta^{\mu\beta}\langle T^{\nu\alpha}(x_1)T^{\rho\sigma}(x_2)\rangle\bigg]\partial_\nu\d(x_1,x_3)\, .
\end{align}
In order to move to momentum space we fix some conventions. The Fourier transform of the correlators is defined as  
\begin{align}
\braket{T^{\mu_1\nu_1}(p_1)\,T^{\m_2\n_2}(p_2)\,T^{\m_3\n_3}(p_3)}=\int d^d x_1 d^d x_2 d^d x_3 e^{i\left( p_1\cdot x_1 + p_2\cdot x_2 + p_3\cdot x_3\right)}\braket{T^{\mu_1\nu_1}(x_1)\,T^{\m_2\n_2}(x_2)\,T^{\m_3\n_3}(x_3)}
\end{align} 
and similarly for the 2-point function. Translational invariance  introduces an overall $\delta(P)$  with 
$P$ being the sum of all the (incoming) momenta, with the generation of derivative terms $\delta'(P)$, after the action of the special conformal transformations on the integrand.
Such terms can be investigated rigorously using the theory of tempered distributions, formulated using a symmetric basis. The analysis has been presented in \cite{Coriano:2018bbe} for a Gaussian basis, to which we refer for more details. In our conventions, we have chosen $p_3$ as the dependent momentum $ {p_3}\to -p_1 - p_2$. Eq. \eqref{WI3PFcoordinate} becomes
\begin{align}
p_{1\n_1}\braket{T^{\mu_1\nu_1}(p_1)\,T^{\m_2\n_2}(p_2)\,T^{\m_3\n_3}({p_3})}&=-p_2^{\m_1}\braket{T^{\m_2\n_2}(p_1+p_2)T^{\m_3\n_3}({p_3})}-{p_3^{\m_1}}\braket{T^{\m_2\n_2}(p_2)T^{\m_3\n_3}(p_1+{p_3})}\notag\\
&\hspace{-1.5cm}+p_{2\a}\left[\d^{\m_1\n_2}\braket{T^{\mu_2\a}(p_1+p_2)T^{\m_3\n_3}({p_3})}+\d^{\m_1\m_2}\braket{T^{\nu_2\a}(p_1+p_2)T^{\m_3\n_3}({p_3})}\right]\notag\\
&\hspace{-1.5cm}+{p_{3\a}}\left[\d^{\m_1\n_3}\braket{T^{\mu_3\a}(p_1+{p_3})T^{\m_2\n_2}(p_2)}+\d^{\m_1\m_3}\braket{T^{\nu_3\a}(p_1+{p_3})T^{\m_2\n_2}(p_2)}\right].
\label{long}
\end{align}
In the next section, in order to clarify that differentiation in $p_3$ has to be performed with the chain rule, we will denote with 
$\bar{p}_3^\mu\equiv -p_1^\mu - p_2^\mu$, the dependent momentum, and the independent 4-momenta will be $p_1^\mu$ and $p_2^\mu$. 
Concerning the naive identity \eqref{trace}, it generates the non-anomalous condition
\begin{equation}
g_{\m_1\n_1}\braket{T^{\mu_1\nu_1}(p_1)\,T^{\m_2\n_2}(p_2)\,T^{\m_3\n_3}(p_3)}=0
\end{equation}
valid in the $d\ne4$ case. \\
After renormalization this equation is modified by the contribution of the conformal 
anomaly, given by the general expression
\bea \label{TraceAnomaly}
g_{\mu\nu}(z)\langle T^{\mu\nu}(z) \rangle
&=&
\sum_{I=F,S,G}n_I \bigg[\beta_a(I)\, C^2(z) + \beta_b(I)\, E(z)\bigg]  
+ \frac{\kappa}{4}n_G F^{a\,\mu\nu}\,F^a_{\mu\nu} (z) \nn 
&\equiv& \mathcal{A}(z,g) \, ,
\eea
by considering only the scheme independent terms with
\begin{align}
\b_a(S)&=-\frac{3\p^2}{720}\,,\hspace{1cm}\b_b(S)=\frac{\p^2}{720}\,,\nn
\b_a(F)&=-\frac{9\p^2}{360}\,,\hspace{1cm}\b_b(F)=\frac{11\p^2}{720}\,\nn
\b_a(G)&=-\frac{18\p^2}{360}\,,\hspace{1.0cm}\b_b(G)=\frac{31\p^2}{360}\,\label{choiceparm}
\end{align}
being the contributions to the $\beta$ functions coming from scalars $(S)$, fermions $(F)$ and vectors $(G)$. We have defined the 
two tensors  
\begin{align}
C^2&=R_{abcd}R^{abcd}-\sdfrac{4}{d-2}R_{ab}R^{ab}+\sdfrac{2}{(d-2)(d-1)}R^2,\hspace{0.7cm}
&E=R_{abcd}R^{abcd}-4R_{ab}R^{ab}+R^2
\end{align}
being the square of the Weyl conformal tensor and the Euler-Poincar\'e density respectively, while $R_{abcd}$ is the Riemann curvature tensor and $R_{ab}$ and $R$ are the Ricci tensor and the Ricci scalar, respectively. Then we get the anomalous WI
\begin{align}
&g_{\mu_1\nu_1}\braket{ T^{\mu_1\nu_1}(p_1)T^{\mu_2\nu_2}(p_2)T^{\mu_3\nu_3}(p_3)}\notag\\
&\hspace{2cm}=
4 \, \mathcal A^{\mu_2\nu_2\mu_3\nu_3}(p_2,p_3)
- 2 \, \braket{ T^{\mu_2\nu_2}(p_1+p_2)T^{\mu_3\nu_3}(p_3)} - 2 \, \braket{ T^{\mu_2\nu_2}(p_2)T^{\mu_3\nu_3}(p_1+p_3)}\nn
&\hspace{2cm}=
4 \, \bigg[ \beta_a\,\big[C^2\big]^{\mu_2\nu_2\mu_3\nu_3}(p_2,p_3)+ \beta_b\, \big[E\big]^{\mu_2\nu_2\mu_3\nu_3}(p_2,p_3) \bigg]\nn
&\hspace{3cm}- 2 \, \braket{ T^{\mu_2\nu_2}(p_1+p_2)T^{\mu_3\nu_3}(p_3)} - 2 \, \braket{ T^{\mu_2\nu_2}(p_2)T^{\mu_3\nu_3}(p_1+p_3)}. \label{munu3PFanomaly}
\end{align}
We just remark that the solutions of all the conformal constraints, in this study, are obtained by working with the non-anomalous expressions of the corresponding CWI's, while the anomaly contributions, as in (\ref{munu3PFanomaly}), are obtained only after taking the $d\to 4$ limit of the general solution and the inclusion of the corresponding counterterms. All these points will be investigated rather thoroughly in the following sections. We briefly pause to comment on the relation between the current and previous analysis \cite{Coriano:2012wp}  of the $TTT$ in free field theory.
The expression for the $TTT$ given in \cite{Coriano:2012wp} has been presented in a complete form only for the gravitational amplitude $g_1(p_1)\to g_2(p_2) + g_3(p_3)$, with $g_2$ and $g_3$ on-shell gravitons, which is quite involved. The expression given in \cite{Coriano:2012wp} breaks the full symmetry of the correlator and requires a basis of 13 form factors, which is nonminimal.
A symmetric and manageable reconstruction of this vertex requires a complete reanalysis of the correlator, with the inclusion also of the special conformal and dilatation constraints, which lower the number of independent form factors to a minimal number, which will be 5. This is the step that we are going to undertake starting from the next section.

\section{Special conformal and dilatation WI's}\label{Skend}
Dilatation and special conformal WI's in position space can be derived in various ways, and the transition to momentum space can be made rigorous by taking suitable distributional limits of the derivative of the Dirac delta functions, as discussed by us in \cite{Coriano:2018bbe}. In coordinate space, for the $TTT$, the special CWI's take the form 
\begin{align}
0={K}^\kappa \braket{T^{\mu_1\nu_1}(x_1)T^{\mu_2\nu_2}(x_2)T^{\mu_3\nu_3}(x_3)} 
&= \sum_{i=1}^{3} {K_i}^{ \kappa}_{scalar}(x_i) \braket{T^{\mu_1\nu_1}(x_1)T^{\mu_2\nu_2}(x_2)T^{\mu_3\nu_3}(x_3)}  \notag\\
& \hspace{-6.5cm}+ 2 \left(  \delta^{\mu_1\kappa} x_{1\rho} - \delta_{\rho}^{\kappa }x_1^{\mu_1 } \right)\braket{T^{\rho\nu_1}(x_1)T^{\mu_2\nu_2}(x_2)T^{\mu_3\nu_3}(x_3)} 
 + 2 \left(  \delta^{\nu_1\kappa} x_{1\rho} - \delta_{\rho}^{\kappa }x_1^{\nu_1 } \right)\braket{T^{\mu_1\rho}(x_1)T^{\mu_2\nu_2}(x_2)T^{\mu_3\nu_3}(x_3)} \nn
& \hspace{-6.5cm}+ 2 \left(  \delta^{\mu_2\kappa} x_{2\rho} - \delta_{\rho}^{\kappa }x_2^{\mu_2 } \right)\braket{T^{\mu_1\nu_1}(x_1)T^{\rho\nu_2}(x_2)T^{\mu_3\nu_3}(x_3)} 
+ 2 \left(  \delta^{\nu_2\kappa} x_{2\rho} - \delta_{\rho}^{\kappa }x_2^{\nu_2 } \right)\braket{T^{\mu_1\nu_1}(x_1)T^{\mu_2\rho}(x_2)T^{\mu_3\nu_3}(x_3)} \nn
& \hspace{-6.5cm}+ 2 \left(  \delta^{\mu_3\kappa} x_{3\rho} - \delta_{\rho}^{\kappa }x_3^{\mu_3 } \right)\braket{T^{\mu_1\nu_1}(x_1)T^{\mu_2\nu_2}(x_2)T^{\rho\nu_3}(x_3)} 
+ 2 \left(  \delta^{\nu_3\kappa} x_{3\rho} - \delta_{\rho}^{\kappa }x_3^{\nu_3 } \right)\braket{T^{\mu_1\nu_1}(x_1)T^{\mu_2\nu_2}(x_2)T^{\mu_3\rho}(x_3)} 
\end{align}
written in terms of a scalar contribution
\begin{equation}
\label{ki}
{{K}_i}^{\kappa}_{scalar}=-x_i^2 \frac{\partial }{\partial x_\kappa} + 2 x_i^\kappa x_i^\tau\frac{\partial}{\partial x_i^\tau} + 2 \Delta_i x_i^\kappa
\end{equation}
and of spin parts. In momentum space this becomes with $\Delta_i$, $i=1,2,3$ being the scaling dimensions of 3 generic rank-2 operators 
- here fixed to be $d$ for the $T^{\m\n}$ -
\begin{align}
&\sum_{j=1}^{2}\left[2(\Delta_j-d)\sdfrac{\partial}{\partial p_j^\k}-2p_j^\a\sdfrac{\partial}{\partial p_j^\a}\sdfrac{\partial}{\partial p_j^\k}+(p_j)_\k\sdfrac{\partial}{\partial p_j^\a}\sdfrac{\partial}{\partial p_{j\a}}\right]\braket{T^{\mu_1\nu_1}(p_1)\,T^{\m_2\n_2}(p_2)\,T^{\mu_3\n_3}(\bar p_3)}\notag\\
&\qquad+2\left(\d^{\k(\mu_1}\sdfrac{\partial}{\partial p_1^{\a_1}}-\delta^{\k}_{\alpha_1}\delta^{\l(\mu_1}\sdfrac{\partial}{\partial p_1^\l}\right)\braket{T^{\nu_1)\alpha_1}(p_1)\,T^{\m_2\n_2}(p_2)\,T^{\mu_3\n_3}(\bar p_3)}\notag\\
&\qquad+2\left(\d^{\k(\mu_2}\sdfrac{\partial}{\partial p_2^{\a_2}}-\delta^{\k}_{\alpha_2}\delta^{\l(\mu_2}\sdfrac{\partial}{\partial p_2^\l}\right)\braket{T^{\nu_2)\alpha_2}(p_2)\,T^{\m_3\n_3}(\bar p_3)\,T^{\mu_1\n_1}( p_1)}=0.\label{SCWTTT}
\end{align} 
Notice that the spin part acts only on two of the three tensors, in this case $T^{\mu_1\nu_1}(p_1)$ and 
$T^{\mu_2\nu_2}(p_2)$, leaving $T^{\mu_3\nu_3}$ as a spin singlet \cite{Coriano:2018bbe}. 
Notice that the Leibnitz rule for the action of the conformal operator $K^\kappa$ is violated and the differentiation respect to the third momentum is performed implicitly. The final result shown above, as explictly discussed in \cite{Coriano:2018bbe}, is a consequence of the Lorentz WI, which has to be used quite extensively. This takes the form  
\begin{equation}
\sum_{j=1}^3 L_{\mu\nu}(x_j) \langle T^{\mu_1\nu_1}(x_1)T^{\mu_2\nu_2}(x_2)T^{\mu_3\nu_3}(x_3)\rangle =0 
\end{equation}
with 
\begin{equation}
L_{\mu\nu}(x)=\left(i(x_\mu\partial_\nu - x_\nu\partial_\mu) +\bar{\Sigma}_{\mu\nu}\right)
\end{equation}
being the generators of the symmetry, separated into the angular momentum component and in the spin part, with
$\bar{\Sigma}$ being the spin generators of $SO(4)$ in the vector representation
\begin{equation}
\left(\bar\Sigma_{\rho\sigma}\right)_{\mu\alpha}=i\left(\delta_{\rho\mu}\delta_{\sigma\alpha}-\delta_{\rho\alpha}\delta_{\sigma\mu}\right).
\end{equation}
 In the case of the $TTT$ this gives
\begin{align}
0&=\sum_{j=1}^3 L_{\mu\nu}(x_j) \braket{ T^{\mu_1\nu_1}(x_1)T^{\mu_2\nu_2}(x_2)T^{\mu_3\nu_3}(x_3)} \notag\\
&=\sum_{j=1}^3 i\left( x_j^\mu \frac{\partial}{\partial {x_j}_\nu} - x_j^\nu \frac{\partial}{\partial {x_j}_\mu}\right) 
\braket{ T^{\mu_1\nu_1}T^{\mu_2\nu_2}T^{\mu_3\nu_3}} + 2 
(\bar{\Sigma}^{\mu\nu})^{(\mu_1}_{\alpha_1}\langle T^{\nu_1)\alpha_1}T^{\mu_2\nu_2}T^{\mu_3\nu_3}\rangle  
\notag\\[1.5ex]
&\hspace{2cm}+ 2 (\bar{\Sigma}^{\mu\nu})^{(\mu_2}_{\alpha_2}\langle T^{\mu_1\nu_1} T^{\nu_2)\alpha_2}T^{\mu_3\nu_3}\rangle  +2 
(\bar{\Sigma}^{\mu\nu})^{(\mu_3}_{\alpha_3}\braket{T^{\mu_1\nu_1}T^{\mu_2\nu_2} T^{\nu_3)\alpha_3}} 
\end{align}
and takes the form in momentum space 
\begin{align}
&\sum_{j=1}^{2}\left[p_j^\nu\sdfrac{\partial}{\partial p_{j\mu}}-p_j^{\mu}\sdfrac{\partial}{\partial p_{j\nu}}\right]\braket{T^{\mu_1\nu_1}(p_1)\,T^{\m_2\n_2}(p_2)\,T^{\mu_3\n_3}(\bar p_3)}\notag\\
&\qquad+2\left(\delta^{\nu}_{\a_1}\d^{\mu(\mu_1}-\delta^{\mu}_{\alpha_1}\delta^{\nu(\mu_1}\right)\braket{T^{\nu_1)\alpha_1}(p_1)\,T^{\m_2\n_2}(p_2)\,T^{\mu_3\n_3}(\bar p_3)}\notag\\
&\qquad+2\left(\delta^{\nu}_{\a_2}\d^{\mu(\mu_2}-\delta^{\mu}_{\alpha_2}\delta^{\nu(\mu_2}\right)\braket{T^{\nu_2)\alpha_2}(p_2)\,T^{\m_3\n_3}(\bar p_3)\,T^{\mu_1\n_1}(p_1)}\notag\\
&\qquad+2\left(\delta^{\nu}_{\a_3}\d^{\mu(\mu_3}-\delta^{\mu}_{\alpha_3}\delta^{\nu(\mu_3}\right)\braket{T^{\nu_3)\alpha_3}(\bar p_3)\,T^{\m_1\n_1}(p_1)\,T^{\mu_2\n_2}(p_2)}=0\label{RotationTTT}.\end{align}
Similarly, the dilatation WI in coordinate space
 \begin{equation}
 \label{ft2}
 \sum_{j=1}^n \left(i\,x_j^\alpha \frac{\partial}{\partial x_j^\alpha} +\Delta_j\right) \braket{T^{\mu_1\nu_1}(x_1)\,T^{\m_2\n_2}(x_2)\,T^{\mu_3\n_3}( x_3)}=0.
 \end{equation}
can be rewritten  in momentum space in the form
\begin{align}
&\left[\sum_{j=1}^3\Delta_j- 2 d-\sum_{j=1}^2\,p_j^\a\sdfrac{\partial}{\partial p_j^\alpha}\right]\braket{T^{\mu_1\nu_1}(p_1)\,T^{\m_2\n_2}(p_2)\,T^{\mu_3\n_3}(\bar p_3)}=0.\label{DilatationTTT}
\end{align}
\section{ Reconstruction in the BMS approach} 
In this section we are going to review the reconstruction method of \cite{Bzowski:2013sza} with the inclusion of extra derivations and details specific to the $TTT$ case, which may illustrate more clearly its formulation. The basic idea of the approach is to introduce a symmetric decomposition of the correlator in terms of its transverse traceless and longitudinal sectors. A second ingredient is that the second order differential equations (primary WI's)  which act on the corresponding form factors separate from the first order ones coming from the conservation Ward identities (secondary WI's). \\
For this one needs the transverse, transverse-traceless and longitudinal projectors
\begin{align}
\pi^{\mu}_{\alpha} & = \delta^{\mu}_{\alpha} - \frac{p^{\mu} p_{\alpha}}{p^2},  \qquad \tilde{\pi}^{\mu}_{\alpha} =\frac{1}{d-1}\pi^{\mu}_{\alpha} \nn
\Pi^{\mu \nu}_{\alpha \beta} & = \frac{1}{2} \left( \pi^{\mu}_{\alpha} \pi^{\nu}_{\beta} + \pi^{\mu}_{\beta} \pi^{\nu}_{\alpha} \right) - \frac{1}{d - 1} \pi^{\mu \nu}\pi_{\alpha \beta}, \nn
\mathcal{I}^{\mu\nu}_\alpha&=\frac{1}{p^2}\left[2 p^{(\mu}\delta^{\nu)}_\alpha - 
\frac{p_\alpha}{d-1}(\delta^{\mu\nu} +(d-2)\frac{p^\mu p^\nu}{p^2})\right]\nn
\mathcal{I}^{\mu\nu}_{\alpha\beta}&=\mathcal{I}^{\m\n}_\alpha p_\beta =\frac{p_{\beta}}{p^2}\left( p^{\mu}\delta^{\nu}_{\alpha} +p^{\nu}\delta^{\mu}_{\alpha} \right)-
\frac{p_{\alpha}p_{\beta}}{p^2}\left( \delta^{\mu\nu} +(d-2)\frac{p^\mu p^\nu}{p^2}\right) \nn
\mathcal{L}^{\mu\nu}_{\alpha\beta}&=\frac{1}{2}\left(\mathcal{I}^{\mu\nu}_{\alpha\beta} +\mathcal{I}^{\mu\nu}_{\beta\alpha}\right) \qquad \tau^{\mu\nu}_{\alpha\beta} =\tilde{\pi}^{\mu \nu}\delta_{\alpha \beta}
\label{projectors}
\end{align}
\begin{align}
\delta^{\mu\nu}_{\alpha\beta}&=\Pi^{\mu \nu}_{\alpha \beta} +\Sigma^{\mu\nu}_{\alpha\beta} \nn
\Sigma^{\mu\nu}_{\alpha\beta}&\equiv\mathcal{L}^{\mu\nu}_{\alpha\beta} +\tau^{\mu\nu}_{\alpha\beta}.
\label{sigmas}
\end{align}
The previous identities allows to decompose a symmetric tensor into its transverse traceless (via $\Pi$), longitudinal (via $\mathcal{L}$) and trace parts (via $\tau$), or on the sum of the combined longitudinal and trace contributions (via $\Sigma$).
Each insertion of stress energy tensor is separated into its longitudinal, transverse traceless and trace parts, in the notation of \cite{Bzowski:2013sza}
\begin{equation}
\label{loca1}
T^{\mu\nu}=t^{\mu\nu} + t_{loc}^{\mu\nu}
\end{equation}
with 
\begin{align}
\label{loca2}
t_{loc}^{\mu\nu}(p)&=\frac{p^{\mu}}{p^2}Q^\nu + \frac{p^{\nu}}{p^2}Q^\mu -
\frac{p^\mu p^\nu}{p^4} Q +\frac{\pi^{\m\nu}}{d-1}(T - \frac{Q}{p^2})\nn
&=\Sigma^{\mu\nu}_{\alpha\beta} T^{\alpha\beta}
\end{align}
and 
\begin{equation}
Q^\mu=p_\nu T^{\mu\nu},\qquad T=\delta_{\mu\nu}T^{\mu\nu}, \qquad Q= p_\nu p_\mu T^{\mu\nu}
\end{equation}
\begin{equation}
t_{loc}^{\mu\nu}=\mathcal{I}^{\mu\nu}_\alpha Q^\alpha +\frac{\pi^{\mu\nu}}{d-1}T.
\end{equation}
We turn to the case of the the 3-graviton vertex. By acting with these projectors on the $TTT$, the 3-point function is divided into two parts: the \emph{transverse-traceless} part and the \emph{local} part (indicated by subscript $loc$) expressible through the transverse and trace Ward Identities. We will be using the suffix "i" in $K_i, \pi_i, \Pi_i$ to indicate operators of momentum $p_i$. In the notation of \cite{Bzowski:2013sza}, the transverse traceless contributions are denoted as
\begin{equation}
\braket{ t^{\mu_1\nu_1}(p_1) t^{\mu_2\nu_2}(p_2) t^{\mu_3\nu_3}(p_3) } ={\Pi_1}^{\mu_1\nu_1}_{\alpha_1\beta_1}{\Pi_2}^{\mu_2\nu_2}_{\alpha_2\beta_2}{\Pi_3}^{\mu_3\nu_3}_{\alpha_3\beta_3} \braket{ T^{\alpha_1\beta_1}(p_1)T^{\alpha_2\beta_2}(p_2)T^{\alpha_3\beta_3}(p_3)}
\end{equation}
while the local contributions, defined by either longitudinal or trace projections, are indicated as 
\begin{align}
\langle t^{\mu_1\nu_1}_{loc}(p_1) T^{\mu_2\nu_2}(p_2) T^{\mu_3\nu_3}(p_3) \rangle &=\Sigma^{\mu_1\nu_1}_{1 \alpha_1\beta_1}\langle T^{\alpha_1\beta_1}(p_1) T^{\mu_2\nu_2}(p_2) T^{\mu_3\nu_3}(p_3)\rangle\notag\\[1.5ex]
\langle t^{\mu_1\nu_1}_{loc}(p_1) t^{\mu_2\nu_2}_{loc}(p_2) T^{\mu_3\nu_3}(p_3) \rangle &=
{\Sigma_1}^{\mu_1\nu_1}_{ \alpha_1\beta_1}{\Sigma_2}^{\mu_2\nu_2}_{ \alpha_2\beta_2}\langle T^{\alpha_1\beta_1}(p_1) T^{\alpha_2\beta_2}(p_2) T^{\mu_3\nu_3}(p_3)\rangle \notag\\[1.5ex]
\langle t^{\mu_1\nu_1}_{loc}(p_1) t^{\mu_2\nu_2}_{loc}(p_2) t_{loc}^{\mu_3\nu_3}(p_3) \rangle &=
{\Sigma_1}^{\mu_1\nu_1}_{ \alpha_1\beta_1}{\Sigma_2}^{\mu_2\nu_2}_{ \alpha_2\beta_2}{\Sigma_3}^{\mu_3\nu_3}_{ \alpha_3\beta_3}\braket{T^{\alpha_1\beta_1}(p_1) T^{\alpha_2\beta_2}(p_2) T^{\alpha_3\beta_3}(p_3)}.
\end{align}
Using the projectors $\Pi$ one can write the most general form of the transverse-traceless part as
\begin{equation}
\braket{t^{\m_1\n_1}(p_1)\,t^{\mu_2\nu_2}(p_2)\,t^{\mu_3\nu_3}(p_3)}=\Pi^{\mu_1\nu_1}_{\alpha_1\beta_1}(p_1)\Pi^{\mu_2\n_2}_{\alpha_2\beta_2}(p_2)\Pi^{\mu_3\n_3}_{\alpha_3\beta_3}(p_3)\,\,X^{\alpha_1\beta_1\,\alpha_2\beta_2\,\alpha_3\beta_3},
\end{equation}
where $X$ is a general tensor of rank six built from the metric and momenta. One can enumerate all possible tensors that can appear in $X$, and simplify the expansion by observing that whenever a tensor component of $X$ contains at least one of the following tensors
\begin{equation}
\delta^{\alpha_1\beta_1},\ \delta^{\alpha_2\beta_2},\ \delta^{\alpha_3\beta_3},\ p_1^{\alpha_1},\ p_1^{\beta_1},\ p_2^{\alpha_2},\ p_2^{\beta_2},\ p_3^{\alpha_3},\ p_3^{\beta_3}
\end{equation}
it will vanish after contraction with the projectors, if these carry the same momentum dependence of each of the $p_i$'s . The expansion of $X$ is chosen to be symmetric respect to the $p_i$. In this way one has only to consider the tensors
\begin{equation}
p_2^{\alpha_1},\ p_2^{\beta_1},\ p_3^{\alpha_2},\ p_3^{\beta_2},\ p_1^{\alpha_3},\ p_1^{\beta_3},\ \delta^{\alpha_2\alpha_3}, \  \delta^{\alpha_1\alpha_2},\  \delta^{\alpha_1\alpha_3},\dots
\end{equation}
and the similar ones with the other indices, and write the most general form of the transverse traceless part as
\begin{align}
&\braket{t^{\m_1\n_1}(p_1)t^{\m_2\n_2}(p_2)t^{\m_3\n_3}(p_3)}
= \Pi^{\m_1\n_1}_{\a_1\b_1}(p_1)
\Pi^{\m_2\n_2}_{\a_2\b_2}(p_2)\Pi^{\m_3\n_3}_{\a_3\b_3}(p_3)\notag\\
&\hspace{0.7cm}\times\Big[ A_1\,p_2^{\a_1} p_2^{\b_1} p_3^{\a_2} p_3^{\b_2} p_1^{\a_3} p_1^{\b_3}+ A_2\,\d^{\b_1\b_2} p_2^{\a_1} p_3^{\a_2} p_1^{\a_3} p_1^{\b_3} 
+ A_2\,(p_1 \leftrightarrow p_3)\, \d^{\b_2\b_3}  p_3^{\a_2} p_1^{\a_3} p_2^{\a_1} p_2^{\b_1} \notag\\
&\hspace{1.3cm}+ A_2\,(p_2\leftrightarrow p_3)\, \d^{\b_3\b_1} p_1^{\a_3} p_2^{\a_1}  p_3^{\a_2} p_3^{\b_2}+ A_3\,\d^{\a_1\a_2} \d^{\b_1\b_2}  p_1^{\a_3} p_1^{\b_3} + A_3(p_1\leftrightarrow p_3)\,\d^{\a_2\a_3} \d^{\b_2\b_3}  p_2^{\a_1} p_2^{\b_1} \notag\\
&\hspace{2cm}
+ A_3(p_2\leftrightarrow p_3)\,\d^{\a_3\a_1} \d^{\b_3\b_1}  p_3^{\a_2} p_3^{\b_2} + A_4\,\d^{\a_1\a_3} \d^{\a_2\b_3}  p_2^{\b_1} p_3^{\b_2} + A_4(p_1\leftrightarrow p_3)\, \d^{\a_2\a_1} \d^{\a_3\b_1}  p_3^{\b_2} p_1^{\b_3} \notag\\
&\hspace{3.5cm}+ A_4(p_2\leftrightarrow p_3)\, \d^{\a_3\a_2} \d^{\a_1\b_2}  p_1^{\b_3} p_2^{\b_1} + A_5  \d^{\a_1\b_2}  \d^{\a_2\b_3}  \d^{\a_3\b_1}\Big] \label{DecompTTT}
\end{align}
where we have used the symmetry properties of the projectors, and the coefficients $A_i\  i=1,\dots,5$, are the {form factors}, scalar functions of the momentum magnitudes $p_i^2$. \\
Using the decompositions (\ref{loca1}) and (\ref{loca2}) the $TTT$ can be re-expressed in terms of longitudinal and transverse traceless operators
\begin{align}
\braket{T^{\mu_1\nu_1}\,T^{\mu_2\n_2}\,T^{\mu_3\n_3}}&=\braket{t^{\mu_1\nu_1}\,t^{\mu_2\nu_2}\,t^{\mu_3\nu_3}}+\braket{t_{loc}^{\mu_1\nu_1}\,t^{\mu_2\nu_2}\,t^{\mu_3\nu_3}}+\braket{t^{\mu_1\nu_1}\,t_{loc}^{\mu_2\nu_2}\,t^{\mu_3\nu_3}}+\braket{t^{\mu_1\nu_1}\,t^{\mu_2\nu_2}\,t_{loc}^{\mu_3\nu_3}}\notag\\
&\hspace{0.8cm} +\braket{t^{\mu_1\nu_1}_{loc}\,t_{loc}^{\mu_2\nu_2}\,t^{\mu_3\nu_3}}+\braket{t_{loc}^{\mu_1\nu_1}\,t_{loc}^{\mu_2\nu_2}\,t^{\mu_3\nu_3}}+\braket{t^{\mu_1\nu_1}\,t_{loc}^{\mu_2\nu_2}\,t_{loc}^{\mu_3\nu_3}}+\braket{t_{loc}^{\mu_1\nu_1}\,t_{loc}^{\mu_2\nu_2}\,t_{loc}^{\mu_3\nu_3}}
\end{align}
or, equivalently, as 
\begin{align}
\label{eqss}
\braket{T^{\mu_1\nu_1}\,T^{\mu_2\n_2}\,T^{\mu_3\n_3}}&=\braket{t^{\mu_1\nu_1}\,t^{\mu_2\n_2}\,t^{\mu_3\n_3}}+\braket{T^{\mu_1\nu_1}\,T^{\mu_2\n_2}\,t_{loc}^{\mu_3\n_3}}+\braket{T^{\mu_1\nu_1}\,t_{loc}^{\mu_2\n_2}\,T^{\mu_3\n_3}}\notag\\
&+\braket{t_{loc}^{\mu_1\nu_1}\,T^{\mu_2\n_2}\,T^{\mu_3\n_3}}-\braket{T^{\mu_1\nu_1}\,t_{loc}^{\mu_2\n_2}\,t_{loc}^{\mu_3\n_3}}-\braket{t_{loc}^{\mu_1\nu_1}\,t_{loc}^{\mu_2\n_2}\,T^{\mu_3\n_3}}\notag\\
&-\braket{t_{loc}^{\mu_1\nu_1}\,T^{\mu_2\n_2}\,t_{loc}^{\mu_3\n_3}}+\braket{t_{loc}^{\mu_1\nu_1}\,t_{loc}^{\mu_2\n_2}\,t_{loc}^{\mu_3\n_3}}.
\end{align}
All the terms on the right-hand side, apart from the first one, may be computed by means of transverse and trace WI's. The exact form of the WI's varies with the exact definition of the operators involved, but all these terms depend uniquely on 2-point functions. \\
The main goal of the approach, after introducing the $A_i$'s, is to find the solution of the corresponding scalar equations that they satisfy. These are obtained by acting with the special conformal transformations and the projectors on the representation of the $TTT$ as given by (\ref{eqss}).  \\
The action of $K$ on the $TTT$, after the projection on the transverse traceless component gets simplified. Using the explicit expression
\begin{align}
 \Sigma^{\alpha\beta}_{\rho\sigma}=2 \frac{p_\sigma}{p^2} p^{(\alpha}\delta^{\beta)}_{\rho} 
 -\frac{1}{(d-1)}\frac{p_\rho p_\sigma}{p^2} \delta^{\alpha \beta} -\frac{d-2}{(d-1)}\frac{p^\alpha p^\beta p_\rho p_\sigma}{(p^2)^2} +
 \frac{1}{d-1}\delta^{\alpha\beta}\delta_{\rho\sigma}-\frac{1}{d-1}\delta_{\rho\sigma}\frac{p^\alpha p^\beta}{p^2}
 \end{align}
and the relation 
\begin{equation}
\Pi^{\mu\nu}_{\alpha\beta}(p) K(p)^\kappa_{scalar} p^\alpha p^\beta=0 
\end{equation}
one can show that terms containing two ${t_{loc}}$ operators simplify 
\begin{align}
\label{kks}
&{\Pi_1}^{\mu_1\nu_1}_{\alpha_1\beta_1}{\Pi_2}^{\mu_2\nu_2}_{\alpha_2\beta_2}{\Pi_3}^{\mu_3\nu_3}_{\alpha_3\beta_3} 
{K_1}^\kappa_{scalar}\left( {\Sigma_1}^{\alpha_1\beta_1}_{\rho_1\sigma_1}{\Sigma_3}^{\alpha_3\beta_3}_{\rho_3\sigma_3} 
{\Pi_2}^{\alpha_2\beta_2}_{\rho_2\sigma_2}\langle T^{\rho_1\sigma_1}T^{\rho_2\sigma_2}T^{\rho_3\sigma_3}\rangle \right)\nn
&={\Pi_1}^{\mu_1\nu_1}_{\alpha_1\beta_1}{\Pi_2}^{\mu_2\nu_2}_{\alpha_2\beta_2}{\Pi_3}^{\mu_3\nu_3}_{\alpha_3\beta_3} {K_1}^\kappa_{scalar}\left[\left( 2 \frac{{p_1}_{\sigma_1}}{p_1^2} p_1^{(\alpha_1}\delta^{\beta_1)}_{\rho_1}\right)
\left( 2 \frac{{p_3}_{\sigma_3}}{{p_3}^2} {p_3}^{(\alpha_3}\delta^{\beta_3)}_{\rho_3}\right) {\Pi_2}^{\alpha_2\beta_2}_{\rho_2\sigma_2} \braket{ T^{\rho_1\sigma_1} T^{\rho_2\sigma_2} T^{\rho_3\sigma_3}}\right]
 \end{align}
The same identities, applied to the spin part give
\begin{align}
\label{kksp}
&{\Pi_1}^{\mu_1\nu_1}_{\alpha_1\beta_1}{\Pi_2}^{\mu_2\nu_2}_{\alpha_2\beta_2}{\Pi_3}^{\mu_3\nu_3}_{\alpha_3\beta_3} 
{K_1}^\kappa_{spin}\left( {\Sigma_1}^{\alpha_1\beta_1}_{\rho_1\sigma_1}{\Sigma_3}^{\alpha_3\beta_3}_{\rho_3\sigma_3} 
{\Pi_2}^{\alpha_2\beta_2}_{\rho_2\sigma_2}\langle T^{\rho_1\sigma_1}T^{\rho_2\sigma_2}T^{\rho_3\sigma_3}\rangle \right)\nn
&=4\, {\Pi_1}^{\mu_1\nu_1}_{\alpha_1\beta_1}{\Pi_2}^{\mu_2\nu_2}_{\alpha_2\beta_2}{\Pi_3}^{\mu_3\nu_3}_{\alpha_3\beta_3} \left( \delta^\kappa_{\alpha_1}\frac{\partial}{\partial p_1^\lambda} -\delta^{\kappa\lambda}
\frac{\partial }{\partial p_1^{\alpha_1}}\right)\left( ( 2 \frac{{p_1}_{\sigma_1}}{p_1^2} p_1^{(\lambda}\delta^{\beta_1)}_{\rho_1})
( 2 \frac{{p_3}_{\sigma_3}}{{p_3}^2} {p_3}^{(\alpha_3}\delta^{\beta_3)}_{\rho_3})\times\right.\nn
&\,\,\,\,\,\,\,\,\,\,\,\,\,\left.\times {\Pi_3}^{\alpha_3\beta_3}_{\rho_3\sigma_3} \langle T^{\rho_1\sigma_1}(p_1) T^{\rho_2\sigma_2}(p_2) T^{\rho_3\sigma_3}(p_3)\rangle \right)
\end{align}
where we have explicitly indicated with large round brackets the expression on which the operator acts. As already mentioned, in the equations above the differentiation is performed also on the third momentum $p_3$, which is not independent from $p_1$ and $p_2$. Adding (\ref{kks}) and (\ref{kksp}) one can show that 
 \be
 {\Pi_1}^{\mu_1\nu_1}_{\alpha_1\beta_1}{\Pi_2}^{\mu_2\nu_2}_{\alpha_2\beta_2}{\Pi_3}^{\mu_3\nu_3}_{\alpha_3\beta_3} 
{K_1}^\kappa \langle t_{loc}^{\alpha_1\beta_1}t^{\alpha_2\beta_2}t_{loc}^{\alpha_3\beta_3}\rangle =0
\ee
which can be easily extended to the entire $K^\kappa$ operator 
 \be
 {\Pi_1}^{\mu_1\nu_1}_{\alpha_1\beta_1}{\Pi_2}^{\mu_2\nu_2}_{\alpha_2\beta_2}{\Pi_3}^{\mu_3\nu_3}_{\alpha_3\beta_3} 
{K}^\kappa \langle t_{loc}^{\alpha_1\beta_1}t^{\alpha_2\beta_2}t_{loc}^{\alpha_3\beta_3}\rangle =0
\ee
A similar relation can be shown to hold also for the term with three $t_{loc}$ contributions.
Proceeding with the reconstruction program, we will be needing the action of the special conformal transformations on correlators with a single $t_{loc}$, after projection on the transverse traceless sector. In this case the treatment of momenta $p_1$ and $p_2$ is similar, while in the case of the third (dependent) momentum $p_3$ the operatorial action on the tensor structures gets very involved. In order to proceed with the derivation of the scalar equations for the form factors $A_i$ we use the Lorentz identity
\begin{align}
&{\Pi_1}^{\mu_1\nu_1}_{\alpha_1\beta_1}{\Pi_2}^{\mu_2\nu_2}_{\alpha_2\beta_2}{\Pi_3}^{\mu_3\nu_3}_{\alpha_3\beta_3} 
{K}^\kappa_{scalar} \langle t^{\alpha_1\beta_1}t^{\alpha_2\beta_2}t_{loc}^{\alpha_3\beta_3}\rangle \notag\\
&\hspace{2cm}={\Pi_1}^{\mu_1\nu_1}_{\alpha_1\beta_1}{\Pi_2}^{\mu_2\nu_2}_{\alpha_2\beta_2}{\Pi_3}^{\mu_3\nu_3}_{\alpha_3\beta_3} 4\left[ (d-1)\,\delta^{\kappa\alpha_3}+{L_1}^{\kappa\,\alpha_3} +{L_2}^{\kappa\,\alpha_3}\right]
\left(\frac{{p_3}_{\rho_3}}{p_3^2}\langle t^{\alpha_1\beta_1}t^{\alpha_2\beta_2}T^{\rho_3\beta_3}\rangle\right)
\end{align}
where 
\begin{equation}
{L_i}^{\kappa\,\alpha_3} = p_i^{\alpha_3}\frac{\partial}{\partial {p_i}_\kappa}- p_i^\kappa\frac{\partial}{\partial {p_i}_{\alpha_3}}  \qquad \qquad i=1,2 
\end{equation}
and
\begin{align}
&{\Pi_1}^{\mu_1\nu_1}_{\alpha_1\beta_1}{\Pi_2}^{\mu_2\nu_2}_{\alpha_2\beta_2}{\Pi_3}^{\mu_3\nu_3}_{\alpha_3\beta_3} 
{K}^\kappa_{spin} \langle t^{\alpha_1\beta_1}t^{\alpha_2\beta_2}t_{loc}^{\alpha_3\beta_3}\rangle \notag\\
&\hspace{0.2cm}={\Pi_1}^{\mu_1\nu_1}_{\alpha_1\beta_1}{\Pi_2}^{\mu_2\nu_2}_{\alpha_2\beta_2}{\Pi_3}^{\mu_3\nu_3}_{\alpha_3\beta_3}4 \left\{ \left(\bar{\Sigma}^{\kappa\alpha_3}\right)_{\rho_1}^{\ \alpha_1}\left[\frac{{p_3}_{\rho_3}}{p_3^2}\braket{ t^{\rho_1\beta_1}t^{\alpha_2\beta_2}T^{\rho_3\beta_3}}\right] +\left(\bar{\Sigma}^{\kappa\alpha_3}\right)_{\rho_2}^{\ \alpha_2}\left[\frac{{p_3}_{\rho_3}}{p_3^2}\braket{ t^{\alpha_1\beta_1}t^{\rho_2\beta_2}T^{\rho_3\beta_3}}\right]\right\}
\end{align}
which allows to derive the projected relations
\begin{align}
\Pi^{\rho_1\sigma_1}_{\mu_1\nu_1}\Pi^{\rho_2\sigma_2}_{\mu_2\nu_2}\Pi^{\rho_3\sigma_3}_{\mu_3\nu_3}\ {K}^\k
\braket{t_{loc}^{\mu_1\nu_1}\,t^{\mu_2\nu_2}\,t^{\mu_3\nu_3}}&=\Pi^{\rho_1\sigma_1}_{\mu_1\nu_1}\Pi^{\rho_2\sigma_2}_{\mu_2\nu_2}\Pi^{\rho_3\sigma_3}_{\mu_3\nu_3}\ \left[\sdfrac{4d}{p_1^2}\,\delta^{\kappa\mu_1}\,p_{1\alpha_1}\,\braket{\braket{T^{\alpha_1\nu_1}T^{\mu_2\nu_2}T^{\mu_3\nu_3}}}\right]\notag\\
\Pi^{\rho_1\sigma_1}_{\mu_1\nu_1}\Pi^{\rho_2\sigma_2}_{\mu_2\nu_2}\Pi^{\rho_3\sigma_3}_{\mu_3\nu_3}\ {K}^\k
\braket{t^{\mu_1\nu_1}\,t_{loc}^{\mu_2\nu_2}\,t^{\mu_3\nu_3}}&=\Pi^{\rho_1\sigma_1}_{\mu_1\nu_1}\Pi^{\rho_2\sigma_2}_{\mu_2\nu_2}\Pi^{\rho_3\sigma_3}_{\mu_3\nu_3}\ \left[\sdfrac{4d}{p_2^2}\,\delta^{\kappa\mu_2}\,p_{2\alpha_2}\,\braket{\braket{T^{\mu_1\nu_1}T^{\alpha_2\nu_2}T^{\mu_3\nu_3}}}\right]\notag\\
\Pi^{\rho_1\sigma_1}_{\mu_1\nu_1}\Pi^{\rho_2\sigma_2}_{\mu_2\nu_2}\Pi^{\rho_3\sigma_3}_{\mu_3\nu_3}{K}^\k
\braket{t^{\mu_1\nu_1}\,t^{\mu_2\nu_2}\,t_{loc}^{\mu_3\nu_3}}&=\Pi^{\rho_1\sigma_1}_{\mu_1\nu_1}\Pi^{\rho_2\sigma_2}_{\mu_2\nu_2}\Pi^{\rho_3\sigma_3}_{\mu_3\nu_3}\ \left[\sdfrac{4d}{p_3^2}\,\delta^{\kappa\mu_3}\,p_{3\alpha_3}\,\braket{\braket{T^{\mu_1\nu_1}T^{\mu_2\nu_2}T^{\alpha_3\nu_3}}}\right]\label{properties}
\end{align}
Using this expression, Eq. \eqref{SCWTTT} takes the form
\begin{align}
0&=\Pi^{\rho_1\sigma_1}_{\mu_1\nu_1}(p_1)\Pi^{\rho_2\sigma_2}_{\mu_2\nu_2}(p_2)\Pi^{\rho_3\sigma_3}_{\mu_3\nu_3}(p_3)\  \bigg({K}^\k\,\braket{T^{\mu_1\nu_1}(p_1)\,T^{\m_2\n_2}(p_2)\,T^{\mu_3\nu_3}( p_3)}\bigg)\notag\\
&=\Pi^{\rho_1\sigma_1}_{\mu_1\nu_1}(p_1)\Pi^{\rho_2\sigma_2}_{\mu_2\nu_2}(p_2)\Pi^{\rho_3\sigma}_{\mu_3\nu_3}(p_3)\times\notag\\
& \bigg\{{K}^\kappa\,\braket{t^{\mu_1\nu_1}(p_1)\,t^{\mu_2\nu_2}(p_2)\,t^{\mu_3\nu_3}(p_3)}+\sdfrac{4d}{p_1^2}\,\delta^{\kappa\mu_1}\,p_{1\alpha_1}\,\braket{T^{\alpha_1\nu_1}(p_1)T^{\mu_2\nu_2}(p_2)T^{\mu_3\nu_3}(p_3)}\notag\\
&\quad+\sdfrac{4d}{p_2^2}\,\delta^{\kappa\mu_2}\,p_{2\alpha_2}\,\braket{T^{\mu_1\nu_1}(p_1)T^{\alpha_2\nu_2}(p_2)T^{\mu_3\nu_3}(p_3)}+\sdfrac{4d}{p_3^2}\,\delta^{\kappa\mu_3}\,p_{3\alpha_3}\,\braket{T^{\mu_1\nu_1}(p_1)T^{\mu_2\nu_2}(p_2)T^{\alpha_3\nu_3}(p_3)}\bigg\}.\label{StrucGenSWIS}
\end{align}
The last three terms in the equation above may be re-expressed in terms of 2-point functions via the transverse Ward identities, while the first term in \eqref{StrucGenSWIS} can be written as
\begin{align}
&\Pi^{\mu_1\nu_1 }_{\alpha_1 \beta_1}(p_1) \Pi^{\mu_2\nu_2 }_{\alpha_2 \beta_2} (p_2)\Pi^{\mu_3\nu_3}_{\alpha_3\beta_3}(p_3){K}^\kappa\,\braket{t^{\mu_1\nu_1}(p_1)\,t^{\mu_2\nu_2}(p_2)\,t^{\mu_3\nu_3}(p_3)}\notag\\
&= \Pi^{\mu_1\nu_1 }_{\alpha_1 \beta_1} \Pi^{\mu_2\nu_2 }_{\alpha_2 \beta_2} \Pi^{\mu_3\nu_3}_{\alpha_3\beta_3}\times \Bigl\{\, p_1^{\kappa} \,\Bigl( C_{1,1} \, p_2^{\alpha_1} p_2^{\beta_1} p_3^{\alpha_2} p_3^{\beta_2}p_1^{\alpha_3} p_1^{\beta_3} + C_{1,2}\, p_2^{\a_1} p_1^{\alpha_3} p_1^{\beta_3} p_3^{\alpha_2}\delta^{\b_1 \beta_2} \notag\\
&\quad+ C_{1,3}\, p_2^{\beta_1} p_2^{\alpha_1} p_1^{\alpha_3} p_3^{\a_2}\delta^{\b_2 \beta_3}+ C_{1,4}\, p_2^{\a_1} p_3^{\alpha_2} p_1^{\alpha_3} p_3^{\beta_2}\delta^{\b_1 \beta_3}+ C_{1,5}\, p_1^{\beta_3} p_2^{\beta_1} \delta^{\alpha_1 \alpha_2}\delta^{\alpha_3 \beta_2}\notag\\
&\quad+ C_{1,6}\, p_1^{\beta_3} p_1^{\alpha_3}\delta^{\alpha_1 \a_2}\delta^{\b_1 \beta_2}+ C_{1,7}\, p_1^{\beta_3} p_3^{\alpha_2}\delta^{\alpha_3 \a_1}\delta^{\b_1 \beta_2}+ C_{1,8}\, p_2^{\beta_1} p_2^{\alpha_1}\delta^{\alpha_2 \a_3}\delta^{\b_2 \beta_3}\notag\\
&\quad+ C_{1,9}\, p_3^{\beta_2} p_3^{\alpha_2}\delta^{\alpha_1 \a_3}\delta^{\b_1 \beta_3}+ C_{1,10}\, \delta^{\alpha_1 \beta_2}\delta^{\alpha_2 \beta_3}\delta^{\alpha_3 \beta_1}\Bigr)+ \big[p_1^\kappa\leftrightarrow p_2^\kappa; \ C_{1,j}\leftrightarrow C_{2,j}\big] \notag\\
& \quad\  + \delta^{\kappa \alpha_1} \Bigl( C_{3,1}\, p_{3}^{\alpha_2}p_{3}^{\beta_2} p_{1}^{\alpha_3}p_{1}^{\beta_3}p_{2}^{\beta_1} + C_{3,2}\, \delta^{\alpha_2 \beta_3} p_{3}^{\beta_2}  p_{1}^{\alpha_3} p_{2}^{\beta_1} + C_{3,3}\, \delta^{\alpha_2 \beta_1} p_{1}^{\alpha_3}  p_{3}^{\beta_2} p_{1}^{\beta_3}+ C_{3,4}\, \delta^{\alpha_3 \beta_1} p_{3}^{\alpha_2}  p_{3}^{\beta_2} p_{1}^{\beta_3}\notag\\
&\qquad\qquad+ C_{3,5}\, \delta^{\alpha_2 \beta_1} \delta^{\alpha_3 \beta_2} p_{1}^{\beta_3}+C_{3,6}\, \delta^{\alpha_2 \beta_3} \delta^{\alpha_3 \beta_1} p_{1}^{\beta_2}+C_{3,7}\, \delta^{\alpha_2 \beta_3} \delta^{\alpha_3 \beta_2} p_{2}^{\beta_1}\Bigr)\notag\\
&\qquad\qquad+[(\a_1,\b_1,p_2)\leftrightarrow (\a_2,\b_2,p_3); \ C_{3,j}\leftrightarrow C_{4,j}]+[(\a_1,\b_1,p_2)\leftrightarrow (\a_3,\b_3,p_1); \ C_{3,j}\leftrightarrow C_{5,j}]\Bigr\}\label{StrucSWIS} 
\end{align}
where now the $C_{ij}$'s are differential equations involving the form factors $A_1,\ A_2,\ A_3,\ A_4,\ A_5$ of the $\braket{ttt}$ in \eqref{DecompTTT}. The equations to solve are obtained by inserting \eqref{StrucSWIS} into \eqref{StrucGenSWIS} and using \eqref{long}.\\
For any 3-point function, the resulting equations can be divided into two groups, the \emph{primary} and the \emph{secondary} CWI's. The primary are second-order differential equations and appear as the coefficients of transverse or transverse-traceless tensor containing $p_1^\kappa$, $p_2^\kappa$ and $p_3^\k$, where $\kappa$ is the special index related to the conformal operator ${K}^\kappa$. The remaining equations, following from all other transverse or transverse-traceless terms, are then secondary conformal Ward identities and correspond to first-order differential equations. Notice that the action of $K^\kappa$ on the $\braket{t t t}$ is endomorphic on the transverse traceless sector (see \cite{Coriano:2018bbe} for a derivation).\\
Obviously, one could define equivalent sets of secondary Ward identities by working directly with \eqref{long}, the conservation WI's for the stress-energy tensor, but this turns out not to be necessary.\\
For this purpose notice that Eq. \eqref{StrucGenSWIS} is the projection of the special CWI into one specific sector, the transverse traceless part. The remaining sectors are associated with at least one
$\Sigma$ projector given in \eqref{sigmas}. Its action on $K^\kappa$ can be assimilated to the action of the momentum operator $P^\mu$, being the correlator traceless $(\Sigma\sim P)$ in $d$ dimensions. Using the commutation relation 
\be
[{K}^\kappa,P^\nu]=2i(\delta^{\kappa\nu}D + M^{\kappa\nu}),
\ee 
if the correlator satisfies both the dilatation and the Lorentz WI's 
then the action of a $\Sigma$ simplifies as $P^\mu K^\kappa =K^\kappa P^\mu$ on the traceless solution. This implies that the remaining sectors in the action of $K^\kappa$ are equivalent to conservation WI's. These are already taken into account by the action of the $\Pi$ projectors on $K^\kappa$, as clear from the right hand side of \eqref{properties}, which shows that the action of $\Pi K^\kappa$ on the local components of the $TTT$ is still local. Combined this information with the fact that $K^\kappa$ maps the transverse-traceless sector into itself, by solving tensorially Eq. \eqref{StrucGenSWIS} we account for all the conformal constraints, except for the dilatation WI's which needs to be considered separately. Notice also that the Lorentz and the permutational symmetries are satisfied by construction, while the dilatation WI's will be diagonal respect to each of the $A_i$'s, as we are going to shown below.

\subsection{The dilatation WI} 
We illustrate the procedure of deriving scalar equations for the form factors in the simpler case of the dilatation WI's obtained by the decomposition \eqref{DilatationTTT}. In this case it is sufficient to use the decomposition
\begin{align}
 &\Pi^{\mu_1\nu_1}_{1\,\alpha_1\beta_1} \Pi^{\mu_2\nu_2}_{2\,\alpha_2\beta_2} \Pi^{\mu_3\nu_3}_{3\,\alpha_3\beta_3} \left(\sum_{j=1}^3\,\Delta_j-2d-\sum_{j=1}^{2}\,p_j^\alpha\sdfrac{\partial}{\partial p_j^\alpha}\right) \braket{T^{\alpha_1\beta_1}T^{\alpha_2\beta_2}T^{\alpha_3\beta_3}} \notag\\
 &= \Pi^{\mu_1\nu_1}_{1\,\alpha_1\beta_1} \Pi^{\mu_2\nu_2}_{2\,\alpha_2\beta_2} \Pi^{\mu_3\nu_3}_{3\,\alpha_3\beta_3} \left(\sum_{j=1}^3\,\Delta_j-2d-\sum_{j=1}^{2}\,p_j^\alpha\sdfrac{\partial}{\partial p_j^\alpha}\right)\bigg[ \braket{t^{\alpha_1\beta_1}t^{\alpha_2\beta_2}t^{\alpha_3\beta_3}} +\dots + \braket{t_{loc}^{\alpha_1\beta_1}t_{loc}^{\alpha_2\beta_2}t^{\alpha_3\beta_3}_{loc}} \bigg]\notag
\end{align}
taking into account the relations 
\begin{equation}
\begin{split}
&\sum_{j=1}^2p_j^\a\sdfrac{\partial}{\partial p_j^\a}\,\left(\Pi^{\mu_i\nu_i}_{\alpha_i\beta_i}(p_i)\right)=0,\qquad i=1,2,3\\
&\sum_{j=1}^2p_j^\a\sdfrac{\partial}{\partial p_j^\a}\,\left(\Sigma^{\mu_i\nu_i}_{\alpha_i\beta_i}(p_i)\right)=0,\qquad i=1,2,3
\end{split}
\end{equation}
and the orthogonality between $\Pi$ and $\Sigma$ to obtain the relation
\begin{align}
0&=\Pi^{\mu_1\nu_1}_{\alpha_1\beta_1}(p_1)\Pi^{\mu_2\nu_2}_{\alpha_2\beta_2}(p_2)\Pi^{\mu_3\nu_3}_{\alpha_3\beta_3}(p_3)\left(\sum_{j=1}^3\,\Delta_j-2d-\sum_{j=1}^{2}\,p_j^\alpha\sdfrac{\partial}{\partial p_j^\alpha}\right)\Bigg[ A_1\,p_2^{\a_1} p_2^{\b_1} p_3^{\a_2} p_3^{\b_2} p_1^{\a_3} p_1^{\b_3}\notag\\
&\quad+ A_2\,\d^{\b_1\b_2} p_2^{\a_1} p_3^{\a_2} p_1^{\a_3} p_1^{\b_3} 
+ A_2\,(p_1 \leftrightarrow p_3)\, \d^{\b_2\b_3}  p_3^{\a_2} p_1^{\a_3} p_2^{\a_1} p_2^{\b_1} + A_2\,(p_2\leftrightarrow p_3)\, \d^{\b_3\b_1} p_1^{\a_3} p_2^{\a_1}  p_3^{\a_2} p_3^{\b_2}\notag\\[1.5ex]
&\quad+ A_3(p_1\leftrightarrow p_3)\,\d^{\a_2\a_3} \d^{\b_2\b_3}  p_2^{\a_1} p_2^{\b_1}+ A_3(p_2\leftrightarrow p_3)\,\d^{\a_3\a_1} \d^{\b_3\b_1}  p_3^{\a_2} p_3^{\b_2} + A_4\,\d^{\a_1\a_3} \d^{\a_2\b_3}  p_2^{\b_1} p_3^{\b_2} \notag\\[1.5ex]
&\quad+A_4(p_1\leftrightarrow p_3)\, \d^{\a_2\a_1} \d^{\a_3\b_1}  p_3^{\b_2} p_1^{\b_3} + A_4(p_2\leftrightarrow p_3)\, \d^{\a_3\a_2} \d^{\a_1\b_2}  p_1^{\b_3} p_2^{\b_1} + A_5  \d^{\a_1\b_2}  \d^{\a_2\b_3}  \d^{\a_3\b_1}\Bigg]
\end{align}
which is equivalent to the equations
\begin{equation}
\left[2d+N_i-\sum_{j=1}^{3}\Delta_j+\sum_{j=1}^2\,p_j^\alpha\sdfrac{\partial}{\partial p_j^\alpha}\right]\,A_i(p_1,p_2,p_3)=0,\label{DilatationFactor} \quad i=1,2\ldots 5
\end{equation}
where $N_i$ is the tensorial dimension of $A_i$, i.e. the number of momenta multiplying the form factor $A_i$ and the projectors $\Pi$. 

\subsection{Primary CWI's}\label{primary}

From the analysis of \eqref{StrucGenSWIS} and \eqref{StrucSWIS}, one can find the primary CWIs that are equivalent to the vanishing of the coefficients $C_{1j}$, $C_{2j}$ and $C_{3j}$ for $j=1,\dots, 10$.
In order to write such equations in a simpler form, we need to rearrange the $C_{jk}$ using the dilatation Ward identities. We will illustrate the explicit procedure for the first coefficient $C_{11}$, being the others similar.\\
In order to write such equations we perform the change of variables
\begin{align}
\frac{\partial}{\partial p_1^{\mu}}&=\frac{\partial p_1}{\partial p_1^{\mu}}\frac{\partial}{\partial p_1}+\frac{\partial p_2}{\partial p_1^{\mu}}\frac{\partial}{\partial p_2}+\frac{\partial p_3}{\partial p_1^{\mu}}\frac{\partial}{\partial p_3}=\frac{p_{\,1\mu}}{ p_1}\frac{\partial}{\partial p_1}+\frac{p_{1\mu}+p_{2\mu} }{ p_3}\frac{\partial}{\partial p_3}\\
\frac{\partial}{\partial p_2^{\mu}}&=\frac{p_{\,2\mu}}{ p_2}\frac{\partial}{\partial p_2}+\frac{p_{1\mu}+p_{2\mu} }{ p_3}\frac{\partial}{\partial p_3}
\end{align}
where $p_i=\sqrt{p_i^2},\ i=1,2,3$ are momentum magnitudes. 
The explicit form of the coefficient $C_{11}$ is given by
\begin{equation}
C_{11}= - \frac{2}{p_3} \left[ p_1 \frac{\partial^2}{\partial p_1 \partial p_3 } + p_2 \frac{\partial^2}{\partial p_2 \partial p_3 } \right] A_1 + \frac{d-1}{p_1} \frac{\partial}{\partial p_1} A_1 - \frac{\partial^2}{\partial p_1^2} A_1 + \frac{d-9}{p_3} \frac{\partial}{\partial p_3} A_1 - \frac{\partial^2}{\partial p_3^2} A_1\,.\label{C11}
\end{equation}
Differentiating the dilatation Ward identities respect to the momentum magnitude $p_3$ we obtain the relation
\begin{equation}
D_n \frac{\partial}{\partial p_3}  A_n = \left[  \frac{\partial p_1}{\partial p_3} \frac{\partial}{\partial p_1} +  p_1 \frac{\partial^2}{\partial p_1 \partial p_3} + \frac{\partial p_2}{\partial p_3} \frac{\partial}{\partial p_2} +  p_2 \frac{\partial^2}{\partial p_3 \partial p_2} +  \frac{\partial p_3}{\partial p_3} \frac{\partial}{\partial p_3} +  p_3 \frac{\partial^2}{\partial p_3 \partial p_3} \right] A_n \,.
\label{eq:dilWI}
\end{equation}
that can be simplified as
\begin{equation}
D_n \frac{\partial}{\partial p_3}  A_n = \left[  p_1 \frac{\partial^2}{\partial p_1 \partial p_3} +  p_2 \frac{\partial^2}{\partial p_3 \partial p_2} +   p_3 \frac{\partial^2}{\partial p_3 \partial p_3} \right] A_n \,,\label{rela}
\end{equation}

By using \eqref{rela}, we can re-expressed the first term in \eqref{C11} as 
\begin{equation}
-\frac{2}{p_3}  \left[ p_1 \frac{\partial^2}{\partial p_1 \partial p_3} +  p_2 \frac{\partial^2}{\partial p_3 \partial p_2}  \right]  A_1 = \frac{\left( 2 - 2 D_1 \right)}{p_3} \frac{\partial}{\partial p_3} A_1 +2  \frac{\partial^2}{\partial p_3^2} A_1 \,,
\end{equation}
recalling that $D_1$ is the degree of the corresponding form factor $A_1$, and in this case $D_1=\D_3-4$. Inserting this result into \eqref{C11} we simplify the form of the differential equation associated to such coefficient
\begin{equation}
C_{11}= \left[ -\frac{\partial^2}{\partial p_1^2} + \frac{d-1}{p_1} \frac{\partial}{\partial p_1} \right] A_1 + \left[ \frac{\partial^2 }{\partial p_3^2}  + \frac{d+1-2\Delta_3}{p_3} \frac{\partial}{\partial p_3} \right] A_1 \,.  
\label{eq:C11quasi}  
\end{equation}
At this stage, in order to write the primary CWIs in a simple way, we define the differential operators
\begin{subequations}
	\begin{align}
	\textup{K}_i &= \frac{\partial^2}{\partial p_i^2} + \frac{d+1-2\Delta_i}{p_i} \frac{\partial}{\partial p_i}  \qquad i=1,2,3    \\ 
	\textup{K}_{ij} &= \textup{K}_i - \textup{K}_j \,,
	\end{align}\label{Koper}
\end{subequations}
where $\D_j$ is the conformal dimension of the j-th operator in the 3-point function under consideration. Through this definition the $C_{11}$ is re-expressed as
\begin{equation}
C_{11} = (\textup{K}_3 - \textup{K}_1 ) A_1 = \textup{K}_{31}A_1 \,.
\end{equation}

The procedure presented above permits us to obtain a simple second-order differential equations and are applied in the same way for all $C_{1j}$, $j=1,2,3$.

The primary CWIs are obtained, as previously discussed, when the coefficients $C_{1j}$ and $C_{2j}$ are equal to zero. For instance, for the $A_1$ form factor we obtain 
\begin{equation}
K_{31}\,A_1=0,\qquad K_{23}A_1=0.
\end{equation} 
Note that, from the definition \eqref{Koper}, we have
\begin{equation}
K_{ii}=0,\qquad K_{ji}=-K_{ij},\qquad K_{ij}+K_{jk}=K_{ik}
\end{equation}
for any $i,j,k ={1,2,3}$. One can therefore subtract corresponding pairs of equations and obtain the following system of independent partial differential equations
\begin{equation}
\begin{matrix}
K_{13}\,A_1=0,\quad K_{12}A_1=0.
\end{matrix}\label{primaryCWI}
\end{equation} 
Since in the $\braket{TTT}$  $\Delta_1=\D_2=\Delta_3=d$, using the manipulations discussed above one obtains all the primary CWIs for the form factors $A_i$ in the form
\begin{equation}
\begin{aligned}[c]
&K_{13}A_1=0 \\
&K_{13}A_2=8 A_1 \\
&K_{13}A_2(p_1\leftrightarrow p_3)=-8 A_1 \\
&K_{13}A_2(p_2\leftrightarrow p_3)=0\\
& K_{13}A_3=2A_2 \\
& K_{13}A_3(p_1\leftrightarrow p_3)=-2A_2(p_1\leftrightarrow p_3) \\
& K_{13}A_3(p_2\leftrightarrow p_3)=0\\
& K_{13}A_4=-4A_2(p_2\leftrightarrow p_3) \\
& K_{13}A_4(p_1\leftrightarrow p_3)=4A_2(p_2\leftrightarrow p_3) \\
& K_{13}A_4(p_2\leftrightarrow p_3)=4A_2(p_1\leftrightarrow p_3)-4A_2\\
& K_{13}A_5=2\left[A_4-A_4(p_1\leftrightarrow p_3) \right]\\
\end{aligned}
\qquad
\begin{aligned}[c]
&K_{23}  A_1=0 \\
&K_{23}A_2=8 A_1 \\
&K_{23}A_2(p_1\leftrightarrow p_3)=0\\
&K_{23}A_2(p_2\leftrightarrow p_3)=-8A_1\\
&K_{23}A_3=2A_2 \\
& K_{23}A_3(p_1\leftrightarrow p_3)=0\\
& K_{23}A_3(p_2\leftrightarrow p_3)=-2A_2(p_2\leftrightarrow p_3)\\
&K_{23}A_4=-4A_2(p_1\leftrightarrow p_3)\\
&K_{23}A_4(p_1\leftrightarrow p_3)= 4A_2(p_2\leftrightarrow p_3)-4A_2\\
&K_{23}A_4(p_2\leftrightarrow p_3)= 4A_2(p_1\leftrightarrow p_3)\\
&K_{23}A_5=2\left[A_4-A_4(p_2\leftrightarrow p_3)\right] \\
\end{aligned}\label{Primary}
\end{equation}
As already mentioned above, the solutions of these equations can be obtained by mapping them  into an 
hypergeometric system of equations for the Appell function $F_4$. As shown in \cite{Coriano:2013jba} each equation is equivalent to 
a system of two coupled equations with specific indices that we have shown in \cite{Coriano:2018bbe} to be universal. Differently from the case considered in \cite{Coriano:2013jba} here 
the system of equations is far more complicated and it has been discussed in \cite{2014JHEP...03..111B} in terms of 3K integrals, which are integrals of 3 Bessel functions.
As we are going to see, the goal of the next section is to show how it is possible to use a direct method based on the operatorial 
splitting of the hypergeometric differential operators in order to relate inhomogeneous solutions to the homogeneous ones. This is obtained by re-expressing the operators $K_{ij}$ in terms of other operators $\bar{K}_{ij}$, which characterize some homogeneous equations, plus extra operators which are first order in the derivative respect to to the momenta.  The action of the extra operators on each $F_4$ can be rearranged by suitable shifts of the parameters in $F_4$ and using the few known properties of this Appell function. The method follows the simpler case discussed in \cite{Coriano:2018bbe}, that we extend. We illustrate the approach, leaving to \appref{fuchs} the more technical details. One of the difficulties of the system of equations \eqref{Primary} is the presence of exchanged momenta on their right hand side which couple all the constants appearing in the solution in a nontrivial way. The equations for each form factor $A_i$ define a coupled system of two equations, which are inhomogeneous, except for $A_1$. The inhomogeneous equations are solved, for each form factor $A_2,...A_5$ separately, as a superposition of a particular solution of the inhomogeneous equations and the general solution of the homogeneous ones, with the free independent constant identified at the end of the entire procedure. We have recollected below the main points, leaving some of the more technical details to an appendix.

%%%%%%%%%%%%%%%%%%%%%%%%%%%%%%%%%%%%%%%%%%%%%%%%%%%%%%%
\section{Solutions of the primary CWI's by an operatorial method}
\label{fuchs}
The transition to the system of hypergeometric differential equations which characterize the form factors can be obtained in various ways. For scalar correlators this has been discussed in 
\cite{Coriano:2013jba} using in $K^\kappa$ the change of variables 
\bea
\label{onechange}
\frac{\partial}{\partial p_{1}^{\mu}}  &=&   2 (p_{1\, \mu} + p_{2 \, \mu}) \frac{\partial}{\partial p_3^2} + \frac{2}{p_3^2}\left( 
(1- x) p_{1 \, \mu}  - x  \,  p_{2 \, \mu} \right) \frac{\partial}{\partial x} - 2  (p_{1\, \mu} + p_{2 \, \mu}) \frac{y}{p_3^2} 
\frac{\partial}{\partial y} \,, \nn \\
\frac{\partial}{\partial p_{2}^{\mu}}  &=& 2 (p_{1\, \mu} + p_{2 \, \mu}) \frac{\partial}{\partial p_3^2}   -   2  (p_{1\, \mu} + 
p_{2 \, \mu}) \frac{x}{p_3^2} \frac{\partial}{\partial x}   + \frac{2}{p_3^2}\left( (1- y) p_{2 \, \mu}  - y  \,  p_{1 \, \mu} 
\right) \frac{\partial}{\partial y}. \, 
\eea
with 
\bea
\label{xy}
x=\frac{p_1^2}{p_3^2} \qquad y=\frac{p_2^2}{p_3^2}.
\eea
 Here we are taking $p_3$ as "pivot" in the expansion, but we could equivalently choose any of the 3  momentum invariants. The hypergeometric character of the CWI's was recognized independently in \cite{Coriano:2013jba} and in \cite{2014JHEP...03..111B}. Here we are going to briefly overview the derivation of such equations in the case of the $TTT$, before discussing a direct method of solutions that we have developed for the $TJJ$ in \cite{Coriano:2018bbe} and that we are going to generalize.\\
As already mentioned, the method exploits the universality of the Fuchsian points of such equations, a property which holds for all the 3-point functions. It is a general characteristics of the CWI's associated to such correlators, as we have verified in several cases. 
The solutions of such equations take a form given by the product of the Appell function $F_4$ times $x$ and $y$ as given in \eqref{xy}, raised at specific powers $a,b$ (indices), which are universal. An overall extra factor of the momentum ($p_3$) raised to a specific power is introduced in such a way to give the correct scaling behaviour of the solution for each form factor $A_i$. \\
For each system (i.e.~each form factor) we first solve the homogeneous equation, determining the general solution. We then add to this a particular solution of the inhomogeneous equation. The latter is obtained by a split of the differential operator $K_{ij}$, which can be performed in various ways. 
The split that we adopt in this case is different form the one used in \cite{Coriano:2018bbe}. \\
We start by reviewing briefly the case of the scalar correlator $\Phi(p_1,p_2, p_3)$ in order to make our discussion self-contained and define our conventions. In this case the equations are homogeneous, of the form 
\begin{equation}
K_{13}\Phi=0  \qquad K_{23}\Phi=0
\end{equation}
and need to be combined with the scaling equation 
\begin{equation}
\label{scale}
\sum_{i=1}^3 p_i\frac{\partial}{\partial p_i} \Phi=(\Delta-2 d) \Phi. 
\end{equation}
The ansatz is generated by a combination of a single power of the momentum "pivot" $p_3$ and powers of $x$ and $y$  
\begin{equation}
\label{ans}
\Phi(p_1,p_2,p_3)=p_3^{\Delta - 2 d} x^{a}y^{b} F(x,y).
\end{equation}
$\Phi$ is required to be homogenous of degree $\Delta-2 d$ under a scale transformation, according to (\ref{scale}), and in (\ref{ans}) this is taken into account by the factor $p_3^{\Delta - 2 d}$. Inserting the ansatz one derives the equation 
 \begin{align}
	K_{13}\Phi &= 4 (p_3^2)^{\Delta/2 -d -1} x^a y^b
	\left(  x(1-x)\frac{\partial }{\partial x \partial x}  + (A x + \gamma)\frac{\partial }{\partial x} -
	2 x y \frac{\partial^2 }{\partial x \partial y}- y^2\frac{\partial^2 }{\partial y \partial y} + 
	D y\frac{\partial }{\partial y} + \left(E +\frac{G}{x}\right)\right) \notag\\
	& \hspace{3cm}\times F(x,y)=0
	\label{red}
\end{align}
with
\begin{align}
	&A=D=\Delta_1 +\Delta_2 - 1 -2 a -2 b -\frac{3 d}{2} \qquad \gamma(a)=2 a +\frac{d}{2} -\Delta_1 + 1
	\notag\\
	& G=\frac{a}{2}(d +2 a - 2 \Delta_1)
	\notag\\
	&E=-\frac{1}{4}(2 a + 2 b +2 d -\Delta_1 -\Delta_2 -\Delta_3)(2 a +2 b + d -\Delta_1 -\Delta_2 +\Delta_3).
\end{align}
and
\begin{align}
	K_{23}\Phi &= 4 p_3^{\Delta -2 d -2} x^a y^b
	\left(  y(1-y)\frac{\partial }{\partial y \partial y}  + (A' y + \gamma')\frac{\partial }{\partial y} -
	2 x y \frac{\partial^2 }{\partial x \partial y}- x^2\frac{\partial^2 }{\partial x \partial x} + 
	D' x\frac{\partial }{\partial x} + \left(E' +\frac{G'}{y}\right)\right) \notag\\
	& \hspace{3cm}\times F(x,y)=0\label{red2}
\end{align}
with
\begin{align}
	&A'=D'= A   \qquad \qquad \gamma'(b)=2 b +\frac{d}{2} -\Delta_2 + 1
	\notag\\
	& G'=\frac{b}{2}(d +2 b - 2 \Delta_2)
	\notag\\
	&E'= E
\end{align}

Notice that in both equations  we need to set $G/x=0$ and $G'/y=0$ in order to reproduce an hypergeometric system, which sets conditions on the Fuchsian exponents $a$ and $b$. These are  
\begin{equation}
\label{cond1}
a=0\equiv a_0 \qquad \textrm{or} \qquad a=\Delta_1 -\frac{d}{2}\equiv a_1.
\end{equation}
and
\begin{equation}
\label{cond2}
b=0\equiv b_0 \qquad \textrm{or} \qquad b=\Delta_2 -\frac{d}{2}\equiv b_1.
\end{equation}
As we have verified, the four independent solutions of the CWI's are all characterised by the same 4 pairs of indices $(a_i,b_j)$ $(i,j=1,2)$. 
Our conventions for the parametric dependences in $F_4$ are the same of those introduced in \cite{Coriano:2018bbe}
\begin{equation}
\alpha(a,b)= a + b + \frac{d}{2} -\frac{1}{2}(\Delta_1 +\Delta_2 -\Delta_3) \qquad \beta (a,b)=a +  b + d -\Delta_1 -\Delta_2 -\Delta_3). \qquad 
\end{equation}
We also have 
\begin{equation}
E=E'=-\alpha(a,b)\beta(a,b) \qquad A=D=A'=D'=-\left(\alpha(a,b) +\beta(a,b) +1\right).
\end{equation}
The general solution for the scalar correlator, for instance,  takes the form 

\begin{equation}
\Phi(p_1,p_2,p_3)=p_3^{\Delta-2 d-2} \sum_{a,b} c(a,b,\vec{\Delta})\,x^a y^b \,F_4(\alpha(a,b), \beta(a,b); \gamma(a), \gamma'(b); x, y) 
\end{equation}
where the sum runs over the four values $a_i, b_i$ $i=0,1$ with constants $c(a,b,\vec{\Delta})$, with $\vec{\Delta}=(\Delta_1,\Delta_2,\Delta_3)$. We will also define
\begin{align}
	& \alpha(a_0,b_0)=\frac{d}{2}-\frac{\Delta_1 + \Delta_2 -\Delta_3}{2},\, &&  \beta(b_0)=d-\frac{\Delta_1 + \Delta_2 +\Delta_3}{2},  \notag\\
	& \gamma(a_0) =\frac{d}{2} +1 -\Delta_1,\, && \gamma(b_0) =\frac{d}{2} +1 -\Delta_2,
\end{align}
and the 4 independent solutions can be re-expressed in terms of the parameters above. The solution for $a=0$ and $b=0$ is known as Appell's function $F_4$, a generalized hypergeometric function of two variables \cite{APPELL}

\begin{align}
	\label{F4def}
	F_4(\alpha(a,b), \beta(a,b); \gamma(a), \gamma'(b); x, y) = \sum_{i = 0}^{\infty}\sum_{j = 0}^{\infty} \frac{(\alpha(a,b))_{i+j} \, 
		(\beta(a,b))_{i+j}}{(\gamma(a))_i \, (\gamma'(b))_j} \frac{x^i}{i!} \frac{y^j}{j!} 
\end{align}
where we have used the standard notations $(\alpha)_i = \Gamma(\alpha + i)/ \Gamma(\alpha)$ for the Pochammer symbol.  $\alpha\ldots \gamma'$ are the first, second$\ldots$, fourth parameters of $F_4$. The 4 independent solutions of \eqref{red} and \eqref{red2} are then all of the form $x^a y^b F_4$, where the 
hypergeometric functions will take some specific values for its parameters $\alpha(a,b), \beta(a,b)\ldots$ etc, with
$a$ and $b$ fixed by (\ref{cond1}) and (\ref{cond2}). \\
Next, we are going to extend the analysis presented for the $TJJ$  to the $TTT$ using an alternative splitting of the hyergeometric operators $K_{i j}$ in order to deal with the more difficult structure of the global system of equations which should be satisfied by the form factors. 

\subsection{Form factors: the solution for $A_1$}
We start from $A_1$ by solving the two equations from (\ref{Primary}) 
\begin{equation}
K_{13}A_1=0   \qquad K_{23}A_1=0.
\end{equation}
In this case we introduce the ansatz 
\begin{equation}
A_1=p_3^{\Delta-2 d - 6}x^a y^b  F(x,y)
\end{equation}
and derive two hypergeometric equations as previously, which are characterised by the same indices 
$(a_i, b_j)$ as before in (\ref{cond1}) and (\ref{cond2}), but new values of the 4 defining parameters. 
We obtain 
\begin{equation}
\label{A1}
A_1(p_1,p_2,p_3)=p_3^{\Delta-2 d - 6}\sum_{a,b} c^{(1)}(a,b,\vec{\Delta})\,x^a y^b \,F_4(\alpha(a,b) +3, \beta(a,b)+3; \gamma(a), \gamma'(b); x, y) 
\end{equation}
with the expression of $\alpha(a,b),\beta(a,b), \gamma(a), \gamma'(b)$ as given before
\begin{align}
	\label{cons1}
	\alpha(a,b)&= a + b + \frac{d}{2} -\frac{1}{2}(\Delta_2 - \Delta_3 +\Delta_1) \notag\\
	\beta(a,b)&= a + b + d -\frac{1}{2}(\Delta_1 + \Delta_2 +\Delta_3) 
\end{align}
 and 
\begin{align} 
	\label{cons2}
	\gamma(a)& =2 a +\frac{d}{2} -\Delta_1 + 1 \notag\\
	\gamma'(b)&=2 b +\frac{d}{2} -\Delta_2 + 1 .
\end{align}

Expressing the values of the scaling dimensions $\D_1=\D_2=\D_3=d$, then 
\begin{align}
A_1(p_1,p_2,p_3)=p_3^{d - 6}\sum_{a,b} c^{(1)}(a,b)\,x^a y^b \,F_4(\alpha(a,b) +3, \beta(a,b)+3; \gamma(a), \gamma'(b); x, y) 
\end{align}
where now
\begin{align}
&a=0,\frac{d}{2}, &&b=0,\frac{d}{2},\notag\\
&	\alpha(a,b)= a + b,&&\beta(a,b)= a + b -\frac{d}{2},\notag\\
&\gamma(a) =2 a -\frac{d}{2} + 1,&&\gamma'(b)=2 b -\frac{d}{2}+ 1.
\end{align}

One can implement the symmetry condition on the $A_1$ form factor which has to be completely symmetric in the exchange of $(p_1,p_2,p_3)$. The three conditions 
\begin{align}
A_1(p_1,p_3,p_2)&=A_1(p_1,p_2,p_3)\nn
A_1(p_3,p_2,p_1)&=A_1(p_1,p_2,p_3)\nn
A_1(p_2,p_1,p_3)&=A_1(p_1,p_2,p_3)
\end{align}
constrain the coefficient $c^{(1)}(a,b)$ and in particular we obtain
\begin{align}
c^{(1)}\left(\frac{d}{2},0\right)&=c^{(1)}\left(0,\frac{d}{2}\right)\notag\\
c^{(1)}(0,0)&=-\frac{(d-4) (d-2)}{(d+2) (d+4)}c^{(1)}\left(0,\frac{d}{2}\right)\notag\\
c^{(1)}\left(\frac{d}{2},\frac{d}{2}\right)&=\frac{\G\left(-\frac{d}{2}\right)\G\left(d+3\right)}{2\,\Gamma\left(\frac{d}{2}\right)} c^{(1)}\left(0,\frac{d}{2}\right)\label{CondA1}
\end{align}
generating a solution which depends only on one arbitrary constant that we identify as $C_1$ 
\begin{equation}
c^{(1)}\left(0,\frac{d}{2}\right)=C_1.\label{oneconst}
\end{equation}
%%%%%%%%%%%%%%%%%%%%%%%%%
\subsection{The solution for $A_2$ and the operatorial shifts}
%%%%%%%%%%%%%%%%%%%%%%%%
The equation for $A_2$ is inhomogeneous, but the solution can be identified using some properties of the hypergeometric forms of such equations. We recall that in this case they are 
\begin{align}
K_{13}A_2 &= 8 A_1\label{inhom}\\\
	K_{23}A_2 &= 8A_1.\label{inhom1}
\end{align}
The ansatz which is in agreement with the scaling behaviour of $A_2$  in this case is 
\begin{equation}
A_2(p_1,p_2,p_3)=p_3^{d - 4}\,x^a\,y^b\,F(x,y).
\end{equation}
At this stage we proceed with the splitting. We observe that the action of $K_{13}$ and $K_{23}$ on $A_2$ can be rearranged as follows
\begin{align}
	K_{13} A_2&=4 x^a y^b p_3^{d -6}\bigg( \bar{K}_{13}F(x,y) +x\frac{\partial}{\partial x} F(x,y)+y\frac{\partial}{\partial y} F(x,y)+\bar{\b}F(x,y)\bigg)\label{rep}\\[1.5ex]
	K_{23} A_2&=4 x^a y^b p_3^{d -6}\bigg( \bar{K}_{23}F(x,y)  +x\frac{\partial}{\partial x} F(x,y)+y\frac{\partial}{\partial y} F(x,y)+\bar{\b}F(x,y)\bigg)\label{rep2}
\end{align}
where
\begin{align}
	\label{k1bar}
	\bar{K}_{13}F(x,y)&=\bigg\{x(1-x) \frac{\partial^2}{\partial x^2} - y^2 \frac{\partial^2}{\partial y^2} - 2 \, x \, y \frac{\partial^2}{\partial x \partial y} +\big[  \gamma(a)- (\tilde\alpha(a,b) + \bar\beta(a,b) + 1) x \big] \frac{\partial}{\partial x} \notag\\
	&\hspace{5cm}+ \frac{a (a-a_1)}{x} - (\tilde\alpha(a,b) + \bar\beta(a,b) + 1) y \frac{\partial}{\partial y}  - \tilde\alpha(a,b) \, \bar\beta (a,b)\bigg\} F(x,y)
\end{align}
and 
\begin{align}
	\label{k2bar}
	\bar{K}_{23} A_2&=\bigg\{ y(1-y) \frac{\partial^2}{\partial y^2} - x^2 \frac{\partial^2}{\partial x^2} - 2 \, x \, y \frac{\partial^2}{\partial x \partial y} +  \big[ \gamma'(b) - (\tilde\alpha(a,b) +\bar \beta(a,b) + 1) y \big] \frac{\partial}{\partial y}\notag\\
	&\hspace{5cm}+  \frac{b(b- b_1)}{y} - (\tilde\alpha(a,b) + \bar\beta(a,b) + 1) x \frac{\partial}{\partial x}  - \tilde\alpha(a,b) \, \bar\beta(a,b) \bigg\} F(x,y)
\end{align}
with 
\begin{equation}
\tilde\alpha(a,b)=\alpha(a,b) + 3 \qquad \bar\beta(a,b)=\beta(a,b) +2
\end{equation}
At this point we notice that the hypergeometric function  that satisfy the system of equations
\begin{equation}
\left\{
\begin{split}
\bar{K}_{23}F(x,y)&=0\\
\bar{K}_{13}F(x,y)&=0\\
\end{split}
\right.
\end{equation}
can be taken of the form
\begin{equation}
\label{rep1}
\Phi_1^{(2)}(x,y)=\sum_{a,b} c^{(2)}_1(a,b)\,x^a y^b \,F_4(\alpha(a,b) +3, \beta(a,b)+2; \gamma(a), \gamma'(b) ; x, y)
\end{equation}
with $c^{(2)}_1$ constant depending on the parameters $a,b$ fixed at the ordinary values $(a_0,b_0), (a_1,b_0), (a_0,b_1)$ and $(a_1,b_1)$ as in the previous cases (\ref{cond1}) and (\ref{cond2}). 
The convention that we adopt on the indices appearing on the constants is as follows.\\
 The superscript $(i)$ on the constant $c^{(i)}$, is the index of the corresponding form factors and it is used in the homogeneous solution of the corresponding set of equations. On the other hand, the subscript $j$, in the constant $c^{(i)}_j$ instead, specifies the particular (inhomogeneous) solution of the same system of equations for the form factor $A_i$.\\
  For instance, if one considers the particular solution $\Phi_1^{(2)}$ in \eqref{rep1}, the constant $c_1^{(2)}$ is well defined using this convention. In fact it tells us that this solution is the first particular solution of the inhomogeneous set of equations for the $A_2$ form factors. It is worth mentioning that all these constants will be fixed, at the end, just in terms of the homogeneous ones that don't carry any subscript. \\
As previously remarked, the values of the exponents $a$ and $b$ remain the same for any equation involving either a $K_{i,j}$ or a $\bar{K}_{i j}$, as one can verify. \\
At this point, to show that $\Phi_1^{(2)}$  is a solution of  Eqs. (\ref{inhom}) we use the property 
\begin{equation}
\frac{\partial^{p+q} F_4(a,b;c_1,c_2;x,y)}{\partial x^p\partial y^q} =\frac{(a)_{p+q}(b)_{p+q}}{(c_1)_{p}(c_2)_{q}}
F_4(a + p + q,b + p + q; c_1 + p ; c_2 + q;x,y)
\end{equation}
using the Pochammer symbol previously defined, from which one derives the simpler relations 
\begin{align}
	& \frac{\partial F_4(a,b;c_1,c_2;x,y)}{\partial x} =\frac{a b}{c_1}F_4(a+1,b+1,c_1+1,c_2,x,y) \notag\\
	&  \frac{\partial F_4(a,b;c_1,c_2;x,y)}{\partial y} =\frac{a b}{c_2}F_4(a+1,b+1,c_1,c_2 +1,x,y).
\end{align} 
We will be also using the known relation on the shift of one parameter of $F_4$
\begin{equation}
\label{exs}
F_4(a,b,c_1-1,c_2;x,y)=F_4(a,b,c_1,c_2,x,y)+x\frac{\partial}{\partial\,x}\,F_4(a,b,c_1,c_2,x,y)
\end{equation}
that leads to the identity
\begin{equation}
x\,F_4(a+1,b+1,c_1,c_2;x,y)=\frac{(c_1-1)(c_1-2)}{a\,b}\bigg[F_4(a,b,c_1-2,c_2,x,y)-F_4(a,b,c_1-1,c_2,x,y)\bigg],
\end{equation}
and furthermore the symmetry relation 
\begin{align}
\label{transfF4}
F_4(\alpha, \beta; \gamma, \gamma'; x, y) &= \frac{\Gamma(\gamma') \Gamma(\beta - \alpha)}{ \Gamma(\gamma' - \alpha) \Gamma(\beta)} (- y)^{- \alpha} \, F_4\left(\alpha, \alpha -\gamma' +1; \gamma, \alpha-\beta +1; \frac{x}{y}, \frac{1}{y}\right) \nn 
&\qquad+  \frac{\Gamma(\gamma') \Gamma(\alpha - \beta)}{ \Gamma(\gamma' - \beta) \Gamma(\alpha)} (- y)^{- \beta} \, F_4\left(\beta -\gamma' +1, \beta ; \gamma, \beta-\alpha +1; \frac{x}{y}, \frac{1}{y}\right) \,.
\end{align}
already used in \cite{Coriano:2013jba} in the analysis of a scalar case, in order to impose the symmetry under the exchange of two of the three momenta. 
All these relations can be used in order to consider the action of $K_{13}$ and $K_{23}$ on the the $\Phi_2^{(1)}$  in (\ref{rep}) and in \eqref{rep2}, obtaining 
\begin{align}
	&K_{13}\Phi_1^{(2)}(x,y) = 4p_3^{d -6} \sum_{a,b} c^{(2)}_1(a,b)\,x^a y^b\bigg(x\frac{\partial}{\partial x} +y\frac{\partial}{\partial y} +(\b+2)\bigg)\,F_4(\alpha +3, \beta+2; \gamma, \gamma' ; x, y)   \notag\\
	&= 4 p_3^{d -6} \sum_{a,b} c^{(1)}_2(a,b)\,x^a y^b\bigg[ (\alpha+3)\bigg(x\frac{(\beta+2)}{\gamma}F_4(\alpha +4, \beta+3; \gamma+1, \gamma'; x, y)\notag\\
	&\hspace{1cm}+y\frac{(\beta+2)}{\gamma'}F_4(\alpha +4, \beta+3; \gamma, \gamma'+1; x, y)\bigg)+(\b+2)\,F_4(\alpha +3, \beta+2; \gamma, \gamma'; x, y) \bigg]
	\end{align}
where, for simplicity, we have denoted with $\a=\a(a,b)$ and $\b=\b(a,b)$. At this point, using the following properties of  hypergeometric functions \cite{APPELL}
\begin{align}
\frac{b}{c_1}\,x\,F_4(a+1,b+1,c_1+1,c_2,x,y)+\frac{b}{c_2}\,y\,F_4(a+1,b+1,c_1,c_2+1,x,y)&=F_4(a+1,b,c_1,c_2,x,y)\notag\\
-F_4(a,b,c_1,c_2,x,y)
aF_4(a+1,b,c_1,c_2,x,y)-bF_4(a,b+1,c_1,c_2,x,y)&=(a-b)F_4(a,b,c_1,c_2,x,y)
\end{align}
after some algebra, it is simple to verify that 
\begin{align}
K_{13}\Phi_1^{(2)}(x,y) &=4p_3^{ d -6} \sum_{a,b} c^{(2)}_1(a,b)\,x^a y^b\big(\b(a,b)+2\big)F_4(\a+3,\b+3,\g,\g',x,y)
\end{align}
and in the same way
	\begin{align}
	&K_{23}\Phi_1^{(2)}(x,y)= 4p_3^{ d -6} \sum_{a,b} c^{(2)}_1(a,b)\,x^a y^b\big(\b(a,b)+2\big)F_4(\a+3,\b+3,\g,\g',x,y).
\end{align}
The non-zero right-hand-side of both equations are proportional to the form factor $A_1$ given in (\ref{A1}). Once this particular solution is determined, \eqref{A1}, by comparison, gives the conditions on $c_1^{(2)}$ as
\begin{align}
	c_1^{(2)}(a,b)&=\frac{2}{\big(\b(a,b)+2\big)}\,c^{(1)}(a,b)\label{condc2}.
\end{align}
Notice that the coefficient $c_1^{(2)}$ of the first particular solution of the inhomogeneous set of equations for $A_2$ is fixed in terms of the coefficient $c^{(1)}$ of homogeneous one of $A_1$.

Finally, we obtain the general solution for $A_2$ in the $TTT$ case (in which $\g=\g'$ ) by superposing the solution of the homogeneous 
system generated by \eqref{inhom} and \eqref{inhom1} and the inhomogeneous one \eqref{rep1}, with a condition on the constants given by \eqref{condc2}. Therefore, the general expression of the solution for $A_2$ is given by
\begin{align}
	A_2&= p_3^{d - 4}\sum_{a b} x^a y^b\left[c^{(2)}(a,b)\,F_4(\alpha +2, \beta+2; \gamma, \gamma'; x, y)+ \frac{2\,c^{(1)}(a,b)}{\big(\b+2\big)}F_4(\alpha +3, \beta+2; \gamma, \gamma'; x, y)\right]\label{A2}
\end{align}
where also in this case we have used a short-hand notation $\a=\a(a,b)$, $\b=\b(a,b)$, $\g=\g(a)$ and $\g'=\g'(b)$. \\
Let's discuss now the symmetry properties of the $A_2$ form factors. The latter in fact is symmetric under the exchange $p_1\leftrightarrow p_2$, and this condition has to be implemented in the form
\begin{equation}
A_2(p_2,p_1,p_3)=A_2(p_1,p_2,p_3).
\end{equation}
Using the properties of the hypergeometric functions previously written, such symmetry constraint relates  $c^{(1)}$ and $c^{(2)}$ for the 4 indices $a,b$ which label the homogeneous solutions in the form  
\begin{equation}
c^{(2)}\left(\frac{d}{2},0\right)=c^{(2)}\left(0,\frac{d}{2}\right)\label{CondA2}
\end{equation}
\begin{equation}
c^{(1)}\left(\frac{d}{2},0\right)=c^{(1)}\left(0,\frac{d}{2}\right)
\end{equation}
Notice that of the two equations above, the second is redundant since it is already present as a symmetry condition on $A_1$, as clear from \eqref{CondA1}. Only the first condition on $c^{(2)}$ is new. \\
We have already established that $A_1$ can be written in terms of only a single constant $C_1$, as evident from \eqref{CondA1} and \eqref{oneconst}. From the expression of $A_2$ in \eqref{A2} and using the property \eqref{CondA2}, we can deduce that so far this form factors can be written in terms of four constants: $C_1,\ c^{(2)}\left(0,d/2\right),\ c^{(2)}\left(d/2,d/2\right),\ c^{(2)}\left(0,0\right)$. We will see that the symmetry condition on $A_5$ will put additional constraint on the coefficients of $A_2$, by allowing us to write this form factors in terms of only two independent constants. At the end, we will see that this iterative method will allow to identify a rather small set of independent  constants for each form factor and the entire solution.

%%%%%%%%%%%%%%%%%%%%%%%%%%%%
\subsection{The solution for $A_3$}
%%%%%%%%%%%%%%%%%%%%%%%%%%
Also in this case the system of two equations is  inhomogeneous 
\begin{equation}
\left\{
\begin{split}
K_{13} A_3 &=2A_2\\ 
K_{23} A_3&=2A_2 
\end{split}\right.
\label{eqA3}
\end{equation}
Using the same strategy of the previous section,  it is possible to find two particular solutions of such system using an operatorial split as above
\begin{align}
\Phi_1^{(3)}(x,y)&= p_3^{d -2} \sum_{a b} c_1^{(3)}(a,b)\, x^a y^b   F_4(\alpha+2,\beta+1; \gamma,\gamma';x,y)\\
\Phi_2^{(3)}(x,y)&= p_3^{ d -2} \sum_{a b} c_2^{(3)}(a,b) \,x^a y^b   F_4(\alpha+3,\beta+1; \gamma,\gamma';x,y)
\end{align}
and the action of $K_{13}$ and $K_{23}$ on them are respectively
\begin{align}
&\left\{
\begin{matrix}
K_{13}\Phi^{(3)}_1=4p_3^{d-4}\sum_{a b} x^a y^b\, c_1^{(3)}(a,b)\,(\b+1)\,F_4(\alpha+2,\beta+2,\gamma,\gamma',x,y)\\[1.5ex]
K_{23}\Phi^{(3)}_1=4p_3^{d-4}\sum_{a b} x^a y^b\, c_1^{(3)}(a,b)\,(\b+1)\,F_4(\alpha+2,\beta+2,\gamma,\gamma',x,y)
\end{matrix}\right.\\[2.2ex]
&\left\{\begin{matrix}
K_{13}\Phi^{(3)}_2=8p_3^{d-4}\sum_{a b} x^a y^b\, c_2^{(3)}(a,b)\,(\b+1)\,F_4(\alpha+3,\beta+2,\gamma,\gamma',x,y)\\[1.5ex]
K_{23}\Phi^{(3)}_2=8p_3^{d-4}\sum_{a b} x^a y^b\, c_2^{(3)}(a,b)\,(\b+1)\,F_4(\alpha+3,\beta+2,\gamma,\gamma',x,y).
\end{matrix}\right.
\end{align}
These equations have to be equal to the right hand side of \eqref{eqA3}, and this condition fixes the integration constants to be those appearing in $A_2$ as
\begin{align}
c_1^{(3)}(a,b)&=\frac{1}{2(\b+1)}\,c^{(2)}(a,b)\\
c_2^{(3)}(a,b)&=\frac{1}{2(\b+1)(\b+2)}\,c^{(1)}(a,b).
\end{align}
The general solution for $A_3$ can be obtained by adding to the particular solution above the general solution of the homogeneous system \eqref{eqA3}, for which
\begin{align}
A_3&=p_3^{d-2}\,\sum_{ab}\,x^a\,y^b\Big[c^{(3)}(a,b)\,F_4(\a+1,\b+1,\g,\g';x,y)\notag\\
&\hspace{1.5cm}+\frac{1}{2(\b+1)}\,c^{(2)}(a,b)\,F_4(\a+2,\b+1,\g,\g';x,y)\notag\\
&\hspace{1.5cm}+\frac{1}{2(\b+1)(\b+2)}\,c^{(1)}(a,b)F_4(\alpha+3,\beta+1,\gamma,\gamma',x,y)
\Big].
\end{align}
Imposing the symmetry condition on $A_3$ under the change $p_1\leftrightarrow p_2$
\begin{equation}
A_3(p_2,p_1,p_3)=A_3(p_1,p_2,p_3)
\end{equation}
we obtain additional constraints on the homogeneous coefficients $c^{(i)}(a,b)$, $i=1,2,3$ as
\begin{align}
c^{(3)}\left(\frac{d}{2},0\right)=c^{(3)}\left(0,\frac{d}{2}\right)\label{CondA3}\\
c^{(2)}\left(\frac{d}{2},0\right)=c^{(2)}\left(0,\frac{d}{2}\right)\\
c^{(1)}\left(\frac{d}{2},0\right)=c^{(1)}\left(0,\frac{d}{2}\right).
\end{align}
We observe that the last two conditions are already satisfied by the solutions of $A_1$ and $A_2$ and the new information follows from the first of the equations in \eqref{CondA3}. At this stage the independent constants appearing in $A_3$ are seven, but this number will be reduced to three once that the symmetry condition on $A_5$ will be also taken into account. 
%%%%%%%%%%%%%%%%%%%%%%% 
%%%%%%%%%%%%%%%%%%%%%%%
\subsection{The solution for $A_4$}\label{A4section}
%%%%%%%%%%%%%%%%%%%%%%%
%%%%%%%%%%%%%%%%%%%%%%%
The solutions by our method for the $A_4$ and $A_5$ form factors require a special treatment, due to exchanged momenta on the functional dependence of the form factors on the right hand side of the respective equations. This complication is not present in the case of the $TJJ$ \cite{Coriano:2018bbe}.
In particular, the primary WI's
\begin{equation}
\left\{\begin{split}
K_{13} A_4 &=-4 A_2(p_2\leftrightarrow p_3)\\
K_{23} A_4&=-4 A_2 (p_1\leftrightarrow p_3)
\end{split}
\right.\label{eqA4}
\end{equation}
involve the symmetrization $A_2(p_2\leftrightarrow p_3)$, that can be obtained from \eqref{A2} with the exchange $(p_2,\D_2)\leftrightarrow (p_3,\D_3)$ and the replacements
\begin{equation}
\begin{split}
&x\to \tilde x=\sdfrac{x}{y},\qquad y\to \tilde y=\sdfrac{1}{y},\\
&\a(a,b)\to \tilde{\a}(a,b)=a+b+\sdfrac{d}{2}-\sdfrac{\D_1-\D_2+\D_3}{2}=\a(a,b)-(\D_2-\D_3)\\
&\b(a,b)\to \tilde\b(a,b)=a+b+d-\sdfrac{\D_1+\D_2+\D_3}{2}=\b(a,b)\\
&\g(a)\to \tilde\g(a)=2a+\sdfrac{d}{2}-\D_1+1=\g(a)\\
&\g'(b)\to \tilde\g'(b)=2b+\sdfrac{d}{2}-\D_3+1=\g'(b)-(\D_2-\D_3)\\
\end{split}
\end{equation}
in the basic solution $F_4$. In the $TTT$ case, inserting the corresponding scaling dimensions, one has $\tilde\a(a,b)=\a(a,b)$ and $\tilde\g'(a,b)=\g'(a,b)$. \\
The two particular solutions of the inhomogeneous equations \eqref{eqA4} can be expressed in the form
\begin{equation}
\begin{split}
\Phi_1^{(4)}&=p_3^{d -2} \sum_{a b} c_1^{(4)}(a,b)\, x^a y^b   F_4(\alpha+2,\beta+1; \gamma,\gamma';x,y)\\
\Phi_2^{(4)}&=p_3^{d -2} \sum_{a b} c_2^{(4)}(a,b)\, x^a y^b   F_4(\alpha+1,\beta+1; \gamma-1,\gamma'-1;x,y)
\end{split}\label{solA4}
\end{equation}
where the action of $K_{13}, K_{23}$ on them gives
\begin{align}
&\left\{\begin{matrix}
K_{13}\Phi_1^{(4)}=4p_3^{d-4}\sum_{a b} x^a y^b\, c_1^{(4)}(a,b)\,(\b+1)\,F_4(\alpha+2,\beta+2,\gamma,\gamma',x,y)\\[1.5ex]
K_{23}\Phi_1^{(4)}=4p_3^{d-4}\sum_{a b} x^a y^b\, c_1^{(4)}(a,b)\,(\b+1)\,F_4(\alpha+2,\beta+2,\gamma,\gamma',x,y)
\end{matrix}\right.\\[2.2ex]
&\left\{\begin{matrix}
K_{13}\Phi_2^{(4)}=4p_3^{d -4} \sum_{a b} c_2^{(4)}(a,b)\,\frac{(\a+1)(\b+1)}{(\g-1)} \,x^a y^b   F_4(\alpha+2,\beta+2; \gamma,\gamma'-1;x,y)\\[1.5ex]
K_{23}\Phi_2^{(4)}=4p_3^{d -4} \sum_{a b} c_2^{(4)}(a,b)\,\frac{(\a+1)(\b+1)}{(\g'-1)} \,x^a y^b   F_4(\alpha+2,\beta+2; \gamma-1,\gamma';x,y).
\end{matrix}\right.
\end{align}
Writing the previous expressions explicitly for the four fundamental indices $a=0,\,d/2$ and $b=0,\,d/2$, one can compare the two solutions in order to extract information about the corresponding constants introduced in \eqref{solA4}.  
We relate the various constants using the intermediate steps worked out in appendix \ref{fuchs}, to which we refer for further details.\\ 
This  allows us to derive a particular solution of the system of equations of \eqref{eqA4}. The general solution to \eqref{eqA4} is obtained by adding such particular solution to the general homogenous one, and can be written in the form
\begin{align}
A_4&=p_3^{d-2}\sum_{ab}x^ay^b\Bigg\{c^{(4)}(a,b)\,F_4(\a+1,\b+1,\g,\g',x,y)\notag\\
&+c^{(4)}_1(a,b)F_4(\a+2,\b+1,\g,\g',x,y)+\,c^{(4)}_2(a,b)F_4(\a+1,\b+1,\g-1,\g'-1,x,y)\notag\\
&+c^{(4)}_3(a,b)F_4(\a+1,\b+1,\g-1,\g',x,y)+\,c^{(4)}_4(a,b)F_4(\a+1,\b+1,\g,\g'-1,x,y)\Bigg\}
\end{align}
with the constants $c^{(4)}_i,\ i=1,2,3,4$ given in terms of $c^{(1)}(a,b)$ and $c^{(2)}(a,b)$ once we enforce the symmetry constraints, and with $\a=\a(a,b)$, $\b=\b(a,b)$, $\g=\g(a)$ and $\g'=\g'(b)$, for simplicity.
The form factor $A_4$ is symmetric under the exchange $p_1\leftrightarrow p_2$\begin{equation}
A_4(p_1,p_2,p_3)=A_4(p_2,p_1,p_3)
\end{equation}
and leads to the conditions \eqref{CondA1}, \eqref{CondA2}, \eqref{CondA3} and to
\begin{align}
&c^{(4)}\left(\frac{d}{2},0\right)=c^{(4)}\left(0,\frac{d}{2}\right),&&c_3^{(4)}\left(0,0\right)=c_4^{(4)}\left(0,0\right)\notag\\
&c_3^{(4)}\left(\frac{d}{2},\frac{d}{2}\right)=c_4^{(4)}\left(\frac{d}{2},\frac{d}{2}\right)&&c_3^{(4)}\left(\frac{d}{2},0\right)=c_4^{(4)}\left(0,\frac{d}{2}\right)\notag\\
&c_3^{(4)}\left(0,\frac{d}{2}\right)=c_4^{(4)}\left(\frac{d}{2},0\right)
\end{align}
\begin{align}
c^{(4)}_3\left(\frac{d}{2},0\right)&=\frac{d^2}{ (d+2) (d+4)} c^{(1)}\left(0,\frac{d}{2}\right)+\frac{d}{(d-2) } c^{(2)}(0,0)-\frac{d}{(d+2)} c^{(2)}\left(0,\frac{d}{2}\right)- c_3^{(4)}\left(0,\frac{d}{2}\right).
\end{align}

Using the relation given in the appendix, the general solution can be parameterized as
\begin{align}
&\begin{aligned}[c]
c_2^{(4)}\left(0,0\right)&=-\frac{d^2}{(d+2)(d+4)}C_1\\
c_2^{(4)}\left(0,\frac{d}{2}\right)&=-\frac{d^2}{(d+2)(d+4)}C_1\\
c_2^{(4)}\left(\frac{d}{2},0\right)&=c_2^{(4)}\left(0,\frac{d}{2}\right)\\
c_2^{(4)}\left(\frac{d}{2},\frac{d}{2}\right)&=\frac{d\sec\left(\frac{\pi\,d}{2}\right)\,\G\left(1-\frac{d}{2}\right)^2}{(d+2)(d+4)\,\G(-d)}\,C_1
\end{aligned}
\label{C41}
\end{align}

%%%%%%%%%%%%%%%%%%%%%%%%
%%%%%%%%%%%%%%%%%%%%%%%%
\subsection{The $A_5$ solution}
%%%%%%%%%%%%%%%%%%%%%%%%
%%%%%%%%%%%%%%%%%%%%%%%%
Also in the case of $A_5$ we have to repeat the approach presented in \secref{A4section}. In particular the primary Ward identities for $A_5$ is given by
\begin{equation}
\left\{
\begin{split}
K_{13} A_5 &=2[A_4- A_4(p_1\leftrightarrow p_3)] \\
K_{23} A_5&=2[A_4- A_4 (p_2\leftrightarrow p_3)]
\end{split}\right.
\label{eqA5}
\end{equation}
and this system of equations admit seven particular solutions. Combined with the homogeneous solution they give
\begin{align}
A_5&=p_3^{d}\sum_{ab}x^ay^b\Bigg\{\frac{1}{\b}\bigg[+c_1^{(5)}(a,b)\,F_4(\a+1,\b,\g-1,\g'-1,x,y)\notag\\
&\hspace{2cm}+c^{(5)}_2(a,b)F_4(\a+1,\b,\g,\g'-1,x,y)+\,c^{(5)}_3(a,b)F_4(\a+1,\b,\g,\g',x,y)\notag\\
&\hspace{-0.7cm}+c^{(5)}_4(a,b)F_4(\a+1,\b,\g-1,\g',x,y)+c^{(5)}_5(a,b)F_4(\a,\b,\g-1,\g',x,y)+c^{(5)}_6(a,b)F_4(\a,\b,\g,\g'-1,x,y)\bigg]\notag\\
&+\frac{1}{\a\,\b}\bigg[c^{(5)}_7(a,b)F_4(\a,\b,\g-1,\g'-1,x,y)+c^{(5)}(a,b)\,F_4(\a,\b,\g,\g',x,y)\bigg]\Bigg\}. 
\end{align}
In particular the coefficients $c^{(5)}_i$, $i=1,\dots,4$ are fixed by the use \eqref{eqA5}, and imposing the symmetry conditions on $A_5$
\begin{align}
A_5(p_3,p_2,p_1)&=A_5(p_1,p_2,p_3) \nn
A_5(p_2,p_1,p_3)&=A_5(p_1,p_2,p_3) \nn
A_5(p_1,p_3,p_2)&=A_5(p_1,p_2,p_3).
\end{align}
We have left to Appendix \ref{fuchs} more details on the identification of the independent constants which characterize this solution and the analogous solution for $A_4$. There are 5 constants overall for the system of primary WI's, in agreement with the result presented in  \cite{Bzowski:2013sza}, which reduce to 3 after imposing the constraints derived by secondary WI's. Such additional reduction can be performed as discussed in \cite{Bzowski:2013sza}. We have listed such secondary CWI's in appendix \ref{secondary}.
%%%%%%%%%%%%%%%%%%%%%%%%%%%%%%%%%%%%%%%%%%%%%%%%%
%%%%%%%%%%%%%%%%%%%%%%%%
{\subsection{Summary}
%%%%%%%%%%%%%%%%%%%%%%%%
%%%%%%%%%%%%%%%%%%%%%%%%%%%%%%%%%%%%%%%%%%%%%%%%%
In this section we will briefly summarize the final solutions obtained for all the form factors. 

We obtain \\
\begin{equation}
A_1=p_3^{d - 6}\sum_{a,b} C_1\,f_1(a,b)\,x^a y^b \,F_4(\alpha(a,b) +3, \beta(a,b)+3; \gamma(a), \gamma'(b); x, y) 
\end{equation}

\begin{align}
f_1\left(0,\frac{d}{2}\right)&=f_1\left(\frac{d}{2},0\right)=1\notag\\
f_1(0,0)&=-\frac{(d-4) (d-2)}{(d+2) (d+4)} \notag\\
f_1\left(\frac{d}{2},\frac{d}{2}\right)&=\frac{\G\left(-\frac{d}{2}\right)\G\left(d+3\right)}{2\,\Gamma\left(\frac{d}{2}\right)} \notag \\
\label{CondA2}
\end{align}

where $f_1(a,b)$ takes four values for the four Fuchsian indices. In this case the function $f_1$ can be read from the expressions \eqref{CondA1};
%Then, after imposing the symmetry conditions on $A_5$, the number of independent constants in $A_i$, $i=2,3,4,5$ decrease, and finally $A_2$ depends on two independent constants $C_1$ and $C_2$ as
\begin{align}
A_2&= p_3^{d - 4}\sum_{a b} x^a y^b\bigg[C_2\,f_2(a,b)\,F_4(\alpha +2, \beta+2; \gamma, \gamma'; x, y)\notag\\
&\hspace{5.5cm}+ \frac{2\,C_1}{\big(\b+2\big)}\,f_1(a,b)\,F_4(\alpha +3, \beta+2; \gamma, \gamma'; x, y)\bigg].
\end{align}

\begin{align} 
f_2\left(0,0\right)&=\frac{d-2}{d+2}\nn
f_2\left(\frac{d}{2},0\right)&=f_2\left(0,\frac{d}{2}\right)=1 \nn
f_2\left(\frac{d}{2},\frac{d}{2}\right)&=\frac{\Gamma(-d/2)\Gamma(d+2)}{\Gamma(d/2)}
\end{align}

In the same way we write the explicit form of $A_3$ using the results in \appref{condition} as
\begin{align}
A_3&=p_3^{d-2}\,\sum_{ab}\,x^a\,y^b\Big[C_3\,f_3(a,b,d)\,F_4(\a+1,\b+1,\g,\g';x,y)\notag\\
&\hspace{3.5cm}+\frac{C_2}{2(\b+1)}\,f_2(a,b,d)\,F_4(\a+2,\b+1,\g,\g';x,y)\notag\\
&\hspace{5.3cm}+\frac{C_1}{2(\b+1)(\b+2)}\,f_1(a,b,d)F_4(\alpha+3,\beta+1,\gamma,\gamma',x,y)
\Big],
\end{align}
where 
\begin{align} 
f_3\left(\frac{d}{2},0\right)&=f_3\left(0,\frac{d}{2}\right)=1\notag\\
f_3\left(0,0\right)&=-1\notag\\
f_3\left(\frac{d}{2},\frac{d}{2}\right)&=\frac{\Gamma(-d/2)\Gamma(d+1)}{\Gamma(d/2)}
\end{align} 

\begin{align}
A_4&=p_3^{d-2}\sum_{ab}x^ay^b\Bigg\{c^{(4)}(a,b)\,F_4(\a+1,\b+1,\g,\g',x,y)\notag\\
&+c^{(4)}_1(a,b)F_4(\a+2,\b+1,\g,\g',x,y)+\,c^{(4)}_2(a,b)F_4(\a+1,\b+1,\g-1,\g'-1,x,y)\notag\\
&+c^{(4)}_3(a,b)F_4(\a+1,\b+1,\g-1,\g',x,y)+\,c^{(4)}_4(a,b)F_4(\a+1,\b+1,\g,\g'-1,x,y)\Bigg\}
\end{align}
\begin{align}
c^{(4)}\left(\frac{d}{2},0\right)&=c^{(4)}\left(0,\frac{d}{2}\right)=C_4\notag\\
c^{(4)}\left(0,0\right)&=-C_4\notag\\
c^{(4)}\left(\frac{d}{2},\frac{d}{2}\right)&=\frac{\G\left(-\frac{d}{2}\right)\G(d+1)}{\G\left(\frac{d}{2}\right)}C_4-\frac{d^2\,\G\left(-\frac{d}{2}\right)\G(d+2)}{(d+1)(d+2)(d+4)\G\left(\frac{d}{2}\right)}C_1\\[4ex]
c_1^{(4)}\left(\frac{d}{2},0\right)&=c^{(4)}_1\left(0,\frac{d}{2}\right)=-C_2\notag\\
c_1^{(4)}\left(0,0\right)&=\frac{2}{(d+2)}\,C_2\notag\\
c_1^{(4)}\left(\frac{d}{2},\frac{d}{2}\right)&=-\frac{2\,\G\left(-\frac{d}{2}\right)\,\G\left(d+2\right)}{(d+2)\,\G\left(\frac{d}{2}\right)}C_2\\[4ex]
c_2^{(4)}\left(0,0\right)&=-\frac{d^2}{(d+2)(d+4)}C_1\notag\\
c_2^{(4)}\left(0,\frac{d}{2}\right)&=c_2^{(4)}\left(0,\frac{d}{2}\right)=-\frac{d^2}{(d+2)(d+4)}C_1\notag\\
c_2^{(4)}\left(\frac{d}{2},\frac{d}{2}\right)&=-\frac{d^2\,\G\left(d+2\right)\G\left(-\frac{d}{2}\right)}{(d+1)(d+2)(d+4)\,\G\left(\frac{d}{2}\right)}\,C_1\\[4ex]
c_3^{(4)}\left(0,0\right)&=c_4^{(4)}\left(0,0\right)=0\notag\\
c_3^{(4)}\left(\frac{d}{2},0\right)&=c_4^{(4)}\left(0,\frac{d}{2}\right)=0\notag\\
c_3^{(4)}\left(0,\frac{d}{2}\right)&=c_4^{(4)}\left(\frac{d}{2},0\right)=\frac{d^2}{(d+2)(d+4)}C_1\notag\\
c_3^{(4)}\left(\frac{d}{2},\frac{d}{2}\right)&=c_4^{(4)}\left(\frac{d}{2},\frac{d}{2}\right)=\frac{d^2\,\G\left(d+2\right)\G\left(-\frac{d}{2}\right)}{(d+1)(d+2)(d+4)\,\G\left(\frac{d}{2}\right)}\,C_1
\end{align}

 Finally we give the form of the $A_5$ form factors as
\begin{align}
A_5&=p_3^{d}\sum_{ab}x^ay^b\Bigg\{\frac{1}{\b}\bigg[+c_1^{(5)}(a,b)\,F_4(\a+1,\b,\g-1,\g'-1,x,y)\notag\\
&\hspace{2cm}+c^{(5)}_2(a,b)F_4(\a+1,\b,\g,\g'-1,x,y)+\,c^{(5)}_3(a,b)F_4(\a+1,\b,\g,\g',x,y)\notag\\
&\hspace{-0.5cm}+\,c^{(5)}_4(a,b)F_4(\a+1,\b,\g-1,\g',x,y)+c^{(5)}_5(a,b)F_4(\a,\b,\g-1,\g',x,y)+\,c^{(5)}_6(a,b)F_4(\a,\b,\g,\g'-1,x,y)\bigg]\notag\\
&+\frac{1}{\a\,\b}\bigg[c^{(5)}_7(a,b)F_4(\a,\b,\g-1,\g'-1,x,y)+c^{(5)}(a,b)\,F_4(\a,\b,\g,\g',x,y)\bigg]\Bigg\}. 
\end{align}
where the coefficients are summarized in Appendix \ref{condition}. The global solution is fixed up to five independent constants. 
\section{Lagrangian realizations and reconstruction}
%%%%%%%%%%%%%%%%%%%%%%%%%%%%%%%%%%%%%%%%%%%%%%%%%
In this section we turn to the central aspect of our analysis, which will allow us to extend the results of the $TJJ$ correlator presented in \cite{Coriano:2018bbe} to the $TTT$. We will be using the free field theory realizations of such correlator in order to study the structure of the conformal Ward identities in momentum space and, in particular, the form of the anomalous Ward identities once the conformal symmetry is broken by the anomaly. We will work in DR and adopt the $\overline{MS}$ renormalization scheme. Our analysis hinges on the correspondence between the exact result obtained by solving the CWI's and the perturbative one. \\
It is clear that the general solutions presented in the former section, though derived regardless of any perturbative picture, become completely equivalent to the latter if, for a given spacetime dimension, we have a sufficient number of independent sectors in the Lagrangian realization that allow us to reproduce the general one. For instance, in $d=3,4$ such correspondence is exact, as already mentioned in the introduction, since the number of constants in the solution coincides wth the number of possible independent sectors in the free field theory.\\
In principle, one could proceed with an analysis of the general solutions - such as those presented in the previous section - as $d\to 4$, by going through a very involved process of extraction of the singularities from their general expressions in terms of $F_4$.\\
 However, this can be avoided once the general results for the $A_i$'s are matched to the perturbative ones. As already mentioned, this brings in an important simplification of the final result for the $TTT$, which is expressed in terms of the simple $\log$ present in $B_0$ and the scalar 3-point function $C_0$. The latter is of type $F_4$ in $d=4$, but takes a far simpler form compared to the expressions derived in the previous section
\bea
 &  C_0 ( p_1^2,p_2^2,p_3^2) = \frac{ 1}{p_3^2} \Phi (x,y),
 \eea
where the function $\Phi (x, y)$  is defined as
\cite{Usyukina:1993ch}
\bea
\Phi( x, y) &=& \frac{1}{\lambda} \biggl\{ 2 [Li_2(-\rho  x) + Li_2(- \rho y)]  +
\ln \frac{y}{ x}\ln \frac{1+ \rho y }{1 + \rho x}+ \ln (\rho x) \ln (\rho  y) + \frac{\pi^2}{3} \biggr\},
\label{Phi}
\eea
with
\bea
 \lambda(x,y) = \sqrt {\Delta},
 \qquad  \qquad \Delta=(1-  x- y)^2 - 4  x  y,
\label{lambda} \\
\rho( x,y) = 2 (1-  x-  y+\lambda)^{-1},
  \qquad  \qquad x=\frac{p_1^2}{p_3^3} \, ,\qquad \qquad y= \frac {p_2^2}{p_3^2}\, .
\eea

 This has the important implication that the study of the specific unitarity cuts in the diagrammatic expansions of the correlator \cite{Giannotti:2008cv, Coriano:2014gja} which are held responsible for the emergence of the anomaly, acquire a simple particle interpretation and are not an artifact of perturbation theory. Once the general correspondence between Lagrangian and non-Lagrangian solutions is established, we will concentrate on showing how renormalization is responsible for the emergence of specific anomaly poles 
 in this correlators. We anticipate that the vertex will separate, after renormalization, into a traceless part and in an anomaly part, following the same pattern of the $TJJ$ \cite{Coriano:2018zdo}. From that point on, one can use just the Feynman expansion to perform complete further analysis of this vertex at one loop, with no loss of generality whatsoever. 
 \subsection{Perturbative sectors}
 In this section we define our conventions used for the perturbative sectors.\\
The quantum actions for the scalar and fermion field are respectively
\begin{align}
S_{scalar}&=\sdfrac{1}{2}\int\, d^dx\,\sqrt{-g}\left[g^{\m\n}\nabla_\m\phi\nabla_\n\phi-\c\, R\,\phi^2\right]\\
S_{fermion}&=\sdfrac{i}{2}\int\, d^dx\,e\,e^{\m}_a\left[\bar{\psi}\g^a(D_\m\psi)-(D_\m\bar{\psi})\g^a\psi\right],
\end{align}
where $\c =(d-2)/(4d-4)$ for a conformally coupled scalar in $d$ dimensions, and $R$ is the Ricci scalar. $e_\m^a$ is the vielbein and $e$ its determinant,  with the covariant derivative $D_\m$ given by 
\begin{equation}
D_\m=\partial_\m+\G_\m=\partial_\m+\sdfrac{1}{2}\Sigma^{ab}\,e^\s_a\nabla_\m\,e_{b\,\s}.
\end{equation}
The $\Sigma^{ab}$ are the generators of the Lorentz group in the spin $1/2$ representation. The Latin indices are related to the flat space-time and the Greek indices to the curved space-time. 
For $d=4$ there is an additional conformal field theory described in terms of free abelian vector fields with the action
\begin{equation}
S_{abelian}=S_{M}+S_{gf}+S_{gh}
\end{equation}
where the three contributions are the Maxwell action, the gauge fixing contribution and the ghost action
\begin{align}
S_M&=-\sdfrac{1}{4}\int d^4x\,\sqrt{-g}\,F^{\m\n}F_{\m\n},\\
S_{gf}&=-\sdfrac{1}{\xi}\int d^4x\,\sqrt{-g}\,(\nabla_\m A^\m)^2,\\
S_{gh}&=\int d^4x\,\sqrt{-g}\,\,\partial^\m\bar c\,\partial_\m \,c.
\end{align}
The computation of the vertices of each theory can be done by taking (at most) two functional derivatives of the action with respect to the metric, since the vacuum expectation values of the third derivatives correspond to massless tadpoles, which are zero in DR. 
They are given in \figref{vertices} and their explicit expressions have been collected in the Appendix \ref{Appendix1}.

\begin{figure}[t]
	\centering
	\vspace{-2cm}
	\subfigure{\includegraphics[scale=0.14]{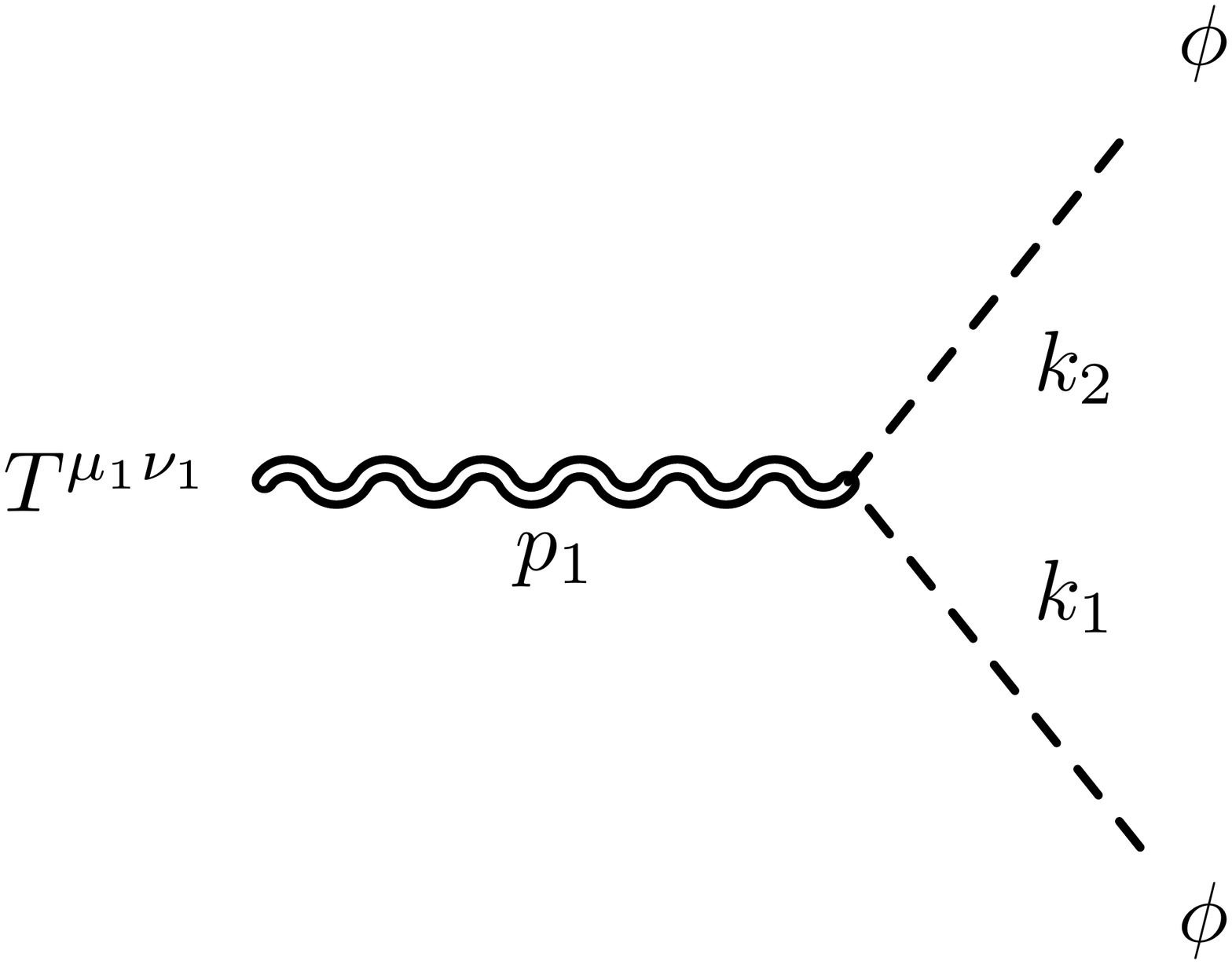}} \hspace{.3cm}
	\subfigure{\includegraphics[scale=0.14]{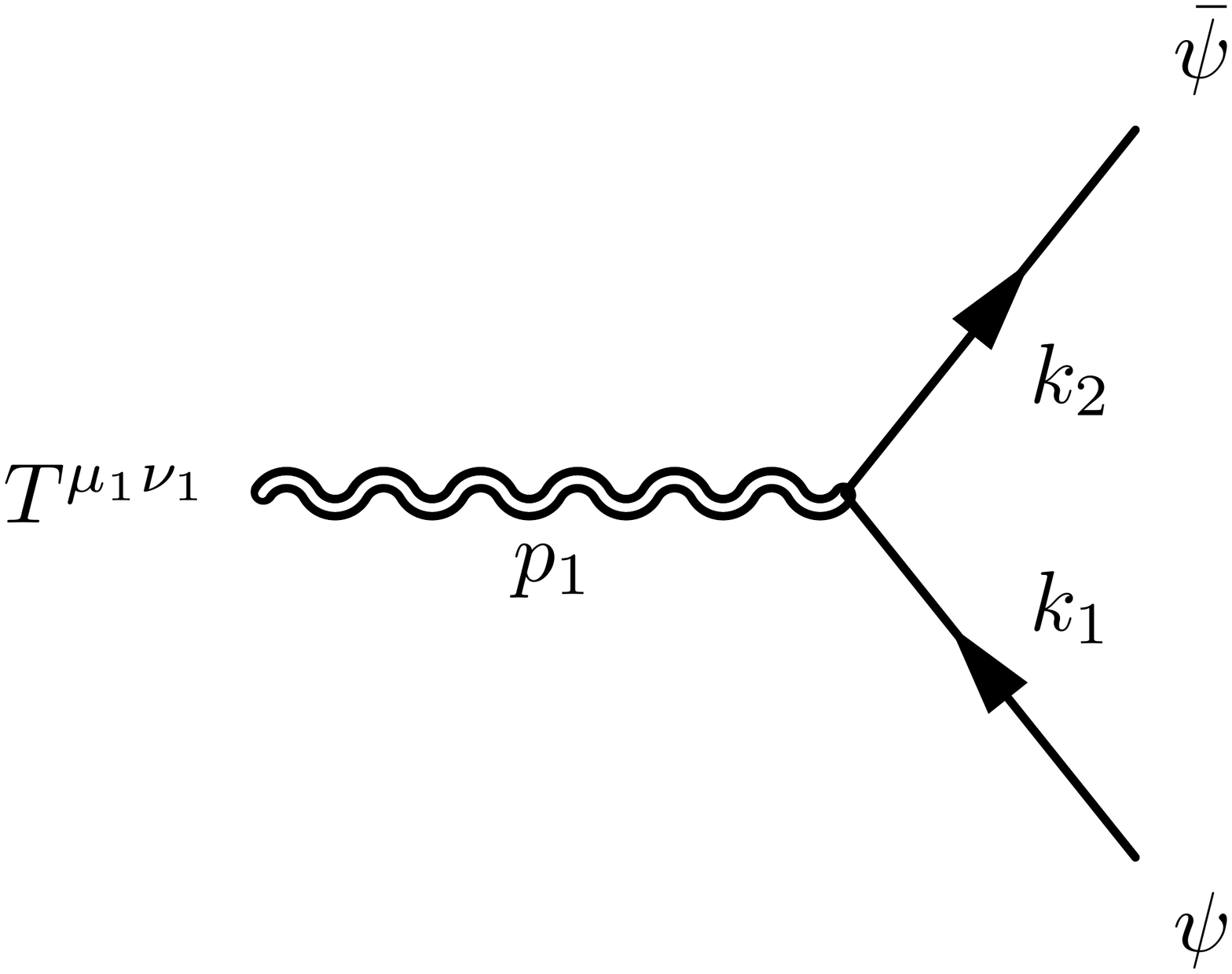}} \hspace{.3cm}
	\subfigure{\includegraphics[scale=0.14]{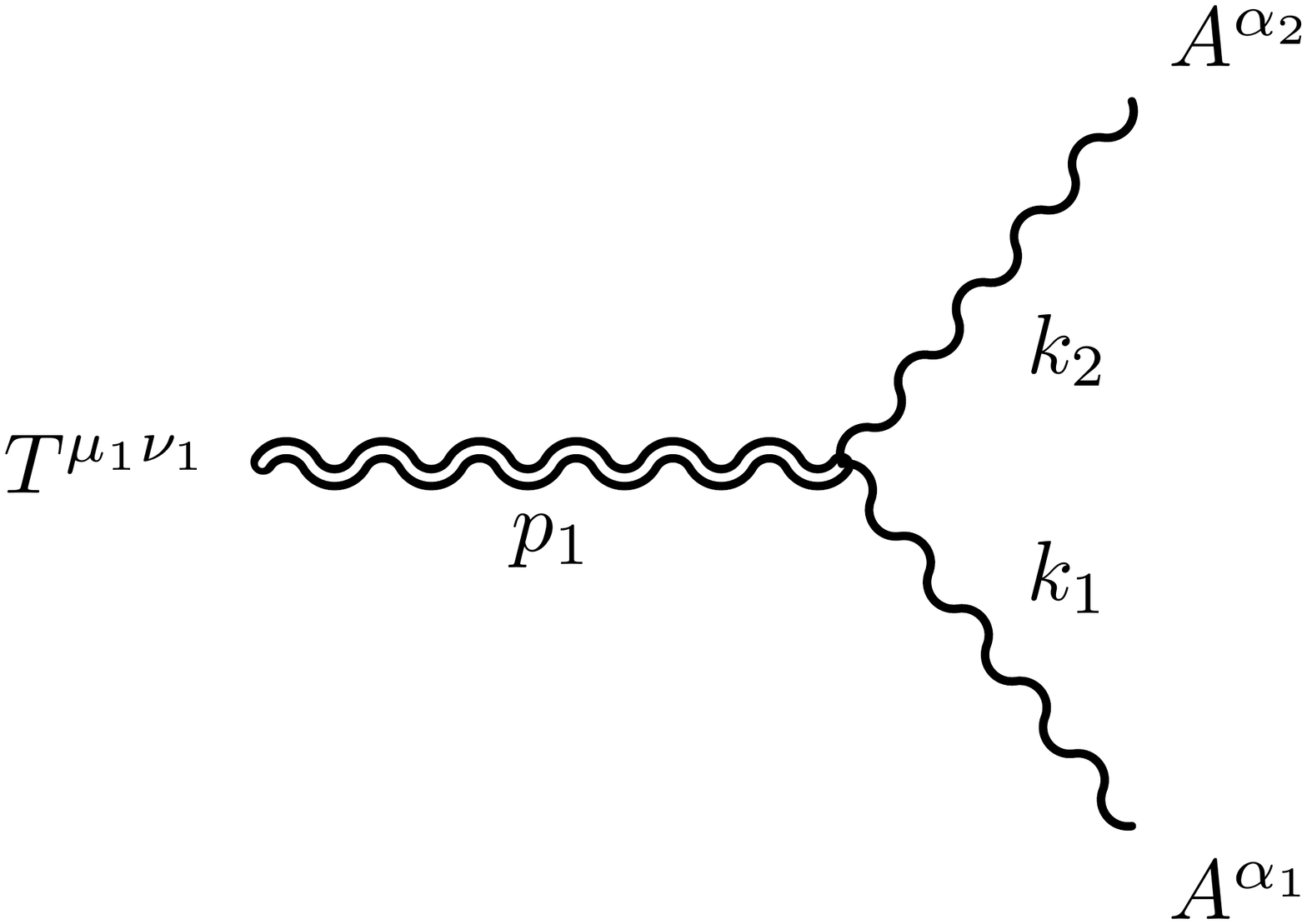}} \hspace{.3cm}
	\subfigure{\includegraphics[scale=0.14]{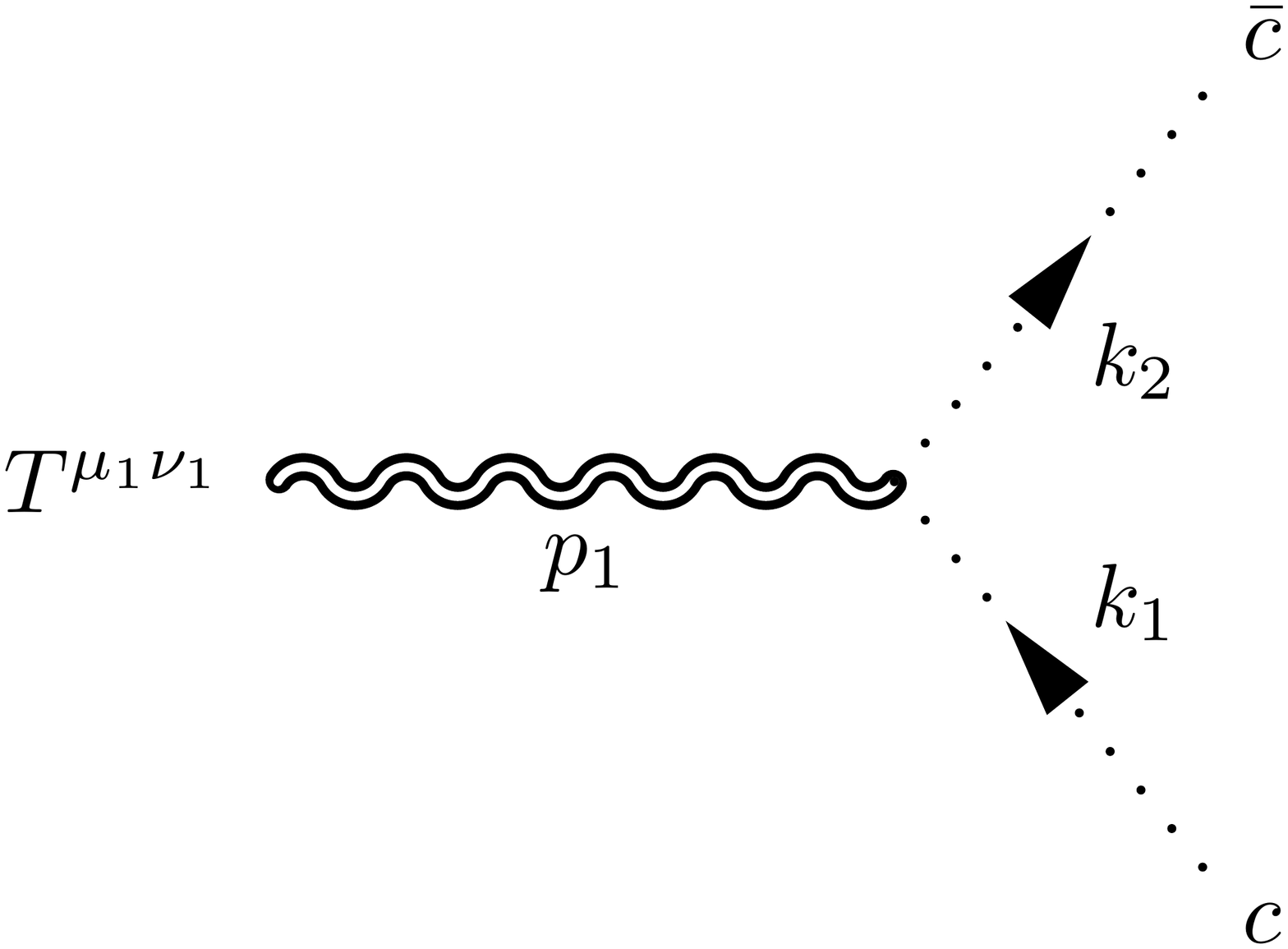}}
	\\
	\vspace{-1.5cm}
	\subfigure{\includegraphics[scale=0.14]{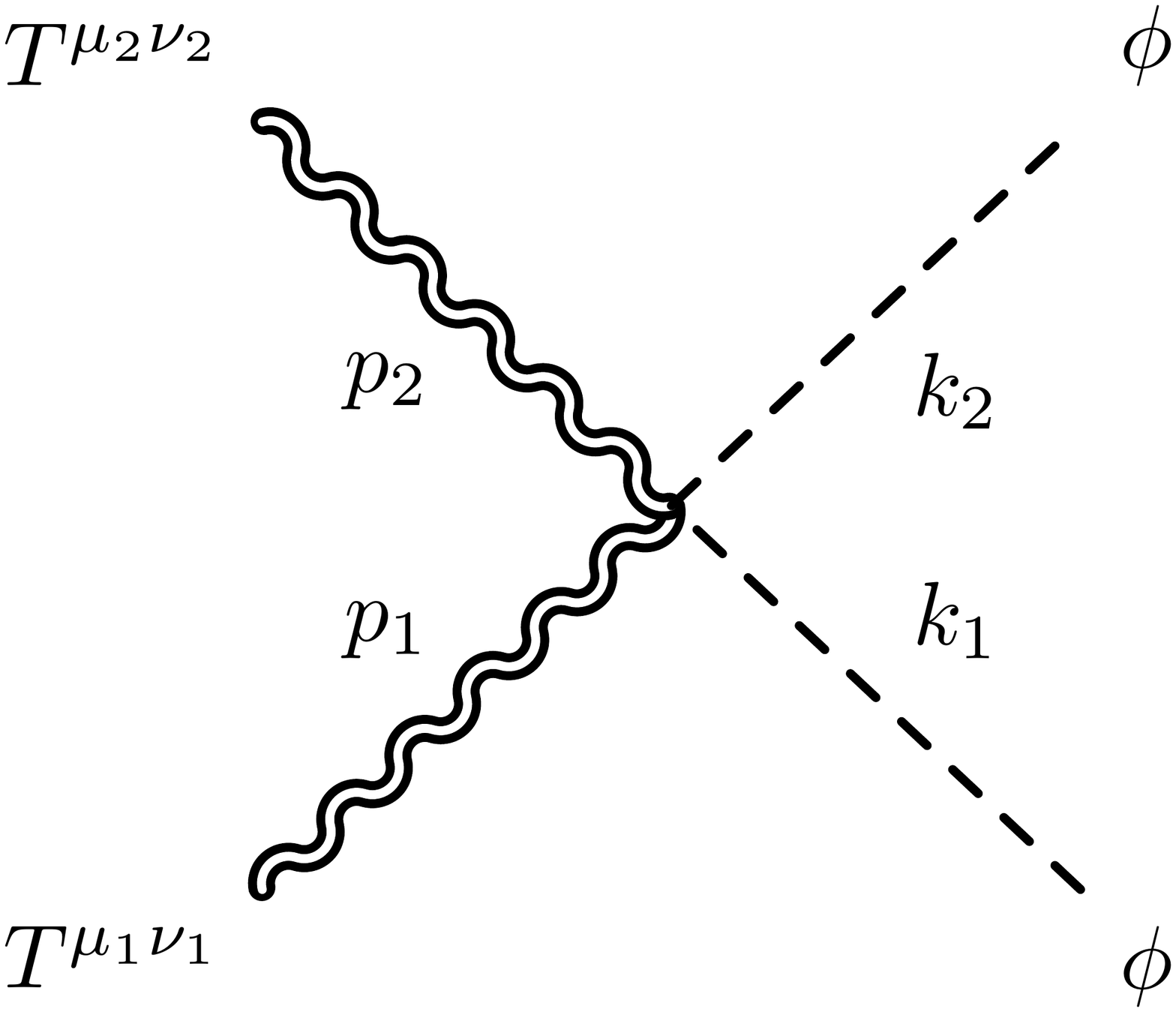}} \hspace{.3cm}
	\subfigure{\includegraphics[scale=0.14]{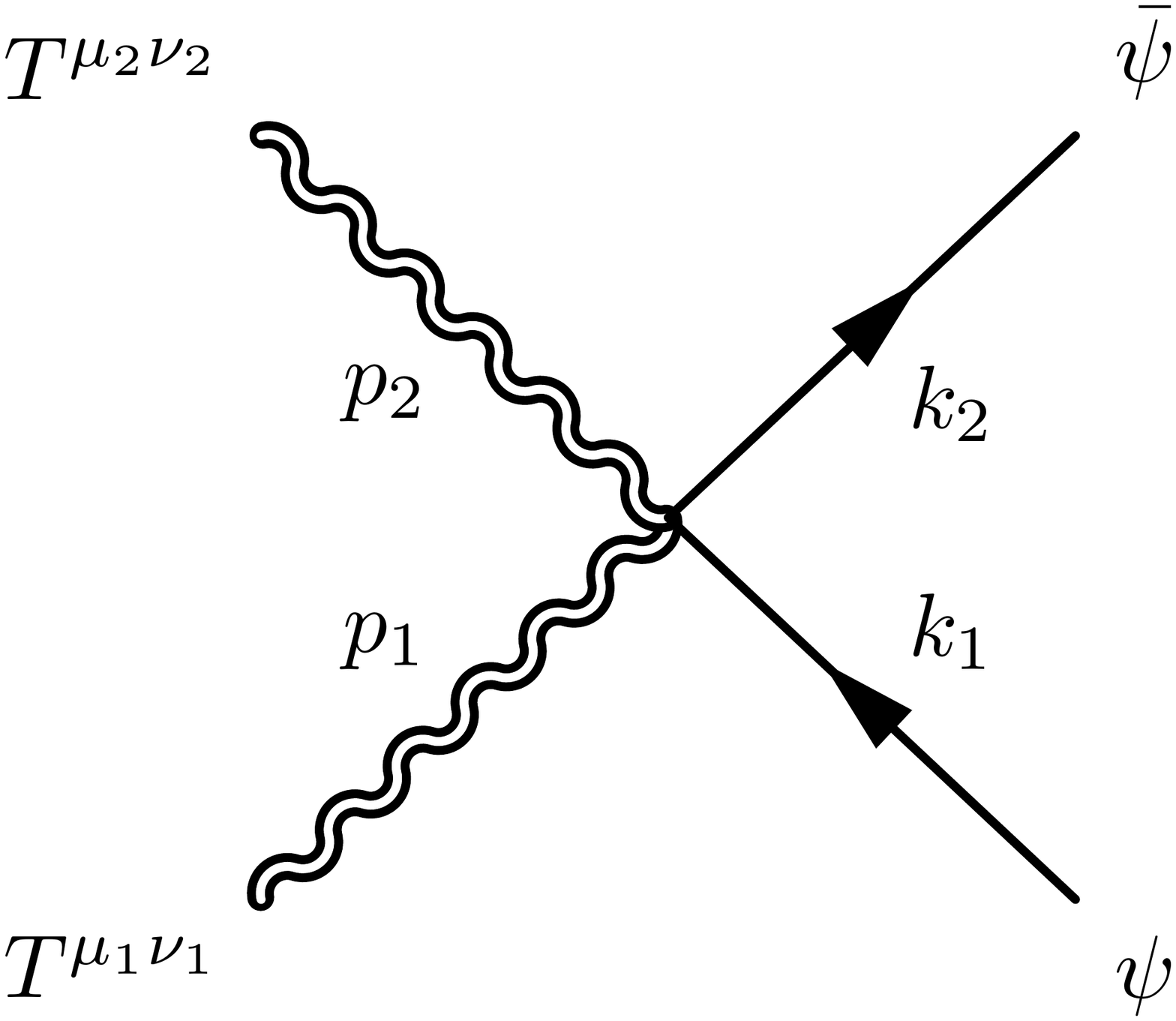}} \hspace{.3cm}
	\subfigure{\includegraphics[scale=0.14]{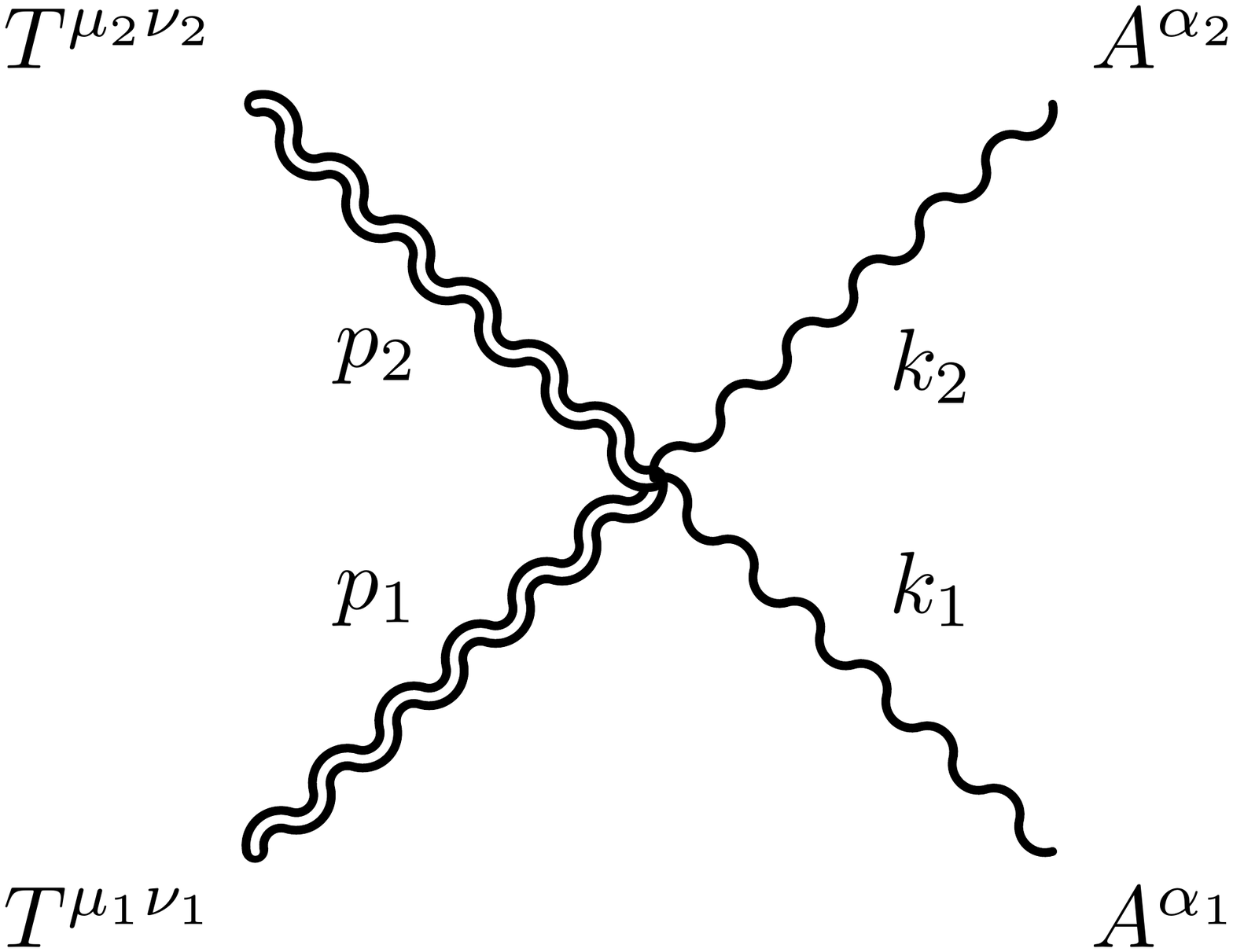}} \hspace{.3cm}
	\subfigure{\includegraphics[scale=0.14]{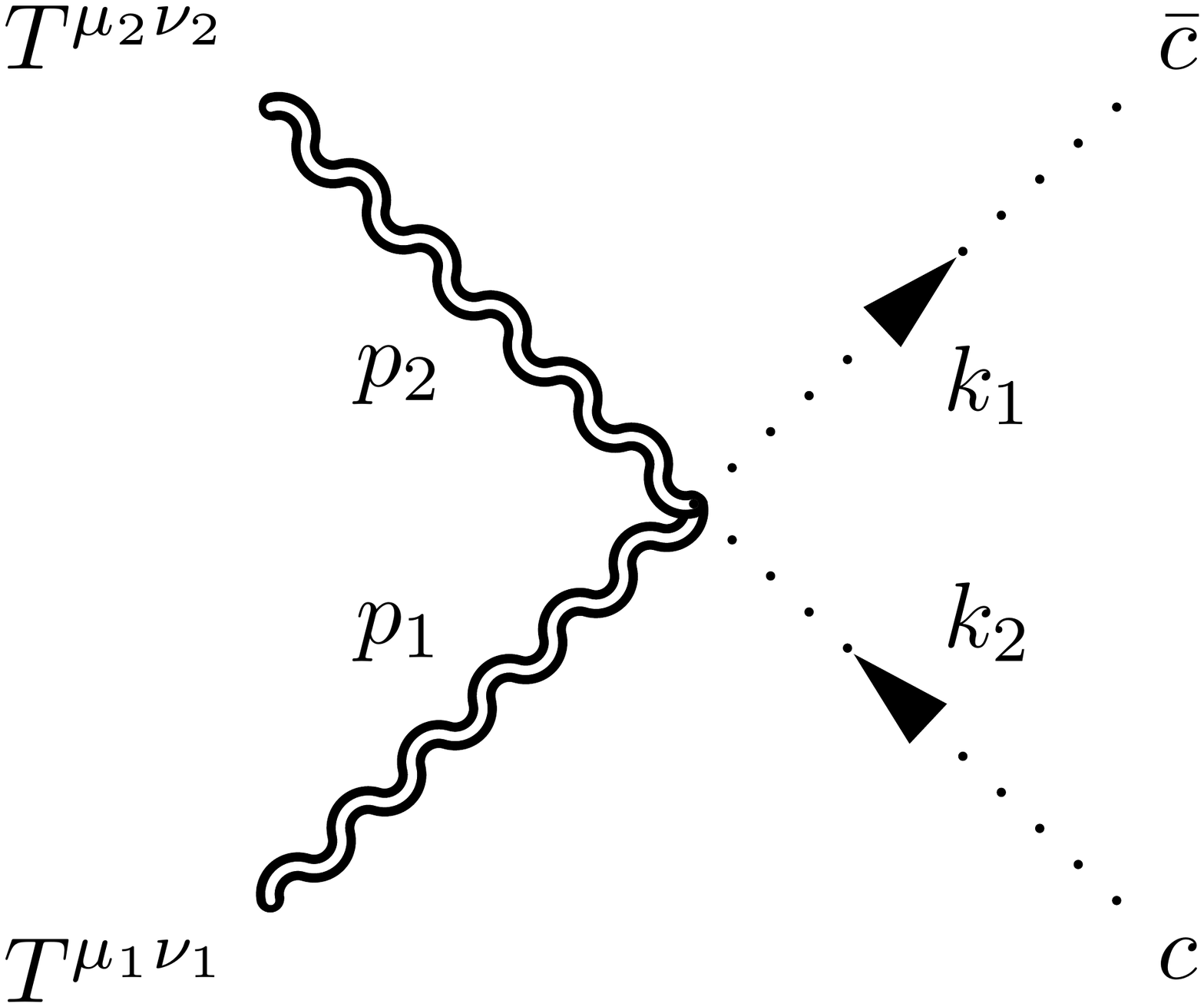}}
	\caption{Vertices used in the Lagrangian realization of the $TTT$ correlator.\label{vertices}}
\end{figure}
\noindent Since we are interested in the most general Lagrangian realization of the $\braket{TTT}$ correlator in the conformal case, this can be obtained only by considering the scalar and fermion sectors in general $d$ dimensions.  

%%%%%%%%%%%%%%%%%%%%%%%%%%%%%%%%%%%%%
\subsection{Scalar sector}
%%%%%%%%%%%%%%%%%%%%%%%%%%%%%%%%%%%%%

\begin{figure}[t]
	\centering
	\subfigure{\includegraphics[scale=0.2]{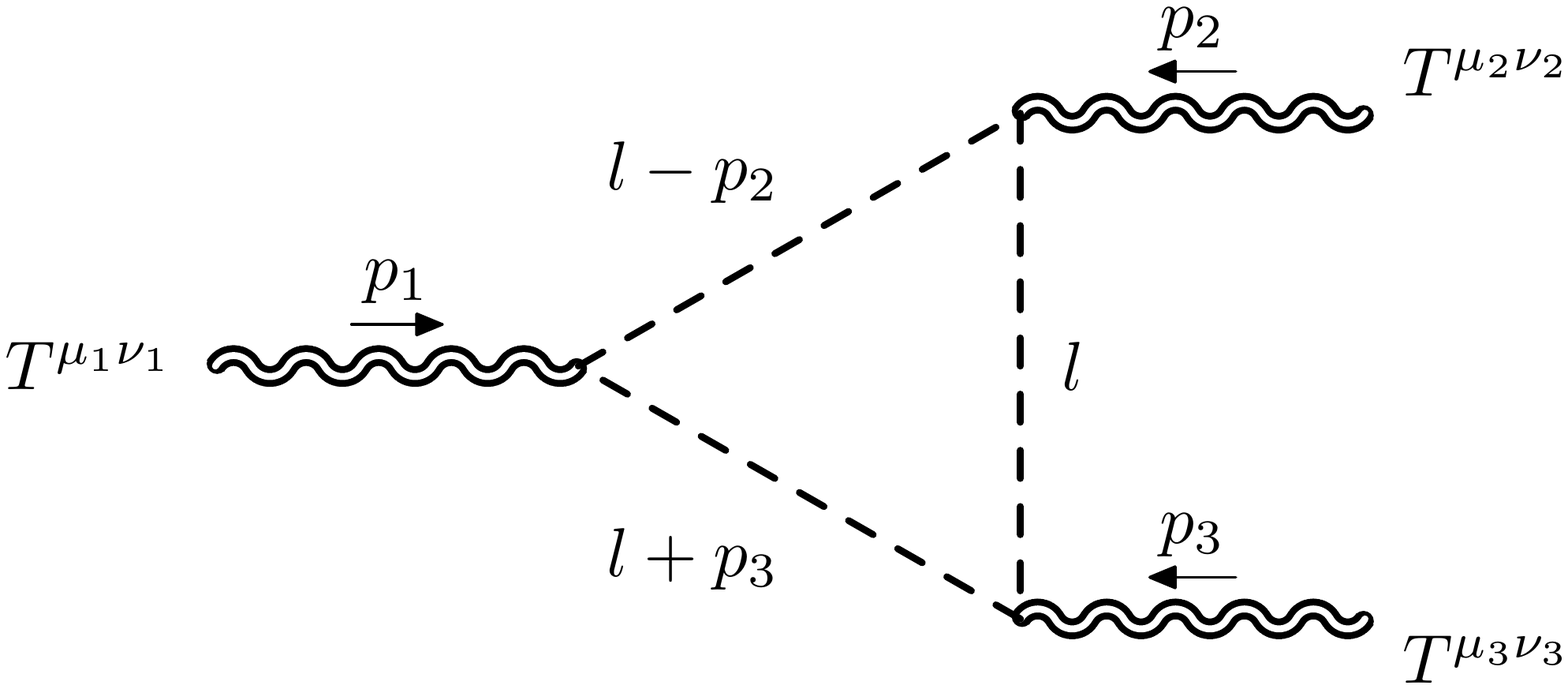}} \hspace{.3cm}
	\subfigure{\includegraphics[scale=0.2]{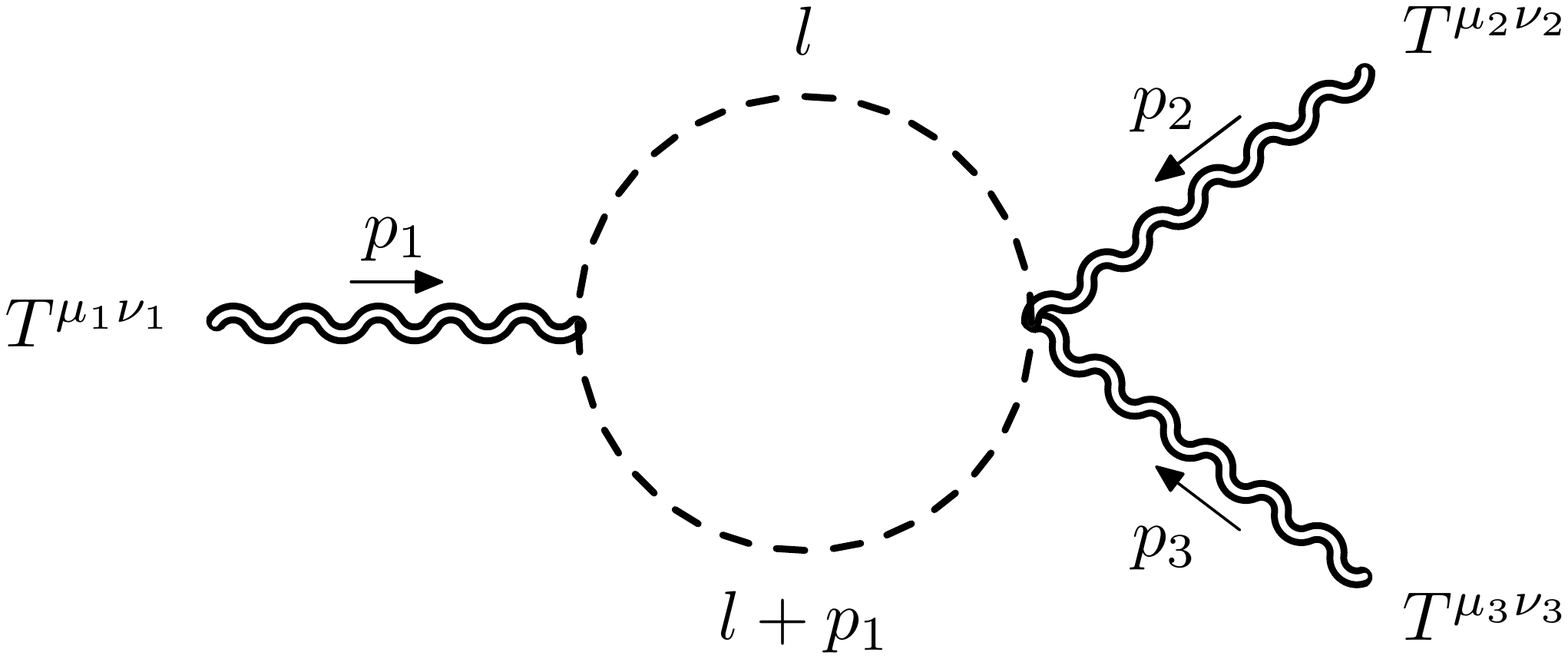}} \hspace{.3cm}
	\raisebox{.12\height}{\subfigure{\includegraphics[scale=0.16]{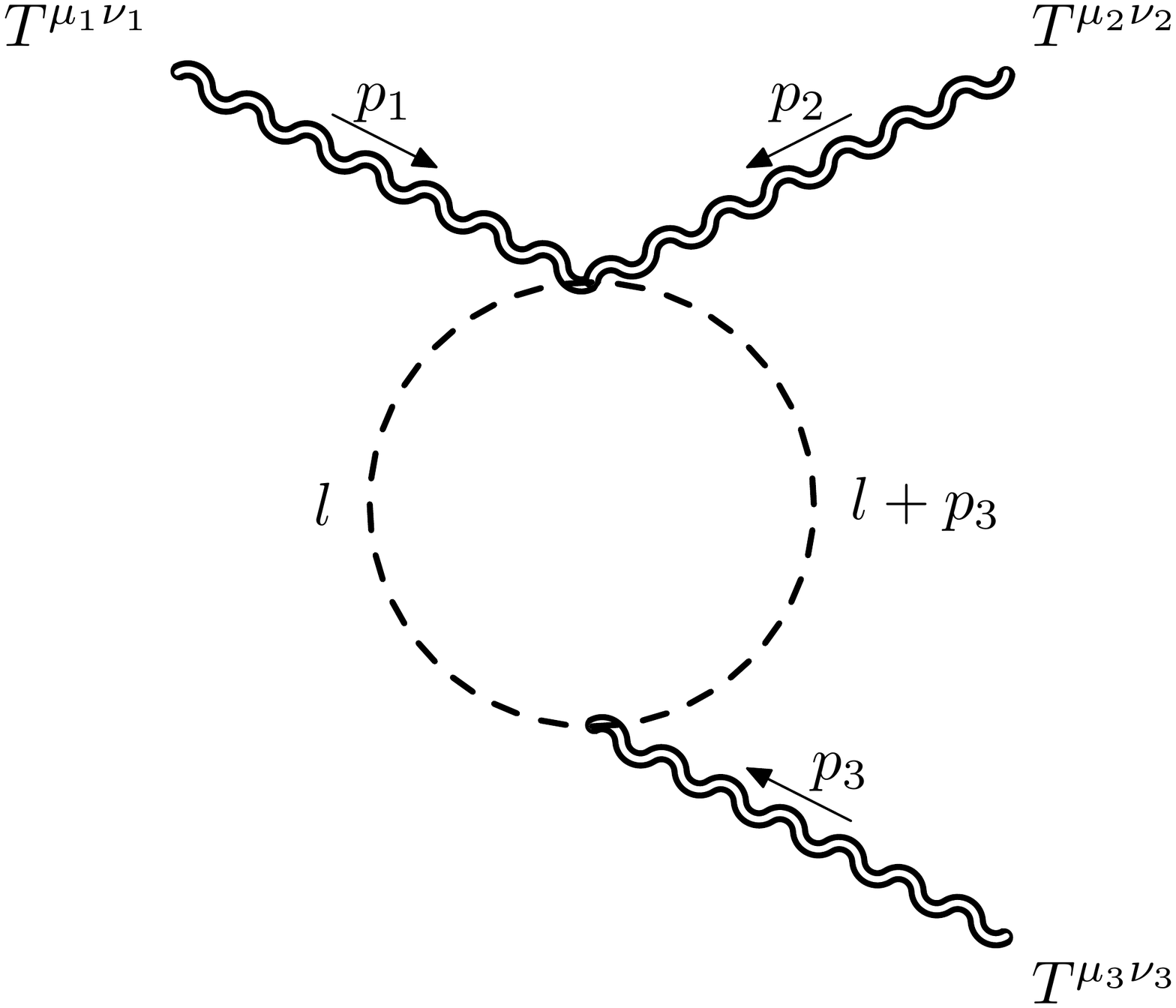}}\hspace{.3cm}}
	\raisebox{.12\height}{\subfigure{\includegraphics[scale=0.16]{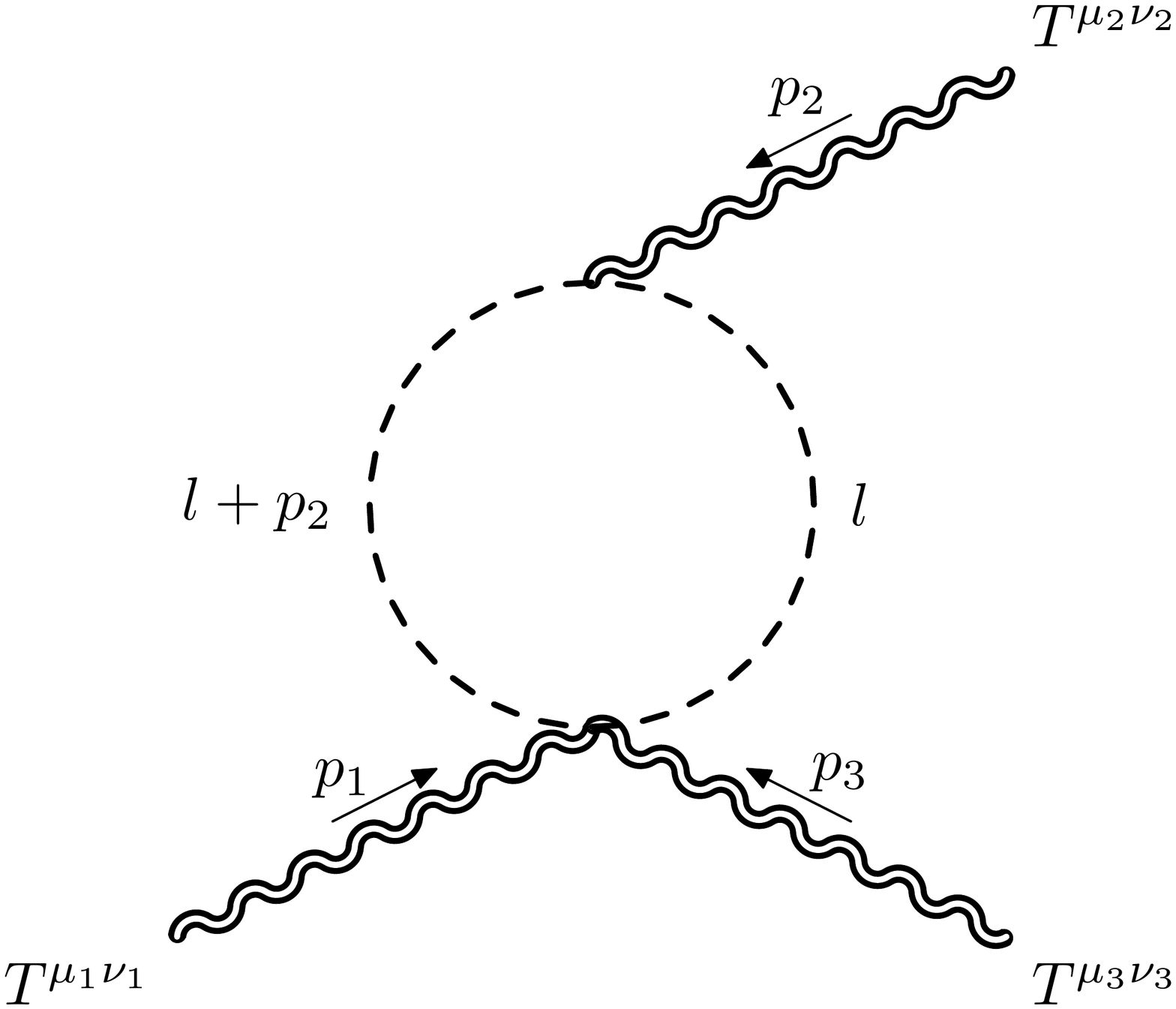}}}
	\vspace{-0.8cm}\caption{One-loop scalar diagrams for the three-graviton vertex.\label{Feynman1}}
\end{figure}

We start from the scalar sector. In the one-loop approximation the contributions to the correlation function are given by the diagrams in \figref{Feynman1}. Using the Feynman rules listed in Appendix \ref{Appendix1}, we calculate all the terms in the defining relation of the $TTT$ Eq. \eqref{expansion} in momentum space, for the scalar sector, as
\begin{align}
\braket{T^{\m_1\n_1}(p_1)T^{\m_2\n_2}(p_2)T^{\m_3\n_3}(p_3)}_S=\, -V_{S}^{\m_1\n_1\m_2\n_2\m_3\n_3}(p_1,p_2,p_3)+\sum_{i=1}^3W_{S,i}^{\m_1\n_1\m_2\n_2\m_3\n_3}(p_1,p_2,p_3)\label{ScalarPart}
\end{align}
where $V_S$ is related to the triangle diagrams in \figref{Feynman1} and $W_{S,i}$ terms are the three bubble contributions labelled by the index $i$, with $i=1,2,3$. These contribution are given by
\begin{align}
V_{S}^{\m_1\n_1\m_2\n_2\m_3\n_3}(p_1,p_2,p_3)& =\int \sdfrac{d^d\ell}{(2\pi)^d}\sdfrac{V^{\m_1\n_1}_{T\phi\phi}(\ell-p_2,\ell+p_3)V^{\m_2\n_2}_{T\phi\phi}(\ell,\ell-p_2)V^{\m_3\n_3}_{T\phi\phi}(\ell,\ell+p_3)}{\ell^2(\ell-p_2)^2(\ell+p_3)^2}\nn
W_{S,1}^{\m_1\n_1\m_2\n_2\m_3\n_3}(p_1,p_2,p_3)& =\sdfrac{1}{2}\int \sdfrac{d^d\ell}{(2\pi)^d}\sdfrac{V^{\m_1\n_1}_{T\phi\phi}(\ell,\ell+p_1)V^{\m_2\n_2\m_3\n_3}_{TT\phi\phi}(\ell,\ell+p_1)}{\ell^2(\ell+p_1)^2}\nn
W_{S,2}^{\m_1\n_1\m_2\n_2\m_3\n_3}(p_1,p_2,p_3)& =\sdfrac{1}{2}\int \sdfrac{d^d\ell}{(2\pi)^d}\sdfrac{V^{\m_3\n_3}_{T\phi\phi}(\ell,\ell+p_3)V^{\m_1\n_1\m_2\n_2}_{TT\phi\phi}(\ell,\ell+p_3)}{\ell^2(\ell+p_3)^2}\nn
W_{S,3}^{\m_1\n_1\m_2\n_2\m_3\n_3}(p_1,p_2,p_3)& =\sdfrac{1}{2}\int \sdfrac{d^d\ell}{(2\pi)^d}\sdfrac{V^{\m_2\n_2}_{T\phi\phi}(\ell,\ell+p_2)V^{\m_1\n_1\m_3\n_3}_{TT\phi\phi}(\ell,\ell+p_2)}{\ell^2(\ell+p_2)^2}
\end{align}
where we have included a symmetry factor $1/2$. The calculation of the integral can be simplified by acting with the projectors $\Pi$ on \eqref{ScalarPart} in order to write the form factors of the transverse and traceless part of the correlator, as in \eqref{DecompTTT}\begin{align}
\braket{t^{\m_1\n_1}(p_1)t^{\m_2\n_2}(p_2)t^{\m_3\n_3}(p_3)}_S&=\,\Pi^{\m_1\n_1}_{\a_1\b_1}(p_1)\Pi^{\m_2\n_2}_{\a_2\b_2}(p_2)\Pi^{\m_3\n_3}_{\a_3\b_3}(p_3)\notag\\
&\times\bigg[ -V_{S}^{\a_1\b_1\a_2\b_2\a_3\b_3}(p_1,p_2,p_3)+\sum_{i=1}^3W_{S,i}^{\a_1\b_1\a_2\b_2\a_3\b_3}(p_1,p_2,p_3)\bigg]
\end{align}
 
%%%%%%%%%%%%%%%%%%%%%%%%%%%%%%%%%%%%%
\subsection{Fermion sector}
%%%%%%%%%%%%%%%%%%%%%%%%%%%%%%%%%%%%%
As in the scalar sector, also in this case we calculate in the one-loop approximation the contribution to the correlation function of the fermion sector by the diagrams in \figref{Feynman2}. These contributions can be written as 
\begin{align}
\braket{T^{\m_1\n_1}(p_1)T^{\m_2\n_2}(p_2)T^{\m_3\n_3}(p_3)}_F=-\, \sum_{j=1}^2V_{F,j}^{\m_1\n_1\m_2\n_2\m_3\n_3}(p_1,p_2,p_3)+\sum_{j=1}^3W_{F,j}^{\m_1\n_1\m_2\n_2\m_3\n_3}(p_1,p_2,p_3)\label{FermionExpa}
\end{align}
using notations similar to the scalar case. In this case we take into account two possible orientations for the fermion in the loop. 
\begin{figure}[t]
	\centering
	\vspace{-2cm}
	\subfigure{\includegraphics[scale=0.2]{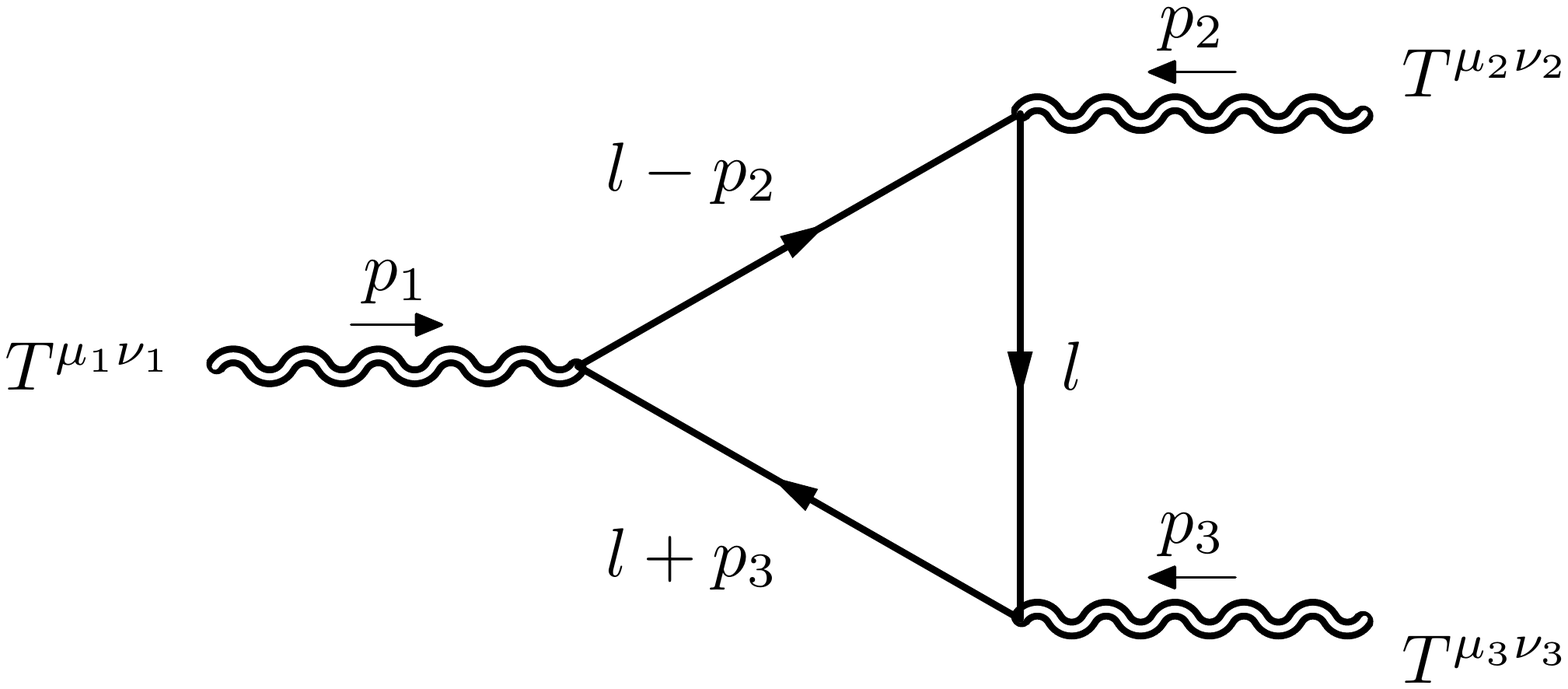}} \hspace{.3cm}
	\subfigure{\includegraphics[scale=0.2]{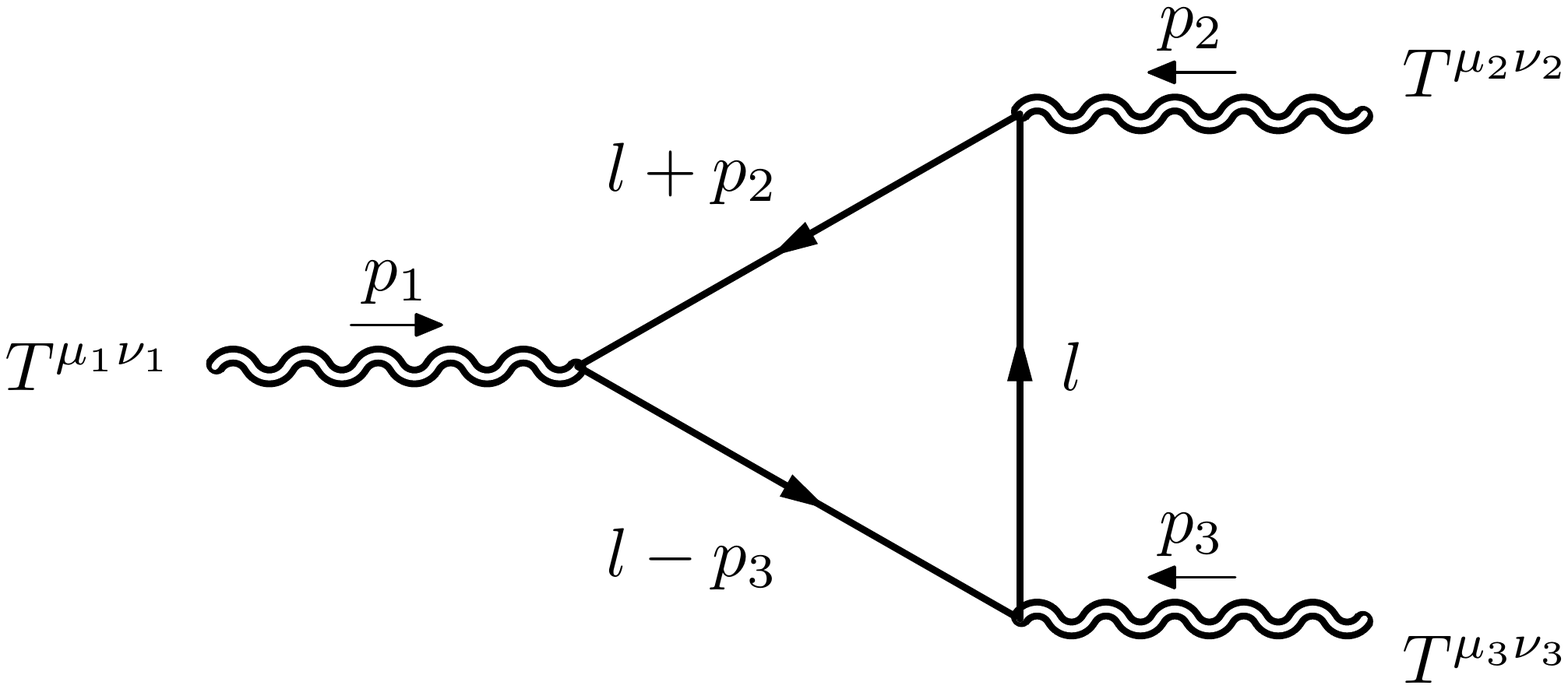}} \hspace{.3cm}
	\subfigure{\includegraphics[scale=0.2]{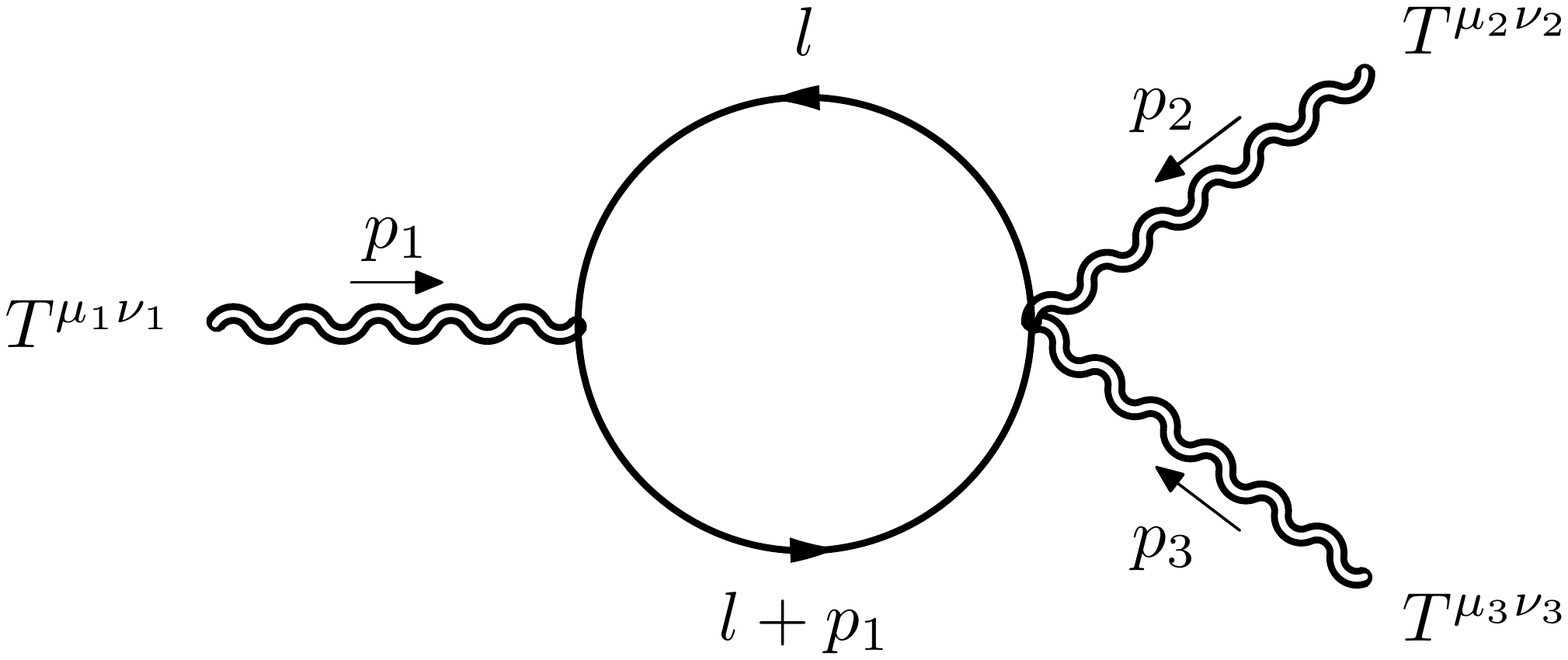}} \hspace{.3cm}\\
	\vspace{-2.5cm}
	\raisebox{.12\height}{\subfigure{\includegraphics[scale=0.16]{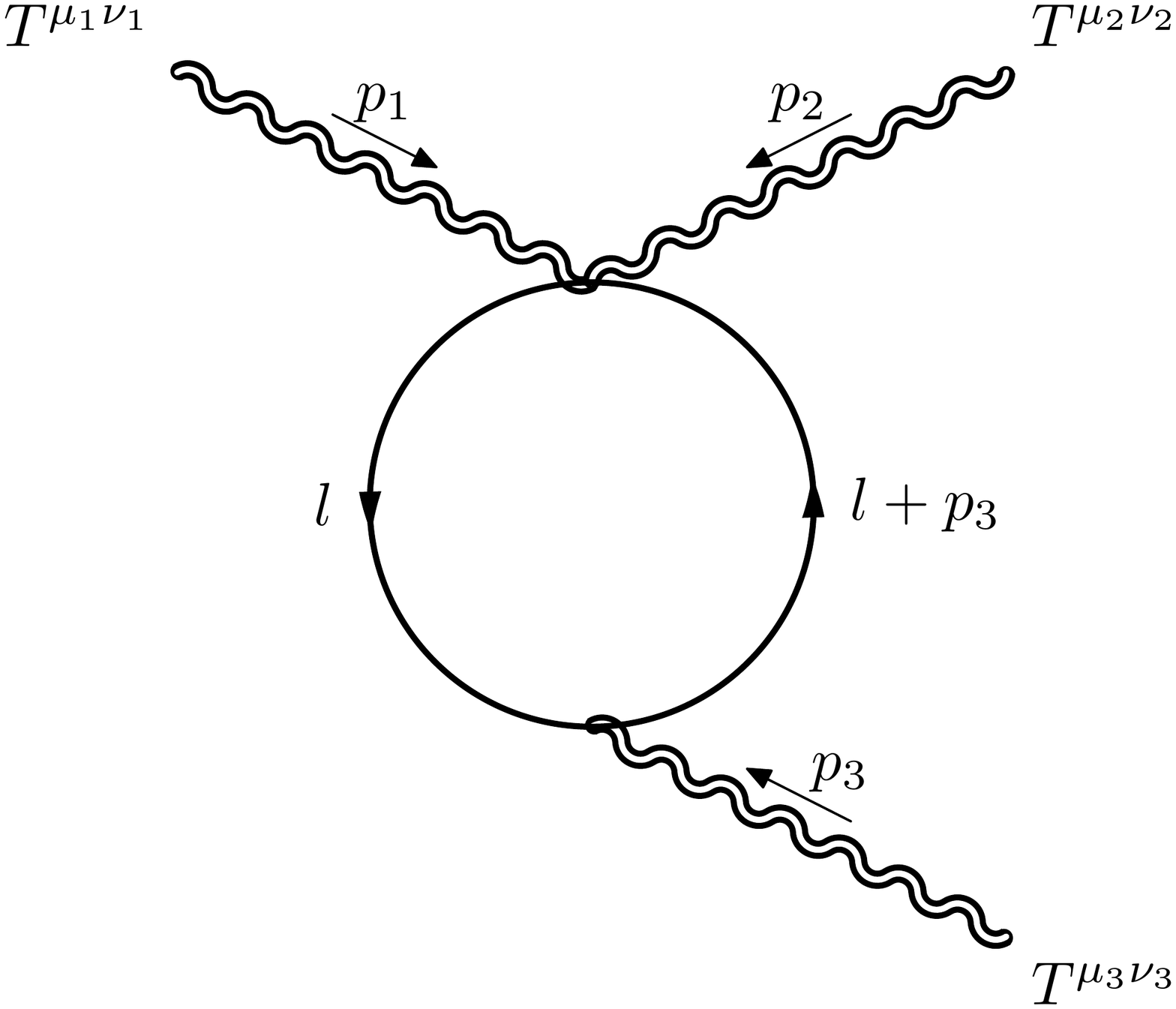}}\hspace{.3cm}}
	\raisebox{.12\height}{\subfigure{\includegraphics[scale=0.16]{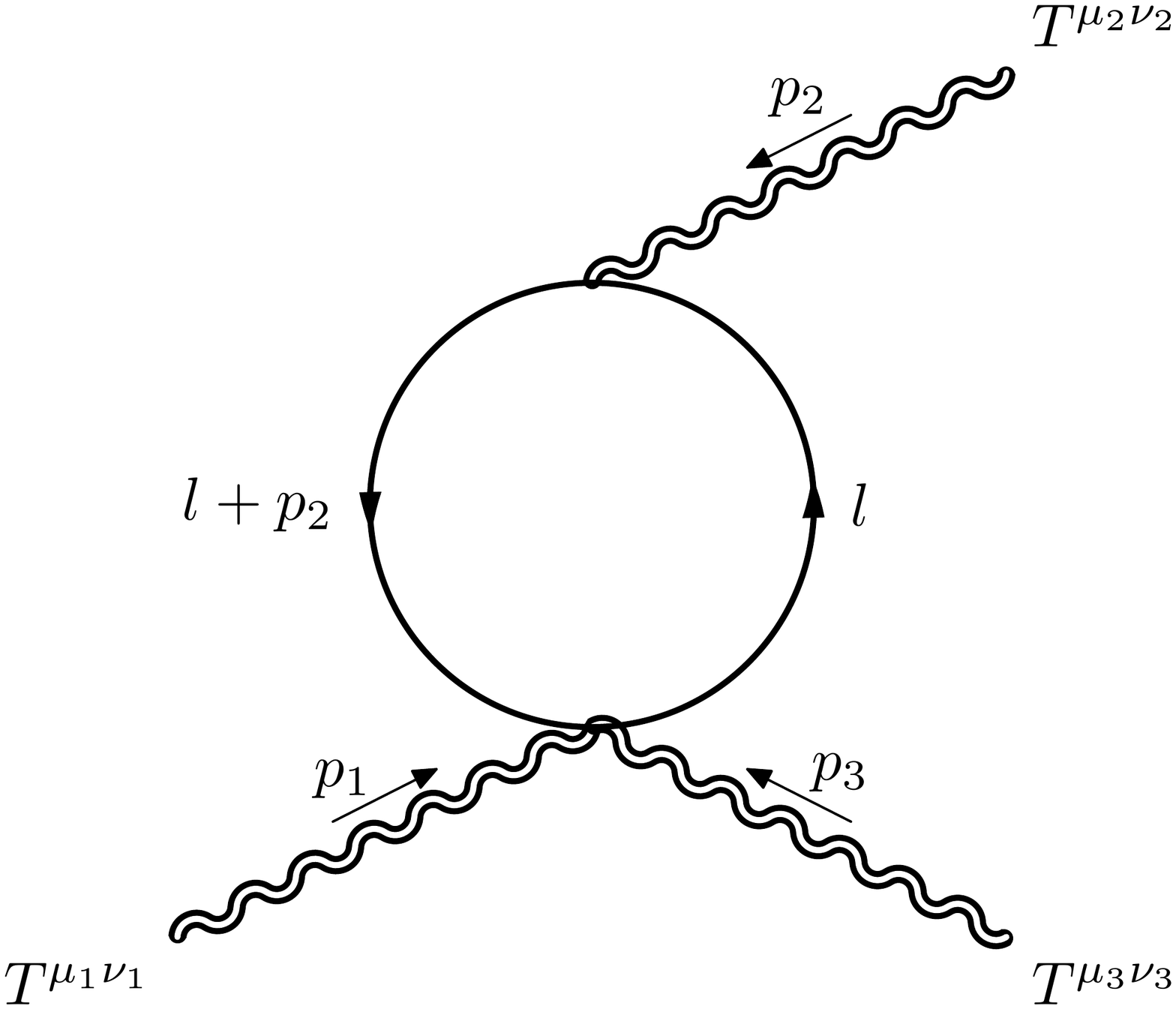}}}
	\vspace{-0.8cm}\caption{One-loop fermion diagrams for the three-graviton vertex.\label{Feynman2}}
\end{figure}
Explicitly the terms in \eqref{FermionExpa} are given by
\begin{align}
V_{F,1}^{\m_1\n_1\m_2\n_2\m_3\n_3}(p_1,p_2,p_3)&=-\int \sdfrac{d^d\ell}{(2\pi)^d}\sdfrac{\Tr\big[V^{\m_1\n_1}_{T\bar\psi\psi}(\ell+p_3,\ell-p_2)(\slashed{\ell}+\slashed{p}_3)V^{\m_3\n_3}_{T\bar\psi\psi}(\ell,\ell+p_3)\slashed{\ell}V^{\m_2\n_2}_{T\bar\psi\psi}(\ell-p_2,\ell)(\slashed{\ell}-\slashed{p}_2)\big]}{\ell^2(\ell-p_2)^2(\ell+p_3)^2}\nn[1ex]
V_{F,2}^{\m_1\n_1\m_2\n_2\m_3\n_3}(p_1,p_2,p_3)&=V_{F,1}^{\m_1\n_1\m_3\n_3\m_2\n_2}(p_1,p_3,p_2)
\end{align}
\begin{align}
W_{F,1}^{\m_1\n_1\m_2\n_2\m_3\n_3}(p_1,p_2,p_3)&=-\int \sdfrac{d^d\ell}{(2\pi)^d}\sdfrac{\Tr\big[V^{\m_1\n_1}_{T\bar\psi\psi}(\ell,\ell+p_1)\ \slashed{\ell}\ V^{\m_2\n_2\m_3\n_3}_{TT\bar\psi\psi}(\ell+p_1,\ell)(\slashed{\ell}+p_1)\big]}{\ell^2(\ell+p_1)^2}\nn[2ex]
W_{F,2}^{\m_1\n_1\m_2\n_2\m_3\n_3}(p_1,p_2,p_3)&=W_{F,1}^{\m_3\n_3\m_1\n_1\m_2\n_2}(p_3,p_1,p_2)\nn[1ex]
W_{F,3}^{\m_1\n_1\m_2\n_2\m_3\n_3}(p_1,p_2,p_3)&=W_{F,1}^{\m_2\n_2\m_1\n_1\m_3\n_3}(p_2,p_1,p_3).
\end{align}
By a direct computation one can verify that the spin part of the two-gravitons and two-fermions vertex does not contribute to the correlation function. \\
Acting with the projectors transverse and traceless $\Pi$ we obtain the form factors $A_i$, $i=1,\dots,5$. For instance, in the fermion sector we obtain
\begin{align}
\braket{t^{\m_1\n_1}(p_1)t^{\m_2\n_2}(p_2)t^{\m_3\n_3}(p_3)}_F&=\,\Pi^{\m_1\n_1}_{\a_1\b_1}(p_1)\Pi^{\m_2\n_2}_{\a_2\b_2}(p_2)\Pi^{\m_3\n_3}_{\a_3\b_3}(p_3)\notag\\
&\times\bigg[ -\sum_{j=1}^2V_{F,j}^{\a_1\b_1\a_2\b_2\a_3\b_3}(p_1,p_2,p_3)+\sum_{j=1}^3W_{F,j}^{\a_1\b_1\a_2\b_2\a_3\b_3}(p_1,p_2,p_3)\bigg]
\end{align}
Also in this case the number of fermion families is kept arbitrary and we will multiply the result by a constant $n_F$ to account for it. It will be essential for matching this contribution to the general non-perturbative one. 

%%%%%%%%%%%%%%%%%%%%%
\section{Comparisons with the conformal solutions in $d=3$ and $d=5$}

\subsection{Normalization of the two point function}
\label{compare}
%%%%%%%%%%%%%%%%%%%%%%%%%%%%%%%%%%%%%%%%%%%%%%%%%%%%
%%%%%%%%%%%%%%%%%%%%%%%%%%%%%%%%%%%%%%%%%%%%%%%%%%%%
In order to investigate the correspondence between the conformal and the perturbative solutions 
we briefly recall the result for the $\braket{TT}$ correlator, that we will need in order to investigate the match between the general conformal solution and its perturbative realization. 
Here we start from a general analysis, based on the CFT solution for this correlator, with a specific application to the case of odd spacetime dimensions, where no renormalization is needed. We will come back to the same correlator in a following section, when we will address the issue of its renormalization in $d=4$.

The $TT$ is fixed by conformal invariance in coordinate space to take the form 
\bea
\label{2PFOsborn}
\langle T^{\mu \nu}(x) \, T^{\alpha \beta} (y)\rangle =
\frac{C_T}{(x-y)^{2d}} \, \mathcal{I}^{\mu\nu ,\alpha\beta}(x-y) \, ,
\eea
with 
\bea \label{Inversion}
\mathcal{I}^{\mu \nu,\alpha \beta} (s) =
I^{\mu\rho}(x-y)I^{\nu\sigma}(x-y) {\epsilon_T}^{\rho\sigma,\alpha\beta} \, ,
\eea
where
\bea
I^{\mu\nu}(x)=
\delta^{\mu\nu} - 2 \frac{x^\mu x^\nu}{x^2}
\eea
and 
\bea\label{epsilon}
{\epsilon_T}^{\mu\nu,\alpha\beta} =
\frac{1}{2} \, (\delta^{\mu\alpha} \delta^{\nu \beta} + \delta^{\mu \beta} \delta^{\nu \alpha} \bigl)
- \frac{1}{d} \, \delta^{\mu \nu} \delta^{\alpha \beta}.
\eea
It is not difficult to check that \eqref{2PFOsborn} can be cast in the form 
\bea
 \langle T^{\mu \nu}(x) \, T^{\alpha \beta} (y)\rangle =
 \frac{C_T}{4 (d-2)^2 d (d+1)} \Delta^{(d) \mu\nu\alpha\beta}(\partial) \frac{1}{((x-y)^2)^{(d-2)}}
\eea
where 
\bea \label{TransverseDeltaCoord}
\hat{\Delta}^{(d)\,\mu\nu\alpha\beta}(\partial)
&=&
\frac{1}{2}\left( \hat{\Theta}^{\mu\alpha} \hat{\Theta}^{\nu\beta} +\hat{ \Theta}^{\mu\beta}\hat{ \Theta}^{\nu\alpha} \right)
- \frac{1}{d-1}\hat{ \Theta^{\mu\nu}} \hat{\Theta}^{\alpha\beta}\, ,
\quad
\text{with}
\quad
\hat{\Theta}^{\mu\nu} = \partial^\mu  \partial^\nu - \delta^{\mu\nu} \, \Box  \nn
\partial_\mu \, \hat{\Delta}^{(d)\,\mu\nu\alpha\beta}(\partial)
&=&
0 \, , \quad
\delta_{\mu\nu} \, \hat{\Delta}^{(d)\,\mu\nu\alpha\beta}(\partial) = 0   \, 
\eea
for any function on which it acts.
Using the representation 
\bea \label{fund}
\frac{1}{(x^2)^\alpha}&=&\equiv C(\alpha) \, \int d^d l \, \frac{e^{i l\cdot x}}{(l^2)^{d/2 - \alpha}} \nonumber \\
C(\alpha) &=&\frac{1}{4^{\alpha}\,\pi^{d/2}} \frac{\Gamma(d/2 - \alpha)}{\Gamma(\alpha)}
\eea
it can be re-expressed in the form 

\bea
\label{firstform}
\braket{T^{\m\n}(p)T^{\alpha\beta}(-p)}&=&\int d^d x e^{i p\cdot x} 
\braket{T^{\m\n}(x)T^{\alpha\beta}(0)}\nn
&=&C_T \frac{ \pi^{d/2} \Gamma(-d/2)}{2^d (d-2)(d+1)\Gamma(d-2) }p^d 
\Pi^{\mu\nu\alpha\beta}(p).  
\eea
Using the expression of the scalar (Euclidean) 2-point function 
\begin{equation}
{B}_0(p_1^2)=\sdfrac{1}{\p^\frac{d}{2}}\int\,d^d \ell\ \frac{l}{\ell^2(\ell-p_1)^2}=\frac{ \left[\G\left(\frac{d}{2}-1\right)\right]^2\G\left(2-\frac{d}{2}\right)}{\G\left(d-2\right)(p_1^2)^{2-\frac{d}{2}}}\label{B0ex}
\end{equation}
which is divergent for $d=2 k $, $k=1,2,3...$, it can also be rewritten in the form 
\be
\braket{T^{\m\n}(p)T^{\alpha\beta}(-p)}=4 C_T \left(\frac{\pi}{4}\right)^{d/2} \frac{1}{(d-2)^2 d (d+1)\Gamma(d/2-1)^2}\Pi^{\mu\nu\alpha\beta}(p)\, p^4  B_0(p^2).
\ee
The singular nature of \eqref{firstform} in even dimensions emerges in DR from the appearance of the
$\Gamma(-d/2)$ factor, which can be regulated by an ordinary shift $d\to d- \epsilon$. After a redefinition of the constant $C_T\to c_T$ which absorbs the $d$-dependent prefactors, it takes the form 
\begin{align}
\label{general}
\braket{T^{\m\n}(p)T^{\alpha\beta}(-p)}=c_T\,\Pi^{\m\n\alpha\beta}(p)\,\G\left(-\frac{d}{2}+\frac{\epsilon}{2}\right)\,p^{d- \epsilon}
\end{align}
where the constant $c_T$ is regular and arbitrary. Using the Lagrangian realization of the $TT$ in terms of the two free field theory sectors available in odd dimensions, it can be written as
\begin{align}
\label{fform}
\braket{T^{\mu_1\nu_1}(p)T^{\mu_2\nu_2}(-p)}&=\frac{\p^\frac{d}{2}\big(n_S+2(d-1)n_F\big)}{4(d-1)(d+1)}\,\Pi^{\mu_1\nu_1\mu_2\nu_2}(p)\,{B}_0(p^2)\,p^d\notag\\
&=\frac{\p^\frac{d}{2}\,\big(n_S+2(d-1)n_F\big)\,d(d-2)}{16(d+1)(d-1)\G\left(d-2\right)}\,\left[\G\left(\frac{d}{2}-1\right)\right]^2\,\Pi^{\mu_1\nu_1\mu_2\nu_2}(p)\,\G\left(-\frac{d}{2}\right)\,p^d
\end{align}
with $c_T$ matched in odd dimensions ($d>1$) according to the expression
\begin{equation}
c_T=\frac{\p^\frac{d}{2}\,\big(n_S+2(d-1)n_F\big)\,d(d-2)}{16(d+1)(d-1)\G\left(d-2\right)}\,\left[\G\left(\frac{d}{2}-1\right)\right]^2.
\end{equation}
In even dimension we have a third (gauge) sector available and therefore it will be necessary to extend \eqref{fform} in order to perform a complete matching. We will address its renormalization in Section \ref{renorm}.\\
In the case of $d=3$ and $d=5$ we get
\begin{equation}
c_T\ \ \mathrel{\stackrel{\makebox[0pt]{\mbox{\normalfont\tiny $d=3$}}}{=}}\ \  \frac{3\,\p^\frac{3}{2}\,\big(n_S+4 n_F\big)}{32(3+1)}\,\left[\G\left(\frac{1}{2}\right)\right]^2=\frac{3\p^\frac{5}{2}}{128}\,\big(n_S+4n_F\big)\,
\end{equation}

\begin{equation}
c_T\ \ \mathrel{\stackrel{\makebox[0pt]{\mbox{\normalfont\tiny $d=5$}}}{=}}\ \  \frac{15\,\p^\frac{5}{2}\,\big(n_S+2(5-1)n_F\big)}{64(5+1)\G\left(5-2\right)}\,\left[\G\left(\frac{3}{2}\right)\right]^2=\frac{5\p^\frac{7}{2}}{1024}\,\big(n_S+8n_F\big)\,
\end{equation}
We will be using these two expressions of $c_T$ in order to perform a comparison with the result of the transverse traceless $A_i$ given in \cite{Bzowski:2013sza} for odd dimensions. In such specific cases there are simplifications both from the exact and the perturbative solutions. In particular, the exact solutions turn into rational functions of the momenta, and, as we are going to show, they can be matched with the perturbative ones that we present below. \\
 In $d=5$, as for all odd dimensions larger than 3, the general solution involves 3 independent constant, as we have mentioned, and we are short of 1 sector in order to match the general result. Nevertheless it is still possible to perform a matching between the two solutions, even though not in the most general case. \\
 Clearly, in this case the perturbative results given below for $d=5$ correspond to a specific choice of the 3 constants of the solution of the conformal constraints. This, obviously, leaves open the issue whether arbitrary choices of all the three constants which appear in odd spacetime dimensions in the general solution correspond to a unitary theory or not, or whether it is possible to formulate, for odd values of values of $d > 3$, CFT's which do not have a free field theory realization. These may correspond to interacting CFT's.

\subsection{Explicit results}
\label{d35}
%%%%%%%%%%%%%%%%%%%%%%%%%%

\subsection{$d=3$ case}
In $d=3$ the scalar integrals ${B}_0$ and ${C}_0$ can be computed in a very simple way. In fact in $d=3$ we get
\begin{equation}
{B}_0(p_1^2)=\frac{\p^{3/2}}{\,p_1}\label{B0}
\end{equation}
where $p_1=|p_1|=\sqrt{p_1^2}$, and analogous relations hold for $p_2$ and $p_3$. The explicit expression of ${C}_0$ can be obtained using the star-triangle relation for which
\begin{equation}
\int\frac{d^{d}x}{[(x-x_1)^2]^{\a_1}\,[(x-x_2)^2]^{\a_2}\,[(x-x_3)^2]^{\a_3}}\ \  \mathrel{\stackrel{\makebox[0pt]{\mbox{\normalfont\tiny $\sum_i\a_i=d$}}}{=}}\ \  \frac{i\pi^{d/2}\n(\a_1)\n(\a_2)\n(\a_3)}{[(x_2-x_3)^2]^{\frac{d}{2}-\a_1}\,[(x_1-x_2)^2]^{\frac{d}{2}-\a_3}\,[(x_1-x_3)^2]^{\frac{d}{2}-\a_2}}\label{startriangle}
\end{equation}
where
\begin{equation}
\n(x)=\frac{\G\left(\frac{d}{2}-x\right)}{\G(x)}
\end{equation}
that holds only if the condition $\sum_i\a_i=d$ is satisfied. In the case $d=3$ the LHS of \eqref{startriangle} is proportional to the three point scalar integral, and in particular
\begin{align}
{C}_0(p_1^2,p_2^2,p_3^2)&=\int\,\frac{d^d\ell}{\p^\frac{d}{2}}\frac{1}{\ell^2(\ell-p_2)^2(\ell+p_3)^2}=\int\,\frac{d^dk}{\p^\frac{d}{2}}\frac{1}{(k-p_1)^2(k+p_3)^2(k+p_3-p_2)^2}\notag\\[2ex]
&=\frac{\left[\G\left(\frac{d}{2}-1\right)\right]^3}{\,(p_1^2)^{\frac{D}{2}-1}(p_2^2)^{\frac{d}{2}-1}(p_3^2)^{\frac{d}{2}-1}}\ \mathrel{\stackrel{\makebox[0pt]{\mbox{\normalfont\tiny $d=3$}}}{=}}\ \frac{\p^{3/2}}{\,p_1\,p_2\,p_3}.\label{C0}
\end{align}

The explicit expression of the form factors in $d=3$, using the perturbative approach to one loop order, can be obtained by taking the limit $d\to3$ of \eqref{B0} and \eqref{C0} derived from the general diagrammatic expansion. We obtain
\begin{align}
A_{1}^{d=3}(p_1,p_2,p_3)&=\frac{\p^3(n_S-4n_F)}{60(p_1+p_2+p_3)^6}\Big[p_1^3+6p_1^2(p_3+p_2)+(6p_1+p_2+p_3)\big((p_2+p_3)^2+3 p_2 p_3\big)\Big]
\end{align}
\begin{align}
A_{2}^{d=3}(p_1,p_2,p_3)&=\frac{\p^3(n_S-4n_F)}{60(p_1+p_2+p_3)^6}\Big[4p_3^2\big(7(p_1+p_2)^2+6p_1 p_2\big)+20p_3^3(p_1+p_2)+4p_3^4\notag\\
&\hspace{4.4cm}+3(5p_3+p_1+p_2)(p_1+p_2)\big((p_1+p_2)^2+p_1 p_2\big)\Big]\notag\\
&+\frac{\p^3\,n_F}{3(p_1+p_2+p_3)^4}\Big[p_1^3+4p_1^2(p_2+p_3)+(4p_1+p_2+p_3)\big((p_2+p_3)^2+p_2 p_3\big)\Big]\nn
\end{align}
\begin{align}
A_{3}^{d=3}(p_1,p_2,p_3)&=\frac{\p^3(n_S-4n_F)\,p_3^2}{240(p_1+p_2+p_3)^4}\Big[28p_3^2(p_1+p_2)+3p_3\big(11(p_1+p_2)^2+6 p_1\,p_2\big)+7p_3^3\notag\\
&+12(p_1+p_2)\big((p_1+p_2)^2+p_1p_2\big)\Big]\notag\\
&+\frac{\p^3n_F\,p_3^2}{6(p_1+p_2+p_3)^3}\Big[3p_2(p_1+p_2)+2\big((p_1+p_2)^2+p_1p_2\big)+p_3^2\Big]\notag\\
&-\frac{\p^3(n_s+4n_F)}{16(p_1+p_2+p_3)^2}\Big[p_1^3+2p_1^2(p_2+p_3)+(2p_1+p_2+p_3)\big((p_2+p_3)^2-p_2 p_3\big)\Big]
\end{align}
\begin{align}
A_{4}^{d=3}(p_1,p_2,p_3)&=\frac{\p^3(n_S-4n_F)}{120(p_1+p_2+p_3)^4}\Big[(4p_3+p_1+p_2)\big(3(p_1+p_2)^4-3(p_1+p_2)^2p_1p_2+4p_1^2p_2^2\big)\notag\\
&+9p_3^2(p_1+p_2)\big((p_1+p_2)^2-3p_1 p_2\big)-3p_3^5-12p_3^4(p_1+p_2)-9p_3^3\big((p_1+p_2)^2+2p_1 p_2\big)\Big]\notag\\
&+\frac{\p^3\,n_F}{6(p_1+p_2+p_3)^3}\Big[(p_1+p_2)\big((p_1+p_2)^2-p_1 p_2\big)(p_1+p_2+3p_3)-p_3^4-3p_3^3(p_1+p_2)\notag\\
&-6p_1p_2p_3^2\Big]-\frac{\p^3(n_s+4n_F)}{8(p_1+p_2+p_3)^2}\Big[p_1^3+2p_1^2(p_2+p_3)+(2p_1+p_2+p_3)\big((p_2+p_3)^2-p_2 p_3\big)\Big]\nn
\end{align}
\begin{align}
A_{5}^{d=3}(p_1,p_2,p_3)&=\frac{\p^3(n_S-4n_F)}{240(p_1+p_2+p_3)^3}\Big[-3(p_1+p_2+p_3)^6+9(p_1+p_2+p_3)^4(p_1 p_2+p_2 p_3+p_1 p_3)\notag\\
&+12(p_1+p_2+p_3)^2(p_1 p_2+p_2p_3+p_3 p_1)^2-33(p_1+p_2+p_3)^2p_1p_2p_3\notag\\
&+12(p_1+p_2+p_3)(p_1p_2+p_2p_3+p_1p_3)p_1p_2p_3+8p_1^2p_2^2p_3^2\Big]\notag\\
&+\frac{\p^3n_F}{12(p_1+p_2+p_3)^2}\Big[-(p_1+p_2+p_3)^5+3(p_1+p_2+p_3)^3(p_1p_2+p_2p_3+p_1p_3)\notag\\
&+4(p_1+p_2+p_3)(p_1p_2+p_2p_3+p_1p_3)^2-11(p_1+p_2+p_3)^2p_1p_2p_3\notag\\
&+4(p_1p_2+p_2p_3+p_1p_3)p_1p_2p_3\Big]-\frac{\p^3(n_S+4n_F)}{16}\Big[p_1^3+p_2^3+p_3^3\Big]
\end{align}

This is in agreement with the expression given in \cite{2014JHEP...03..111B} in terms of he constant $\a_1,\a_2$ and $c_T$ if we choose 
(see \cite{2014JHEP...03..111B})
\bea
\a_1=\frac{\p^3(n_S-4n_F)}{480}, \qquad \a_2=\frac{\p^3\,n_F}{6}, \qquad c_T=\frac{3\p^{5/2}}{128}(n_S+4n_F), \qquad c_g=0 
\eea
Notice that $c_g$ is a constant appearing in \cite{2014JHEP...03..111B} related to the possibility of having a nonzero functional variation of the stress energy tensor respect to the metric ($\sim \delta T^{\mu\nu}(x)/\delta g_{\alpha\beta}(y))$ which is an extra contact term not included in our discussion.

%%%%%%%%%%%%%%%%%%%%%%%%%%
\subsection{$d=5$ case}
%%%%%%%%%%%%%%%%%%%%%%%%%%
In this case we have
\begin{equation}
{C}_0(p_1^2,p_2^2,p_3^2)=\frac{\p^{3/2}}{p_1+p_2+p_3}.
\end{equation}
From \eqref{B0ex} the ${B}_0$ is calculated in $d=5$ as
\begin{equation}
{B}_0(p_1^2)=-\frac{\p^{3/2}}{4} p_1.
\end{equation}
In the $d\to 5$ limit the $A_1$ form factor becomes, for instance, 
\begin{align}
A_{1}^{d=5}(p_1,p_2,p_3)&=\frac{\p^4(n_S-4n_F)}{560(p_1+p_2+p_3)^7}\Big[(p_1+p_2+p_3)^2\big((p_1+p_2+p_3)^4+(p_1+p_2+p_3)^2(p_1p_2+p_2p_3+p_1p_3)\notag\\
&\hspace{-1.9cm}+(p_1p_2+p_2p_3+p_1p_3)^2\big)+(p_1+p_2+p_3)\big((p_1+p_2+p_3)^2+5(p_1p_2+p_2p_3+p_1p_3)\big)p_1p_2p_3+10p_1^2p_2^2p_3^2\Big].
\end{align}
The remaining form factors are given in Appendix \ref{aaa5}. 
Their expressions are in agreement with those given in \cite{2014JHEP...03..111B} when the 
corresponding constants (denoted by $\a_1$ and $\a_2$) are matched by the relations
\bea
\a_1=\frac{\p^4(n_S-4n_F)}{560 \times 72}, \qquad \a_2=\frac{\p^4\,n_F}{240}, \qquad c_T=\frac{5\p^{7/2}}{1024}(n_S+8n_F).
\eea
The case that we have analysed and their correspondence shows that we can safely move to $d=4$. In this case we will not 
attempt a comparison with the results of \cite{2014JHEP...03..111B} which are far more involved and 
require the implementation of some recursion relations on the renormalized 3K integrals. In our case we will have to specialize our computation to $d=4$ with the inclusion of a third free field theory sector and extract the $A_i$'s after addressing their renormalization.

%%%%%%%%%%%%%%%%%%%%%%%%%%%%%%%%%%%%%%%%%%%%%%%%%%%%
%%%%%%%%%%%%%%%%%%%%%%%%%%%%%%%%%%%%%%%%%%%%%%%%%%%%
%%%%%%%%%%%%%%%%%%%%%%%%%%%%%
%%%%%%%%%%%%%%%%%%%%%%%%%%%%%
\section{The correlator in $d=4$ and the trace anomaly}
%%%%%%%%%%%%%%%%%%%%%%%%%%%%%
%%%%%%%%%%%%%%%%%%%%%%%%%%%%%
\begin{figure}[t]
\centering
\vspace{-2cm}
\subfigure{\includegraphics[scale=0.2]{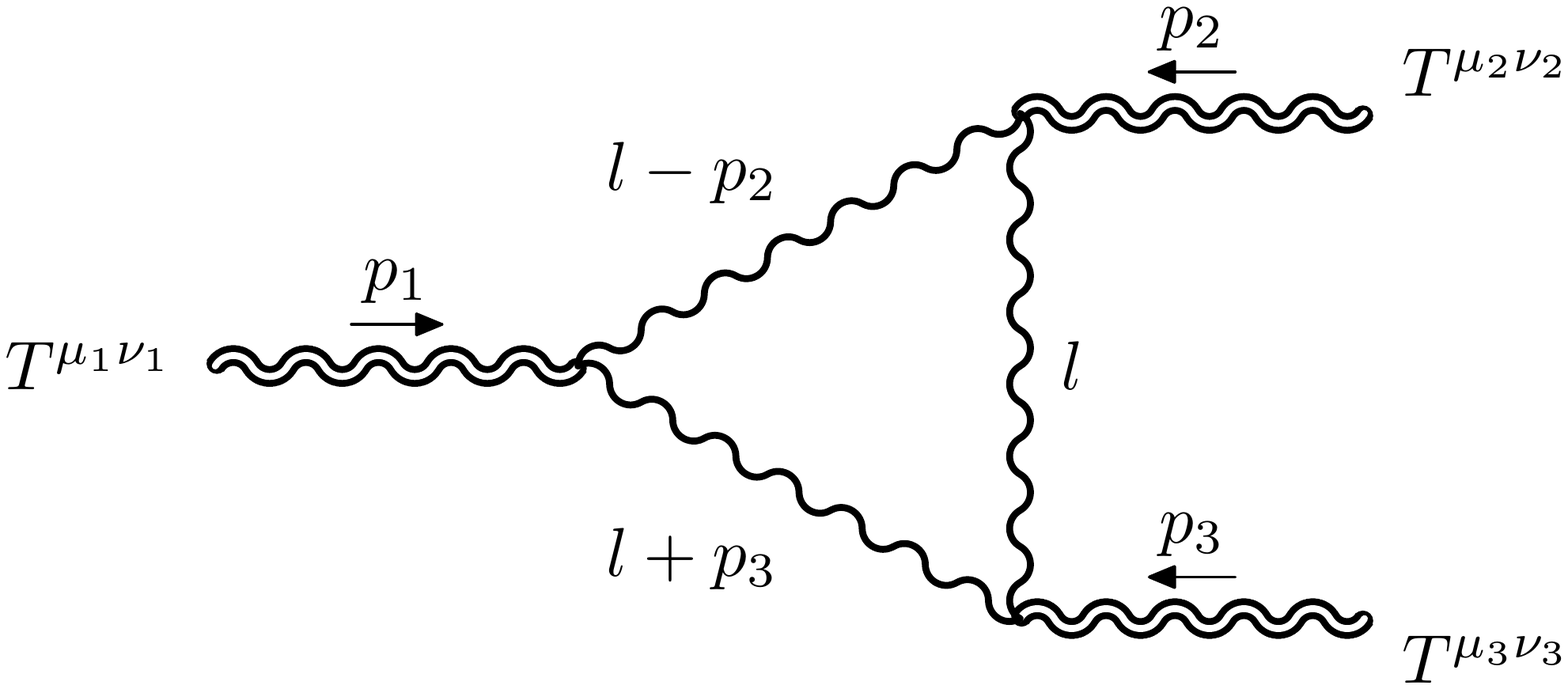}} \hspace{.3cm}
\subfigure{\includegraphics[scale=0.2]{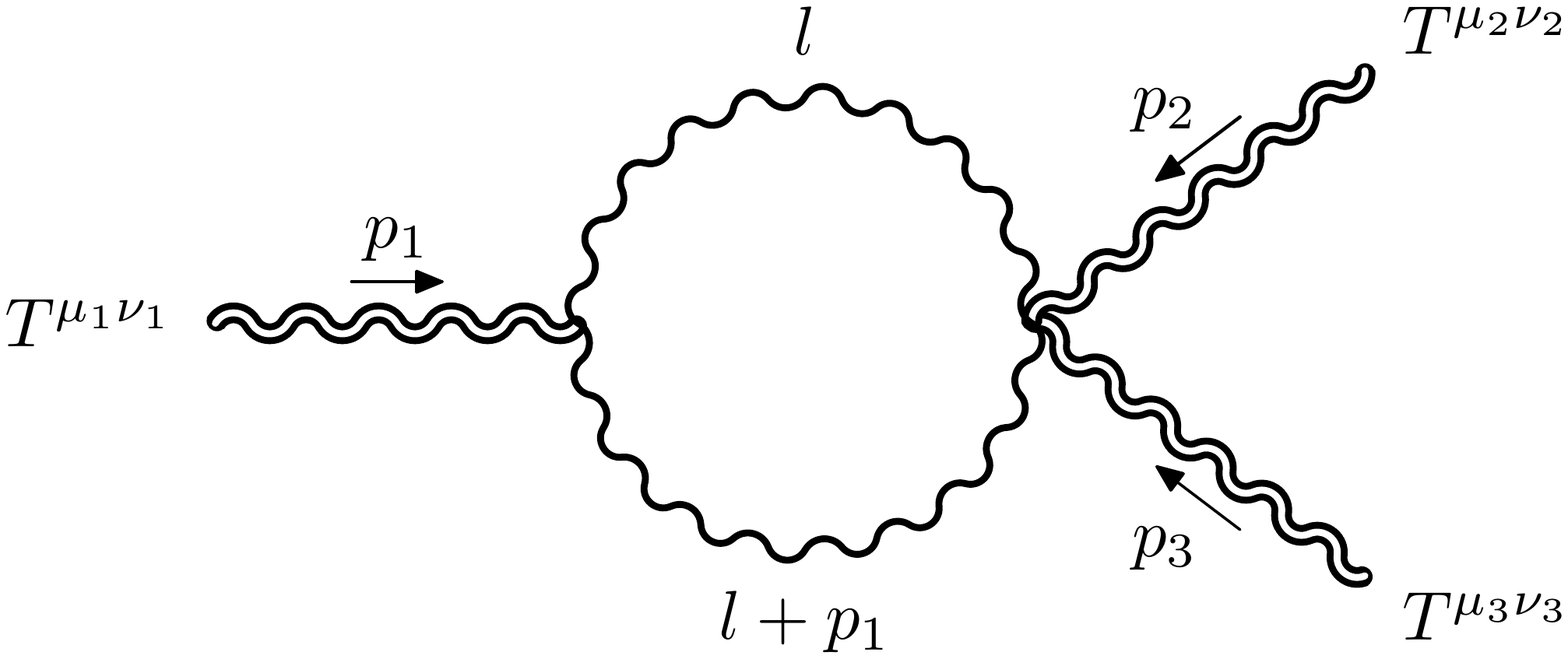}} \hspace{.3cm}
\raisebox{.12\height}{\subfigure{\includegraphics[scale=0.16]{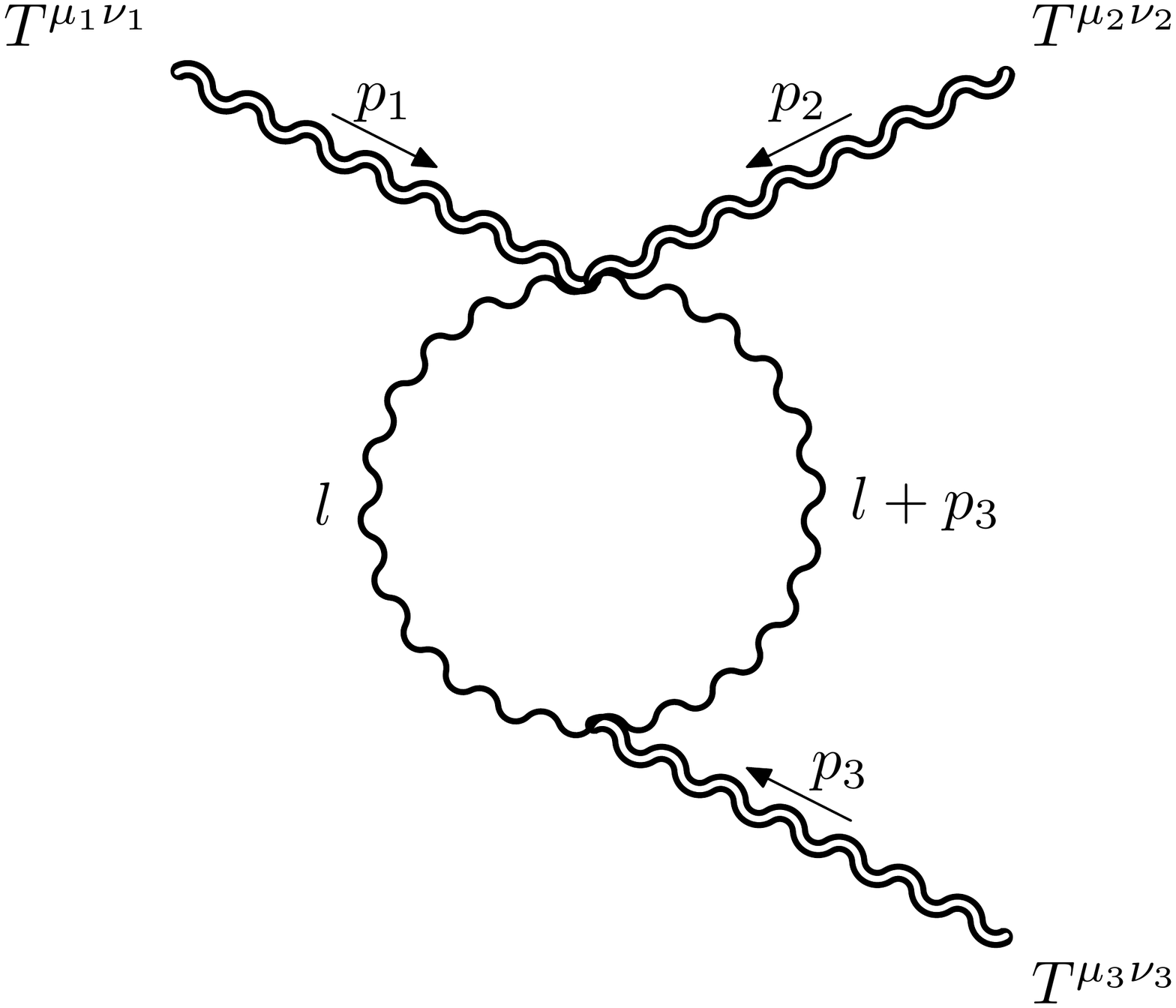}}\hspace{.3cm}}
\raisebox{.12\height}{\subfigure{\includegraphics[scale=0.16]{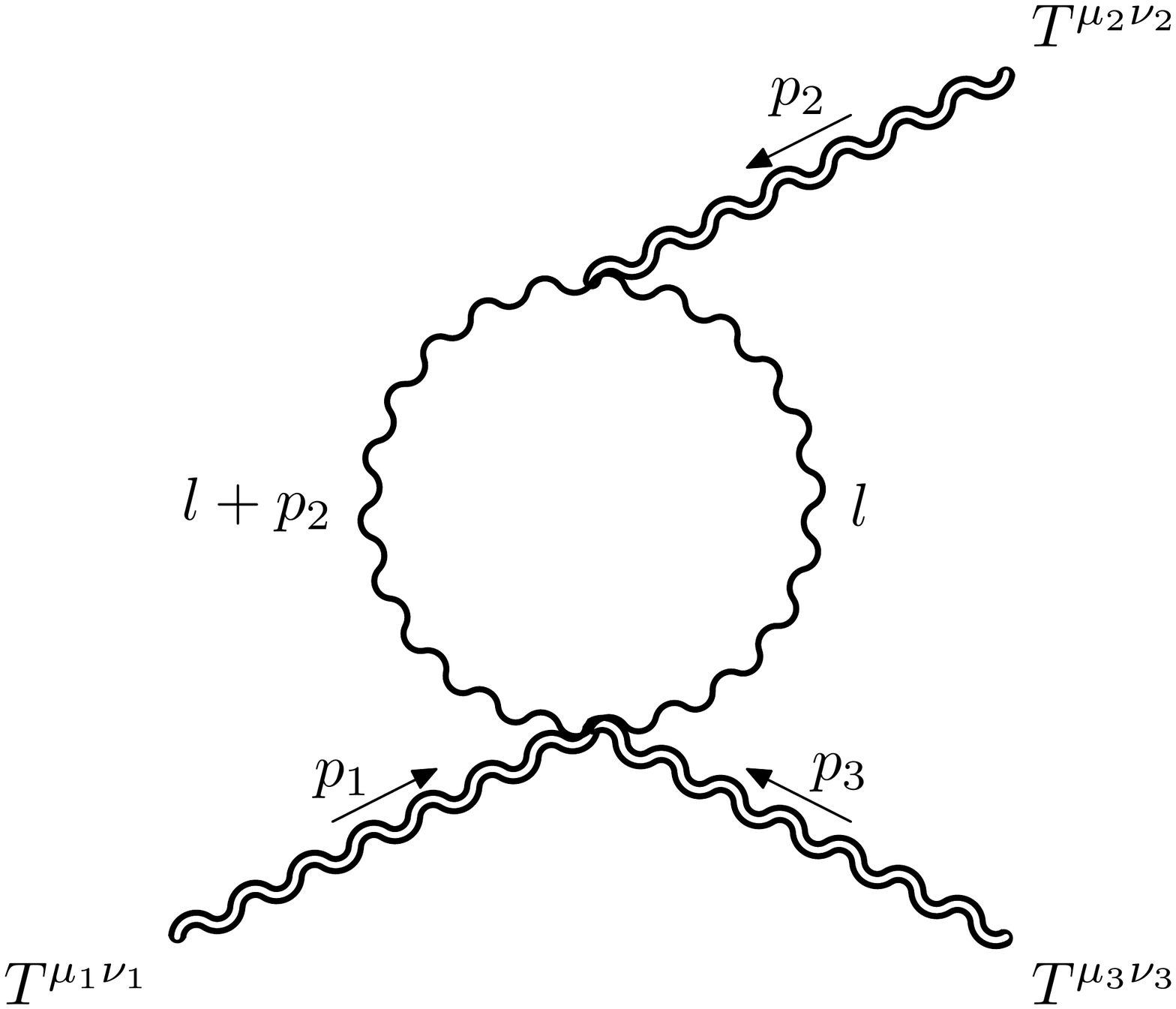}}}
\vspace{-0.8cm}\caption{One-loop gauge diagrams for the three-graviton vertex.\label{Feynman3}}
\end{figure}
%%%%%%%%%%%%%%%%%%%%%%%%%%%%%
\subsection{Gauge and Ghost sectors}
%%%%%%%%%%%%%%%%%%%%%%%%%%%%%

We have to consider, as already mentioned, the contributions coming from the spin-1 sector, and in the one-loop approximation they correspond to the diagrams in \figref{Feynman3}. We have also to consider the contributions from the ghost, which can be calculated from the same type of diagrams given in \figref{Feynman3} but now with a ghost field running in the loop. A direct computation shows that the ghost and the gauge fixing contributions cancel.
Therefore we calculate in the one-loop approximation the contribution to the correlation function of the gauge sector, given by the diagrams in \figref{Feynman3}. These contributions can be written as 
\begin{align}
\braket{T^{\m_1\n_1}(p_1)T^{\m_2\n_2}(p_2)T^{\m_3\n_3}(p_3)}_G=\, -V_{G}^{\m_1\n_1\m_2\n_2\m_3\n_3}(p_1,p_2,p_3)+\sum_{i=1}^3W_{G,i}^{\m_1\n_1\m_2\n_2\m_3\n_3}(p_1,p_2,p_3)\label{GaugeExpa}
\end{align}
for the triangle and the bubble topologies respectively. They are given by
\begin{align}
&V_{G}^{\m_1\n_1\m_2\n_2\m_3\n_3}(p_1,p_2,p_3)=\notag\\
&\qquad=\int \sdfrac{d^d\ell}{(2\pi)^d}\sdfrac{V^{\m_1\n_1\a_1\b_1}_{TAA}(\ell+p_3,\ell-p_2)\,\d_{\a_1\b_2}\,V^{\m_3\n_3\a_2\b_2}_{TAA}(\ell,\ell+p_3)\,\d_{\a_2\b_3}\,V^{\m_2\n_2\a_3\b_3}_{TAA}(\ell-p_2,\ell)\d_{\a_3\b_1}}{\ell^2(\ell-p_2)^2(\ell+p_3)^2}
\end{align}
\begin{align}
W_{G,1}^{\m_1\n_1\m_2\n_2\m_3\n_3}(p_1,p_2,p_3)&=\frac{1}{2}\int \sdfrac{d^d\ell}{(2\pi)^d}\sdfrac{V^{\m_1\n_1\a_1\b_1}_{TAA}(\ell,\ell+p_1)\,\d_{\b_1\a_2}\,V^{\m_2\n_2\m_3\n_3\a_2\b_2}_{TTAA}(\ell+p_1,\ell)\,\d_{\a_1\b_2}}{\ell^2(\ell+p_1)^2}\\[2ex]
W_{G,2}^{\m_1\n_1\m_2\n_2\m_3\n_3}(p_1,p_2,p_3)&=W_{G,1}^{\m_3\n_3\m_1\n_1\m_2\n_2}(p_3,p_1,p_2)\\[1ex]
W_{G,3}^{\m_1\n_1\m_2\n_2\m_3\n_3}(p_1,p_2,p_3)&=W_{G,1}^{\m_2\n_2\m_1\n_1\m_3\n_3}(p_2,p_1,p_3).
\end{align}
One can show that the spin part of the two-graviton/two-fermion vertex does not contribute to the correlation function. \\
By acting with the transverse-traceless $\Pi$ projectors, we obtain the form factors $A_i$, $i=1,\dots,5$ in the fermion sector, in particular
\begin{align}
\braket{t^{\m_1\n_1}(p_1)t^{\m_2\n_2}(p_2)t^{\m_3\n_3}(p_3)}_G&=\,\Pi^{\m_1\n_1}_{\a_1\b_1}(p_1)\Pi^{\m_2\n_2}_{\a_2\b_2}(p_2)\Pi^{\m_3\n_3}_{\a_3\b_3}(p_3)\notag\\
&\times\bigg[ -V_{G}^{\a_1\b_1\a_2\b_2\a_3\b_3}(p_1,p_2,p_3)+\sum_{i=1}^3W_{G,i}^{\a_1\b_1\a_2\b_2\a_3\b_3}(p_1,p_2,p_3)\bigg]
\end{align}
Also in this case the number of gauge fields are kept arbitrary by the inclusion of an overall factor $n_G$.
%%%%%%%%%%%%%%%%%%%
\subsection{Divergences}
%%%%%%%%%%%%%%%%%%%

In $d=4$ the complete correlation function can be written as
\begin{equation}
\braket{T^{\m_1\n_1}(p_1)T^{\m_2\n_2}(p_2)T^{\m_3\n_3}(p_3)}=\sum_{I=F,G,S}\,n_I\,\braket{T^{\m_1\n_1}(p_1)T^{\m_2\n_2}(p_2)T^{\m_3\n_3}(p_3)}_I
\end{equation}
also valid for the transverse traceless part of the correlator. In this case we encounter divergenes in the forms of single poles in $1/\epsilon$  ($\epsilon=(4-d)/2$). In this section we discuss the structures of such divergences and their elimination in DR using the two usual gravitational counterterms.\\
As a first remark, it is easy to realize for dimensional reasons and power counting that $A_{1}$ is UV finite. All other form factors have divergent parts explicitly given as
\begin{subequations}
\begin{align}
A_2^{Div}&=\frac{\p^2}{45\,\varepsilon}\,\big[26n_G-7n_F-2n_S\big]\\[1ex]
A_3^{Div}&=\frac{\p^2}{90\,\varepsilon}\,\big[3(s+s_1)\big(6n_F+n_S+12n_G\big)+s_2(11n_F+62n_G+n_S)\big]\\[1ex]
A_4^{Div}&=\frac{\p^2}{90\,\varepsilon}\,\big[(s+s_1)\big(29n_F+98n_G+4n_S\big)+s_2(43n_F+46n_G+8n_S)\big]\\[1ex]
A_5^{Div}&=\frac{\p^2}{180\,\varepsilon}\bigg\{n_F \left(43 s^2-14 s (s_1+s_2)+43 s_1^2-14 s_1 s_2+43 s_2^2\right)\notag\\
&\quad+2 \big[n_G \left(23 s^2+26 s (s_1+s_2)+23 s_1^2+26 s_1 s_2+23s_2^2\right)+2n_S\left(2 s^2-s (s_1+s_2)+2 s_1^2-s_1 s_2+2 s_2^2\right)\big]\bigg\}
\end{align}\label{diverg}
\end{subequations}
and at this point we can proceed with their renormalization. 
\section{Renormalization of the $TTT$}
\label{renorm}
The renormalization of the 3-graviton vertex is obtained by the addition of 2 counterterms in the defining Lagrangian. In perturbation theory the one loop counterterm Lagrangian is
\begin{equation}
S_{count}=-\sdfrac{1}{\varepsilon}\,\sum_{I=F,S,G}\,n_I\,\int d^dx\,\sqrt{-g}\bigg(\,\b_a(I)\,C^2+\b_b(I)\,E\bigg)
\label{scount}
\end{equation}
corresponding to the Weyl tensor squared and the Euler density, omitting the extra 
$R^2$ operator which is responsible for the $\square R$ term in \eqref{TraceAnomaly}, having choosen the local part of anomaly 
$(\sim \beta_c\square R)$ vanishing ($\beta_c=0$). We refer to \cite{Coriano:2012wp} for a more detailed discussion of this point and of the finite renormalization needed to get from the general $\beta_c\neq 0$ to the $\beta_c=0$ case. The corresponding vertex counterterms are
\begin{align}
&\braket{T^{\m_1\n_1}(p_1)T^{\m_2\n_2}(p_2)T^{\m_3\n_3}(p_3)}_{count}=\notag\\
&\hspace{2cm}=-\sdfrac{1}{\varepsilon}\sum_{I=F,S,G}n_I\bigg(\b_a(I)\,V_{C^2}^{\m_1\n_1\m_2\n_2\m_3\n_3}(p_1,p_2,p_3)+\b_b(I)\,V_{E}^{\m_1\n_1\m_2\n_2\m_3\n_3}(p_1,p_2,p_3)\bigg)
\end{align} 
where
\begin{align}
V_{C^2}^{\m_1\n_1\m_2\n_2\m_3\n_3}(p_1,p_2,p_3)&=8\int\,d^dx_1\,\,d^dx_2\,\,d^dx_3\,\,d^dx\,\bigg(\sdfrac{\d^3(\sqrt{-g}C^2)(x)}{\d g_{\m_1\n_1}(x_1)\d g_{\m_2\n_2}(x_2)\d g_{\m_3\n_3}(x_3)}\bigg)_{flat}\,e^{-i(p_1\,x_1+p_2\,x_2+p_3\,x_3)}\notag\\
&\equiv 8\big[\sqrt{-g}\,C^2\big]^{\m_1\n_1\m_2\n_2\m_3\n_3}(p_1,p_2,p_3)\\[2ex]
V_{E}^{\m_1\n_1\m_2\n_2\m_3\n_3}(p_1,p_2,p_3)&=8\int\,d^dx_1\,\,d^dx_2\,\,d^dx_3\,\,d^dx\,\bigg(\sdfrac{\d^3(\sqrt{-g}E)(x)}{\d g_{\m_1\n_1}(x_1)\d g_{\m_2\n_2}(x_2)\d g_{\m_3\n_3}(x_3)}\bigg)_{flat}\,e^{-i(p_1\,x_1+p_2\,x_2+p_3\,x_3)}\notag\\
&\equiv 8\big[\sqrt{-g}\,E\big]^{\m_1\n_1\m_2\n_2\m_3\n_3}(p_1,p_2,p_3).\label{count}
\end{align}

These vertices satisfy the relations
\begin{align}
\d_{\mu_1\nu_1}\,V_{C^2}^{\mu_1\nu_1\mu_2\nu_2\mu_3\nu_3}(p_1,p_2,p_3)&=4(d-4)\big[C^2\big]^{\mu_2\nu_2\mu_3\nu_3}(p_2,p_3)\notag\\
&-8\,\bigg([C^2]^{\mu_2\nu_2\mu_3\nu_3}(p_1+p_2,p_3)+[C^2]^{\mu_2\nu_2\mu_3\nu_3}(p_2,p_1+p_3)\bigg)\\[2ex]
\d_{\mu_1\nu_1}\,V_{E}^{\mu_1\nu_1\mu_2\nu_2\mu_3\nu_3}(p_1,p_2,p_3)&=4(d-4)\big[E\big]^{\mu_2\nu_2\mu_3\nu_3}(p_2,p_3)\\[2ex]
p_{1\mu_1}\,V_{C^2}^{\mu_1\nu_1\mu_2\nu_2\mu_3\nu_3}(p_1,p_2,p_3)&=-4\,\bigg(p_2^{\nu_1}[C^2]^{\mu_2\nu_2\mu_3\nu_3}(p_1+p_2,p_3)+p_3^{\nu_1}[C^2]^{\mu_2\nu_2\mu_3\nu_3}(p_2,p_1+p_3)\bigg)\notag\\
&\hspace{-0.7cm}+4\,p_{2\a}\bigg(\d^{\mu_2\nu_1}[C^2]^{\a\nu_2\mu_3\nu_3}(p_1+p_2,p_3)+\d^{\nu_2\nu_1}[C^2]^{\a\mu_2\mu_3\nu_3}(p_1+p_2,p_3)\bigg)\notag\\
&\hspace{-0.7cm}+4\,p_{3\a}\bigg(\d^{\mu_3\nu_1}[C^2]^{\mu_2\nu_2\a\nu_3}(p_2,p_1+p_3)+\d^{\nu_3\nu_1}[C^2]^{\mu_2\mu_2\mu_3\a}(p_2,p_1+p_3)\bigg)\\[2ex]
p_{1\mu_1}\,V_{E}^{\mu_1\nu_1\mu_2\nu_2\mu_3\nu_3}(p_1,p_2,p_3)&=0.
\end{align}

\section{Divergences of the two-point function: a worked out example}
%%%%%%%%%%%%%%%%%%%%%%%%%%%%%%%%%%
Before coming to a discussion of the $TTT$, in this section we illustrate 
in some detail the way the generation of the extra tensor structures for such correlators takes place after renormalization. 
We will work out some of the intermediate steps first of the $TT$, for simplicity, presenting enough details which will be then applied to the $TTT$. 

 We start by extending the analysis of Section \ref{compare} in the perturbative sector by including all the three sectors (scalar, fermion and gauge) in $d$ dimensions. This choice obviously violates conformal symmetry since the spin 1 contribution is not conformally invariant and it is responsible for an extra trace term proportional to $n_G$. One obtains
\begin{align}
\braket{T^{\mu_1\nu_1}(p)T^{\mu_2\nu_2}(-p)}&=-\frac{\p^2\,p^4}{4(d-1)(d+1)}\,B_0(p^2)\,\P^{\mu_1\nu_1\mu_2\nu_2}(p)\Big[2(d-1)n_F+(2d^2-3d-8)n_G+n_S\Big]\notag\\
&\hspace{1cm}+\frac{\p^2\,p^4\,n_G}{8(d-1)^2}(d-4)^2(d-2)\p^{\mu_1\nu_1}(p)\p^{\mu_2\nu_2}(p)\,B_0(p^2)\label{TTddim}
\end{align}
with a second contributon proportional to $n_G$. This term vanishes in $d=4$, as clear from the discussion below. \\
For this purpose, we recall that around $d=4$, the projectors are expanded according to the relation 
\begin{equation}
\label{pexp}
\P^{\,\mu_1\nu_1\mu_2\nu_2}(p)=\P^{(4)\,\mu_1\nu_1\mu_2\nu_2}(p)-\frac{2}{9}\varepsilon\,\pi^{\mu_1\nu_1}(p)\,\pi^{\mu_2\nu_2}(p)+O(\varepsilon^2). 
\end{equation}
This equations requires some clarification and we pause for a moment in order to illustrate its correct use. 

A consistent approach to the calculation is to perform all the tensor contractions in $d-$dimensions and only at the end move to $d=4$ in the limit of $\epsilon\to 0$. In this way one is reassured that the contraction of a metric tensor (in this case $\delta_{\mu}^{\mu}$, being us in the Euclidean case) gives $d$ and not $4$. The use of \eqref{pexp} is possible only if we are sure that there will not be any trace of the metric to perform. If these conditions are satisfied, then two methods of computation are equivalent and do not generate any ambiguity. \\
We illustrate this for the $TT$. Using \eqref{pexp} in \eqref{TTddim}, the latter takes the form

\begin{align}
\label{result}
\braket{T^{\mu_1\nu_1}(p)T^{\mu_2\nu_2}(-p)}&=-\frac{\p^2\,p^4}{4}\,\bigg(\frac{1}{\varepsilon}+\bar{B}_0(p^2)\bigg)\,\bigg(\P^{(4)\,\mu_1\nu_1\mu_2\nu_2}(p)-\frac{2}{9}\varepsilon\,\pi^{\mu_1\nu_1}(p)\,\pi^{\mu_2\nu_2}(p)+O(\varepsilon^2)\bigg)\notag\\
&\hspace{-2cm}\times\Bigg[\bigg(\frac{2}{5}+\frac{4}{25}\varepsilon+O(\varepsilon^2)\bigg)n_F+\bigg(\frac{4}{5}-\frac{22}{25}\varepsilon+O(\varepsilon^2)\bigg)n_G+\bigg(\frac{1}{15}+\frac{16}{225}\varepsilon+O(\varepsilon^2)\bigg)n_S\Bigg]\notag\\
&\hspace{-1cm}+\frac{\p^2\,p^4\,n_G}{8}\p^{\mu_1\nu_1}(p)\p^{\mu_2\nu_2}(p)\,\bigg(\frac{1}{\varepsilon}+\bar{B}_0(p^2)\bigg)\bigg[\frac{8}{9}\varepsilon^2+\frac{8}{27}\varepsilon^3+O(\varepsilon^4)\bigg]
\end{align}
where $\P^{ (4)\,\,\mu_1\nu_1\mu_2\nu_2}(p)$ is the transverse and traceless projector in $d=4$ and $\bar{B}_0(p^2)= 2 + \log(\mu^2/p^2)$ is the finite part in $d=4$ of the scalar integral in the $\overline{MS}$ scheme. As anticipated above, the last term of \eqref{result}, generated by the addition of a non-conformal sector ($\sim n_G$) vanishes separately as $\epsilon\to 0$. Finally, combining all the terms we obtain the regulated ($reg$) expression of the $TT$ around $d=4$ in the form

\begin{align}
\label{regular}
\braket{T^{\mu_1\nu_1}(p)T^{\mu_2\nu_2}(-p)}_{reg}&=-\frac{\p^2\,p^4}{60\,\varepsilon}\Pi^{(4)\,\mu_1\nu_1\mu_2\nu_2}(p)\left(6 n_F + 12 n_G + n_S\right)\notag\\
&\hspace{-3cm}+\frac{\p^2\,p^4}{270}\p^{\mu_1\nu_1}(p)\p^{\mu_2\nu_2}(p)\left(6 n_F + 12 n_G + n_S\right)-\frac{\p^2\,p^4}{300}\bar{B}_0(p^2)\Pi^{\mu_1\nu_1\mu_2\nu_2}(p)\left(30n_F+60n_G+5n_S\right)\notag\\
&\hspace{-2cm}-\frac{\p^2\,p^4}{900}\Pi^{\mu_1\nu_1\mu_2\nu_2}(p)\left(36n_F-198 n_G+16n_S\right)+O(\varepsilon)
\end{align}
The divergence in the previous expression can be removed through the one loop counterterm Lagrangian \eqref{scount}. In fact, the second functional derivative of $S_{count}$ with respect to the background metric gives 
\begin{align}
\braket{T^{\mu_1\nu_1}(p)T^{\mu_2\nu_2}(-p)}_{count}&\equiv -\sdfrac{1}{\varepsilon}\sum_{I=F,S,G}\bigg(4\b_a(I)\,\big[\sqrt{-g}\,C^2\big]^{\m_1\n_1\m_2\n_2}(p,-p)\bigg)\notag\\
&=-\frac{8}{\varepsilon}\frac{(d-3)\,}{(d-2)}p^4\P^{(d)\,\mu_1\nu_1\mu_2\nu_2}(p)\,\bigg(n_S\,\b_a(S)+n_F\,\b_a(F)+n_G\,\b_a(G)\,\bigg)
\end{align}
having used the relation $V_{E}^{\m_1\n_1\m_2\n_2}(p,-p)=0$. In particular, expanding around $d=4$ and using again \eqref{pexp} we obtain
\begin{align}
\braket{T^{\mu_1\nu_1}(p)T^{\mu_2\nu_2}(-p)}_{count}&=-\frac{8\,p^4}{\varepsilon}\bigg(\P^{(4)\,\mu_1\nu_1\mu_2\nu_2}(p)-\frac{2}{9}\varepsilon\,\pi^{\mu_1\nu_1}(p)\,\pi^{\mu_2\nu_2}(p)+O(\varepsilon^2)\bigg)\,\bigg(\frac{1}{2}-\frac{\varepsilon}{2}+O(\varepsilon^2)\bigg)\notag\\
&\hspace{1cm}\times\bigg(n_S\,\b_a(S)+n_F\,\b_a(F)+n_G\,\b_a(G)\,\bigg)\notag\\
&\hspace{-3cm}=-\sdfrac{4}{\varepsilon}p^4\bigg(n_S\,\b_a(S)+n_F\,\b_a(F)+n_G\,\b_a(G)\,\bigg)\,\P^{(4)\,\mu_1\nu_1\mu_2\nu_2}(p)\notag\\
&\hspace{-2cm}+4\, p^4\bigg(n_S\,\b_a(S)+n_F\,\b_a(F)+n_G\,\b_a(G)\,\bigg)\bigg[\P^{(4)\,\mu_1\nu_1\mu_2\nu_2}(p)+\frac{2}{9}\p^{\mu_1\nu_1}(p)\p^{\mu_2\nu_2}(p)\bigg]+O(\varepsilon)
\end{align}
which cancels the divergence arising in the two point function, if one chooses the parameters as in \eqref{choiceparm}.  
The renormalized 2-point function using  \eqref{choiceparm} then takes the form  \begin{align}
\braket{T^{\mu_1\nu_1}(p)T^{\mu_2\nu_2}(-p)}_{Ren}&=\braket{T^{\mu_1\nu_1}(p)T^{\mu_2\nu_2}(-p)}+\braket{T^{\mu_1\nu_1}(p)T^{\mu_2\nu_2}(-p)}_{count}\notag\\
&=-\frac{\p^2\,p^4}{60}\bar{B}_0(p^2)\Pi^{\mu_1\nu_1\mu_2\nu_2}(p)\left(6n_F+12n_G+n_S\right)\notag\\
&\quad-\frac{\p^2\,p^4}{900}\P^{\mu_1\nu_1\mu_2\nu_2}(p)\big(126n_F-18n_G+31n_S\big)
\label{tren}
\end{align}
Notice that the choice of the choice $\beta_c=0$ takes us to 
a final expression which is transverse and traceless. 
The same choice of parameters $\b_a,\,\b_b$ given in \eqref{choiceparm} removes the divergences in the three point function, as we are going to discuss below.

%%%%%%%%%%%%%%%%%%%%%%%%%%%%%%%%%%%%%%%%%%%%%%%%%%%%%%%
%%%%%%%%%%%%%%%%%%%%%%%%%%%%%%%%%%%%%%%%%%%%%%%%%%%%%%%
\section{Anomalous Conformal Ward Identities in $d=4$ and free field content}
%%%%%%%%%%%%%%%%%%%%%%%%%%%%%%%%%%%%%%%%%%%%%%%%%%%%%%%
%%%%%%%%%%%%%%%%%%%%%%%%%%%%%%%%%%%%%%%%%%%%%%%%%%%%%%%
The divergences arising in the form factors in $d=4$ and their renormalization induce a breaking of the conformal symmetry, thereby generating a set of anamalous CWI's.  In this section we will give the explicit form of the such identities in the presence of a trace anomaly.

%%%%%%%%%%%%%%%%%%%%%%%%%%%%%%%%%
\subsection{Primary anomalous CWI's and free field content}
%%%%%%%%%%%%%%%%%%%%%%%%%%%%%%%%%
The equations for the anomalous primary CWI's are generated after renormalization, starting from the $d$-dimensional expressions of the $A_i$'s  given in Appendix \ref{renexp}. The renormlization procedure will involve only $B_0$.
 The primary anomalous CWI's take the form
\begin{equation}
\begin{split}
& \textup{K}_{13}A^{Ren}_3=2A^{Ren}_2-\sdfrac{2\p^2}{45}\left(7n_F-26n_G+2n_S\right) \\
&\textup{K}_{23}A^{Ren}_3=2A^{Ren}_2-\sdfrac{2\p^2}{45}\left(7n_F-26n_G+2n_S\right) \\[1.1ex]
& \textup{K}_{13}A^{Ren}_4=-4A^{Ren}_2(p_2\leftrightarrow p_3)+\sdfrac{4\p^2}{45}\left(7n_F-26n_G+2n_S\right) \\
&\textup{K}_{23}A^{Ren}_4=-4A^{Ren}_2(p_1\leftrightarrow p_3)+\sdfrac{4\p^2}{45}\left(7n_F-26n_G+2n_S\right) \\[1.1ex]
& \textup{K}_{13}A^{Ren}_5=2\left[A^{Ren}_4-A^{Ren}_4(p_1\leftrightarrow p_3) \right]-\sdfrac{4\p^2}{9}(s-s_2)\left(5n_F+2n_G+n_s\right)\\
&\textup{K}_{23}A^{Ren}_5=2\left[A^{Ren}_4-A^{Ren}_4(p_2\leftrightarrow p_3)\right] -\sdfrac{4\p^2}{9}(s_1-s_2)\left(5n_F+2n_G+n_s\right)
\end{split}\label{PrimaryAnom}
\end{equation}
where now the differential operators $K_i$ take the form
\begin{equation}
K_i=\frac{\partial^2}{\partial p_i^2}-\frac{3}{p_i}\frac{\partial}{\partial p_i}=4s_{i-1}\frac{\partial^2}{\partial s_{i-1}^2}-4\frac{\partial}{\partial s_{i-1}},\qquad i=1,2,3
\end{equation}
with the identification $s_0=s$. The $(p_1\leftrightarrow p_3)$ and $(p_2\leftrightarrow p_3)$ versions of the anomalous Ward identities can be obtained from \eqref{PrimaryAnom}. 
%It is worth noticing that the primary CWIs of $A_1$ and $A_2$ do not change at this stage. 
Using the expressions given in the \appref{Mvc}, we can identify the corresponding counterterms for the $A_i$, extracted from the 
transverse traceless parts of the vertices generated by the counterterm Lagrangian in \eqref{scount}, obtaining
\begin{align}
A_2^{count}&=-\frac{16}{\varepsilon}\,\sum_{I=F,S,G}\,n_I\,\big[\b_a(I)+\b_b(I)\big]\nn
A_3^{count}&=-\frac{8}{\varepsilon}\,\sum_{I=F,S,G}\,n_I\,\big[s_2\,\b_b(I)-(s+s_1)\b_a(I)\big]+o(\epsilon)\nn
A_4^{count}&=-\frac{8}{\varepsilon}\,\sum_{I=F,S,G}\,n_I\,\big[(s+s_1-s_2)\,\b_b(I)-(s+s_1+3s_2)\b_a(I)\big]+o(\epsilon)\nn
A_5^{count}&=-\frac{4}{\varepsilon}\,\sum_{I=F,S,G}\,n_I\,\big[-\big(s^2-2s(s_1+s_2)+(s_1-s_2)^2\big)\,\b_b(I)\nn
&\hspace{3cm}-\big(3s^2-2s(s_1+s_2)+3s_1^2-2s_1s_2+3s_2^2\big)\b_a(I)\big]+o(\epsilon)
\end{align}
In order to cancel the divergences arising from the form factors, we need to choose the coefficient $\b_b(I)$ and $\b_a(I)$ as in \eqref{choiceparm}. The renormalized form factors can then be written as
\begin{align}
A_2^{Ren}&=A_2^{Reg}\nn
A_3^{Ren}&=A_3^{Reg}-8\,(s+s_1+s_2)\sum_{I=F,S,G}n_I\,\b_a(I)\nn
A_4^{Ren}&=A_4^{Reg}-16\,(s+s_1+s_2)\sum_{I=F,S,G}n_I\,\b_a(I)\nn
A_5^{Ren}&=A_5^{Reg}-8\,(s^2+s_1^2+s_2^2)\sum_{I=F,S,G}n_I\,\b_a(I)
\end{align}
where with ``reg'' we indicate those form factors which remain unmodified by the procedure, being finite. Such are $A_1$ and $A_2$.
\subsection{Secondary anomalous CWI's from free field theory }
The derivation of the secondary anomalous CWI's has been discussed within the general formalism in \cite{Bzowski:2013sza}  and in the perturbative approach in \cite{Coriano:2018bbe} in the case of the $TJJ$ correlator. The details of this analysis, which has been discussed at length in our previous work \cite{Coriano:2018bbe}, also in this case remain similar. We refer to Appendix \ref{secondary} for a definiton of the corresponding operators appearing in such equations and to \cite{Coriano:2018bbe}.  A lengthy computation gives
\begin{align}
&L_6 A^{Ren}_1+R A^{Ren}_2-R A^{Ren}_2(p_2\leftrightarrow p_3)=0\nn 
&L_4\,A^{Ren}_2+2p_1^2\,A^{Ren}_2+4RA_3-2RA^{Ren}_4(p_1\leftrightarrow p_3)=\frac{4\p^2\,p_1^2}{45}\left(7n_F-26n_G+2n_S\right)\\[1.5ex]\nn
&L_4\,A^{Ren}_2(p_1\leftrightarrow p_3)-R\,A^{Ren}_4+RA^{Ren}_4(p_2\leftrightarrow p_3)+2p_1^2(A^{Ren}_2(p_2\leftrightarrow p_3)-A^{Ren}_2)=\frac{2\p^2\,p_1^2}{45}\left(7n_F-26n_G+2n_S\right)\nn
&L_4\,A^{Ren}_2(p_2\leftrightarrow p_3)-4R\,A^{Ren}_3(p_2\leftrightarrow p_3)+2RA^{Ren}_4(p_1\leftrightarrow p_3)-2p_1^2A^{Ren}_2(p_2\leftrightarrow p_3)=0\nn
&L_2\,A^{Ren}_3(p_1\leftrightarrow p_3)+p_1^2(A^{Ren}_4-A^{Ren}_4(p_2\leftrightarrow p_3)=\frac{30\p^2}{225}(6n_F+12n_G+n_S)\big(s_2^2 \bar{B}_0(s_2)-s_1^2B_0^{Reg}(s_1)\big)\nonumber
\end{align}
\begin{align}
&\hspace{+1cm}-\frac{\p^2}{225}\bigg[n_F\big(55s^2+5s(29s_1+7s_2)+252(s_1^2-s_2^2)\big)+2n_G\big(155s^2+245s\,s_1-65s\,s_2-18(s_1^2-s_2^2)\big)\notag\nn
&\hspace{+1.5cm}+n_S\big(5s^2+10s(2s_1+s_2)+62(s_1^2-s_2^2)\big)\bigg]\\[1.5ex]
&L_2\,A^{Ren}_4+2R\,A^{Ren}_5+8p_1^2A^{Ren}_3(p_2\leftrightarrow p_3)-2p_1^2(A^{Ren}_4+A^{Ren}_4(p_1\leftrightarrow p_3))=\notag\nn
&\hspace{+1cm}-\frac{120\p^2\,s_1^2}{225}B_0(s_1)\big(6n_F+12n_G+n_sS\big)-\frac{4\p^2}{225}\Bigg[15s^2(6n_F+12n_G+n_S)+5s\,s_1(11n_F+62n_G)\notag\nn
&\hspace{+1.5cm}+2s_1^2(126n_F-18n_G+31n_S)\bigg]\\[1.5ex]
&L_2\,A^{Ren}_4(p_2\leftrightarrow p_3)-2R\,A^{Ren}_5-8p_1^2A^{Ren}_3+2p_1^2(A^{Ren}_4(p_2\leftrightarrow p_3)+A^{Ren}_4(p_1\leftrightarrow p_3))=\notag\\
&\hspace{+1cm}+\frac{2\p^2}{225} \bigg[60 s_2^2(6 n_F+12 n_G+n_S)B_0^{Reg}(s_2) +5 s \bigg(s (7 n_F-26 n_G+2 n_S)-s_1 (43 n_F+46 n_G+8 n_S)\bigg)\notag\\
&\hspace{+1.5cm}-5 s s_2 (7 n_F-26 n_G+2 n_S)+4 s_2^2 (126 n_F-18 n_G+31 n_S)\bigg].
\end{align}
The most involved part of this analysis involves a rewriting of the differential action of the $L$ operators on $B_0$ and $C_0$. 
We have explicitly verified that the renormalized $A_i$ satisfy such equations confirming the consistency of the entire approach. 

\section{Reconstruction of the $\braket{TTT}$ in $d=4$}

%%%%%%%%%%%%%%%%%%%%%%%%%%%%%%%%%%%%%%%%%%%%%%%%%%%%%%%
%%%%%%%%%%%%%%%%%%%%%%%%%%%%%%%%%%%%%%%%%%%%%%%%%%%%%%%
In this section we will illustrate the reconstruction procedure for the $TTT$ using the perturbative realization of this correlator. In this case our goal will be to show how the separation of the vertex into a traceless part and an anomaly contribution takes place after renormalization. As already remarked in the introduction, the advantage of using a direct perturbative approach is to present for the 
transverse traceless sector of this vertex the simplest explicit form, in terms of the renormalized scalar 2- and 3-point functions. \\
The approach is obviously the standard one, where the renormalization is obtained by the addition to the bare vertex of the counterterms worked out in the previous two sections, but we will try to illustrate in some detail how the generation of the anomaly poles in the trace part takes place in these types of correlators. \\
We start from the bare local contributions in $d$ dimensions which take the form
\begin{align}
\label{loc}
\braket{t_{loc}^{\mu_1\nu_1}T^{\mu_2\nu_2}T^{\mu_3\nu_3}}&=\Big(\mathcal{I}^{\mu_1\nu_1}_{\a_1}(p_1)\,p_{1\b_1}+\frac{\p^{\mu_1\nu_1}(p_1)}{(d-1)}\d_{\a_1\b_1}\Big)\braket{T^{\a_1\b_1}T^{\mu_2\nu_2}T^{\mu_3\nu_3}}\notag\\
&=-\frac{2\,\p^{\mu_1\nu_1}(p_1)}{(d-1)}\Big[\braket{T^{\mu_2\nu_2}(p_1+p_2)T^{\mu_3\nu_3}(p_3)}+\braket{T^{\mu_2\nu_2}(p_2)T^{\mu_3\nu_3}(p_1+p_3)}\Big]
\notag\\
&+\mathcal{I}^{\mu_1\nu_1}_{\a_1}(p_1)\Big\{-p_2^{\a_1}\braket{T^{\mu_2\nu_2}(p_1+p_2)T^{\mu_3\nu_3}(p_3)}-p_3^{\a_1}\braket{T^{\mu_2\nu_2}(p_2)T^{\mu_3\nu_3}(p_1+p_3)}\notag\\
&+p_{2\b}\Big[\d^{\a_1\mu_2}\braket{T^{\b\mu_2}(p_1+p_2)T^{\mu_3\nu_3}(p_3)}+\d^{\a_1\nu_2}\braket{T^{\b\mu_2}(p_1+p_2)T^{\mu_3\nu_3}(p_3)}\Big]\notag\\
&+p_{3\b}\Big[\d^{\a_1\mu_3}\braket{T^{\nu_2\mu_2}(p_2)T^{\b_3\nu_3}(p_1+p_3)}+\d^{\a_1\nu_3}\braket{T^{\mu_2\nu_2}(p_2)T^{\mu_3\b}(p_1+p_3)}\Big]\Big\}
\end{align}
which develop a singularity for $\epsilon\to 0$, with $\epsilon=(4-d)/2$, just like all the other contributions appearing in \eqref{DecompTTT}. We pause for a moment to describe the structure of this expression and comment on the general features of the regularization procedure.
\\ We perform all the tensor contractions in $d$ dimensions and in the final expression we set $d= 4 +\epsilon$. For example, if a projector such as $\Pi^{(d)}$ appears, we will be using Eq.~\eqref{pexp}, which relates $\Pi^{(d)}$ to $\Pi^{(4)}$, and so on. For instance, a projector such as $\pi^{\mu_1\nu_1}$ with open indices remains unmodified since it has no explicit $d$-dependence, unless it is contracted with a $\delta^{\mu\nu}$. It is then clear, from a cursory look at the right hand side of \eqref{loc} that the regulated expression of this expression involves a prefactor $1/(d-1)$, which is expanded around $d=4$ and the replacements of all the two point functions with the regulated expression given by Eq. \eqref{regular}, with the insertion of the appropriate momenta.\\
The corresponding counterterm is given by 
\begin{align}
\label{locco}
\braket{t_{loc}^{\mu_1\nu_1}T^{\mu_2\nu_2}T^{\mu_3\nu_3}}_{count}&=\Big(\mathcal{I}^{\mu_1\nu_1}_{\a_1}(p_1)\,p_{1\b_1}+\frac{\p^{\mu_1\nu_1}(p_1)}{(d-1)}\d_{\a_1\b_1}\Big)\braket{T^{\a_1\b_1}T^{\mu_2\nu_2}T^{\mu_3\nu_3}}_{(count)}\notag\\
&\hspace{-3cm}=-\frac{1}{\varepsilon}\frac{(d-4)}{(d-1)}\p^{\mu_1\nu_1}(p_1)\bigg(4[E]^{\mu_2\nu_2\mu_3\nu_3}(p_2,p_3)+4[C^2]^{\mu_2\nu_2\mu_3\nu_3}(p_2,p_3)\bigg)\notag\\
&\hspace{-3cm}+\frac{1}{\varepsilon}\frac{2}{(d-1)}\p^{\mu_1\nu_1}(p_1)\bigg(4[C^2]^{\mu_2\nu_2\mu_3\nu_3}(p_1+p_2,p_3)+4[C^2]^{\mu_2\nu_2\mu_3\nu_3}(p_2,p_1+p_3)\bigg)\notag\\
&\hspace{-3cm}-\frac{1}{\varepsilon}\mathcal{I}^{\mu_1\nu_1}_{\a_1}(p_1)\bigg\{-4p_2^{\a_1}[C^2]^{\mu_2\nu_2\mu_3\nu_3}(p_1+p_2,p_3)-p_3^{\a_1}[C^2]^{\mu_2\nu_2\mu_3\nu_3}(p_2,p_1+p_3)\notag\\
&\hspace{-3cm}+4p_{2\b}\Big[\d^{\a_1\mu_2}[C^2]^{\b\nu_2\mu_3\nu_3}(p_1+p_2,p_3)+\d^{\a_1\nu_2}[C^2]^{\mu_2\b\mu_3\nu_3}(p_1+p_2,p_3)\Big]\notag\\
&\hspace{-3cm}+4p_{3\b}\Big[\d^{\a_1\mu_3}[C^2]^{\mu_2\nu_2\b\nu_3}(p_2,p_1+p_3)+\d^{\a_1\nu_3}[C^2]^{\mu_2\nu_2\mu_3\b}(p_2,p_1+p_3)\Big]\bigg\}.
\end{align}
where, for simplicity, we have absorbed the dependence on the total contributions to the beta functions $\beta_a$ and $\beta_b$ 
\begin{equation}
\beta_{a,b}\equiv\sum_{I=f,s,G} \beta_{a,b} (I)
\end{equation}
into $[E]$ and $[C^2]$.\\
It is worth mentioning that all the divergent parts of the local term given in \eqref{loc} above are cancelled by the local parts of the counterterm \eqref{locco}. For its renormalized expression we obtain
\begin{align}
\braket{t_{loc}^{\mu_1\nu_1}T^{\mu_2\nu_2}T^{\mu_3\nu_3}}_{Ren}&=\braket{t_{loc}^{\mu_1\nu_1}T^{\mu_2\nu_2}T^{\mu_3\nu_3}}+\braket{t_{loc}^{\mu_1\nu_1}T^{\mu_2\nu_2}T^{\mu_3\nu_3}}_{(count)}\notag\\
&=\mathcal{V}_{loc\,0\,0 } +\braket{t_{loc}^{\mu_1\nu_1}T^{\mu_2\nu_2}T^{\mu_3\nu_3}}^{(4)}_{extra}
\end{align}
where
\begin{align}
&\mathcal{V}_{loc\, 0 \, 0 }=-\frac{2\,\p^{\mu_1\nu_1}(p_1)}{3}\Big[\braket{T^{\mu_2\nu_2}(p_1+p_2)T^{\mu_3\nu_3}(-p_1-p_2)}_{Ren}+\braket{T^{\mu_2\nu_2}(p_2)T^{\mu_3\nu_3}(-p_2)}_{Ren}\Big]
\notag\\
&\quad+\mathcal{I}^{(4)\,\mu_1\nu_1}_{\a_1}(p_1)\Big\{-p_2^{\a_1}\braket{T^{\mu_2\nu_2}(p_1+p_2)T^{\mu_3\nu_3}(-p_1-p_2)}_{Ren}-p_3^{\a_1}\braket{T^{\mu_2\nu_2}(p_2)T^{\mu_3\nu_3}(-p_2)}_{Ren}\notag\\
&\quad+p_{2\b}\Big[\d^{\a_1\mu_2}\braket{T^{\b\mu_2}(p_1+p_2)T^{\mu_3\nu_3}(-p_1-p_2)}_{Ren}+\d^{\a_1\nu_2}\braket{T^{\b\mu_2}(p_1+p_2)T^{\mu_3\nu_3}(-p_1-p_2)}_{Ren}\Big]\notag\\
&\quad+p_{3\b}\Big[\d^{\a_1\mu_3}\braket{T^{\nu_2\mu_2}(p_2)T^{\b\nu_3}(-p_2)}_{Ren}+\d^{\a_1\nu_3}\braket{T^{\mu_2\nu_2}(p_2)T^{\mu_3\b}(-p_2)}_{Ren}\Big]\Big\}
\end{align}
with $\braket{TT}_{ren}$ given by \eqref{tren}. Notice the presence of an extra contribution coming from the local parts of counterterms that takes the explicit form
\begin{align}
&\braket{t_{loc}^{\mu_1\nu_1}T^{\mu_2\nu_2}T^{\mu_3\nu_3}}^{(4)}_{extra}=\frac{\hat{\p}^{\mu_1\nu_1}(p_1)}{3 \, p_1^2}\bigg(4[E]^{\mu_2\nu_2\mu_3\nu_3}(p_2,p_3)+4[C^2]^{\mu_2\nu_2\mu_3\nu_3}(p_2,p_3)\bigg),
\end{align}
having defined 
\bea
\hat{\p}^{\mu\nu}(p)=(\delta^{\mu_1\nu_1}p^2 - p^\mu p^\nu)
\eea
which shows the emergence of an anomaly pole, similarly to the $TJJ$ cases  \cite{Giannotti:2008cv,Armillis:2009pq,Coriano:2018zdo}. \\
The renormalization of the other local contributions follows a similar pattern. In particular, the correlator with two $t_{loc}$ projections takes the form
\begin{align}
&\braket{t_{loc}^{\mu_1\nu_1}t_{loc}^{\mu_2\nu_2}T^{\mu_3\nu_3}}_{Ren}= \mathcal{V}^{\mu_1\nu_1\mu_2\nu_2\mu_3\nu_3}_{loc\, loc \, 0} +\braket{t_{loc}^{\mu_1\nu_1}t_{loc}^{\mu_2\nu_2}T^{\mu_3\nu_3}}^{(4)}_{extra}
\end{align}
where
\begin{align}
&\mathcal{V}^{\mu_1\nu_1\mu_2\nu_2\mu_3\nu_3}_{loc\, loc \, 0}=\Big(\mathcal{I}^{(4)\,\mu_2\nu_2}_{\a_2}(p_2)\,p_{2\b_2}+\frac{\p^{\mu_2\nu_2}(p_2)}{3}\d_{\a_2\b_2}\Big)\notag\\
&\times\Bigg\{-\frac{2\,\p^{\mu_1\nu_1}(p_1)}{3}\Big[\braket{T^{\a_2\b_2}(p_1+p_2)T^{\mu_3\nu_3}(-p_1-p_2)}_{Ren}+\braket{T^{\a_2\b_2}(p_2)T^{\mu_3\nu_3}(-p_2)}_{Ren}\Big]
\notag\\
&\quad+\mathcal{I}^{(4)\,\mu_1\nu_1}_{\a_1}(p_1)\Big[-p_2^{\a_1}\braket{T^{\a_2\b_2}(p_1+p_2)T^{\mu_3\nu_3}(-p_1-p_2)}_{Ren}-p_3^{\a_1}\braket{T^{\a_2\b_2}(p_2)T^{\mu_3\nu_3}(-p_2)}_{Ren}\notag\\
&\quad+p_{2\b}\Big(\d^{\a_1\a_2}\braket{T^{\b\b_2}(p_1+p_2)T^{\mu_3\nu_3}(-p_1-p_2)}_{Ren}+\d^{\a_1\b_2}\braket{T^{\b\a_2}(p_1+p_2)T^{\mu_3\nu_3}(-p_1-p_2)}_{Ren}\Big)\notag\\
&\quad+p_{3\b}\Big(\d^{\a_1\mu_3}\braket{T^{\b_2\a_2}(p_2)T^{\b\nu_3}(-p_2)}_{Ren}+\d^{\a_1\nu_3}\braket{T^{\a_2\b_2}(p_2)T^{\mu_3\b}(-p_2)}_{Ren}\Big)\Big]\Bigg\}
\end{align}
in which we define 
\begin{align}
\mathcal{I}^{(4)\,\mu\nu}_{\a}(p)\equiv \frac{1}{p^2}\left[2 p^{(\mu}\delta^{\nu)}_\alpha - 
\frac{p_\alpha}{3}\left(\delta^{\mu\nu} +2\,\frac{p^\mu p^\nu}{p^2}\right)\right]
\end{align}
and with the presence of an extra term of the form
\begin{align}
\braket{t_{loc}^{\mu_1\nu_1}t_{loc}^{\mu_2\nu_2}T^{\mu_3\nu_3}}^{(4)}_{extra}=\frac{\p^{\mu_1\nu_1}(p_1)}{3}\frac{\p^{\mu_2\nu_2}(p_2)}{3}\d_{\a_2\b_2}\bigg(4[E]^{\a_2\b_2\mu_3\nu_3}(p_2,p_3)+4[C^2]^{\a_2\b_2\mu_3\nu_3}(p_2,p_3)\bigg)
\end{align}
and finally the term with three insertions of $t_{loc}$
\begin{align}
&\braket{t_{loc}^{\mu_1\nu_1}t_{loc}^{\mu_2\nu_2}t_{loc}^{\mu_3\nu_3}}_{Ren}= \mathcal{V}^{\mu_1\nu_1\mu_2\nu_2\mu_3\nu_3}_{loc\,loc\,loc} +\braket{t_{loc}^{\mu_1\nu_1}t_{loc}^{\mu_2\nu_2}t_{loc}^{\mu_3\nu_3}}^{(4)}_{extra}
\end{align}
with
\begin{align}
&\mathcal{V}^{\mu_1\nu_1\mu_2\nu_2\mu_3\nu_3}_{loc\,loc\,loc}=\Big(\mathcal{I}^{(4)\,\mu_2\nu_2}_{\a_2}(p_2)\,p_{2\b_2}+\frac{\p^{\mu_2\nu_2}(p_2)}{3}\d_{\a_2\b_2}\Big)\Big(\mathcal{I}^{(4)\,\mu_3\nu_3}_{\a_3}(p_3)\,p_{3\b_3}+\frac{\p^{\mu_3\nu_3}(p_3)}{3}\d_{\a_3\b_3}\Big)\notag\\
&\times\Bigg\{-\frac{2\,\p^{\mu_1\nu_1}(p_1)}{3}\Big[\braket{T^{\a_2\b_2}(p_1+p_2)T^{\a_3\b_3}(-p_1-p_2)}_{Ren}+\braket{T^{\a_2\b_2}(p_2)T^{\a_3\b_3}(-p_2)}_{Ren}\Big]
\notag\\
&\quad+\mathcal{I}^{(4)\,\mu_1\nu_1}_{\a_1}(p_1)\Big[-p_2^{\a_1}\braket{T^{\a_2\b_2}(p_1+p_2)T^{\a_3\b_3}(-p_1-p_2)}_{Ren}-p_3^{\a_1}\braket{T^{\a_2\b_2}(p_2)T^{\a_3\b_3}(-p_2)}_{Ren}\notag\\
&\quad+p_{2\b}\Big(\d^{\a_1\a_2}\braket{T^{\b\b_2}(p_1+p_2)T^{\a_3\b_3}(-p_1-p_2)}_{Ren}+\d^{\a_1\b_2}\braket{T^{\b\a_2}(p_1+p_2)T^{\a_3\b_3}(-p_1-p_2)}_{Ren}\Big)\notag\\
&\quad+p_{3\b}\Big(\d^{\a_1\a_3}\braket{T^{\b_2\a_2}(p_2)T^{\b\b_3}(-p_2)}_{Ren}+\d^{\a_1\b_3}\braket{T^{\a_2\b_2}(p_2)T^{\a_3\b}(-p_2)}_{Ren}\Big)\Big]\Bigg\}
\end{align}
where
\begin{align}
\braket{t_{loc}^{\mu_1\nu_1}t_{loc}^{\mu_2\nu_2}t_{loc}^{\mu_3\nu_3}}^{(4)}_{extra}=\frac{\p^{\mu_1\nu_1}(p_1)\,\p^{\mu_2\nu_2}(p_2)\,\p^{\mu_3\nu_3}(\bar{p}_3)}{27}\d_{\a_2\b_2}\d_{\a_3\b_3}\bigg(4[E]^{\a_2\b_2\a_3\b_3}(p_2,\bar{p}_3)+4[C^2]^{\a_2\b_2\a_3\b_3}(p_2,\bar{p}_3)\bigg).
\end{align}
In summary, the counterterm cancels all the divergences arising in the 3-point function and from the local part of the counterterms there are extra contributions in the final renormalized $\braket{TTT}$ of the form
\begin{align}
\braket{T^{\mu_1\nu_1}T^{\mu_2\nu_2}T^{\mu_3\nu_3}}^{(4)}_{extra}&=\left(\frac{\p^{\mu_1\nu_1}(p_1)}{3}\bigg(4[E]^{\mu_2\nu_2\mu_3\nu_3}(p_2,\bar{p}_3)+4[C^2]^{\mu_2\nu_2\mu_3\nu_3}(p_2,\bar{p}_3)\bigg)+(\text{perm.})\right)\notag\\
&\hspace{-1cm}-\left(\frac{\p^{\mu_1\nu_1}(p_1)}{3}\frac{\p^{\mu_2\nu_2}(p_2)}{3}\d_{\a_2\b_2}\bigg(4[E]^{\a_2\b_2\mu_3\nu_3}(p_2,\bar{p}_3)+4[C^2]^{\a_2\b_2\mu_3\nu_3}(p_2,\bar{p}_3)\bigg) +(\text{perm.})\right)\notag\\
&\hspace{-1cm}+\frac{\p^{\mu_1\nu_1}(p_1)}{3}\frac{\p^{\mu_2\nu_2}(p_2)}{3}\frac{\p^{\mu_3\nu_3}(\bar{p}_3)}{3}\d_{\a_2\b_2}\d_{\a_3\b_3}\Big(4[E]^{\a_2\b_2\a_3\b_3}(p_2,\bar{p}_3)+4[C^2]^{\a_2\b_2\a_3\b_3}(p_2,\bar{p}_3)\Big).
\end{align}
This extra contribution is exactly the anomalous part of the $TTT$, which in the flat limit becomes
\begin{align}
\braket{T(p_1)T^{\mu_2\nu_2}(p_2)T^{\mu_3\nu_3}(\bar{p}_3)}_{anomaly}^{(4)}&=\big(4[E]^{\m_2\n_2\m_3\n_3}(p_2,p_3)+4[C^2]^{\m_2\n_2\m_3\n_3}(p_2,\bar{p}_3)\big)\\
\braket{T(p_1)T(p_2)T^{\mu_3\nu_3}(\bar{p}_3)}_{anomaly}^{(4)}&=\d_{\a_2\b_2}\big(4[E]^{\a_2\b_2\m_3\n_3}(p_2,p_3)+4[C^2]^{\a_2\b_2\m_3\n_3}(p_2,\bar{p}_3)\big)\notag\\
\braket{T(p_1)T(p_2)T(\bar{p}_3)}_{anomaly}^{(4)}&=\d_{\a_2\b_2}\d_{\a_3\b_3}\big(4[E]^{\a_2\b_2\a_3\b_3}(p_2,p_3)+4[C^2]^{\a_2\b_2\a_3\b_3}(p_2,\bar{p}_3)\big)
\end{align}
(with $T(p)\equiv \delta_{\m\n}T^{\m\nu}$).
The second order functional derivatives of the anomaly can be reconstructed using the expressions 
\begin{align}
\big[E\big]^{\m_i\nu_i\m_j\nu_j}(p_i,p_j) &=\big[R_{\m\a\n\b}\,R^{\m\a \n\b}\big]^{\m_i\nu_i\m_j\nu_j}
-4\,\big[R_{\m\n}R^{\m\n}\big]^{\m_i\nu_i\m_j\nu_j}
+\big[ R^2\big]^{\m_i\nu_i\m_j\nu_j}\notag\\
&\hspace{-2cm}=\bigg\{\big[R_{\m\a\n\b}\big]^{\m_i\nu_i}(p_i)\big[R^{\m\a \n\b}\big]^{\m_j\nu_j}(p_j)
-4\,\big[R_{\m\n}\big]^{\m_i\nu_i}(p_i)\big[R^{\m\n}\big]^{\m_j\nu_j}(p_j)
+\big[ R\big]^{\m_i\nu_i}(p_i)\big[R\big]^{\m_j\nu_j}(p_j)\bigg\}\notag\\[1.2ex]
&\hspace{2cm}+\{(\mu_i,\nu_i,p_i)\leftrightarrow (\mu_j,\nu_j,p_j)\}\\[2ex]
\big[C^2\big]^{\m_i\nu_i\m_j\nu_j}(p_i,p_j) &= \big[R_{\m\a\n\b}R^{\m\a \n\b}\big]^{\m_i\nu_i\m_j\nu_j}
-2\,\big[R_{\m\n}R^{\m\n}\big]^{\m_i\nu_i\m_j\nu_j}
+ \sdfrac{1}{3}\,\big[R^2\big]^{\m_i\nu_i\m_j\nu_j}\notag\\
&\hspace{-2cm}=\bigg\{\big[R_{\m\a\n\b}\big]^{\m_i\nu_i}(p_i)\big[R^{\m\a \n\b}\big]^{\m_j\nu_j}(p_j)
-2\,\big[R_{\m\n}\big]^{\m_i\nu_i}(p_i)\big[R^{\m\n}\big]^{\m_j\nu_j}(p_j)
+\frac{1}{3}\big[ R\big]^{\m_i\nu_i}(p_i)\big[R\big]^{\m_j\nu_j}(p_j)\bigg\}\notag\\[1.2ex]
&\hspace{2cm}+\{(\mu_i,\nu_i,p_i)\leftrightarrow (\mu_j,\nu_j,p_j)\},
\end{align}
for which we obtain
\begin{align}
\braket{T^{\mu_1\nu_1}(p_1)T^{\mu_2\nu_2}(p_2)T^{\mu_3\nu_3}(\bar{p}_3)}^{(4)}_{extra}&=\left(\frac{\p^{\mu_1\nu_1}(p_1)}{3}\braket{T(p_1)T^{\mu_2\nu_2}(p_2)T^{\mu_3\nu_3}(\bar{p}_3)}^{(4)}_{anomaly}+(\text{perm.})\right)\notag\\
&\hspace{-1cm}-\left(\frac{\p^{\mu_1\nu_1}(p_1)}{3}\frac{\p^{\mu_2\nu_2}(p_2)}{3}\braket{T(p_1)T(p_2)T^{\mu_3\nu_3}(\bar{p}_3)}^{(4)}_{anomaly}+(\text{perm.})\right)\notag\\
&\hspace{-1cm}+\frac{\p^{\mu_1\nu_1}(p_1)}{3}\frac{\p^{\mu_2\nu_2}(p_2)}{3}\frac{\p^{\mu_3\nu_3}(\bar{p}_3)}{3}\braket{T(p_1)T(p_2)T(\bar{p}_3)}^{(4)}_{anomaly}.
\end{align}
\subsection{Summary}
To summarize, the full renormalized $\braket{TTT}$ in $d=4$ can be constructed using the renormalized transverse traceless and the local terms. In particular we find
\begin{align}
&\braket{T^{\mu_1\nu_1}(p_1)T^{\mu_2\nu_2}(p_2)T^{\mu_3\nu_3}(\bar{p}_3)}^{(4)}=\braket{t^{\mu_1\nu_1}(p_1)t^{\mu_2\nu_2}(p_2)t^{\mu_3\nu_3}(\bar{p}_3)}^{(4)}_{Ren}\notag\\
&\hspace{1cm}+\left(\braket{t_{loc}^{\mu_1\nu_1}(p_1)T^{\mu_2\nu_2}(p_2)T^{\mu_3\nu_3}(\bar{p}_3)}^{(4)}_{Ren}+(\text{cyclic perm.})\right)\notag\\
&-\left(\braket{t_{loc}^{\mu_1\nu_1}(p_1)t_{loc}^{\mu_2\nu_2}(p_2)T^{\mu_3\nu_3}(\bar{p}_3)}^{(4)}_{Ren}+(\text{cyclic perm.})\right)+\braket{t_{loc}^{\mu_1\nu_1}(p_1)t_{loc}^{\mu_2\nu_2}(p_2)t_{loc}^{\mu_3\nu_3}(\bar{p}_3)}^{(4)}_{Ren}
\end{align} 
where the transverse and traceless parts are expressed as
\begin{align}
&\braket{t^{\mu_1\nu_1}(p_1)t^{\mu_2\nu_2}(p_2)t^{\mu_3\nu_3}(\bar{p}_3)}^{(4)}_{Ren}=\Pi^{(4)\,\mu_1\nu_1}_{\a_1\b_1}(p_1)\Pi^{(4)\,\mu_2\nu_2}_{\a_2\b_2}(p_2)\Pi^{(4)\,\mu_3\nu_3}_{\a_3\b_3}(\bar{p}_3)\notag\\
&\times\Big\{A_1^{Ren}\,p_2^{\a_1} p_2^{\b_1} \bar{p}_3^{\a_2} p_3^{\b_2} p_1^{\a_3} p_1^{\b_3}+ A_2^{Ren}\,\d^{\b_1\b_2} p_2^{\a_1} p_3^{\a_2} p_1^{\a_3} p_1^{\b_3} 
+ A_2^{ren}\,(p_1 \leftrightarrow p_3)\, \d^{\b_2\b_3}  p_3^{\a_2} p_1^{\a_3} p_2^{\a_1} p_2^{\b_1} \notag\\
&\hspace{0.8cm}+ A_2^{Ren}\,(p_2\leftrightarrow p_3)\, \d^{\b_3\b_1} p_1^{\a_3} p_2^{\a_1}  p_3^{\a_2} p_3^{\b_2}+ A_3^{Ren}\,\d^{\a_1\a_2} \d^{\b_1\b_2}  p_1^{\a_3} p_1^{\b_3} + A_3^{Ren}(p_1\leftrightarrow p_3)\,\d^{\a_2\a_3} \d^{\b_2\b_3}  p_2^{\a_1} p_2^{\b_1} \notag\\
&\hspace{1.2cm}
+ A_3^{Ren}(p_2\leftrightarrow p_3)\,\d^{\a_3\a_1} \d^{\b_3\b_1}  p_3^{\a_2} p_3^{\b_2} + A_4^{Ren}\,\d^{\a_1\a_3} \d^{\a_2\b_3}  p_2^{\b_1} p_3^{\b_2} + A_4^{Ren}(p_1\leftrightarrow p_3)\, \d^{\a_2\a_1} \d^{\a_3\b_1}  p_3^{\b_2} p_1^{\b_3} \notag\\
&\hspace{3.5cm}+ A_4^{Ren}(p_2\leftrightarrow p_3)\, \d^{\a_3\a_2} \d^{\a_1\b_2}  p_1^{\b_3} p_2^{\b_1} + A_5 ^{Ren} \d^{\a_1\b_2}  \d^{\a_2\b_3}  \d^{\a_3\b_1}\Big\}
\end{align}
with the renormalized form factors given in \appref{renexp}. It can be further simplified in the form 
\bea
\braket{T^{\m_1\n_1}T^{\m_2\n_2}T^{\m_3\n_3}}_{Ren}= \braket{t^{\m_1\n_1} t^{\m_2\n_2} t^{\m_3\n_3}}_{Ren} + 
\braket{T^{\m_1\n_1}T^{\m_2\n_2}T^{\m_3\n_3}}_{Ren\, l\,t} + \braket{T^{\m_1\n_1}T^{\m_2\n_2}T^{\m_3\n_3}}_{anomaly}\nn
\eea
 and with the renormalized longitudinal traceless contribution ($l \, t$ ) given by 
\begin{align}
\braket{ T^{\m_1\n_1}T^{\m_2\n_2}T^{\m_3\n_3}}_{Ren\,l \, t}&\equiv\left(\mathcal{V}^{\mu_1\nu_1\mu_2\nu_2\mu_3\nu_3}_{loc\,0\, 0} + \mathcal{V}^{\mu_1\nu_1\mu_2\nu_2\mu_3\nu_3}_{0\, loc\, 0}  + \mathcal{V}^{\mu_1\nu_1\mu_2\nu_2\mu_3\nu_3}_{0\, 0 \, loc} \right) \nn
&-
\left(\mathcal{V}^{\mu_1\nu_1\mu_2\nu_2\mu_3\nu_3}_{loc\,loc\, 0} + \mathcal{V}^{\mu_1\nu_1\mu_2\nu_2\mu_3\nu_3}_{0\, loc\, loc}  + \mathcal{V}^{\mu_1\nu_1\mu_2\nu_2\mu_3\nu_3}_{loc\, 0 \, loc} \right) + \mathcal{V}^{\mu_1\nu_1\mu_2\nu_2\mu_3\nu_3}_{loc\,loc\,loc}
\end{align}
and 
\begin{align}
\label{expans}
& \lla T^{\mu_1 \nu_1}({p}_1) T^{\mu_2 \nu_2}({p}_2) T^{\mu_3 \nu_3}({p}_3) \rra_{anomaly} = \frac{\hat\pi^{\mu_1 \nu_1}({p}_1)}{3\, p_1^2} \lla T({p}_1) T^{\mu_2 \nu_2}({p}_2) T^{\mu_3 \nu_3}({p}_3) \rra_{anomaly} \nn
& + \frac{\hat\pi^{\mu_2 \nu_2}({p}_2)}{3\, p_2^2} \lla T^{\mu_1 \nu_1}({p}_1) T({p}_2) T^{\mu_3 \nu_3}({p}_3) \rra_{anomaly} + \frac{\hat\pi^{\mu_3 \nu_3}({p}_3)}{3\, p_3^2} \lla T^{\mu_1 \nu_1}({p}_1) T^{\mu_2 \nu_2}({p}_2)  T({p}_3)\rra_{anomaly} \nn
& - \: \frac{\hat\pi^{\mu_1 \nu_1}({p}_1) \hat\pi^{\mu_2 \nu_2}({p}_2)}{9\, p_1^2 p_2^2}\lla T({p}_1)T({p}_2)T^{\mu_3 \nu_3}({p}_3)\rra_{anomaly} - \: \frac{\hat\pi^{\mu_2 \nu_2}({p}_2) \hat\pi^{\mu_3 \nu_3}({p}_2)}{9 p_2^2 p_3^2}\lla T^{\mu_1 \nu_1}({p}_1)T(p_2)T({p}_3)\rra_{anomaly} \nn
& - \: \frac{\hat\pi^{\mu_1 \nu_1}({p}_1)\hat \pi^{\mu_3 \nu_3}(\bar{p}_3)}{9 p_1^2 p_3^2}\lla T({p}_1) T^{\mu_2 \nu_2}({p}_2)T({p}_3)\rra_{anomaly}  + \frac{\hat\pi^{\mu_1 \nu_1}({p}_1)\hat\pi^{\mu_2 \nu_2}({p}_2) \hat\pi^{\mu_3 \nu_3}(\bar{p}_3)}{27 p_1^2 p_2^2 p_3^2}\lla T({p}_1)T({p}_2)T(\bar{p}_3)\rra_{anomaly}  .
\end{align}
As a final step, it is convenient to collect together the two renormalized contributions, the transverse and the longitudinal one, which are both traceless, into a single contribution 
\bea
\mathcal{V}^{\mu_1\nu_1\mu_2\nu_2\mu_3\nu_3}_{ traceless}\equiv \braket{t^{\m_1\n_1} t^{\m_2\n_2} t^{\m_3\n_3}}_{Ren} + \braket{ T^{\m_1\n_1}T^{\m_2\n_2}T^{\m_3\n_3}}_{Ren\,l \, t}
\eea
in order to cast the entire vertex in the form 
\bea
\braket{T^{\m_1\n_1}T^{\m_2\n_2}T^{\m_3\n_3}}_{Ren}=\mathcal{V}^{\mu_1\nu_1\mu_2\nu_2\mu_3\nu_3}_{traceless} + 
\braket{T^{\m_1\n_1}T^{\m_2\n_2}T^{\m_3\n_3}}_{anomaly}.
\eea
We are going to comment briefly on the implications of these results at diagrammatic level.

\begin{figure}[t]
\centering
\subfigure[]{\includegraphics[scale=0.49]{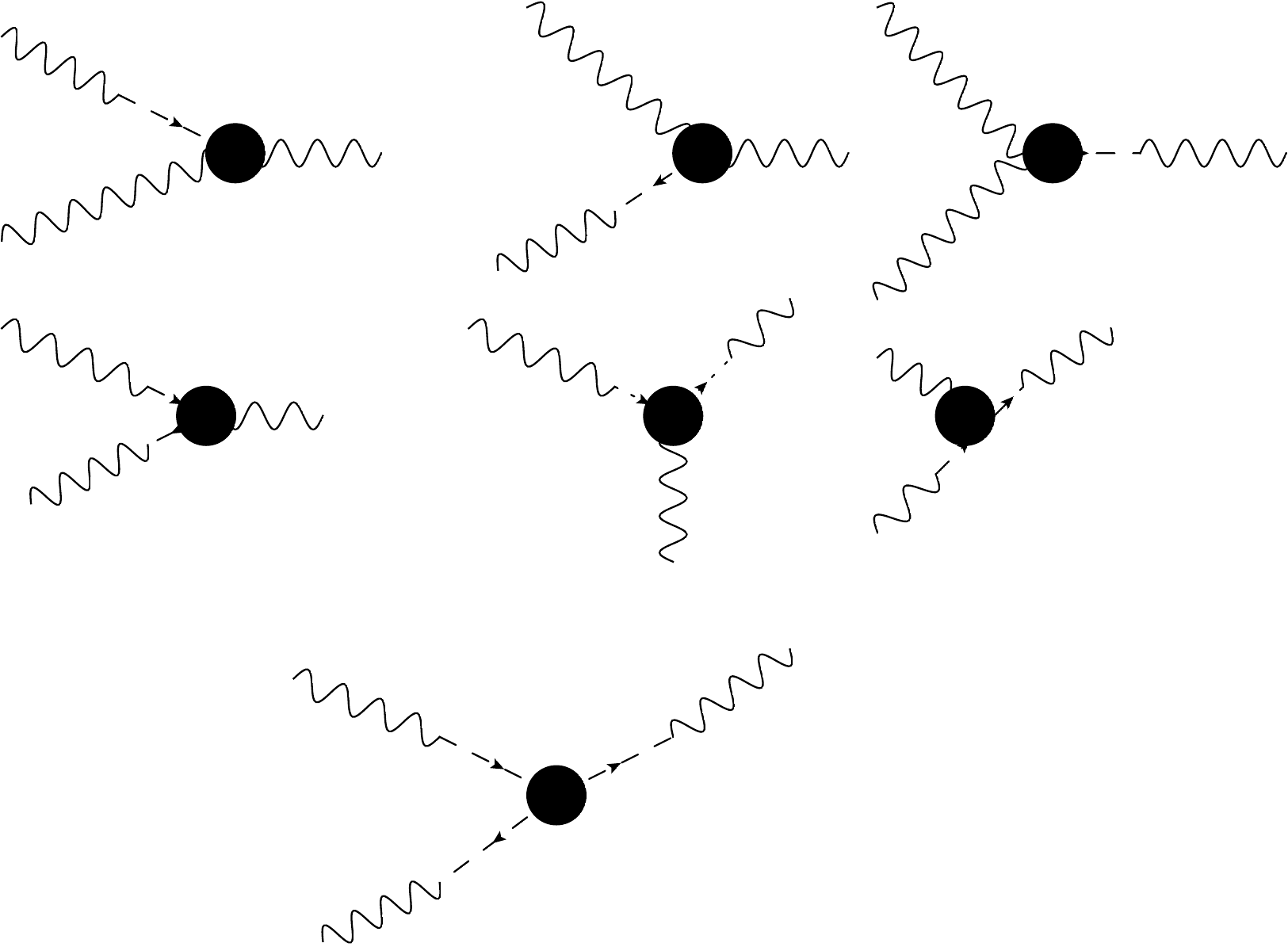}} 
\caption{Anomaly interactions mediated by the exchange of one, two or three poles. The poles are generated by the renormalization of the longitudinal sector of the $TTT$.}
\label{dec}
\end{figure}

\subsection{The perturbative structure of the $TTT$ and the poles separation} 
The structure of the poles in the $TTT$ is summarized in Fig. \ref{dec} where we have denoted with a dashed line the exchange of one or more massless $(\sim 1/p_i^2)$ interactions. In configuration space such extra terms, related to the renormalization of the correlator, are the natural generalization of the typical anomaly pole interaction found, for instance, in the case of the $TJJ$, where the effect of the anomaly is in the generation of a nonlocal interaction of the form \cite{Giannotti:2008cv,Armillis:2009pq,Armillis:2010qk}
 \bea
\mathcal{S}_{an}\sim \beta(e) \int d^4 x\, d^4 y R^{(1)}(x)  \left(\frac{1}{\square}\right)(x,y) F F(y)
 \eea
with $F$ being the QED field strength and $\beta(e)$ the corresponding beta function of the gauge coupling. In the $TTT$ case, as one can immediately figure out from \eqref{expans} such contributions can be rewritten as contribution to the anomaly action in the form
 \bea
\mathcal{S}_{an}\sim \int d^4 x\, d^4 y R^{(1)}(x)  \left(\frac{1}{\square}\right)(x,y) \left( \beta_b E^{(2)}(y) + \beta_a (C^2)^{(2)}(y)\right)
 \eea
and similar for the other contributions extracted from \eqref{expans}. Notice that each $\hat\pi$ projector in \eqref{expans} is accompanied by a corresponding anomaly (single) pole of the external invariants, generating contributions of the form $1/p_i^2$,  $1/(p_i^2 p_j^2) (i\neq j)$ and $1/(p_1^2 p_2^2 p_3^2)$, where multiple poles are connected to separate external graviton lines. Each momentum invariant appears as a single pole. One can use the correspondence 
\be
\frac{1}{p^2} \hat\pi^{\mu\nu} \leftrightarrow R^{(1)}\frac{1}{\square} 
\ee
to include such nonzero trace contributions into the anomaly action. This involves a multiplication of the vertex by the external fields together with an integration over all the internal points. As shown in \cite{Coriano:2017mux} such nonzero trace contributions are automatically generated by the nonlocal conformal anomaly action, which accounts for the entire expression \eqref{expans}. \\
The diagrammatic interpretation suggests a possible generalization of this result also to higher point functions, as one can easily guess, in a combination similar to that shown in Fig. \ref{dec}.\\
 Notice that the numerators of such decompositions, which correspond to single, double and triple traces are, obviously, purely polynomial in the external invariants, being derived from the anomaly functional, which is local in momentum space.\\

\section{Conclusions and perspectives}
We have presented a comparative study of the 3-graviton vertex $TTT$ in CFT's in momentum space. The analysis of conformal correlators is relatively knew, beyond the standard Lagrangian approach, though the interest in this approach is growing \cite{Isono:2018rrb,Gillioz:2018mto}. Our analysis extends a previous work on the $TJJ$ correlator \cite{Coriano:2018bbe} and on the same $TTT$ vertex given in \cite{Coriano:2012wp}, based on similar approaches. Building on the analysis presented in \cite{Bzowski:2013sza, Bzowski:2017poo}, we have gone over the reconstruction program proposed in those works from a perturbative perspective. We have also presented an independent analysis of the solutions of the CWI's. This is based on a new approach which exploits some properties of the solutions of the hypergeometric systems of differential equations associated with the CWI's and equivalent to them. The method is alternative to the approach presented in \cite{Bzowski:2013sza}, which requires a rather complex analysis of the singularities of the solutions, given in terms of 3-K integrals. 
The method has been extended by us also to 4-point functions, in the search for special solutions of such systems, which are controlled by a larger class of hypergeometrics, respect to the simpler solutions found for 3-point functions discussed here, as we will show in a forthcoming work.

The comparison with perturbation theory allows to provide drastic simplifications of the results for the vertex, while keeping, for specific dimensions, the generality of the conformal 
(non-Lagrangian) solution. At the same time, having established a direct link between the two 
- i.e. the perturbative and the non-Lagrangian formulations - this opens the way to several independent analysis of this vertex - entirely based on the Feynman's expansion. \\
This would allow to identify the singularities - and henceforth the anomaly poles - present in such correlator, from a simple and physical perspective based on the analysis of the Landau conditions of the basic (1-loop) diagrams generated by the matching, as done in the 
 simpler case of \cite{Coriano:2014gja} in a supersymmetric context.
The expression that we have presented of such vertex is the simplest one that can be written and down and in $d=4$ keeps its generality. Obviously, it is possible to extend our analysis to higher (even) dimensions by the inclusion of antisymmetric forms, building on the analysis of \cite{Bastianelli:2000rs} as a third (beside scalar and fermions) sector, which would provide an extension of the approach presented in our work. 

Finally, we have also discussed the organization of the result fo the $TTT$ - for its renormalized expression - in terms of a homogenous (zero trace) contribution and of an anomaly part. The anomaly (nonzero trace) part, is generated by the renormalization of 
the local components of the $TTT$.  
Our detailed analysis shows that such contributions are not an artifact of the parameterization of the form factors or can be attributed to a 
specific decomposition but are a general feature of CFT's and is related to renormalization. This is in agreement with the analysis \cite{Bzowski:2017poo} and, at the same time, with the predictions - limitedly to the anomaly part - coming from the nonlocal anomaly action \cite{Coriano:2017mux}.

\vspace{0.5cm}
\centerline{\bf Acknowledgements} 
We thank Emil Mottola for discussion and collaboration on a related project. We thank Kostas Skenderis, Paul McFadden, Fiorenzo Bastianelli, Loriano Bonora, Maxim Chernodub, Olindo Corradini, Luigi Delle Rose and Mirko Serino for discussions.  C.C. thanks Fiorenzo Bastianelli and Olindo Corradini for hospitality at the University of Bologna and Giacomo Cacciapaglia and 
Aldo Deandrea for hospitality at the Codyce 2018 workshop at the University of Lyon. This work is partially supported by INFN  Iniziativa specifica QFT-HEP. 

\appendix
\section{Secondary CWI's}\label{secondary}

The secondary conformal Ward identities are first-order partial differential equations and involve the semi-local information contained in $t^{\m\n}_{loc}$. In order to write them compactly, following \cite{2014JHEP...03..111B} we define the two differential operators
\begin{align}
\textup{L}_N&= p_1(p_1^2 + p_2^2 - p_3^2) \frac{\partial}{\partial p_1} + 2 p_1^2 p_2 \frac{\partial}{\partial p_2} + \left[ (2d - \Delta_1 - 2\Delta_2 +N)p_1^2 + (2\Delta_1-d)(p_3^2-p_2^2)  \right]\label{Ldef} \\
\textup{R} &= p_1 \frac{\partial}{\partial p_1} - (2\Delta_1-d) \label{Rdef}\,. 
\end{align}
as well as their symmetric versions
\begin{align}
&L'_N=L_N,\quad\text{with}\ p_1\leftrightarrow p_2\ \text{and}\ \D_1\leftrightarrow\D_2,\\
&R'=R,\qquad\text{with}\ p_1\mapsto p_2\ \text{and}\ \D_1\mapsto\D_2.
\end{align}
These operators depend on the conformal dimensions of the operators involved in the 3-point function under consideration, and additionally on a single parameter $N$ determined by the Ward identity in question. In the $\braket{TTT}$ case one finds considering the structure \eqref{StrucSWIS} one can find for $C_{3j}$, $j=1,\dots,7$
\begingroup\makeatletter\def\f@size{9}\check@mathfonts
\def\maketag@@@#1{\hbox{\m@th\large\normalfont#1}}%
\begin{equation}
\begin{split}
C_{31}&=-\sdfrac{2}{p_1^2}\left[L_6 A_1+R A_2-R A_2(p_2\leftrightarrow p_3)\right]\\
C_{32}&=-\sdfrac{1}{p_1^2}\left[L_4\,A_2+2p_1^2\,A_2+4RA_3-2RA_4(p_1\leftrightarrow p_3)\right]\\
C_{33}&=-\sdfrac{2}{p_1^2}\left[L_4\,A_2(p_1\leftrightarrow p_3)-R\,A_4+RA_4(p_2\leftrightarrow p_3)+2p_1^2(A_2(p_2\leftrightarrow p_3)-A_2)\right]\\
C_{34}&=-\sdfrac{1}{p_1^2}\left[L_4\,A_2(p_2\leftrightarrow p_3)-4R\,A_3(p_2\leftrightarrow p_3)+2RA_4(p_1\leftrightarrow p_3)-2p_1^2A_2(p_2\leftrightarrow p_3)\right]\\
C_{35}&=-\sdfrac{2}{p_1^2}\left[L_2\,A_3(p_1\leftrightarrow p_3)+p_1^2(A_4-A_4(p_2\leftrightarrow p_3)\right]\\
C_{36}&=-\sdfrac{1}{p_1^2}\left[L_2\,A_4+2R\,A_5+8p_1^2A_3(p_2\leftrightarrow p_3)-2p_1^2(A_4+A_4(p_1\leftrightarrow p_3))\right]\\
C_{37}&=-\sdfrac{1}{p_1^2}\left[L_2\,A_4(p_2\leftrightarrow p_3)-2R\,A_5-8p_1^2A_3+2p_1^2(A_4(p_2\leftrightarrow p_3)+A_4(p_1\leftrightarrow p_3))\right]\\
\end{split}
\end{equation}
and for $C_{4,j}$, $j=1,\dots, 7$
\begin{equation}
\begin{split}
C_{41}&=\sdfrac{2}{p_2^2}\left[L'_6 A_1+R' A_2-R' A_2(p_1\leftrightarrow p_3)\right]\\
C_{42}&=\sdfrac{1}{p_2^2}\left[L'_4\,A_2+2p_2^2\,A_2+4R'A_3-2R'A_4(p_2\leftrightarrow p_3)\right]\\
C_{43}&=\sdfrac{1}{p_2^2}\left[L'_4\,A_2(p_1\leftrightarrow p_3)-4R'\,A_3(p_1\leftrightarrow p_3)+2RA_4(p_2\leftrightarrow p_3)-2p_2^2A_2(p_1\leftrightarrow p_3)\right]\\
C_{44}&=\sdfrac{2}{p_2^2}\left[L'_4\,A_2(p_2\leftrightarrow p_3)-R'\,A_4+R'A_4(p_1\leftrightarrow p_3)-2p_2^2(A_2-A_2(p_1\leftrightarrow p_3))\right]\\
C_{45}&=\sdfrac{2}{p_2^2}\left[L'_2\,A_3(p_2\leftrightarrow p_3)+p_2^2(A_4-A_4(p_1\leftrightarrow p_3)\right]\\
C_{46}&=\sdfrac{1}{p_2^2}\left[L'_2\,A_4+2R'\,A_5+8p_2^2A_3(p_1\leftrightarrow p_3)-2p_2^2(A_4+A_4(p_2\leftrightarrow p_3))\right]\\
C_{47}&=\sdfrac{1}{p_2^2}\left[L'_2\,A_4(p_1\leftrightarrow p_3)-2R'\,A_5-8p_2^2A_3+2p_2^2(A_4(p_2\leftrightarrow p_3)+A_4(p_1\leftrightarrow p_3))\right]\\
\end{split}
\end{equation}
and finally for $C_{5j}$, $j=1,\dots, 7$
\begin{equation}
\begin{split}
C_{51}&=\sdfrac{2}{p_3^2}\bigg\{(L_6 -L'_6) A_1+(R+R')\big[A_2(p_2\leftrightarrow p_3)-A_2(p_1\leftrightarrow p_3)\big]+2(d+2)\big[A_2(p_2\leftrightarrow p_3)-A_2(p_1\leftrightarrow p_3)\big]\bigg\}\\[1.5ex]
C_{52}&=\sdfrac{2}{p_3^2}\bigg\{(L_4-L'_4)\,A_2+2p_3^2\,\big[A_2(p_2\leftrightarrow p_3)-A_2(p_1\leftrightarrow p_3)\big]\notag\\
&\hspace{2cm}+\big(R+R'+2(d+1)\big)\big[A_4(p_1\leftrightarrow p_3)-A_4(p_2\leftrightarrow p_3)\big]\bigg\}\\[1.5ex]
C_{53}&=\sdfrac{1}{p_3^2}\bigg\{(L_4-L'_4)\,A_2(p_1\leftrightarrow p_3)+2p_3^2A_2(p_1\leftrightarrow p_3)-2\big(R'+R+2(d+1)\big)\,\big[2A_3(p_1\leftrightarrow p_3)-A_4\big]\bigg\}\\
C_{54}&=\sdfrac{1}{p_3^2}\bigg\{(L_4-L'_4)\,A_2(p_2\leftrightarrow p_3)-2p_3^2A_2(p_2\leftrightarrow p_3)+2\big(R'+R+2(d+1)\big)\,\big[2A_3(p_2\leftrightarrow p_3)-A_4\big]\bigg\}\\
C_{55}&=\sdfrac{2}{p_3^2}\bigg\{(L_2-L'_2)\,A_3-p_3^2\big[A_4(p_2\leftrightarrow p_3)-A_4(p_1\leftrightarrow p_3)\big]\bigg\}\\
C_{56}&=\sdfrac{1}{p_3^2}\bigg\{(L_2-L'_2)\,A_4(p_1\leftrightarrow p_3)+2p_3^2\big[4A_3(p_2\leftrightarrow p_3)-A_4-A_4(p_1\leftrightarrow p_3)\big]-2\big(R+R'+2d\big)A_5\bigg\}\\
C_{57}&=\sdfrac{1}{p_3^2}\bigg\{(L_2-L'_2)\,A_4(p_2\leftrightarrow p_3)-2p_3^2\big[4A_3(p_1\leftrightarrow p_3)-A_4-A_4(p_2\leftrightarrow p_3)\big]+2\big(R+R'+2d\big)A_5\bigg\}
\end{split}
\end{equation}
\endgroup

%%%%%%%%%%%%%%%%%%%%%%%%%%%%%%%

\section{Fuchsian solutions of the primary CWI's}
\label{fuchs}
\subsection{$A_4$ solution}
Under the exchange of two momenta, $A_2(p_2\leftrightarrow p_3)$ becomes

\begin{align}
A_2(p_2\leftrightarrow p_3)&= p_3^{d - 4}\sum_{a b} x^a y^{\frac{d}{2} - 2-a-b}\bigg[c^{(2)}(a,b)\,F_4\left(\alpha +2, \beta+2; \gamma, \gamma'; \sdfrac{x}{y}, \sdfrac{1}{y}\right)\notag\\
&\hspace{-0.2cm}+ \frac{2\,c^{(1)}(a,b)}{\big(\b+2\big)}F_4(\alpha +3, \beta+2; \gamma, \gamma'; \sdfrac{x}{y}, \sdfrac{1}{y}\bigg)\bigg]. 
\end{align}
In order to solve the equation \eqref{eqA4} we will be needing the transformation property of $F_4$ given by \eqref{transfF4}.
Once plugged into the explicit expression on $A_4(p_2\leftrightarrow p_3)$, and separating its 4 indicial components $(a_i, b_i)$ we obtain
\begingroup\makeatletter\def\f@size{8.5}\check@mathfonts
\def\maketag@@@#1{\hbox{\m@th\large\normalfont#1}}%
\begin{align}
A_2(p_2\leftrightarrow p_3)&=p_3^{d-4}\Bigg\{\ c^{(2)}\left(0,\frac{d}{2}\right)\frac{(d-2)}{(d+2)}F_4\left(2-\frac{d}{2},2,1-\frac{d}{2},1-\frac{d}{2},x,y\right)\notag\\
&\hspace{4cm}+\ c^{(1)}\left(0,\frac{d}{2}\right)\,\frac{d(d-2)}{(d+2)(d+4)}\,F_4\left(2-\frac{d}{2},2,1-\frac{d}{2},-\frac{d}{2},x,y\right)\Bigg\}\notag\\
&\hspace{-2cm}+p_3^{d-4}x^{d/2}\Bigg\{\ c^{(2)}\left(\frac{d}{2},0\right)\,F_4\left(\frac{d}{2}+2,2,\frac{d}{2}+1,1-\frac{d}{2},x,y\right)+\ c^{(1)}\left(\frac{d}{2},0\right)\,\frac{d}{(d+4)}\,F_4\left(\frac{d}{2}+2,2,\frac{d}{2}+1,-\frac{d}{2},x,y\right)\Bigg\}\notag\\
&\hspace{-2cm}+p_3^{d-4}y^{d/2}\Bigg\{\ c^{(2)}\left(0,0\right)\frac{(d+2)}{(d-2)}F_4\left(2,\frac{d}{2}+2,1-\frac{d}{2},\frac{d}{2}+1,x,y\right)\notag\\
&\hspace{4cm}+ c^{(1)}\left(0,0\right)\,\frac{4(d+4)y}{(d-4)(d-2)}F_4\left(3,\frac{d}{2}+3,1-\frac{d}{2},\frac{d}{2}+2,x,y\right)\Bigg\}\notag\\
&\hspace{-2cm}+p_3^{d-4}y^{d/2}x^{d/2}(-1)^d\Bigg\{-\ c^{(2)}\left(\frac{d}{2},\frac{d}{2}\right)F_4\left(d+2,\frac{d}{2}+2,\frac{d}{2}+1,\frac{d}{2}+1,x,y\right)\,\notag\\
&\hspace{4cm}+\ 4c^{(1)}\left(\frac{d}{2},\frac{d}{2}\right)\frac{y}{d+2}F_4\left(d+3,\frac{d}{2}+3,\frac{d}{2}+1,\frac{d}{2}+2,x,y\right)\Bigg\}
\end{align}
\endgroup
and with a similar one for $A_4(p_1\leftrightarrow p_3)$, that we omit. At this stage, using the property of $F_4$ given in \eqref{exs} we can rearrange the expression of $A_2(p_2\leftrightarrow p_3)$ as
\begingroup\makeatletter\def\f@size{8.5}\check@mathfonts
\def\maketag@@@#1{\hbox{\m@th\large\normalfont#1}}%
\begin{align}
A_2(p_2\leftrightarrow p_3)&=p_3^{d-4}\Bigg\{\ c^{(2)}\left(0,\frac{d}{2}\right)\frac{(d-2)}{(d+2)}F_4\left(2-\frac{d}{2},2,1-\frac{d}{2},1-\frac{d}{2},x,y\right)\notag\\
&\hspace{-1.5cm}+\ c^{(1)}\left(0,\frac{d}{2}\right)\,\frac{d(d-2)}{(d+2)(d+4)}\,F_4\left(2-\frac{d}{2},2,1-\frac{d}{2},-\frac{d}{2},x,y\right)\Bigg\}+p_3^{d-4}x^{d/2}\Bigg\{c^{(2)}\left(\frac{d}{2},0\right)\notag\\
 &\hspace{-1.5cm}\times\,F_4\left(\frac{d}{2}+2,2,\frac{d}{2}+1,1-\frac{d}{2},x,y\right)+\ c^{(1)}\left(\frac{d}{2},0\right)\,\frac{d}{(d+4)}\,F_4\left(\frac{d}{2}+2,2,\frac{d}{2}+1,-\frac{d}{2},x,y\right)\Bigg\}\notag\\
&\hspace{1.5cm}+p_3^{d-4}y^{d/2}\Bigg\{\ c^{(2)}\left(0,0\right)\frac{(d+2)}{(d-2)}F_4\left(2,\frac{d}{2}+2,1-\frac{d}{2},\frac{d}{2}+1,x,y\right)\notag
\end{align}
\begin{align}
&\hspace{1.5cm}+c^{(1)}\left(0,0\right)\,\frac{d(d+2)}{(d-4)(d-2)} \Bigg[F_4\left(2,\frac{d}{2}+2,1-\frac{d}{2},\frac{d}{2},x,y\right)-F_4\left(2,\frac{d}{2}+2,1-\frac{d}{2},\frac{d}{2}+1,x,y\right)\Bigg]\Bigg\}\notag\\
&\hspace{2cm}+p_3^{d-4}y^{d/2}x^{d/2}(-1)^d\Bigg\{-\ c^{(2)}\left(\frac{d}{2},\frac{d}{2}\right)F_4\left(d+2,\frac{d}{2}+2,\frac{d}{2}+1,\frac{d}{2}+1,x,y\right)\,\notag\\
&\hspace{1cm}+\ c^{(1)}\left(\frac{d}{2},\frac{d}{2}\right)\frac{2d}{(d+4)(d+2)}\Bigg[F_4\left(d+2,\frac{d}{2}+2,\frac{d}{2}+1,\frac{d}{2},x,y\right)-F_4\left(d+2,\frac{d}{2}+2,\frac{d}{2}+1,\frac{d}{2}+1,x,y\right)\Bigg]\Bigg\}.
\end{align}
\endgroup
A similar expression can be derived for $A_2(p_1\leftrightarrow p_3)$.
The  constants appearing in the solution for the form factor $A_4$ can be related as
\begingroup\makeatletter\def\f@size{8.5}\check@mathfonts
\def\maketag@@@#1{\hbox{\m@th\large\normalfont#1}}%
\begin{align}
&\begin{aligned}
c_1^{(4)}\left(0,0\right)&=\frac{2}{(d+2)}c^{(2)}\left(0,\frac{d}{2}\right)+\frac{2}{(d+2)}c_3^{(4)}\left(0,0\right)\\
c_1^{(4)}\left(0,\frac{d}{2}\right)&=-c^{(2)}\left(0,\frac{d}{2}\right)-\frac{(d+2)}{d}c_3^{(4)}\left(\frac{d}{2},0\right)\\
c_1^{(4)}\left(\frac{d}{2},0\right)&=c_1^{(4)}\left(0,\frac{d}{2}\right)\\
c_1^{(4)}\left(\frac{d}{2},\frac{d}{2}\right)&=\frac{2\sec\left(\frac{\p\,d}{2}\right)\left[\G\left(1-\frac{d}{2}\right)\right]^2}{(d^2+6d+8)\G(-d-1)}c^{(1)}\left(0,\frac{d}{2}\right)+\frac{2(-1)^d}{d+2}c^{(2)}\left(\frac{d}{2},\frac{d}{2}\right)\notag\\
&-\frac{2(d+1)}{d}c_3^{(4)}\left(\frac{d}{2},\frac{d}{2}\right)-\frac{2(-1)^{d/2}\G\left(1-\frac{d}{2}\right)\G\left(-\frac{d}{2}\right)}{(d+2)\G(-d-1)}c^{(2)}\left(0,\frac{d}{2}\right)\\[2ex]
\end{aligned}\notag\\
\end{align}
\endgroup
\subsection{Relating the constants in the $A_5$ solution}
 \label{condition}
The constants in the expression of $A_5$ are fixed as follows

\begingroup\makeatletter\def\f@size{8.5}\check@mathfonts
\def\maketag@@@#1{\hbox{\m@th\large\normalfont#1}}%

\begin{align}
\begin{aligned}[c]
c^{(2)}\left(\frac{d}{2},0\right)&=c^{(2)}\left(0,\frac{d}{2}\right)\equiv C_2\\
c^{(2)}\left(0,0\right)&=\frac{(d-2)}{(d+2)}\,C_2\\
c^{(2)}\left(\frac{d}{2},\frac{d}{2}\right)&=\frac{\G\left(-\frac{d}{2}\right)\,\G\left(d+2\right)}{\G\left(\frac{d}{2}\right)}C_2\\
\end{aligned}
\hspace{2cm}
\begin{aligned}[c]
c^{(3)}\left(\frac{d}{2},0\right)&=c^{(3)}\left(0,\frac{d}{2}\right)\equiv C_3\\[0.8ex]
c^{(3)}\left(0,0\right)&=-C_3\\[1ex]
c^{(3)}\left(\frac{d}{2},\frac{d}{2}\right)&=\frac{\G\left(-\frac{d}{2}\right)\,\G\left(d+1\right)}{\G\left(\frac{d}{2}\right)}C_3\\
\end{aligned}
\end{align}

\begin{align}
\begin{aligned}[c]
c^{(5)}\left(\frac{d}{2},0\right)&=c^{(5)}\left(0,\frac{d}{2}\right)=C_5=c^{(5)}\left(0,0\right)\\
c^{(5)}\left(\frac{d}{2},\frac{d}{2}\right)&=-\frac{d^2\G\left(-\frac{d}{2}-1\right)\G(d+1)}{8\G\left(\frac{d}{2}\right)}C_2+\frac{\G\left(-\frac{d}{2}\right)\,\G(d+1)}{2\G\left(\frac{d}{2}+1\right)}C_5
\end{aligned}
\end{align}
\begin{align}
\begin{aligned}[c]
c_1^{(5)}\left(\frac{d}{2},0\right)&=c_1^{(5)}\left(0,\frac{d}{2}\right)=-\frac{d^2}{2(d+2)(d+4)}C_1=c_1^{(5)}\left(0,0\right)\\
c_1^{(5)}\left(\frac{d}{2},\frac{d}{2}\right)&=-\frac{d\,\G\left(1-\frac{d}{2}\right)\G(d+1)}{(d+2)(d+4)\G\left(\frac{d}{2}\right)}C_1
\end{aligned}
\end{align}
\begin{align}
\begin{aligned}[c]
c_2^{(5)}\left(0,0\right)&=c_4^{(5)}\left(0,0\right)=-\frac{d}{2(d+2)}C_2\\
c_2^{(5)}\left(\frac{d}{2},0\right)&=c_4^{(5)}\left(0,\frac{d}{2}\right)=\frac{d^2}{2(d+2)(d+4)}C_1-\frac{d}{d+2}C_2\\
c_2^{(5)}\left(0,\frac{d}{2}\right)&=c_4^{(5)}\left(\frac{d}{2},0\right)=\frac{d}{d+2}C_2\notag\\
c_2^{(5)}\left(\frac{d}{2},\frac{d}{2}\right)&=c_4^{(5)}\left(\frac{d}{2},\frac{d}{2}\right)=\frac{d\,\G\left(-\frac{d}{2}-1\right)\G(d+2)}{4(d+1)\G\left(\frac{d}{2}\right)}C_2-\frac{d^2\,\G\left(d+2\right)\G\left(-\frac{d}{2}-1\right)}{4(d+1)(d+4)\,\G\left(\frac{d}{2}\right)}C_1
\end{aligned}
\end{align}
\begin{align}
\begin{aligned}[c]
c^{(5)}_3\left(0,0\right)&=0\\
c^{(5)}_3\left(0,\frac{d}{2}\right)&=c^{(5)}_3\left(\frac{d}{2},0\right)=-C_2\frac{d}{d+2}\\
c^{(5)}_3\left(\frac{d}{2},\frac{d}{2}\right)&=-\frac{\pi\,d^2\csc\left(\frac{\pi\,d}{2}\right)\G(d+1)}{8\G\left(\frac{d}{2}+3\right)\G\left(\frac{d}{2}\right)}C_1-\frac{\pi\,d(d+4)\csc\left(\frac{\pi\,d}{2}\right)\G(d+1)}{4\G\left(\frac{d}{2}+3\right)\G\left(\frac{d}{2}\right)}C_2
\end{aligned}
\end{align}
\begin{align}
\begin{aligned}[c]
c_5^{(5)}\left(0,0\right)&=c_6^{(5)}\left(0,0\right)=0\\
c_5^{(5)}\left(\frac{d}{2},0\right)&=c_6^{(5)}\left(0,\frac{d}{2}\right)=0\notag\\
c_5^{(5)}\left(0,\frac{d}{2}\right)&=c_6^{(5)}\left(\frac{d}{2},0\right)=-\frac{d^2}{2(d+2)}C_2\notag\\
c_5^{(5)}\left(\frac{d}{2},\frac{d}{2}\right)&=c_6^{(5)}\left(\frac{d}{2},\frac{d}{2}\right)=\frac{d^2\,\G\left(-\frac{d}{2}-1\right)\,\G(d+1)}{8\,\G\left(\frac{d}{2}\right)}C_2
\end{aligned}
\end{align}
\begin{align}
\begin{aligned}[c]
c_7^{(5)}\left(0,0\right)&=-\frac{d^2}{2(d+2)}C_2\notag\\
c_7^{(5)}\left(\frac{d}{2},0\right)&=c_7^{(5)}\left(0,\frac{d}{2}\right)=\frac{d^2}{2(d+2)}C_2\notag\\
c_7^{(5)}\left(\frac{d}{2},\frac{d}{2}\right)&=-\frac{d^2\,\G\left(-\frac{d}{2}-1\right)\,\G(d+1)}{8\,\G\left(\frac{d}{2}\right)}C_2
\end{aligned}
\end{align}

\endgroup

\section{Summary: Reconstructions for odd dimensions} 
We summarize the main steps collecting all the equations needed for the reconstruction in odd dimensions using the matched solutions in $d =3$ and 5. The vertex is reconstructed from the relation
\begin{align}
\label{eqss1}
\braket{T^{\mu_1\nu_1}\,T^{\mu_2\n_2}\,T^{\mu_3\n_3}}&=\braket{t^{\mu_1\nu_1}\,t^{\mu_2\n_2}\,t^{\mu_3\n_3}}+\braket{T^{\mu_1\nu_1}\,T^{\mu_2\n_2}\,t_{loc}^{\mu_3\n_3}}+\braket{T^{\mu_1\nu_1}\,t_{loc}^{\mu_2\n_2}\,T^{\mu_3\n_3}}\notag\\
&+\braket{t_{loc}^{\mu_1\nu_1}\,T^{\mu_2\n_2}\,T^{\mu_3\n_3}}-\braket{T^{\mu_1\nu_1}\,t_{loc}^{\mu_2\n_2}\,t_{loc}^{\mu_3\n_3}}-\braket{t_{loc}^{\mu_1\nu_1}\,t_{loc}^{\mu_2\n_2}\,T^{\mu_3\n_3}}\notag\\
&-\braket{t_{loc}^{\mu_1\nu_1}\,T^{\mu_2\n_2}\,t_{loc}^{\mu_3\n_3}}+\braket{t_{loc}^{\mu_1\nu_1}\,t_{loc}^{\mu_2\n_2}\,t_{loc}^{\mu_3\n_3}}
\end{align}
The transverse traceless sector $\braket{ttt}$ is constructed using the corresponsing form  factors $A_i$ in the respective dimensions 
using \eqref{DecompTTT}. Their explicit expressions are found in Section \ref{d35}. For the local terms in (\ref{eqss}) we use eq. \eqref{loc} 
\begingroup\makeatletter\def\f@size{10}\check@mathfonts
\def\maketag@@@#1{\hbox{\m@th\large\normalfont#1}}%
\begin{align}
&\braket{t_{loc}^{\mu_1\nu_1}(p_1)T^{\mu_2\nu_2}(p_2)T^{\mu_3\nu_3}(\bar{p}_3)}=\Big(\mathcal{I}^{\mu_1\nu_1}_{\a_1}(p_1)\,p_{1\b_1}+\frac{\p^{\mu_1\nu_1}(p_1)}{(d-1)}\d_{\a_1\b_1}\Big)\braket{T^{\a_1\b_1}(p_1)T^{\mu_2\nu_2}(p_2)T^{\mu_3\nu_3}(\bar{p}_3)}\notag\\
&\hspace{1.5cm}=-\frac{2\,\p^{\mu_1\nu_1}(p_1)}{(d-1)}\Big[\braket{T^{\a_2\b_2}(p_1+p_2)T^{\mu_3\nu_3}(-p_1-p_2)}+\braket{T^{\a_2\b_2}(p_2)T^{\mu_3\nu_3}(-p_2)} \Big]
\notag\\
&\hspace{1.8cm}+\mathcal{I}^{\mu_1\nu_1}_{\a_1}(p_1)\Big[-p_2^{\a_1}\braket{T^{\a_2\b_2}(p_1+p_2)T^{\mu_3\nu_3}(-p_1-p_2)} -p_3^{\a_1}\braket{T^{\a_2\b_2}(p_2)T^{\mu_3\nu_3}(-p_2)} \notag\\
&\hspace{1.8cm}+p_{2\b}\Big(\d^{\a_1\a_2}\braket{T^{\b\b_2}(p_1+p_2)T^{\mu_3\nu_3}(-p_1-p_2)} +\d^{\a_1\b_2}\braket{T^{\b\a_2}(p_1+p_2)T^{\mu_3\nu_3}(-p_1-p_2)} \Big)\notag\\
&\hspace{1.8cm}+p_{3\b}\Big(\d^{\a_1\mu_3}\braket{T^{\b_2\a_2}(p_2)T^{\b\nu_3}(-p_2)} +\d^{\a_1\nu_3}\braket{T^{\a_2\b_2}(p_2)T^{\mu_3\b}(-p_2)} \Big)\Big]
\end{align}
\endgroup
with the two-point functions defined as in 
\be
\braket{T^{\m\n}(p)T^{\alpha\beta}(-p)}=c_T\,\Pi^{\m\n\alpha\beta}(p)\,\G\left(-\frac{d}{2}+\frac{\epsilon}{2}\right)\,p^{d- \epsilon}
\end{equation}
with 
\begin{equation}
c_T=\frac{3\p^\frac{5}{2}}{128}\,\big(n_S+4n_F\big)\,
\end{equation}
in $d=3$ and 
\begin{equation}
c_T=\frac{5\p^\frac{7}{2}}{1024}\,\big(n_S+8n_F\big)\,
\end{equation}
in $d=5$. In $d=3$ we generate this way the most general CFT solution which matches the analogous case presented in \cite{2014JHEP...03..111B}. \\
The $\braket{t_{loc} t_{loc} T}$ and $\braket{t_{loc} t_{loc} t_{loc}}$ terms are obtained from  
\begingroup\makeatletter\def\f@size{8.5}\check@mathfonts
\def\maketag@@@#1{\hbox{\m@th\large\normalfont#1}}%
\begin{align}
&\braket{t_{loc}^{\mu_1\nu_1}(p_1)t_{loc}^{\mu_2\nu_2}(p_2)T^{\mu_3\nu_3}(\bar{p}_3)}=\Big(\mathcal{I}^{\mu_2\nu_2}_{\a_2}(p_2)\,p_{2\b_2}+\frac{\p^{\mu_2\nu_2}(p_2)}{(d-1)}\d_{\a_2\b_2}\Big)\notag\\
&\hspace{1.5cm}\times\Bigg\{-\frac{2\,\p^{\mu_1\nu_1}(p_1)}{(d-1)}\Big[\braket{T^{\a_2\b_2}(p_1+p_2)T^{\mu_3\nu_3}(-p_1-p_2)}+\braket{T^{\a_2\b_2}(p_2)T^{\mu_3\nu_3}(-p_2)} \Big]
\notag\\
&\hspace{1.8cm}+\mathcal{I}^{\mu_1\nu_1}_{\a_1}(p_1)\Big[-p_2^{\a_1}\braket{T^{\a_2\b_2}(p_1+p_2)T^{\mu_3\nu_3}(-p_1-p_2)} -p_3^{\a_1}\braket{T^{\a_2\b_2}(p_2)T^{\mu_3\nu_3}(-p_2)} \notag\\
&\hspace{1.8cm}+p_{2\b}\Big(\d^{\a_1\a_2}\braket{T^{\b\b_2}(p_1+p_2)T^{\mu_3\nu_3}(-p_1-p_2)} +\d^{\a_1\b_2}\braket{T^{\b\a_2}(p_1+p_2)T^{\mu_3\nu_3}(-p_1-p_2)} \Big)\notag\\
&\hspace{1.8cm}+p_{3\b}\Big(\d^{\a_1\mu_3}\braket{T^{\b_2\a_2}(p_2)T^{\b\nu_3}(-p_2)} +\d^{\a_1\nu_3}\braket{T^{\a_2\b_2}(p_2)T^{\mu_3\b}(-p_2)} \Big)\Big]\Bigg\}
\end{align}
\begin{align}
&\braket{t_{loc}^{\mu_1\nu_1}(p_1)t_{loc}^{\mu_2\nu_2}(p_2)t_{loc}^{\mu_3\nu_3}(\bar{p}_3)}=\left(\mathcal{I}^{\mu_3\nu_3}_{\a_3}(\bar{p}_3)\,\bar{p}_{3\b_3}+\frac{\p^{\mu_3\nu_3}(\bar{p}_3)}{(d-1)}\d_{\a_3\b_3}\right)\left(\mathcal{I}^{\mu_2\nu_2}_{\a_2}(p_2)\,p_{2\b_2}+\frac{\p^{\mu_2\nu_2}(p_2)}{(d-1)}\d_{\a_2\b_2}\right)\notag\\
&\hspace{1.5cm}\times\Bigg\{-\frac{2\,\p^{\mu_1\nu_1}(p_1)}{(d-1)}\Big[\braket{T^{\a_2\b_2}(p_1+p_2)T^{\a_3\b_3}(-p_1-p_2)}+\braket{T^{\a_2\b_2}(p_2)T^{\a_3\b_3}(-p_2)} \Big]
\notag\\
&\hspace{1.5cm}\quad+\mathcal{I}^{\mu_1\nu_1}_{\a_1}(p_1)\Big[-p_2^{\a_1}\braket{T^{\a_2\b_2}(p_1+p_2)T^{\a_3\b_3}(-p_1-p_2)} -p_3^{\a_1}\braket{T^{\a_2\b_2}(p_2)T^{\a_3\b_3}(-p_2)} \notag\\
&\hspace{1.5cm}\quad+p_{2\b}\Big(\d^{\a_1\a_2}\braket{T^{\b\b_2}(p_1+p_2)T^{\a_3\b_3}(-p_1-p_2)} +\d^{\a_1\b_2}\braket{T^{\b\a_2}(p_1+p_2)T^{\a_3\b_3}(-p_1-p_2)} \Big)\notag\\
&\hspace{1.5cm}\quad+p_{3\b}\Big(\d^{\a_1\a_3}\braket{T^{\b_2\a_2}(p_2)T^{\b\b_3}(-p_2)} +\d^{\a_1\b_3}\braket{T^{\a_2\b_2}(p_2)T^{\a_3\b}(-p_2)} \Big)\Big]\Bigg\}
\end{align}
\endgroup
by inserting the corresponding values of $d$. The matching with the analogous solution given in \cite{2014JHEP...03..111B}
is obtained, in $d=3$, for 
\bea
\a_1=\frac{\p^3(n_S-4n_F)}{480}, \qquad \a_2=\frac{\p^3\,n_F}{6}, \qquad c_T=\frac{3\p^{5/2}}{128}(n_S+4n_F), \qquad c_g=0 
\eea
and in $d=5$ we use similar matching conditions 
\bea
\a_1=\frac{\p^4(n_S-4n_F)}{560 \times 72}, \qquad \a_2=\frac{\p^4\,n_F}{240}, \qquad c_T=\frac{5\p^{7/2}}{1024}(n_S+8n_F).
\eea

\section{Vertices}\label{Appendix1}

We have shown in \figref{vertices} a list of all the vertices which are needed for the momentum space computation of the $TTT$ correlator. We list them below and they are computed by taking functional derivatives of the action in order to allows to keep multi-graviton correlators symmetric. We use the letter $V$ to indicate the vertex, the subscript is referred to the fields involved and the Greek indices are linked to the Lorentz structure of the space-time. Furthermore about the momenta convention, we consider the graviton momenta incoming as well as $k_1$, instead of the out coming $k_2$ momentum as pictured in \figref{vertices}. In order to simplify the notation, we introduce the tensor components
\begin{align}
A^{\m_1\n_1\m\n}&\equiv\d^{\mu_1\nu_1}\d^{\m\n}-2\d^{\m(\m_1}\d^{\n_1)\n}\nn
B^{\m_1\n_1\m\n}&\equiv\d^{\mu_1\nu_1}\d^{\m\n}-\d^{\m(\m_1}\d^{\n_1)\n}\nn
C^{\m_1\n_1\m_2\n_2\m\n}&\equiv\d^{\m(\m_1}\d^{\n_1)(\m_2}\d^{\n_2)\n}+\d^{\m(\m_2}\d^{\n_2)(\m_1}\d^{\n_1)\n}\nn
\tilde{C}^{\m_1\n_1\m_2\n_2\m\n}&\equiv\d^{\m(\m_1}\d^{\n_1)(\m_2}\d^{\n_2)\n}\nn
D^{\m_1\n_1\m_2\n_2\m\n}&\equiv\d^{\m_1\n_1}\d^{\m(\m_2}\d^{\n_2)\n}+\d^{\m_2\n_2}\d^{\m(\m_1}\d^{\n_1)\n}\nn
E^{\m_1\n_1\m_2\n_2\m\n}&\equiv\d^{\m_1\n_1}B^{\m_2\n_2\m\n}+C^{\m_1\n_1\m_2\n_2\m\n},\nn
F^{\a_1\a_2\m\n}&\equiv\d^{\a_1[\m}\d^{\n]\a_2}\nn
\tilde{F}^{\a_1\a_2\m\n}&\equiv\d^{\a_1(\n}\d^{\m)\a_2}\nn
\tilde{F}^{\a_1\a_2}_{\m\n}&\equiv\d^{(\a_1}_{\n}\d_{\m}^{\a_2)}\nn
G^{\m_1\n_1\a_1\a_2\m\n}&\equiv\d^{\m[\n}\d^{\a_2](\m_1}\d^{\n_1)\a_1}+\d^{\a_1[\a_2}\d^{\n](\m_1}\d^{\n_1)\m}\nn
H^{\m_1\n_1\m_2\n_1\a_1\a_2\m\n}&\equiv A^{\m_1\n_1\m\a_1}\tilde{F}^{\m_2\n_2\n\a_2}-A^{\m_2\n_2\m\a_1}\tilde{F}^{\m_1\n_1\n\a_2}\nn
I^{\m_1\n_1\m_2\n_2\a_1\a_1\m\n}&\equiv\d^{\m_1\n_1}D^{\m\a_1\n\a_2\m_2\n_2}-\sdfrac{1}{2}\d^{\a_1\m}\d^{\a_2\n}A^{\m_1\n_1\m_2\n_2}
\end{align}
where we indicate with the circle brackets the symmetrization of the indices and the square brackets the anti-symmetrization of the indices, as follows
\begin{align}
\d^{\m(\m_1}\d^{\n_1)\n}&\equiv\sdfrac{1}{2}\bigg(\d^{\m\m_1}\d^{\n_1\n}+\d^{\m\n_1}\d^{\m_1\n}\bigg)\nn
\d^{\m[\m_1}\d^{\n_1]\n}&\equiv\sdfrac{1}{2}\bigg(\d^{\m\m_1}\d^{\n_1\n}-\d^{\m\n_1}\d^{\m_1\n}\bigg)
\end{align}
In the scalar sectors we obtain
\begin{align}
V^{\mu_1\nu_1}_{T\phi\phi}(k_1,k_2)&=\sdfrac{1}{2}A^{\m_1\n_1\m\n}\,k_{1\nu}\,k_{2\m}+\c\,B^{\m_1\n_1\m\n}\,(k_1-k_2)_\mu\,(k_1-k_2)_\nu\\[2ex]
V^{\mu_1\nu_1\mu_2\nu_2}_{TT\phi\phi}(p_2,k_1,k_2)&=\left(\sdfrac{1}{4}A^{\m_1\n_1\m_2\n_2}\d^{\m\n}+C^{\m_1\n_1\m_2\m_2\m\n}-\sdfrac{1}{2}D^{\m_1\n_1\m_2\n_2\m\n}\right)\,k_{1\nu}\,k_{2\m}\notag\\
&\hspace{-2cm}+\sdfrac{\c}{2}\bigg[\sdfrac{1}{2}\,\Big(E^{\m_1\n_1\m_2\n_2\m\n}-D^{\m_2\n_2\m\n\m_1\n_1}\Big)\,p_{2\m}p_{2\n}+\sdfrac{1}{2}\,\Big(E^{\m_1\n_1\m_2\n_2\m\n}-D^{\m_1\n_1\m\n\m_2\n_2}\Big)\,p_{2\m}(k_2-k_1)_\n\notag\\
&\hspace{1cm}+\Big(C^{\m_1\n_1\m_2\n_2\m\n}-D^{\m_1\n_1\m\n\m_2\n_2}\Big)(k_2-k_1)_\m(k_2-k_1)_\n\bigg].
\end{align}
 
In the fermion sector
\begin{align}
V^{\mu_1\nu_1}_{T\bar\psi\psi}(k_1,k_2)&=\sdfrac{1}{4}\,B^{\m_1\n_1\m\n}\,\g_\n\,(k_1+k_2)_\m\\[2ex]
V^{\mu_1\nu_1\m_2\n_2}_{TT\bar\psi\psi}(p_2,k_1,k_2)&=\sdfrac{1}{8}\bigg[\d^{\m\n}A^{\m_1\n_1\m_2\n_2}-D^{\m_1\n_1\m_2\n_2\m\n}+C^{\m_1\n_1\m_2\n_2\m\n}+\tilde{C}^{\m_2\n_2\m_1\n_1\m\n}\bigg]\,\g_\n\,(k_1+k_2)_\m\notag\\
&\hspace{1cm}+\sdfrac{1}{32}\tilde{C}^{\m_1\n_1\m_2\n_2\n\m}\,p_2^\sigma\,\left\{\g_\n,\left[\g_\m,\g_\sigma\right]\,\right\}
\end{align}
where we notice that the spin connection contribute to the two gravitons and two fermions vertex. However one can prove that this term does not contribute to the bubble diagrams. 

In the gauge sector we split the contribution of the Maxwell action with respect to the gauge fixing contribution. We labelled the first type of contribution to the vertices with a subscript $M$, instead of the second with $GF$. Then in this case we obtain for the pure gauge term
\begin{align}
V^{\m_1\n_1\a_1\a_2}_{TAA,\,M}(k_1,k_2)&=\bigg(\d^{\m_1\n_1}F^{\a_1\m\n\a_2}+2\,G^{\m_1\n_1\a_1\a_2\m\n}\bigg)\,k_{1\m}\,k_{2\n}\\[2ex]
V^{\m_1\n_1\m_2\n_2\a_1\a_2}_{TTAA,\,M}(k_1,k_2)&=\bigg[-\sdfrac{1}{2}A^{\m_1\n_1\m_2\n_2}F^{\m\a_1\n\a_2}+\d^{\m_2\n_2}\,G^{\m_1\n_1\a_1\a_2\m\n}+\d^{\m_1\n_1}\,G^{\m_2\n_2\a_1\a_2\m\n}\notag\\
&\hspace{-2cm}-\big(\d^{\a_1\a_2}C^{\m_1\n_1\m_2\n_2\m\n}+\d^{\m\n}C^{\m_1\n_1\m_2\n_2\a_1\a_2}-\d^{\a_1\n}C^{\m_1\n_1\m_2\n_2\a_2\m}-\d^{\a_2\m}C^{\m_1\n_1\m_2\n_2\a_1\n}\big)\notag\\
&\hspace{-2cm}-\big(\tilde{F}^{\m\n\m_1\n_1}\tilde{F}^{\m_2\n_2\a_1\a_2}+\tilde{F}^{\m\n\m_2\n_2}\tilde{F}^{\m_1\n_1\a_1\a_2}-\tilde{F}^{\m\a_2\m_1\n_1}\tilde{F}^{\m_2\n_2\a_1\n}-\tilde{F}^{\m\a_2\m_2\n_2}\tilde{F}^{\m_1\n_1\n\a_1}\big)\,\bigg]\,k_{1\m}\,k_{2\n}
\end{align}
and for the gauge fixing term
\begin{align}
V^{\m_1\n_1\a_1\a_2}_{TAA,\,GF}(k_1,k_2)&=-\sdfrac{1}{2\xi}\bigg[-\d^{\m_1\n_1}\d^{\a_1\m}\d^{\a_2\n}\,k_{1\m}\,k_{2\n}+\big(\d^{\m_1\n_1}\tilde{F}^{\a_1\a_2\m\n}-2\tilde{C}^{\m_1\n_1\m\n\a_1\a_2}\big)\,k_{2\m}\,k_{2\n}\nn
&\hspace{2cm}+\big(\d^{\m_1\n_1}\tilde{F}^{\a_1\a_2\m\n}-2\tilde{C}^{\m_1\n_1\m\n\a_2\a_1}\big)\,k_{1\m}\,k_{1\n}\,\bigg].
\end{align}
\begin{align}
V^{\m_1\n_1\m_2\n_2\a_1\a_2}_{TTAA,\,GF}(p_2,k_1,k_2)&=-\sdfrac{1}{2\xi}\bigg[I^{\m_1\n_1\m_2\n_2\a_1\a_2\m\n}\,k_{1\m}\,k_{2\n}+H^{\m_2\n_2\m_1\n_1\a_2\a_1\m\n}\,p_{2\m}k_{1\n}\notag\\
&\hspace{-3cm}+H^{\m_1\n_1\m_2\n_2\a_1\a_2\m\n}\,p_{2\m}k_{2\n}-\big(I^{\m_1\n_1\m_2\n_2\a_1\a_2\m\n}+A^{\m_2\n_2\a_2\n}\tilde{F}^{\m_1\n_1\m\a_1}-2\d^{\a_2\n}C^{\m_1\n_1\m_2\n_2\m\a_1}\big)\,k_{2\m}\,k_{2\n}\notag\\
&\hspace{-3cm}-\big(I^{\m_1\n_1\m_2\n_2\a_2\a_1\m\n}+A^{\m_2\n_2\a_1\n}\tilde{F}^{\m_1\n_1\m\a_2}-2\d^{\a_1\n}C^{\m_1\n_1\m_2\n_2\m\a_2}\big)\,k_{1\m}\,k_{1\n}+\big(4\,\tilde{F}^{\m_1\n_1\n(\a_1}\tilde{F}^{\a_2)\m\m_2\n_2}\notag\\
&\hspace{-2cm}-2\d^{\m_1\n_1}\tilde{C}^{\a_1\a_2\m_2\n_2\n\m}-2\d^{\m_2\n_2}\tilde{C}^{\a_1\a_2\m_1\n_1\m\n}+\d^{\m_1\n_1}\d^{\m_2\n_2}\tilde{F}^{\m\n\a_1\a_2}\big)\,p_{2\m}\,(p_2-k_2+k_1)_\n
\bigg]
\end{align}
For the ghost sector the vertices are
\begin{align}
V^{\m_1\n_1}_{T\bar cc}(k_1,k_2)&=\sdfrac{1}{2}\,A^{\m_1\n_1\m\n}\,k_{1\m}\,k_{2\n}\nn
V^{\m_1\n_1\m_2\n_2}_{TT\bar cc}(k_1,k_2)&=\bigg(\sdfrac{1}{4}\d^{\m_2\n_2}\,A^{\m_1\n_1\m\n}-\sdfrac{1}{2}\,D^{\m_1\n_1\m\n\m_2\n_2}+C^{\m_1\n_1\m_2\n_2\m\n}\bigg)\,k_{1\m}\,k_{2\n}
\end{align}

Note that in order to find the vertices, we have to consider a $i$ complex factor coming from the generating functional and the Fourier transformation is conventionally set in these expressions with the exponential factor $\exp[-i(px-qy)]$ if $p$ is an incoming momentum and $q$ is outgoing. 

\section{Metric variations of the counterterms}\label{Mvc}
%%%%%%%%%%%%%%%%%%%%%%%%%%%%%%%%%%%%%%%%%%%%%%%%%%%%
%%%%%%%%%%%%%%%%%%%%%%%%%%%%%%%%%%%%%%%%%%%%%%%%%%%%
In this appendix we list the metric variations of the counterterms. In particular we give them directly in the momentum using the definition of the Fourier transform in \eqref{count}. The metric variation are consider in the flat space-time limit and the first variation of the square of the metric, Riemann, Ricci and the scalar curvature are given as
\begingroup\makeatletter\def\f@size{9.5}\check@mathfonts
\def\maketag@@@#1{\hbox{\m@th\large\normalfont#1}}%
\begin{align}
\big[\sqrt{-g}\big]^{\m_i\n_i}&=\sdfrac{1}{2}\d^{\m_i\n_i}\nn
%%%%%%%%%%%%%%%%%%%%%%%%%%%%%%%%%%%%%%
%%%%%%%%%%%%%%%%%%%%%%%%%%%%%%%%%%%%%%
\big[R_{\m\a\n\b}\big]^{\m_i\n_i}(p_i)&=\frac{1}{2}\,\bigg(\d_\a^{(\m_i}\d^{\n_i)}_\b\,p_{i\m}\,p_{i\n}
+\d_\m^{(\m_i}\d^{\n_i)}_\n\,p_{i\a}\,p_{i\b}-\d_\m^{(\m_i}\d^{\n_i)}_\b\,p_{i\a}\,p_{i\n}-\d_\a^{(\m_i}\d^{\n_i)}_\n\,p_{i\m}\,p_{i\b}\bigg)\nn
%%%%%%%%%%%%%%%%%%%%%%%%%%%%%%%%%%%%%%
%%%%%%%%%%%%%%%%%%%%%%%%%%%%%%%%%%%%%%
\big[R^{\m\a\n\b}\big]^{\m_i\n_i}(p_i)&=\frac{1}{2}\,\bigg(\d^{\a(\m_i}\d^{\n_i)\b}\,p_i^\m\,p_i^\n
+\d^{\m(\m_i}\d^{\n_i)\n}\,p_i^\a\,p_i^\b-\d^{\m(\m_i}\d^{\n_i)\b}\,p_i^\a\,p_i^\n-\d^{\a(\m_i}\d^{\n_i)\n}\,p_i^\m\,p_i^\b\bigg)\nn
%%%%%%%%%%%%%%%%%%%%%%%%%%%%%%%%%%%%%%
%%%%%%%%%%%%%%%%%%%%%%%%%%%%%%%%%%%%%%
\big[R_{\m\n}\big]^{\m_i\n_i}(p_i)&=\frac{1}{2}\,\bigg(\d_\m^{(\m_i}\d^{\n_i)}_\n\,p_{i}^2
+\d^{\m_i\n_i}\,p_{i\m}\,p_{i\n}-p_i^{(\m_i}\d^{\n_i)}_\m\,p_{i\n}-p_i^{(\m_i}\d^{\n_i)}_\n\,p_{i\m}\bigg)\nn
%%%%%%%%%%%%%%%%%%%%%%%%%%%%%%%%%%%%%%
%%%%%%%%%%%%%%%%%%%%%%%%%%%%%%%%%%%%%%
\big[R^{\m\n}\big]^{\m_i\n_i}(p_i)&=\frac{1}{2}\,\bigg(\d^{\m(\m_i}\d^{\n_i)\n}\,p_{i}^2
+\d^{\m_i\n_i}\,p_i^\m\,p_i^\n-p_i^{(\m_i}\d^{\n_i)\m}\,p_i^\n-p_i^{(\m_i}\d^{\n_i)\n}\,p_i^\m\bigg)\nn
%%%%%%%%%%%%%%%%%%%%%%%%%%%%%%%%%%%%%%
%%%%%%%%%%%%%%%%%%%%%%%%%%%%%%%%%%%%%%
\big[R\big]^{\m_i\n_i}(p_i)&=\bigg(\d^{\m_i\n_i}\,p_{i}^2-p_i^{(\m_i}p_i^{\n_i)}\bigg)\nn
%%%%%%%%%%%%%%%%%%%%%%%%%%%%%%%%%%%%%%
%%%%%%%%%%%%%%%%%%%%%%%%%%%%%%%%%%%%%%
\big[\square R\big]^{\m_i\n_i}(p_i)&=p_i^2\,\bigg(p_i^{(\m_i}p_i^{\n_i)}-\d^{\m_i\n_i}\,p_{i}^2\bigg)
\end{align}
\endgroup
and the second variations of these object can be calculated in order to obtain
\begingroup\makeatletter\def\f@size{9.5}\check@mathfonts
\def\maketag@@@#1{\hbox{\m@th\large\normalfont#1}}%
\begin{align}
\big[R^{\,\b}_{\ \ \n\r\s}\big]^{\m_1\n_1\m_2\n_2}(p_1,p_2)&=\Big[-\frac{1}{2}\,\tilde F^{\m_1\n_1\b\epsilon}p_{1\s}\big(\tilde F^{\m_2\n_2}_{\e\n}p_{2\r}+\tilde F^{\m_2\n_2}_{\e\r}p_{2\n}-\tilde F^{\m_2\n_2}_{\n\r}p_{2\e}\big)\notag\nn
&\hspace{-1.5cm}-\frac{1}{2}\big(\tilde C^{\m_1\n_1\m_2\n_2\b}_{\hspace{1.4cm}\r}\ p_{2\n}-\tilde F^{\m_1\n_1\b\e}\,\tilde F^{\m_2\n_2}_{\n\r}\,p_{2\e}\big)p_{2\s}\notag\nn
&\hspace{-1.5cm}-\frac{1}{4}\big(\tilde F^{\m_1\n_1}_{\a\n}p_{1\s}+\tilde F^{\m_1\n_1}_{\a\s}p_{1\n}-\tilde F^{\m_1\n_1}_{\s\n}p_{1\a}\big)\big(\tilde F^{\m_2\n_2\b\a}\,p_{2\r}+\tilde F^{\m_2\n_2\b}_{\qquad\r}\,p_{2}^\a-\tilde F^{\m_2\n_2\a}_{\qquad\r}\,p_{2}^{\b}\big)\Big]-(\s\leftrightarrow \r)\nn
%%%%%%%%%%%%%%%%%%%%%%%%%%%
\big[R_{\m\n\r\s}\big]^{\m_1\n_1\m_2\n_2}(p_1,p_2)&=\d^{(\m_1}_\m\d^{\n_1)}_\b\big[R^{\,\b}_{\ \ \n\r\s}\big]^{\m_2\n_2}(p_2)+\d_{\m\b}\big[R^{\,\b}_{\ \ \n\r\s}\big]^{\m_1\n_1\m_2\n_2}(p_1,p_2)\nn
%%%%%%%%%%%%%%%%%%%%%%%%%%%
\end{align}
\begin{align}
\big[R^{\m\n\r\s}\big]^{\m_1\n_1\m_2\n_2}(p_1,p_2)&=\d^{\a\n}\d^{\b\r}\d^{\s\g}\big[R^{\,\m}_{\ \ \a\b\g}\big]^{\m_1\n_1\m_2\n_2}(p_1,p_2)\notag\nn
&\hspace{-1.5cm}-\big(\d^{\a(\m_1}\d^{\n_1)\n}\d^{\b\r}\d^{\s\g}+\d^{\a\n}\d^{\b(\m_1}\d^{\n_1)\r}\d^{\s\g}+\d^{\a\n}\d^{\b\r}\d^{\s(\m_1}\d^{\n_1)\g}\big)\big[R^{\,\b}_{\ \ \n\r\s}\big]^{\m_2\n_2}(p_2)
%%%%%%%%%%%%%%%%%%%%%%%%%%%
\end{align}
\begin{align}
\big[R_{\n\s}\big]^{\m_1\n_1\m_2\n_2}(p_1,p_2)&=-\frac{1}{2}\tilde F^{\m_1\n_1\m_2\n_2}\left(p_{1\s}p_{2\n}-\sdfrac{1}{2}p_{1\n}p_{2\s}+p_{2\n}p_{2\s}\right)-\frac{1}{4}\d^{\m_2\n_2}\left(\tilde F^{\m_1\n_1}_{\a\n}\,p_{1\s}+\tilde F^{\m_1\n_1}_{\a\s}\,p_{1\n}\right)\,p_2^\a\notag\\
&\hspace{-1.5cm}+\frac{1}{2}\big(\tilde C^{\m_1\n_1\m_2\n_2\e}_{\hspace{1.4cm}\n}\,p_{2\s}+\tilde C^{\m_1\n_1\m_2\n_2\e}_{\hspace{1.4cm}\s}\,p_{2\n}\big)\,(p_1+p_2)_{\e}+\frac{1}{2}F^{\m_2\n_2}_{\a\s}\tilde F^{\m_1\n_1}_{\b\n}\,p_1^\a\,p_2^\b\notag\\
&\hspace{-1.5cm}-\frac{1}{2}\tilde F^{\m_2\n_2}_{\n\s}\,\tilde F^{\m_1\n_1\a\b}(p_1+p_2)_\a\,p_{2\b}-\frac{1}{2}\left(\tilde C^{\m_1\n_1\m_2\n_2}_{\hspace{1.25cm}\n\s}-\frac{1}{2}\d^{\m_2\n_2}\,\tilde F^{\m_1\n_1}_{\n\s}\right)\,p_1\cdot p_2\nn
\end{align}
\begin{align}
%%%%%%%%%%%%%%%%%%%%%%%%%%%
\big[R^{\n\s}\big]^{\m_1\n_1\m_2\n_2}(p_1,p_2)&=\d^{\n\a}\d^{\s\b}\big[R_{\a\b}\big]^{\m_1\n_1\m_2\n_2}(p_1,p_2)-\big(\d^{\n(\m_1}\d^{\n_1)\a}\d^{\s\b}+\d^{\n\a}\d^{\s(\m_1}\d^{\n_1)\b}\big)\big[R_{\a\b}\big]^{\m_2\n_2}(p_2)\\
%%%%%%%%%%%%%%%%%%%%%%%%%%%
\big[R\big]^{\m_1\n_1\m_2\n_2}(p_1,p_2)&=-\left(p_2^2+\sdfrac{1}{4}p_1\cdot p_2\right)\,\tilde F^{\m_1\n_1\m_2\n_2}+\frac{1}{4}\,A^{\m_1\n_1\m_2\n_2}\,p_1\cdot p_2\notag\\
&\hspace{-1.5cm}+\tilde C^{\m_1\n_1\m_2\n_2\a\b}\,(p_{1\a}+2p_{2\a})p_{2\b}-\d^{\m_2\n_2}\tilde F^{\m_1\n_1\a\b}\,(p_{1\a}+p_{2\a})p_{2\b}+\sdfrac{1}{2}\tilde C^{\m_2\n_2\m_1\n_1\a\b}\,p_{1\a}p_{2\b}
\end{align}
\begin{align}
%%%%%%%%%%%%%%%%%%%%%%%%%%%
\big[\square R\big]^{\m_1\n_1\m_2\n_2}(p_1,p_2)&=\tilde F^{\m_1\n_1\m_2\n_2}\,\bigg[p_2^2(p_1+p_2)^2+\sdfrac{3}{2}(p_2^2+p_1\cdot p_2)\bigg]+\sdfrac{1}{2}\d^{\m_1\n_1}\tilde F^{\m_2\n_2\a\b}(p_1\cdot p_2)\,p_{2\a}p_{2\b}\notag\\
&\hspace{-1.5cm}-\sdfrac{1}{2}\d^{\m_1\n_1}\d^{\m_2\n_2}(p_1\cdot p_2)\bigg[(p_1+p_2)^2-p_1\cdot p_2\bigg]+\d^{\m_2\n_2}F^{\m_1\n_1\a\b}p_{2\a}(p_1+p_2)_\b\bigg[(p_1+p_2)^2+p_2^2\bigg]\notag\\
&\hspace{-1.5cm}-\tilde F^{\m_2\n_2\a\b}p_{2\a}p_{2\b}\,\tilde F^{\m_1\n_1\g\d}p_{2\g}(p_1+p_2)_\d-(p_1+p_2)^2\,\tilde C^{\m_1\n_1\m_2\n_2\a\b}\bigg[2p_{2\a}p_{2\b}+p_{1\a}p_{2\b}+\sdfrac{1}{2}p_{2\a}p_{1\b}\bigg]
\end{align}
\endgroup
remembering that the order of indices of variation is important, because these second variation are not symmetrized. 

Using these relations it is possible to find the third variation of the counterterms. For instance the Weyl tensor counterterm is expressed as
\begingroup\makeatletter\def\f@size{9.5}\check@mathfonts
\def\maketag@@@#1{\hbox{\m@th\large\normalfont#1}}%
\begin{align}
 \big[\sqrt{-g}\,C^2\big]^{\m_1\n_1\m_2\n_2\m_3\n_3}(p_1,p_2,p_3)&=\Bigg\{[\sqrt{-g}]^{\m_1\n_1}\bigg([R_{abcd}]^{\m_2\n_2}(p_2)[R^{abcd}]^{\m_3\n_3}(p_3)-\sdfrac{4}{d-2}[R_{ab}]^{\m_2\n_2}(p_2)[R^{ab}]^{\m_3\n_3}(p_3)\notag\\
 &\hspace{-3cm}+\sdfrac{2}{(d-2)(d-1)}[R]^{\m_2\n_2}(p_2)[R]^{\m_3\n_3}(p_3)\bigg)+\bigg([R_{abcd}]^{\m_1\n_1\m_2\n_2}(p_1,p_2)[R^{abcd}]^{\m_3\n_3}(p_3)\notag\\
 &\hspace{-3cm}+[R_{abcd}]^{\m_2\n_2}(p_2)[R^{abcd}]^{\m_1\n_1\m_3\n_3}(p_1,p_3)-\sdfrac{4}{d-2}[R_{ab}]^{\m_1\n_1\m_2\n_2}(p_1,p_2)[R^{ab}]^{\m_3\n_3}(p_3)\notag\\
 &\hspace{-3cm}-\sdfrac{4}{d-2}[R_{ab}]^{\m_2\n_2}(p_2)[R^{ab}]^{\m_1\n_1\m_3\n_3}(p_1,p_3)+\sdfrac{2}{(d-2)(d-1)}[R]^{\m_1\n_1\m_2\n_2}(p_1,p_2)[R]^{\m_3\n_3}(p_3)\notag\\
 &\hspace{-3cm}+\sdfrac{2}{(d-2)(d-1)}[R]^{\m_2\n_2}(p_2)[R]^{\m_1\n_1\m_3\n_3}(p_1,p_3)\bigg)\Bigg\}+\text{permutations}
\end{align}
\endgroup
where ``permutation'' indicate all the possible permutations of the indices $(\m_i,\n_i)$. In a same way the Euler density counterterm is given as
\begingroup\makeatletter\def\f@size{9.5}\check@mathfonts
\def\maketag@@@#1{\hbox{\m@th\large\normalfont#1}}%

\begin{align}
\big[\sqrt{-g}\,E\big]^{\m_1\n_1\m_2\n_2\m_3\n_3}(p_1,p_2,p_3)&=\Bigg\{[\sqrt{-g}]^{\m_1\n_1}\bigg([R_{abcd}]^{\m_2\n_2}(p_2)[R^{abcd}]^{\m_3\n_3}(p_3)-4[R_{ab}]^{\m_2\n_2}(p_2)[R^{ab}]^{\m_3\n_3}(p_3)\notag\\
&\hspace{-4.7cm}+[R]^{\m_2\n_2}(p_2)[R]^{\m_3\n_3}(p_3)\bigg)+\bigg([R_{abcd}]^{\m_1\n_1\m_2\n_2}(p_1,p_2)[R^{abcd}]^{\m_3\n_3}(p_3)+[R_{abcd}]^{\m_2\n_2}(p_2)[R^{abcd}]^{\m_1\n_1\m_3\n_3}(p_1,p_3)\notag\\
&\hspace{-4.7cm}-4[R_{ab}]^{\m_1\n_1\m_2\n_2}(p_1,p_2)[R^{ab}]^{\m_3\n_3}(p_3)-4[R_{ab}]^{\m_2\n_2}(p_2)[R^{ab}]^{\m_1\n_1\m_3\n_3}(p_1,p_3)+[R]^{\m_1\n_1\m_2\n_2}(p_1,p_2)[R]^{\m_3\n_3}(p_3)\notag\\
&\hspace{-3cm}+[R]^{\m_2\n_2}(p_2)[R]^{\m_1\n_1\m_3\n_3}(p_1,p_3)\bigg)\Bigg\}+\text{permutations}.
\end{align}
\endgroup

\section{Form Factors in $d=5$}
\label{aaa5}
\begingroup\makeatletter\def\f@size{9.5}\check@mathfonts
\def\maketag@@@#1{\hbox{\m@th\large\normalfont#1}}%

\begin{align}
A_{2}^{D=5}(p_1,p_2,p_3)&=\frac{\p^4(n_S-4\,n_F)}{1680(p_1+p_2+p_3)^6}\Big[-24p_3^5\big(8(p_1+p_2)^2-p_1 p_2\big)-24p_3^4(p_1+p_2)\big(13(p_1+p_2)-6p_1p_2\big)\notag\\
&-8 p_3^3 \left(42 (p_1+p_2)^4-p_1 p_2 \left(33 (p_1+p_2)^2+8 p_1 p_2\right)\right)\notag\\
&-5 (6p_3+p_1+p_2) (p_1+p_2)^2 \big(3 (p_1+p_2)^4-p_1 p_2 \left(3 (p_1+p_2)^2+p_1 p_2\right)\big)\notag\\
&-3 p_3^2 (p_1+p_2) \left(77 (p_1+p_2)^4-p_1 p_2 \left(63 (p_1+p_2)^2+43 p_1 p_2\right)\right)-72 p_3^6 (p_1+p_2)-12 p_3^7\Big]\notag\\
&+\frac{\pi ^4\,n_F}{40 (p_1+p_2+p_3)^5} \Big[-3 p_1^5 (p_1+5p_2+5p_3)-4 p_1^4 \left(8 (p_2+p_3)^2-p_2 p_3\right)\notag\\
&-20 p_1^3 (p_2+p_3) \left(2 (p_2+p_3)^2- p_2 p_3\right)-4 p_1^2 \left(8 (p_2+p_3)^4-p_2 p_3 \left(7 (p_2+p_3)^2+2 p_2 p_3\right)\right)\notag\\
&-(p_2+p_3)(5p_1+p_2+p_3) \left(3(p_2+p_3)^4-3p_2p_3(p_2+p_3)^2-p_2^2p_3^2\right)\Big]]
\end{align}
\begin{align}
A_{3}^{D=5}(p_1,p_2,p_3)&=\frac{\pi ^4  (n_S-4 n_F)p_3^2}{20160 (p_1+p_2+p_3)^5} \Big[-8 p_3^2 \left(-13 p_1^2 p_2^2+87 (p_1+p_2)^4-63 p_1 p_2 (p_1+p_2)^2\right)\notag\\
&\hspace{-2cm}-100 p_3 (p_1+p_2) \left(-p_1^2 p_2^2+3 (p_1+p_2)^4-3 p_1 p_2 (p_1+p_2)^2\right)-20 (p_1+p_2)^2 \big(-p_1^2 p_2^2+3 (p_1+p_2)^4\notag\\
&\hspace{-2cm}-3 p_1 p_2 (p_1+p_2)^2\big)-405 p_3^5 (p_1+p_2)-3 p_3^4 \left(281 (p_1+p_2)^2-22 p_1 p_2\right)-15 p_3^3 (p_1+p_2) \big(65 (p_1+p_2)^2\notag\\
&\hspace{-2cm}-22 p_1 p_2\big)-81 p_3^6\Big]+\frac{\pi ^4 \,n_F\,p_3^2}{240 (p_1+p_2+p_3)^4} \Big[-8 p_3 \left(-p_1^2 p_2^2+3 (p_1+p_2)^4-3 p_1 p_2 (p_1+p_2)^2\right)\notag\\
&\hspace{-2cm}-2 (p_1+p_2) \left(-p_1^2 p_2^2+3 (p_1+p_2)^4-3 p_1 p_2 (p_1+p_2)^2\right)-36 p_3^4 (p_1+p_2)-3 p_3^3 \left(19 (p_1+p_2)^2-2 p_1 p_2\right)\notag\\
&\hspace{-2cm}-24 p_3^2 (p_1+p_2) \left(2 (p_1+p_2)^2-p_1 p_2\right)-9 p_3^5\Big]+\frac{\pi ^4 (n_S+8 n_F)}{576 (p_1+p_2+p_3)^3} \Big[2 p_1^2 p_2^2 p_3^2\notag\\
&\hspace{-2cm}+3 p_1 p_2 p_3 (p_1+p_2+p_3) \left((p_1+p_2+p_3)^2+p_1 p_2+p_1 p_3+p_2 p_3\right)+(p_1 p_2+p_1 p_3+p_2 p_3)^2\notag\\
&\hspace{-2cm}+3 (p_1+p_2+p_3)^2 \left((p_1+p_2+p_3)^4-3 (p_1 p_2+p_1 p_3+p_2 p_3) (p_1+p_2+p_3)^2\right)\Big]
\end{align}
\begin{align}
A_4^{D=5}(p_1,p_2,p_3)&=\frac{\pi ^4  (n_S-4 n_F)p_3^2}{10080 (p_1+p_2+p_3)^5} \Big[45 p_3^8+225p_3^7(p_1+p_2)+15p_3^6\big(29(p_1+p_2)^2+2p_1p_2\big)\notag\\
&\hspace{-2cm}+75p_3^5(p_1+p_2)\big(5(p_1+p_2)^2+2p_1p_2\big)+8p_3^4\big(75(p_1+p_2)^2-23p_1p_2\big)p_1p_2\notag\\
&\hspace{-2cm}-5p_3^3(p_1+p_2)\big(75(p_1+p_2)^4-255(p_1+p_2)^4-255(p_1+p_2)^2p_1p_2+79p_1^2p_2^2\big)\notag\\
&\hspace{-2cm}-p_3^2\big(435(p_1+p_2)^6-1335(p_1+p_2)^4p_1p_2+343(p_1+p_2)^2p_1^2p_2^2-96p_1^3p_2^3\big)\notag\\
&\hspace{-2cm}-3(p_1+p_2)(5p_3+(p_1+p_2))(15(p_1+p_2)^6-45(p_1+p_2)^4p_1p_2+11(p_1+p_2)^2p_1^2p_2^2-4p_1^3p_2^3)\Big]\notag\\
&\hspace{-2cm}+\frac{\p^4\,n_F}{240(p_1+p_2+p_3)^4}\Big[9p_3^7+36\,p_3^6(p_1+p_2)+3p_3^5\big(17(p_1+p_2)^2+2p_1p_2\big)\notag\\
&\hspace{-2cm}+24p_3^4(p_1+p_2)\big((p_1+p_2)^2+p_1p_2\big)-8p_3^3\big(3(p_1+p_2)^4-12(p_1+p_2)^2p_1p_2+5p_1^2p_2^2\big)\notag\\
&\hspace{-2cm}-p_3^2(p_1+p_2)\big(51(p_1+p_2)^4-159(p_1+p_2)^2p_1p_2+55p_1^2p_2^2\big)\notag\\
&\hspace{-2cm}-9(p_1+p_2)^2(4p_3+p_1+p_2)\big((p_1+p_2)^4-3(p_1+p_2)^2p_1p_2+p_1^2p_2^2\big)\Big]+\frac{\p^4(n_s+8n_F)}{288(p_1+p_2+p_3)^3}\Big[2p_1^2p_2^2p_3^2\notag\\
&\hspace{-2cm}+3(p_1+p_2+p_3)^2\big((p_1+p_2+p_3)^4-3(p_1+p_2+p_3)^2(p_1p_3+p_2p_1+p_3p_2)+(p_1p_3+p_2p_1+p_3p_2)^2\big)\notag\\
&\hspace{-2cm}+3(p_1+p_2+p_3)\big((p_1+p_2+p_3)^2+p_1p_2+p_2p_3+p_1p_3\big)p_1p_2p_3\Big]
\end{align}
\begin{align}
A_5^{D=5}(p_1,p_2,p_3)&=\frac{\pi ^4 (n_S-4 n_F)}{6720 (p_1+p_2+p_3)^4} \Big[8 p_1^3 p_2^3 p_3^3-4 p_1^2 p_2^2 p_3^2 (p_1+p_2+p_3) \big((p_1+p_2+p_3)^2\notag\\
&\hspace{-2cm}-4 (p_1 p_2+p_1 p_3+p_2 p_3)\big)+5 p_1 p_2 p_3 (p_1+p_2+p_3)^2 \big(23 (p_1+p_2+p_3)^4\notag\\
&\hspace{-2cm}-23 (p_1 p_2+p_1 p_3+p_2 p_3) (p_1+p_2+p_3)^2+4 (p_1 p_2+p_1 p_3+p_2 p_3)^2\big)\notag\\
&\hspace{-2cm}+5 (p_1+p_2+p_3)^3 \left((p_1+p_2+p_3)^2-4 (p_1 p_2+p_1 p_3+p_2 p_3)\right) \big(3 (p_1+p_2+p_3)^4\notag\\
&\hspace{-2cm}-3 (p_1 p_2+p_1 p_3+p_2 p_3) (p_1+p_2+p_3)^2-(p_1 p_2+p_1 p_3+p_2 p_3)^2\big)\Big]\notag\\
&\hspace{-2cm}+\frac{\pi ^4 n_F }{480 (p_1+p_2+p_3)^3}\Big[-4 p_1^2 p_2^2 p_3^2 \left((p_1+p_2+p_3)^2-2 (p_1 p_2+p_1 p_3+p_2 p_3)\right)\notag\\
&\hspace{-2cm}+3 p_1 p_2 p_3 (p_1+p_2+p_3) \big(23 (p_1+p_2+p_3)^4-23 (p_1 p_2+p_1 p_3+p_2 p_3) (p_1+p_2+p_3)^2\notag\\
&\hspace{-2cm}+4 (p_1 p_2+p_1 p_3+p_2 p_3)^2\big)+3 (p_1+p_2+p_3)^2 \left((p_1+p_2+p_3)^2-4 (p_1 p_2+p_1 p_3+p_2 p_3)\right) \times\notag\\
&\hspace{-2cm}\times\left(3 (p_1+p_2+p_3)^4-3 (p_1 p_2+p_1 p_3+p_2 p_3) (p_1+p_2+p_3)^2-(p_1 p_2+p_1 p_3+p_2 p_3)^2\right)\Big]\notag\\
&\hspace{-2cm}+\frac{\pi ^4 (n_S+8 n_F) }{192} \left(p_1^5+p_2^5+p_3^5\right)
\end{align}
\endgroup

\section{Renormalized Form Factors in $d=4$}
\label{renexp}
%%%%%%%%%%%%%%%%%%%%%%%%%%%%%%%%%%
%%%%%%%%%%%%%%%%%%%%%%%%%%%%%%%%%%
We collect the explicit expressions of the renormalized form factors in $d=4$.\\
 We define $\s=s^2-2 s (s_1+s_2)+(s_1-s_2)^2$ 
%%%%%%%%%%%%%%%%%%%%%%
%%%%%%%%%%%%%%%%%%%%%%
\begingroup\makeatletter\def\f@size{8}\check@mathfonts
\def\maketag@@@#1{\hbox{\m@th\large\normalfont#1}}%
\begin{align}
A_1^{Ren}&=\pi ^2 (4 n_F-2 n_G-n_S)\bigg\{\frac{1}{45 \s^5}\bigg[s^9-13 s^8 (s_1+s_2)+2 s^7 \left(25 s_1^2+77 s_1 s_2+25 s_2^2\right)-2 s^6 (s_1+s_2) \left(41 s_1^2-9 s_1 s_2+41 s_2^2\right)\notag\\
&\hspace{-3ex}+s^5 \left(44 s_1^4-922 s_1^3 s_2+5088 s_1^2 s_2^2-922 s_1 s_2^3+44 s_2^4\right)+2 s^4 (s_1+s_2) \left(22 s_1^4+823 s_1^3 s_2-3360 s_1^2 s_2^2+823 s_1 s_2^3+22 s_2^4\right)\notag\\
&\hspace{-3ex}-2 s^3 \left(41 s_1^6+461 s_1^5 s_2+2537 s_1^4 s_2^2-8598 s_1^3 s_2^3+2537 s_1^2 s_2^4+461 s_1 s_2^5+41 s_2^6\right)+2 s^2 (s_1-s_2)^2 (s_1+s_2) \big(25 s_1^4-7 s_1^3 s_2\notag\\
&\hspace{-3ex}+2562 s_1^2 s_2^2-7 s_1 s_2^3+25 s_2^4\big)-s (s_1-s_2)^4 \left(13 s_1^4-102 s_1^3 s_2-422 s_1^2 s_2^2-102 s_1 s_2^3+13 s_2^4\right)+(s_1-s_2)^6 (s_1+s_2) \left(s_1^2-8 s_1 s_2+s_2^2\right)\bigg]\notag\\
&\hspace{-3ex}-\frac{4s^2 }{15 \s^6} \bigg[s_2^5 \left(-35 s^4-2469 s^2 s_1^2+2428 s s_1^3+726 s_1^4\right)+s_2^6 \big(35 s^3+135 s^2 s_1+448 s s_1^2-1052 s_1^3\big)+s_2^2 (s-s_1)^3 \big(s^4-24 s^3 s_1-675 s^2 s_1^2\notag\\
&\hspace{-3ex}-1348 s s_1^3-300 s_1^4\big)+3 s_2^4 \big(7 s^5-45 s^4 s_1+581 s^3 s_1^2+925 s^2 s_1^3-1850 s s_1^4+242 s_1^5\big)-s_2^3 (s-s_1) \big(7 s^5-101 s^4 s_1-705 s^3 s_1^2\notag\\
&\hspace{-3ex}+4151 s^2 s_1^3+1376 s s_1^4-1052 s_1^5\big)-27 s_1^2 s_2 (s-s_1)^5 (s+s_1)+s_1^2 (s-s_1)^7+s_2^8 (7 s+27 s_1)-3 s_2^7 (s-2 s_1) (7 s+50 s_1)-s_2^9\bigg]\,\bar{B}_0(s)\notag\\
&\hspace{-3ex}-\frac{4 s_1^2}{15 \s^6} \bigg[-s^9+s^8 (7 s_1+27 s_2)-3 s^7 (s_1-2 s_2) (7 s_1+50 s_2)+s^6 \left(35 s_1^3+135 s_1^2 s_2+448 s_1 s_2^2-1052 s_2^3\right)+s^5 \big(-35 s_1^4\notag\\
&\hspace{-3ex}-2469 s_1^2 s_2^2+2428 s_1 s_2^3+726 s_2^4\big)+3 s^4 \left(7 s_1^5-45 s_1^4 s_2+581 s_1^3 s_2^2+925 s_1^2 s_2^3-1850 s_1 s_2^4+242 s_2^5\right)-s^3 (s_1-s_2) \big(7 s_1^5-101 s_1^4 s_2\notag\\
&\hspace{-3ex}-705 s_1^3 s_2^2+4151 s_1^2 s_2^3+1376 s_1 s_2^4-1052 s_2^5\big)+s^2 (s_1-s_2)^3 \left(s_1^4-24 s_1^3 s_2-675 s_1^2 s_2^2-1348 s_1 s_2^3-300 s_2^4\right)\notag\\
&\hspace{-3ex}-27 s s_2^2 (s_1-s_2)^5 (s_1+s_2)+s_2^2 (s_1-s_2)^7\bigg]\bar{B}_0(s_1)-\frac{4 s_2^2}{15 \s^6}\bigg[-s^9+s^8 (27 s_1+7 s_2)+3 s^7 (2 s_1-s_2) (50 s_1+7 s_2)+s^6 \big(-1052 s_1^3\notag\\
&\hspace{-3ex}+448 s_1^2 s_2+135 s_1 s_2^2+35 s_2^3\big)+s^5 \left(726 s_1^4+2428 s_1^3 s_2-2469 s_1^2 s_2^2-35 s_2^4\right)+3 s^4 \big(242 s_1^5-1850 s_1^4 s_2+925 s_1^3 s_2^2+581 s_1^2 s_2^3\notag\\
&\hspace{-3ex}-45 s_1 s_2^4+7 s_2^5\big)-s^3 (s_1-s_2) \left(1052 s_1^5-1376 s_1^4 s_2-4151 s_1^3 s_2^2+705 s_1^2 s_2^3+101 s_1 s_2^4-7 s_2^5\right)+s^2 (s_1-s_2)^3 \big(300 s_1^4+1348 s_1^3 s_2\notag\\
&\hspace{-3ex}+675 s_1^2 s_2^2+24 s_1 s_2^3-s_2^4\big)+27 s s_1^2 (s_1-s_2)^5 (s_1+s_2)-s_1^2 (s_1-s_2)^7\bigg]\bar{B}_0(s_2)+\frac{16  s^2 s_1^2 s_2^2 }{\s^6} \bigg[3 s^6-4 s^5 (s_1+s_2)\notag\\
&\hspace{-3ex}+s^4 \left(-11 s_1^2+40 s_1 s_2-11 s_2^2\right)+12 s^3 (2 s_1-s_2) (s_1+s_2) (s_1-2 s_2)-s^2 \left(11 s_1^4+36 s_1^3 s_2-108 s_1^2 s_2^2+36 s_1 s_2^3+11 s_2^4\right)\notag\\
&\hspace{-3ex}-4 s (s_1-s_2)^2 (s_1+s_2) \left(s_1^2-9 s_1 s_2+s_2^2\right)+(s_1-s_2)^4 \left(3 s_1^2+8 s_1 s_2+3 s_2^2\right)\bigg]C_0(s,s_1,s_2)
\end{align}\endgroup
%%%%%%%%%%%%%%%%%%%%%%
%%%%%%%%%%%%%%%%%%%%%%
where $\bar{B}_0(s_i)$ is defined as the regular part of the two point scalar integral.
Then the others form factors in $d=4$ are explicitly given as follows. The 
expression of $A_2$ is given by
%%%%%%%%%%%%%%%%%%%%%%
%%%%%%%%%%%%%%%%%%%%%%
%%%%%%%%%%%%%%%%%%%%%%
\begingroup\makeatletter\def\f@size{7.5}\check@mathfonts
\def\maketag@@@#1{\hbox{\m@th\large\normalfont#1}}%
\begin{align}
A_2^{Ren}&=\frac{\pi ^2 }{450 \s^4}\bigg\{n_F \bigg[7 s^8+4 (21 s_1+41 s_2) s^7-4 \left(161 s_1^2+815 s_2 s_1+211 s_2^2\right) s^6+4 \left(427 s_1^3+2199 s_2 s_1^2+819 s_2^2 s_1+367 s_2^3\right) s^5-2 \big(1155 s_1^4\notag\\
&\hspace{-3ex}+2850 s_2 s_1^3+19914 s_2^2 s_1^2-4430 s_2^3 s_1+455 s_2^4\big) s^4+4 \left(427 s_1^5-1425 s_2 s_1^4+18698 s_2^2 s_1^3-1982 s_2^3 s_1^2-3645 s_2^4 s_1-73 s_2^5\right) s^3+\big(-644 s_1^6\notag\\
&\hspace{-3ex}+8796 s_2 s_1^5-39828 s_2^2 s_1^4-7928 s_2^3 s_1^3+33972 s_2^4 s_1^2+4956 s_2^5 s_1+676 s_2^6\big) s^2+4 (s_1-s_2)^3 \left(21 s_1^4-752 s_2 s_1^3-1500 s_2^2 s_1^2-8 s_2^3 s_1+79 s_2^4\right) s\notag\\
&\hspace{-3ex}+(s_1-s_2)^5 \left(7 s_1^3+199 s_2 s_1^2+81 s_2^2 s_1-47 s_2^3\right)\bigg]+2 n_S \bigg[-4 s^8+(52 s_1+42 s_2) s^7-\left(232 s_1^2+430 s_2 s_1+207 s_2^2\right) s^6+2 \big(262 s_1^3+519 s_2 s_1^2\notag\\
&\hspace{-3ex}+864 s_2^2 s_1+277 s_2^3\big) s^5-\left(680 s_1^4+650 s_2 s_1^3-591 s_2^2 s_1^2+2470 s_2^3 s_1+855 s_2^4\right) s^4+\big(524 s_1^5-650 s_2 s_1^4-4224 s_2^2 s_1^3+6116 s_2^3 s_1^2+460 s_2^4 s_1\notag\\
&\hspace{-3ex}+774 s_2^5\big) s^3+\left(-232 s_1^6+1038 s_2 s_1^5+591 s_2^2 s_1^4+6116 s_2^3 s_1^3-8634 s_2^4 s_1^2+1518 s_2^5 s_1-397 s_2^6\right) s^2+2 (s_1-s_2)^3 \big(26 s_1^4-137 s_2 s_1^3\notag\\
&\hspace{-3ex}+375 s_2^2 s_1^2+327 s_2^3 s_1-51 s_2^4\big) s-(s_1-s_2)^5 \left(4 s_1^3-22 s_2 s_1^2+57 s_2^2 s_1-9 s_2^3\right)\bigg]+2n_G \bigg[487 s^8-4 (1039 s_1+1044 s_2) s^7+2 \big(7598 s_1^2\notag\\
&\hspace{-3ex}+12320 s_2 s_1+7623 s_2^2\big) s^6-4 \left(7793 s_1^3+12216 s_2 s_1^2+11871 s_2^2 s_1+7778 s_2^3\right) s^5+\big(39290 s_1^4+28400 s_2 s_1^3+53202 s_2^2 s_1^2+24760 s_2^3 s_1\notag\\
&\hspace{-3ex}+38940 s_2^4\big) s^4-4 \left(7793 s_1^5-7100 s_2 s_1^4+10482 s_2^2 s_1^3+5312 s_2^3 s_1^2-7655 s_2^4 s_1+7668 s_2^5\right) s^3+2 \big(7598 s_1^6-24432 s_2 s_1^5+26601 s_2^2 s_1^4\notag\\
&\hspace{-3ex}-10624 s_2^3 s_1^3+17376 s_2^4 s_1^2-23952 s_2^5 s_1+7433 s_2^6\big) s^2-4 (s_1-s_2)^3 \left(1039 s_1^4-3043 s_2 s_1^3-375 s_2^2 s_1^2+2853 s_2^3 s_1-1014 s_2^4\right) s\notag\\
&\hspace{-3ex}+(s_1-s_2)^5 \left(487 s_1^3-1741 s_2 s_1^2+1671 s_2^2 s_1-477 s_2^3\right)\bigg]\bigg\}-\frac{\pi ^2 s^2}{45 \s^5} \bigg\{n_F \bigg[7 s^8-70 (s_1+s_2) s^7+\left(294 s_1^2+490 s_2 s_1+270 s_2^2\right) s^6-2 \big(343 s_1^3\notag\\
&\hspace{-3ex}+463 s_2 s_1^2+295 s_2^2 s_1+271 s_2^3\big) s^5+10 \big(98 s_1^4-27 s_2 s_1^3+774 s_2^2 s_1^2-99 s_2^3 s_1+62 s_2^4\big) s^4-2 \big(441 s_1^5-1495 s_2 s_1^4+9090 s_2^2 s_1^3+258 s_2^3 s_1^2\notag\\
&\hspace{-3ex}-1255 s_2^4 s_1+201 s_2^5\big) s^3+10 \big(49 s_1^6-397 s_2 s_1^5+995 s_2^2 s_1^4+2090 s_2^3 s_1^3-1633 s_2^4 s_1^2-157 s_2^5 s_1+13 s_2^6\big) s^2+2 \big(-77 s_1^7+1115 s_2 s_1^6+2505 s_2^2 s_1^5\notag\\
&\hspace{-3ex}-15015 s_2^3 s_1^4+8385 s_2^4 s_1^3+3057 s_2^5 s_1^2+35 s_2^6 s_1-5 s_2^7\big) s+3 (s_1-s_2)^3 \big(7 s_1^5-137 s_2 s_1^4-1832 s_2^2 s_1^3-1352 s_2^3 s_1^2-47 s_2^4 s_1+s_2^5\big)\bigg]-2 n_S \bigg[-s^8\notag\\
&\hspace{-3ex}+10 (s_1+s_2) s^7-\big(42 s_1^2+70 s_2 s_1+45 s_2^2\big) s^6+2 \big(49 s_1^3+79 s_2 s_1^2+100 s_2^2 s_1+58 s_2^3\big) s^5-5 \big(28 s_1^4+18 s_2 s_1^3-9 s_2^2 s_1^2+36 s_2^3 s_1+37 s_2^4\big) s^4\notag\\
&\hspace{-3ex}+2 \big(63 s_1^5-85 s_2 s_1^4-270 s_2^2 s_1^3+834 s_2^3 s_1^2-115 s_2^4 s_1+93 s_2^5\big) s^3-5 \big(14 s_1^6-62 s_2 s_1^5-5 s_2^2 s_1^4-380 s_2^3 s_1^3+652 s_2^4 s_1^2-122 s_2^5 s_1+23 s_2^6\big) s^2\notag\\
&\hspace{-3ex}+2 \big(11 s_1^7-95 s_2 s_1^6+330 s_2^2 s_1^5-1680 s_2^3 s_1^4+1245 s_2^4 s_1^3+399 s_2^5 s_1^2-230 s_2^6 s_1+20 s_2^7\big) s-3 (s_1-s_2)^3 \big(s_1^5-11 s_2 s_1^4+79 s_2^2 s_1^3+319 s_2^3 s_1^2+34 s_2^4 s_1\notag\\
&\hspace{-3ex}-2 s_2^5\big)\bigg]-2n_G \bigg[13 s^8-130 (s_1+s_2) s^7+\big(546 s_1^2+910 s_2 s_1+540 s_2^2\big) s^6-2 \big(637 s_1^3+937 s_2 s_1^2+895 s_2^2 s_1+619 s_2^3\big) s^5+10 \big(182 s_1^4+27 s_2 s_1^3\notag\\
&\hspace{-3ex}+747 s_2^2 s_1^2+9 s_2^3 s_1+173 s_2^4\big) s^4-2 \big(819 s_1^5-2005 s_2 s_1^4+7470 s_2^2 s_1^3+5262 s_2^3 s_1^2-1945 s_2^4 s_1+759 s_2^5\big) s^3+10 \big(91 s_1^6-583 s_2 s_1^5+980 s_2^2 s_1^4\notag\\
&\hspace{-3ex}+950 s_2^3 s_1^3+323 s_2^4 s_1^2-523 s_2^5 s_1+82 s_2^6\big) s^2+2 \big(-143 s_1^7+1685 s_2 s_1^6+525 s_2^2 s_1^5-4935 s_2^3 s_1^4+915 s_2^4 s_1^3+663 s_2^5 s_1^2+1415 s_2^6 s_1-125 s_2^7\big) s\notag\\
&\hspace{-3ex}+3 (s_1-s_2)^3 \big(13 s_1^5-203 s_2 s_1^4-1358 s_2^2 s_1^3+562 s_2^3 s_1^2+157 s_2^4 s_1-11 s_2^5\big)\bigg]\bigg\}\bar{B}_0(s)-\frac{\pi ^2 s_1^2 }{45 \s^5}\bigg\{n_F \bigg[21 s^8-2 (77 s_1+237 s_2) s^7+10 \big(49 s_1^2\notag\\
&\hspace{-3ex}+223 s_2 s_1-420 s_2^2\big) s^6+\big(-882 s_1^3-3970 s_2 s_1^2+5010 s_2^2 s_1+11178 s_2^3\big) s^5+10 \big(98 s_1^4+299 s_2 s_1^3+995 s_2^2 s_1^2-3003 s_2^3 s_1-405 s_2^4\big) s^4\notag\\
&\hspace{-3ex}-2 \big(343 s_1^5+135 s_2 s_1^4+9090 s_2^2 s_1^3-10450 s_2^3 s_1^2-8385 s_2^4 s_1+3123 s_2^5\big) s^3+2 (s_1-s_2)^2 \big(147 s_1^4-169 s_2 s_1^3+3385 s_2^2 s_1^2+6681 s_2^3 s_1+1812 s_2^4\big) s^2\notag\\
&\hspace{-3ex}-10 (s_1-s_2)^4 \big(7 s_1^3-21 s_2 s_1^2-67 s_2^2 s_1-15 s_2^3\big) s+(s_1-s_2)^6 \big(7 s_1^2-28 s_2 s_1-3 s_2^2\big)\bigg]-2 n_S \bigg[-3 s^8+(22 s_1+42 s_2) s^7-5 \big(14 s_1^2+38 s_2 s_1+69 s_2^2\big) s^6\notag\\
&\hspace{-3ex}+2 \big(63 s_1^3+155 s_2 s_1^2+330 s_2^2 s_1-72 s_2^3\big) s^5-5 \big(28 s_1^4+34 s_2 s_1^3-5 s_2^2 s_1^2+672 s_2^3 s_1-405 s_2^4\big) s^4+2 \big(49 s_1^5-45 s_2 s_1^4-270 s_2^2 s_1^3+950 s_2^3 s_1^2\notag\\
&\hspace{-3ex}+1245 s_2^4 s_1-1161 s_2^5\big) s^3-(s_1-s_2)^2 \big(42 s_1^4-74 s_2 s_1^3-235 s_2^2 s_1^2-2064 s_2^3 s_1-633 s_2^4\big) s^2+10 (s_1-s_2)^4 \big(s_1^3-3 s_2 s_1^2+2 s_2^2 s_1+12 s_2^3\big) s\notag\\
&\hspace{-3ex}-(s_1-s_2)^6 \big(s_1^2-4 s_2 s_1+6 s_2^2\big)\bigg]-2n_G \bigg[39 s^8-22 (13 s_1+33 s_2) s^7+10 \big(91 s_1^2+337 s_2 s_1-213 s_2^2\big) s^6-2 \big(819 s_1^3+2915 s_2 s_1^2-525 s_2^2 s_1\notag\\
&\hspace{-3ex}-6021 s_2^3\big) s^5+10 \big(182 s_1^4+401 s_2 s_1^3+980 s_2^2 s_1^2-987 s_2^3 s_1-1620 s_2^4\big) s^4+2 \big(-637 s_1^5+135 s_2 s_1^4-7470 s_2^2 s_1^3+4750 s_2^3 s_1^2+915 s_2^4 s_1+3843 s_2^5\big) s^3\notag\\
&\hspace{-3ex}+2 (s_1-s_2)^2 \big(273 s_1^4-391 s_2 s_1^3+2680 s_2^2 s_1^2+489 s_2^3 s_1-87 s_2^4\big) s^2-10 (s_1-s_2)^4 \big(13 s_1^3-39 s_2 s_1^2-55 s_2^2 s_1+57 s_2^3\big) s+(s_1-s_2)^6 \big(13 s_1^2\notag\\
&\hspace{-3ex}-52 s_2 s_1+33 s_2^2\big)\bigg]\bigg\}\bar{B}_0(s_1)+\frac{\pi ^2 s_2^2}{45 \s^5} \bigg\{n_F \bigg[-45 s^8+s^7 (810 s_1+298 s_2)+10 s^6 \big(144 s_1^2-295 s_1 s_2-85 s_2^2\big)+2 s^5 \big(-6165 s_1^3+5751 s_1^2 s_2\notag\\
&\hspace{-3ex}+1745 s_1 s_2^2+681 s_2^3\big)+10 s^4 \big(2025 s_1^4-885 s_1^3 s_2-2003 s_1^2 s_2^2-59 s_1 s_2^3-134 s_2^4\big)-2 s^3 \big(6165 s_1^5+4425 s_1^4 s_2-16910 s_1^3 s_2^2-354 s_1^2 s_2^3+945 s_1 s_2^4\notag\\
&\hspace{-3ex}-415 s_2^5\big)+2 s^2 (s_1-s_2)^2 \big(720 s_1^4+7191 s_1^3 s_2+3647 s_1^2 s_2^2+457 s_1 s_2^3-159 s_2^4\big)+10 s (s_1-s_2)^4 \big(81 s_1^3+29 s_1^2 s_2-21 s_1 s_2^2+7 s_2^3\big)-(s_1-s_2)^6 \big(45 s_1^2\notag\\
&\hspace{-3ex}-28 s_1 s_2+7 s_2^2\big)\bigg]+2 n_G \bigg[45 s^8-2 s^7 (405 s_1+161 s_2)+s^6 \big(-1440 s_1^2+3550 s_1 s_2+1000 s_2^2\big)+2 s^5 \big(6165 s_1^3-1539 s_1^2 s_2-2855 s_1 s_2^2-879 s_2^3\big)\notag
\end{align}
\begin{align}
&-10 s^4 \big(2025 s_1^4+15 s_1^3 s_2-1232 s_1^2 s_2^2-341 s_1 s_2^3-191 s_2^4\big)+2 s^3 \big(6165 s_1^5-75 s_1^4 s_2-2090 s_1^3 s_2^2-5286 s_1^2 s_2^3+405 s_1 s_2^4-655 s_2^5\big)\notag\\
&-2 s^2 (s_1-s_2)^2 \big(720 s_1^4+2979 s_1^3 s_2-922 s_1^2 s_2^2+463 s_1 s_2^3-276 s_2^4\big)-10 s (s_1-s_2)^4 \big(81 s_1^3-31 s_1^2 s_2-39 s_1 s_2^2+13 s_2^3\big)+(s_1-s_2)^6 \big(45 s_1^2-52 s_1 s_2\notag\\
&+13 s_2^2\big)\bigg]+2n_S s_2 \bigg[4 s^7-25 s^6 (4 s_1+s_2)+s^5 \big(-1404 s_1^2+370 s_1 s_2+66 s_2^2\big)+5 s^4 \big(300 s_1^3+257 s_1^2 s_2-94 s_1 s_2^2-19 s_2^3\big)+4 s^3 \big(375 s_1^4-1235 s_1^3 s_2\notag\\
&+411 s_1^2 s_2^2+45 s_1 s_2^3+20 s_2^4\big)-s^2 (s_1-s_2)^2 \big(1404 s_1^3+1523 s_1^2 s_2-2 s_1 s_2^2+39 s_2^3\big)-10 s (s_1-s_2)^4 \big(10 s_1^2+3 s_1 s_2-s_2^2\big)+(s_1-s_2)^6 (4 s_1-s_2)\bigg]\bigg\}\bar{B}_0(s_2)\notag\\
&-\frac{4 \pi ^2 s^2 s_1^2 s_2^2}{3 \s^5} \bigg\{n_F \bigg[45 s^5-3 s^4 (45 s_1+13 s_2)+2 s^3 \big(45 s_1^2+152 s_1 s_2-63 s_2^2\big)+2 s^2 \big(45 s_1^3-265 s_1^2 s_2+59 s_1 s_2^2+81 s_2^3\big)+s \big(-135 s_1^4+304 s_1^3 s_2\notag\\
&+118 s_1^2 s_2^2-272 s_1 s_2^3-15 s_2^4\big)+3 (s_1-s_2)^3 \big(15 s_1^2+32 s_1 s_2+9 s_2^2\big)\bigg]-2 n_G \bigg[45 s^5-3 s^4 (45 s_1+37 s_2)+s^3 \big(90 s_1^2+256 s_1 s_2+36 s_2^2\big)+2 s^2 \big(45 s_1^3\notag\\
&-145 s_1^2 s_2-64 s_1 s_2^2+54 s_2^3\big)+s \big(-135 s_1^4+256 s_1^3 s_2-128 s_1^2 s_2^2+112 s_1 s_2^3-105 s_2^4\big)+3 (s_1-s_2)^3 \big(15 s_1^2+8 s_1 s_2-9 s_2^2\big)\bigg]-2n_S s_2 \bigg[12 s^4+s^3 (8 s_1\notag\\
&-27 s_2)+s^2 \big(-40 s_1^2+41 s_1 s_2+9 s_2^2\big)+s \big(8 s_1^3+41 s_1^2 s_2-64 s_1 s_2^2+15 s_2^3\big)+3 (s_1-s_2)^3 (4 s_1+3 s_2)\bigg]\bigg\}C_0(s,s_1,s_2).
\end{align}\endgroup
%%%%%%%%%%%%%%%%%%%%%%
%%%%%%%%%%%%%%%%%%%%%%
%%%%%%%%%%%%%%%%%%%%%%
The form factor $A_3$ is given by
\begingroup\makeatletter\def\f@size{7.5}\check@mathfonts
\def\maketag@@@#1{\hbox{\m@th\large\normalfont#1}}%
\begin{align}
A_3^{Ren}&=\frac{\pi ^2}{900 \s^3} \bigg\{s_2^5 \big[s^2 (763 n_F+12166 n_G-22 n_S)-6 s s_1 (17 n_F-2956 n_G+77 n_S)+s_1^2 (763 n_F+12166 n_G-22 n_S)\big]-2 s_2^4 (s+s_1) \big[5 s^2 (24 n_F\notag\\
&\hspace{-3ex}+1988 n_G-31 n_S)+s s_1 (-2414 n_F+2202 n_G-659 n_S)+5 s_1^2 (24 n_F+1988 n_G-31 n_S)\big]+s_2^2 (s-s_1)^2 (s+s_1) \big[s^2 (896 n_F-10678 n_G\notag\\
&\hspace{-3ex}+451 n_S)+4 s s_1 (941 n_F+1112 n_G+196 n_S)+s_1^2 (896 n_F-10678 n_G+451 n_S)\big]-s_2 (s-s_1)^4 \big[s^2 (423 n_F-3214 n_G+188 n_S)+2 s s_1 (551 n_F\notag\\
&\hspace{-3ex}-918 n_G+206 n_S)+s_1^2 (423 n_F-3214 n_G+188 n_S)\big]-s_2^3 \big[5 s^4 (139 n_F-3812 n_G+109 n_S)+2 s^3 s_1 (3556 n_F-2258 n_G+811 n_S)+2 s^2 s_1^2 (641 n_F\notag\\
&\hspace{-3ex}-1888 n_G+1021 n_S)+2 s s_1^3 (3556 n_F-2258 n_G+811 n_S)+5 s_1^4 (139 n_F-3812 n_G+109 n_S)\big]-s_2^6 (s+s_1) (472 n_F+4054 n_G+57 n_S)\notag\\
&\hspace{-3ex}+4 (s-s_1)^6 (s+s_1) (18 n_F-99 n_G+8 n_S)+s_2^7 (99 n_F+568 n_G+19 n_S)\bigg\}+\frac{\pi ^2 s^2}{90 \s^4} \bigg\{-s_2^5 \big[s^2 (443 n_F+386 n_G+103 n_S)+40 s s_1 (26 n_F\notag\\
&\hspace{-3ex}+7 (2 n_G+n_S))+3 s_1^2 (259 n_F+358 n_G-11 n_S)\big]+2 s_2^2 (s-s_1)^3 \big[s^2 (208 n_F+301 n_G+38 n_S)+4 s s_1 (112 n_F+214 n_G+17 n_S)+3 s_1^2 (154 n_F\notag\\
&\hspace{-3ex}-17 n_G+19 n_S)\big]+s_2^4 \big[5 s^3 (146 n_F+152 n_G+31 n_S)+s^2 s_1 (1642 n_F+1384 n_G+347 n_S)+s s_1^2 (-14 n_F+1852 n_G+191 n_S)+15 s_1^3 (22 n_F\notag\\
&\hspace{-3ex}+196 n_G-23 n_S)\big]-s_2^3 (s-s_1) \big[5 s^3 (143 n_F+176 n_G+28 n_S)+5 s^2 s_1 (311 n_F+52 (8 n_G+n_S))+s s_1^2 (181 n_F+292 n_G+386 n_S)+15 s_1^3 (67 n_F\notag\\
&\hspace{-3ex}-164 n_G+22 n_S)\big]+2 s_2^6 (74 n_F s+114 n_F s_1+53 n_G s+33 n_G s_1+19 n_S s+39 n_S s_1)-s_2 (s-s_1)^5 (s (133 n_F+226 n_G+23 n_S)+3 s_1 (59 n_F\notag\\
&\hspace{-3ex}+158 n_G+9 n_S))+3 (s-s_1)^7 (6 n_F+12 n_G+n_S)-3 s_2^7 (7 n_F+4 n_G+2 n_S)\bigg\}\bar{B}_0(s)+\frac{\pi ^2 s_1^2}{90 \s^4} \bigg\{-s_2^5 \big[3 s^2 (259 n_F+358 n_G-11 n_S)+40 s s_1 \notag\\
&\hspace{-3ex}\times(26 n_F+7 (2 n_G+n_S))+s_1^2 (443 n_F+386 n_G+103 n_S)\big]-2 s_2^2 (s-s_1)^3 \big[3 s^2 (154 n_F-17 n_G+19 n_S)+4 s s_1 (112 n_F+214 n_G+17 n_S)\notag\\
&\hspace{-3ex}+s_1^2 (208 n_F+301 n_G+38 n_S)\big]+s_2^4 \big[15 s^3 (22 n_F+196 n_G-23 n_S)+s^2 s_1 (-14 n_F+1852 n_G+191 n_S)+s s_1^2 (1642 n_F+1384 n_G+347 n_S)\notag\\
&\hspace{-3ex}+5 s_1^3 (146 n_F+152 n_G+31 n_S)\big]+s_2^3 (s-s_1) \big[15 s^3 (67 n_F-164 n_G+22 n_S)+s^2 s_1 (181 n_F+292 n_G+386 n_S)+5 s s_1^2 (311 n_F+52 (8 n_G+n_S))\notag\\
&\hspace{-3ex}+5 s_1^3 (143 n_F+176 n_G+28 n_S)\big]+2 s_2^6 (114 n_F s+74 n_F s_1+33 n_G s+53 n_G s_1+39 n_S s+19 n_S s_1)+s_2 (s-s_1)^5 (3 s (59 n_F+158 n_G+9 n_S)\notag\\
&\hspace{-3ex}+s_1 (133 n_F+226 n_G+23 n_S))-3 (s-s_1)^7 (6 n_F+12 n_G+n_S)-3 s_2^7 (7 n_F+4 n_G+2 n_S)\bigg\}\bar{B}_0(s_1)+\frac{\pi ^2 s_2^2}{90 \s^4} \bigg\{s_2^5 \big[5 s^2 (37 n_F+292 n_G+2 n_S)\notag\\
&\hspace{-3ex}+8 s s_1 (19 n_F+238 n_G-n_S)+5 s_1^2 (37 n_F+292 n_G+2 n_S)\big]-2 s_2^4 (s+s_1) \big[5 s^2 (26 n_F+257 n_G+n_S)-4 s s_1 (58 n_F+61 n_G+8 n_S)+5 s_1^2 (26 n_F\notag\\
&\hspace{-3ex}+257 n_G+n_S)\big]+s_2^3 \big[5 s^4 (41 n_F+542 n_G+n_S)-8 s^3 s_1 (103 n_F+196 n_G+8 n_S)-6 s^2 s_1^2 (107 n_F+134 n_G+77 n_S)-8 s s_1^3 (103 n_F+196 n_G\notag\\
&\hspace{-3ex}+8 n_S)+5 s_1^4 (41 n_F+542 n_G+n_S)\big]-s_2^2 (s+s_1) \big[s^4 (86 n_F+1712 n_G+n_S)-24 s^3 s_1 (36 n_F+282 n_G+n_S)-2 s^2 s_1^2 (182 n_F-3916 n_G\notag\\
&\hspace{-3ex}+187 n_S)-24 s s_1^3 (36 n_F+282 n_G+n_S)+s_1^4 (86 n_F+1712 n_G+n_S)\big]-5 s_2^6 (s+s_1) (14 n_F+92 n_G+n_S)+s_2^7 (11 n_F+62 n_G+n_S)\notag\\
&\hspace{-3ex}+15 s_2 (s-s_1)^2 \big[s^4 (n_F+40 n_G)-2 s^3 s_1 (7 n_F+88 n_G)-2 s^2 s_1^2 (47 n_F+32 n_G)-2 s s_1^3 (7 n_F+88 n_G)+s_1^4 (n_F+40 n_G)\big]-90 n_G (s-s_1)^4 (s+s_1) \notag\\
&\hspace{-3ex}\times\big[s^2-8 s s_1+s_1^2\big]\bigg\}\bar{B}_0(s_2)+\frac{2 \pi ^2 s^2 s_1^2 s_2^2}{3 \s^4} \bigg\{-s_2^2 \big[3 s^2 (11 n_F+8 n_G+n_S)-2 s s_1 (n_F-14 n_G-4 n_S)+3 s_1^2 (11 n_F+8 n_G+n_S)\big]\notag\\
&\hspace{-3ex}+3 s_2^3 (s+s_1) (7 n_F -2 n_G+2 n_S)-3 s_2^4 (n_F-2 n_G+n_S)+3 s_2 (5 n_F+14 n_G) (s+s_1) (s-s_1)^2-18 n_G (s-s_1)^4\bigg\}C_0(s,s_1,s_2)\notag\\
&\hspace{-3ex} -\frac{8\p^2}{720}\,(s+s_1+s_2)(n_S+11n_F+62n_G).
\end{align}\endgroup
%%%%%%%%%%%%%%%%%%%%%%
%%%%%%%%%%%%%%%%%%%%%%
%%%%%%%%%%%%%%%%%%%%%%
Then $A_4$ is expressed as
\begingroup\makeatletter\def\f@size{7.5}\check@mathfonts
\def\maketag@@@#1{\hbox{\m@th\large\normalfont#1}}%
\begin{align}
A_4^{Ren}&=\frac{\pi ^2}{900 \s^3} \bigg\{n_F \bigg[151 s^7-s^6 (675 s_1+629 s_2)+s^5 \big(1119 s_1^2-1350 s_1 s_2+743 s_2^2\big)+s^4 \big(-595 s_1^3+7221 s_1^2 s_2+8339 s_1 s_2^2+435 s_2^3\big)-s^3 \big(595 s_1^4\notag\\
&\hspace{-3ex}+10484 s_1^3 s_2+13882 s_1^2 s_2^2+9356 s_1 s_2^3+1875 s_2^4\big)+s^2 \big(1119 s_1^5+7221 s_1^4 s_2-13882 s_1^3 s_2^2+2034 s_1^2 s_2^3+1659 s_1 s_2^4+1849 s_2^5\big)-s (s_1-s_2)^3 \big(675 s_1^3\notag\\
&\hspace{-3ex}+3375 s_1^2 s_2-239 s_1 s_2^2-811 s_2^3\big)+(s_1-s_2)^5 \big(151 s_1^2+126 s_1 s_2-137 s_2^2\big)\bigg]+2 n_G \bigg[91 s^7-s^6 (475 s_1+1689 s_2)+s^5 \big(879 s_1^2+7650 s_1 s_2+7963 s_2^2\big)\notag\\
&\hspace{-3ex}-s^4 \big(495 s_1^3+14439 s_1^2 s_2+15401 s_1 s_2^2+17665 s_2^3\big)+s^3 \big(-495 s_1^4+16956 s_1^3 s_2+8638 s_1^2 s_2^2-596 s_1 s_2^3+21625 s_2^4\big)+s^2 \big(879 s_1^5-14439 s_1^4 s_2\notag\\
&\hspace{-3ex}+8638 s_1^3 s_2^2-4806 s_1^2 s_2^3+24819 s_1 s_2^4-15091 s_2^5\big)-s (s_1-s_2)^3 \big(475 s_1^3-6225 s_1^2 s_2-4699 s_1 s_2^2+5649 s_2^3\big)+(s_1-s_2)^5 \big(91 s_1^2-1234 s_1 s_2+883 s_2^2\big)\bigg]\notag\\
&\hspace{-3ex}+4 n_S \bigg[14 s^7-s^6 (75 s_1+56 s_2)+s^5 \big(141 s_1^2+52 s_2^2\big)+s^4 \big(-80 s_1^3+744 s_1^2 s_2+496 s_1 s_2^2+90 s_2^3\big)-2 s^3 \big(40 s_1^4+688 s_1^3 s_2+124 s_1^2 s_2^2+317 s_1 s_2^3\notag\\
&\hspace{-3ex}+125 s_2^4\big)+s^2 \big(141 s_1^5+744 s_1^4 s_2-248 s_1^3 s_2^2-924 s_1^2 s_2^3+51 s_1 s_2^4+236 s_2^5\big)-s (s_1-s_2)^3 \big(75 s_1^3+225 s_1^2 s_2-46 s_1 s_2^2-104 s_2^3\big)+2 (s_1-s_2)^5 \big(7 s_1^2\notag\\
&\hspace{-3ex}+7 s_1 s_2-9 s_2^2\big)\bigg]\bigg\}+\frac{\pi ^2 s^2}{90 \s^4} \bigg\{n_F \bigg[29 s^7-7 s^6 (29 s_1+27 s_2)+s^5 \big(625 s_1^2+584 s_1 s_2+525 s_2^2\big)-s^4 \big(1095 s_1^3+309 s_1^2 s_2+231 s_1 s_2^2+805 s_2^3\big)\notag\\
&\hspace{-3ex}+s^3 \big(1175 s_1^4-1272 s_1^3 s_2-3822 s_1^2 s_2^2-856 s_1 s_2^3+735 s_2^4\big)+s^2 \big(-769 s_1^5+2665 s_1^4 s_2+978 s_1^3 s_2^2+4526 s_1^2 s_2^3+1159 s_1 s_2^4-399 s_2^5\big)+s \big(283 s_1^6\notag\\
&\hspace{-3ex}-2064 s_1^5 s_2+2625 s_1^4 s_2^2-960 s_1^3 s_2^3+525 s_1^2 s_2^4-528 s_1 s_2^5+119 s_2^6\big)-15 (s_1-s_2)^3 \big(3 s_1^4-30 s_1^3 s_2-94 s_1^2 s_2^2+2 s_1 s_2^3-s_2^4\big)\bigg]+2 n_G \bigg[49 s^7\notag\\
&\hspace{-3ex}-s^6 (343 s_1+369 s_2)+s^5 \big(1025 s_1^2+1384 s_1 s_2+1185 s_2^2\big)-s^4 \big(1695 s_1^3+1329 s_1^2 s_2+1191 s_1 s_2^2+2105 s_2^3\big)+s^3 \big(1675 s_1^4-1032 s_1^3 s_2-822 s_1^2 s_2^2\notag\\
&\hspace{-3ex}-2216 s_1 s_2^3+2235 s_2^4\big)+s^2 \big(-989 s_1^5+2765 s_1^4 s_2-582 s_1^3 s_2^2+646 s_1^2 s_2^3+5099 s_1 s_2^4-1419 s_2^5\big)+s \big(323 s_1^6-1824 s_1^5 s_2+3765 s_1^4 s_2^2-480 s_1^3 s_2^3\notag\\
&\hspace{-3ex}+1365 s_1^2 s_2^4-3648 s_1 s_2^5+499 s_2^6\big)-15 (s_1-s_2)^3 \big(3 s_1^4-18 s_1^3 s_2+94 s_1^2 s_2^2+46 s_1 s_2^3-5 s_2^4\big)\bigg]+4 n_S s \bigg[s^6-s^5 (7 s_1+6 s_2)+s^4 \big(20 s_1^2+16 s_1 s_2+15 s_2^2\big)\notag\\
&\hspace{-3ex}+s^3 \big(-30 s_1^3+24 s_1^2 s_2+6 s_1 s_2^2-20 s_2^3\big)+s^2 \big(25 s_1^4-108 s_1^3 s_2-18 s_1^2 s_2^2-44 s_1 s_2^3+15 s_2^4\big)+s \big(-11 s_1^5+110 s_1^4 s_2+282 s_1^3 s_2^2-116 s_1^2 s_2^3+41 s_1 s_2^4\notag\\
&\hspace{-3ex}-6 s_2^5\big)+2 s_1^6-36 s_1^5 s_2-285 s_1^4 s_2^2+240 s_1^3 s_2^3+90 s_1^2 s_2^4-12 s_1 s_2^5+s_2^6\bigg]\bigg\}\bar{B}_0(s)+\frac{\pi ^2 s_1^2}{90 \s^4} \bigg\{-n_F \bigg[45 s^7-s^6 (283 s_1+585 s_2)+s^5 \big(769 s_1^2+2064 s_1 s_2\notag\\
&\hspace{-3ex}+75 s_2^2\big)-5 s^4 \big(235 s_1^3+533 s_1^2 s_2+525 s_1 s_2^2-573 s_2^3\big)+3 s^3 \big(365 s_1^4+424 s_1^3 s_2-326 s_1^2 s_2^2+320 s_1 s_2^3-1295 s_2^4\big)-s^2 (s_1-s_2)^2 \big(625 s_1^3+941 s_1^2 s_2\notag\\
&\hspace{-3ex}-2565 s_1 s_2^2-1545 s_2^3\big)+s (s_1-s_2)^4 \big(203 s_1^2+228 s_1 s_2-75 s_2^2\big)-(29 s_1-15 s_2) (s_1-s_2)^6\bigg]-2 n_G \bigg[45 s^7-s^6 (323 s_1+405 s_2)+s^5 \big(989 s_1^2\notag\\
&\hspace{-3ex}+1824 s_1 s_2+2355 s_2^2\big)-5 s^4 \big(335 s_1^3+553 s_1^2 s_2+753 s_1 s_2^2+879 s_2^3\big)+3 s^3 \big(565 s_1^4+344 s_1^3 s_2+194 s_1^2 s_2^2+160 s_1 s_2^3+785 s_2^4\big)-s^2 (s_1-s_2)^2 \notag\\
&\hspace{-3ex}\times\big(1025 s_1^3+721 s_1^2 s_2-405 s_1 s_2^2-885 s_2^3\big)+s (s_1-s_2)^4 \big(343 s_1^2-12 s_1 s_2-915 s_2^2\big)-(49 s_1-75 s_2) (s_1-s_2)^6\bigg]+4 n_S s_1 \bigg[2 s^6-s^5 (11 s_1+36 s_2)\notag\\
&\hspace{-3ex}+5 s^4 \big(5 s_1^2+22 s_1 s_2-57 s_2^2\big)-6 s^3 \big(5 s_1^3+18 s_1^2 s_2-47 s_1 s_2^2-40 s_2^3\big)+2 s^2 (s_1-s_2)^2 \big(10 s_1^2+32 s_1 s_2+45 s_2^2\big)-s (s_1-s_2)^4 (7 s_1+12 s_2)\notag\\
&\hspace{-3ex}+(s_1-s_2)^6\bigg]\bigg\}\bar{B}_0(s_1)+\frac{\pi ^2 s_2^2}{90 \s^4} \bigg\{2 n_G \bigg[3 s^7+5 s^6 (s_2-9 s_1)-s^5 \big(243 s_1^2+224 s_1 s_2+93 s_2^2\big)+s^4 \big(285 s_1^3-1469 s_1^2 s_2+1211 s_1 s_2^2+285 s_2^3\big)\notag\\
&\hspace{-3ex}+s^3 \big(285 s_1^4+2896 s_1^3 s_2+5082 s_1^2 s_2^2-1704 s_1 s_2^3-415 s_2^4\big)-s^2 (s_1-s_2)^2 \big(243 s_1^3+1955 s_1^2 s_2-1415 s_1 s_2^2-327 s_2^3\big)-s (s_1-s_2)^4 \big(45 s_1^2\notag\\
&\hspace{-3ex}+404 s_1 s_2+135 s_2^2\big)+(s_1-s_2)^6 (3 s_1+23 s_2)\bigg]-4n_S \bigg[3 s^7-10 s^6 (3 s_1+2 s_2)+s^5 \big(162 s_1^2+116 s_1 s_2+57 s_2^2\big)-s^4 \big(135 s_1^3+244 s_1^2 s_2+149 s_1 s_2^2\notag\\
&\hspace{-3ex}+90 s_2^3\big)+s^3 \big(-135 s_1^4+416 s_1^3 s_2-18 s_1^2 s_2^2+36 s_1 s_2^3+85 s_2^4\big)+2 s^2 (s_1-s_2)^2 \big(81 s_1^3+40 s_1^2 s_2-10 s_1 s_2^2-24 s_2^3\big)-s (s_1-s_2)^4 \big(30 s_1^2+4 s_1 s_2\notag\\
&\hspace{-3ex}-15 s_2^2\big)+(s_1-s_2)^6 (3 s_1-2 s_2)\bigg]-n_F \big(57 s^7-5 s^6 (135 s_1+77 s_2)+s^5 \big(243 s_1^2+2704 s_1 s_2+1113 s_2^2\big)+s^4 \big(375 s_1^3-1511 s_1^2 s_2-3751 s_1 s_2^2\notag\\
&\hspace{-3ex}-1785 s_2^3\big)+s^3 \big(375 s_1^4-3536 s_1^3 s_2+1518 s_1^2 s_2^2+1464 s_1 s_2^3+1715 s_2^4\big)+s^2 (s_1-s_2)^2\big(243 s_1^3-1025 s_1^2 s_2-775 s_1 s_2^2-987 s_2^3\big)-s (s_1-s_2)^4\notag\\
&\hspace{-3ex}\times \big(675 s_1^2-4 s_1 s_2-315 s_2^2\big)+(s_1-s_2)^6 (57 s_1-43 s_2)\big)\bigg\}\bar{B}_0(s_2)-\frac{2 \pi ^2 s^2 s_1^2 s_2^2}{3 \s^4} \bigg\{n_F \bigg[21 s^4-18 s^3 (2 s_1+3 s_2)+s^2 \big(30 s_1^2+70 s_1 s_2+36 s_2^2\big)\notag\\
&\hspace{-3ex}+s \big(-36 s_1^3+70 s_1^2 s_2-40 s_1 s_2^2+6 s_2^3\big)+3 (s_1-s_2)^3 (7 s_1+3 s_2)\bigg]-2 n_G \bigg[3 s^4+18 s^3 s_2-2 s^2 \big(3 s_1^2+7 s_1 s_2+36 s_2^2\big)+2 s s_2 \big(-7 s_1^2-32 s_1 s_2+39 s_2^2\big)\notag\\
&\hspace{-3ex}+3 (s_1-s_2)^3 (s_1+9 s_2)\big]-4 n_S s s_1 \bigg[3 s^2+s (s_2-6 s_1)+3 s_1^2+s_1 s_2-4 s_2^2\bigg]\bigg\}C_0(s,s_1,s_2)-\frac{16\p^2}{720}\,(s+s_1+s_2)\big(n_S+11n_F+62n_G\big).
\end{align}\endgroup
%%%%%%%%%%%%%%%%%%
%%%%%%%%%%%%%%%%%%
%%%%%%%%%%%%%%%%%%
Finally the $A_5$ form factor can be written as
\begingroup\makeatletter\def\f@size{7.5}\check@mathfonts
\def\maketag@@@#1{\hbox{\m@th\large\normalfont#1}}%
\begin{align}
A_5^{Ren}&=\frac{\pi ^2}{1800 \s^2} \bigg\{n_F \bigg[137 s^6-674 s^5 (s_1+s_2)+11 s^4 \big(133 s_1^2+150 s_1 s_2+133 s_2^2\big)-4 s^3 \big(463 s_1^3+244 s_1^2 s_2+244 s_1 s_2^2+463 s_2^3\big)+s^2 \big(1463 s_1^4\notag\\
&\hspace{-3ex}-976 s_1^3 s_2+1234 s_1^2 s_2^2-976 s_1 s_2^3+1463 s_2^4\big)-2 s (s_1-s_2)^2 \big(337 s_1^3-151 s_1^2 s_2-151 s_1 s_2^2+337 s_2^3\big)+(s_1-s_2)^4 \big(137 s_1^2-126 s_1 s_2+137 s_2^2\big)\bigg]\notag\\
&\hspace{-3ex}-2 n_G \bigg[883 s^6-4766 s^5 (s_1+s_2)+s^4 \big(11117 s_1^2+12450 s_1 s_2+11117 s_2^2\big)-4 s^3 \big(3617 s_1^3+1921 s_1^2 s_2+1921 s_1 s_2^2+3617 s_2^3\big)+s^2 \big(11117 s_1^4\notag\\
&\hspace{-3ex}-7684 s_1^3 s_2-194 s_1^2 s_2^2-7684 s_1 s_2^3+11117 s_2^4\big)-2 s (s_1-s_2)^2 \big(2383 s_1^3-1459 s_1^2 s_2-1459 s_1 s_2^2+2383 s_2^3\big)+(s_1-s_2)^4 \big(883 s_1^2-1234 s_1 s_2\notag\\
&\hspace{-3ex}+883 s_2^2\big)\bigg]+4 n_S \bigg[18 s^6-86 s^5 (s_1+s_2)+7 s^4 \big(26 s_1^2+25 s_1 s_2+26 s_2^2\big)-s^3 \big(228 s_1^3+89 s_1^2 s_2+89 s_1 s_2^2+228 s_2^3\big)+s^2 \big(182 s_1^4\notag\\
&\hspace{-3ex}-89 s_1^3 s_2+26 s_1^2 s_2^2-89 s_1 s_2^3+182 s_2^4\big)-s (s_1-s_2)^2 \big(86 s_1^3-3 s_1^2 s_2-3 s_1 s_2^2+86 s_2^3\big)+2 (s_1-s_2)^4 \big(9 s_1^2-7 s_1 s_2+9 s_2^2\big)\bigg]\bigg\}+\frac{\pi ^2 s^2}{180 \s^3} \bigg\{n_F \bigg[43 s^6\notag\\
&\hspace{-3ex}-272 s^5 (s_1+s_2)+s^4 \big(715 s_1^2+928 s_1 s_2+715 s_2^2\big)-20 s^3 \big(50 s_1^3+39 s_1^2 s_2+39 s_1 s_2^2+50 s_2^3\big)+s^2 \big(785 s_1^4-508 s_1^3 s_2-114 s_1^2 s_2^2-508 s_1 s_2^3+785 s_2^4\big)\notag\\
&\hspace{-3ex}-4 s \big(82 s_1^5-251 s_1^4 s_2+109 s_1^3 s_2^2+109 s_1^2 s_2^3-251 s_1 s_2^4+82 s_2^5\big)+3 (s_1-s_2)^2 \big(19 s_1^4-86 s_1^3 s_2+14 s_1^2 s_2^2-86 s_1 s_2^3+19 s_2^4\big)\bigg]+2 n_G \bigg[23 s^6\notag\\
&\hspace{-3ex}-112 s^5 (s_1+s_2)+s^4 \big(215 s_1^2+128 s_1 s_2+215 s_2^2\big)-40 s^3 \big(5 s_1^3-6 s_1^2 s_2-6 s_1 s_2^2+5 s_2^3\big)+s^2 \big(85 s_1^4-368 s_1^3 s_2+6 s_1^2 s_2^2-368 s_1 s_2^3+85 s_2^4\big)\notag\\
&\hspace{-3ex}-8 s \big(s_1^5-8 s_1^4 s_2+82 s_1^3 s_2^2+82 s_1^2 s_2^3-8 s_1 s_2^4+s_2^5\big)-3 (s_1-s_2)^2 \big(s_1^4-14 s_1^3 s_2-94 s_1^2 s_2^2-14 s_1 s_2^3+s_2^4\big)\bigg]+4 n_S \bigg[2 s^6-13 s^5 (s_1+s_2)+s^4 \big(35 s_1^2\notag\\
&\hspace{-3ex}+47 s_1 s_2+35 s_2^2\big)-5 s^3 \big(10 s_1^3+9 s_1^2 s_2+9 s_1 s_2^2+10 s_2^3\big)+s^2 \big(40 s_1^4-17 s_1^3 s_2-36 s_1^2 s_2^2-17 s_1 s_2^3+40 s_2^4\big)+s \big(-17 s_1^5+46 s_1^4 s_2+s_1^3 s_2^2+s_1^2 s_2^3\notag\\
&\hspace{-3ex}+46 s_1 s_2^4-17 s_2^5\big)+3 (s_1-s_2)^6\bigg]\bigg\}\bar{B}_0(s)+\frac{\pi ^2 s_1^2}{180 \s^3} \bigg\{n_F \bigg[57 s^6-4 s^5 (82 s_1+93 s_2)+s^4 \big(785 s_1^2+1004 s_1 s_2+615 s_2^2\big)-4 s^3 \big(250 s_1^3+127 s_1^2 s_2\notag\\
&\hspace{-3ex}+109 s_1 s_2^2+150 s_2^3\big)+s^2 \big(715 s_1^4-780 s_1^3 s_2-114 s_1^2 s_2^2-436 s_1 s_2^3+615 s_2^4\big)-4 s (s_1-s_2)^3 \big(68 s_1^2-28 s_1 s_2-93 s_2^2\big)+(s_1-s_2)^5 (43 s_1-57 s_2)\bigg]\notag\\
&\hspace{-3ex}-2 n_G \bigg[3 s^6+8 s^5 (s_1-6 s_2)-s^4 \big(85 s_1^2+64 s_1 s_2+195 s_2^2\big)+8 s^3 \big(25 s_1^3+46 s_1^2 s_2+82 s_1 s_2^2+60 s_2^3\big)-s^2 \big(215 s_1^4+240 s_1^3 s_2+6 s_1^2 s_2^2-656 s_1 s_2^3\notag\\
&\hspace{-3ex}+195 s_2^4\big)+16 s (s_1-s_2)^3 \big(7 s_1^2+13 s_1 s_2+3 s_2^2\big)-(s_1-s_2)^5 (23 s_1+3 s_2)\big)+4 n_S \bigg[3 s^6-s^5 (17 s_1+18 s_2)+s^4 \big(40 s_1^2+46 s_1 s_2+45 s_2^2\big)+s^3 \big(-50 s_1^3\notag\\
&\hspace{-3ex}-17 s_1^2 s_2+s_1 s_2^2-60 s_2^3\big)+s^2 \big(35 s_1^4-45 s_1^3 s_2-36 s_1^2 s_2^2+s_1 s_2^3+45 s_2^4\big)-s (s_1-s_2)^3 \big(13 s_1^2-8 s_1 s_2-18 s_2^2\big)+(s_1-s_2)^5 (2 s_1-3 s_2)\bigg]\bigg\}\bar{B}_0(s_1)\notag\\
&+\frac{\pi ^2 s_2^2}{180\s^3} \bigg\{n_F \bigg[57 s^6-4 s^5 (93 s_1+82 s_2)+s^4 \big(615 s_1^2+1004 s_1 s_2+785 s_2^2\big)-4 s^3 \big(150 s_1^3+109 s_1^2 s_2+127 s_1 s_2^2+250 s_2^3\big)+s^2 \big(615 s_1^4\notag\\
&\hspace{-3ex}-436 s_1^3 s_2-114 s_1^2 s_2^2-780 s_1 s_2^3+715 s_2^4\big)-4 s (s_1-s_2)^3 \big(93 s_1^2+28 s_1 s_2-68 s_2^2\big)+(s_1-s_2)^5 (57 s_1-43 s_2)\bigg]-2 n_G \bigg[3 s^6+8 s^5 (s_2-6 s_1)\notag\\
&\hspace{-3ex}-s^4 \big(195 s_1^2+64 s_1 s_2+85 s_2^2\big)+8 s^3 \big(60 s_1^3+82 s_1^2 s_2+46 s_1 s_2^2+25 s_2^3\big)-s^2 \big(195 s_1^4-656 s_1^3 s_2+6 s_1^2 s_2^2+240 s_1 s_2^3+215 s_2^4\big)-16 s (s_1-s_2)^3 \notag\\
&\hspace{-3ex}\times\big(3 s_1^2+13 s_1 s_2+7 s_2^2\big)+(s_1-s_2)^5 (3 s_1+23 s_2)\bigg]+4 n_S \bigg[3 s^6-s^5 (18 s_1+17 s_2)+s^4 \big(45 s_1^2+46 s_1 s_2+40 s_2^2\big)+s^3 \big(-60 s_1^3+s_1^2 s_2-17 s_1 s_2^2\notag\\
&\hspace{-3ex}-50 s_2^3\big)+s^2 \big(45 s_1^4+s_1^3 s_2-36 s_1^2 s_2^2-45 s_1 s_2^3+35 s_2^4\big)-s (s_1-s_2)^3 \big(18 s_1^2+8 s_1 s_2-13 s_2^2\big)+(s_1-s_2)^5 (3 s_1-2 s_2)\bigg]\bigg\}\bar{B}_0(s_2)\notag\\
&\hspace{-3ex}+\frac{\pi ^2 s^2 s_1^2 s_2^2}{3 \s^3} \bigg\{n_F \bigg[3 s^3-3 s^2 (s_1+s_2)-s \big(3 s_1^2+2 s_1 s_2+3 s_2^2\big)+3 (s_1-s_2)^2 (s_1+s_2)\bigg]+n_G \bigg[-6 s^3+6 s^2 (s_1+s_2)+s \big(6 s_1^2+28 s_1 s_2+6 s_2^2\big)\notag\\
&\hspace{-3ex}-6 (s_1-s_2)^2 (s_1+s_2)\bigg]-4 n_S s s_1 s_2\bigg\}C_0(s,s_1,s_2)-\frac{8\p^2}{720}\,(s^2+s_1^2+s_2^2)\big(n_S+11n_F+62n_G\big).
\end{align}\endgroup

\bibliographystyle{h-physrev5.bst}
\bibliography{TJJdilatonG}

\begin{thebibliography}{10}

\bibitem{Duff:1993wm}
M.~J. Duff,
\newblock Class. Quant. Grav. {\bf 11}, 1387 (1994), arXiv:hep-th/9308075.
%%CITATION = HEP-TH/9308075;%%

\bibitem{Osborn:1993cr}
H.~Osborn and A.~C. Petkou,
\newblock Ann. Phys. {\bf 231}, 311 (1994), arXiv:hep-th/9307010.
%%CITATION = HEP-TH/9307010;%%

\bibitem{Erdmenger:1996yc}
J.~Erdmenger and H.~Osborn,
\newblock Nucl.Phys. {\bf B483}, 431 (1997), arXiv:hep-th/0103237.

\bibitem{Giannotti:2008cv}
M.~Giannotti and E.~Mottola,
\newblock Phys. Rev. {\bf D79}, 045014 (2009), arXiv:0812.0351.
%%CITATION = 0812.0351;%%

\bibitem{Armillis:2009pq}
R.~Armillis, C.~Corian\`{o}, and L.~Delle~Rose,
\newblock Phys. Rev. {\bf D81}, 085001 (2010), arXiv:0910.3381.
%%CITATION = 0910.3381;%%

\bibitem{Coriano:2018zdo}
C.~Corian\`o and M.~M. Maglio,
\newblock (2018), arXiv:1802.01501.
%%CITATION = ARXIV:1802.01501;%%

\bibitem{Coriano:2018bbe}
C.~Corianò and M.~M. Maglio,
\newblock (2018), arXiv:1802.07675.
%%CITATION = ARXIV:1802.07675;%%

\bibitem{Coriano:2013xua}
C.~Corian\`o, L.~Delle~Rose, C.~Marzo, and M.~Serino,
\newblock (2013), arXiv:1306.4248.
%%CITATION = ARXIV:1306.4248;%%

\bibitem{Coriano:2013nja}
C.~Corian\`o, L.~Delle~Rose, C.~Marzo, and M.~Serino,
\newblock Class. Quant. Grav. {\bf 31}, 105009 (2014), arXiv:1311.1804.
%%CITATION = ARXIV:1311.1804;%%

\bibitem{Bastianelli:2012bz}
F.~Bastianelli, O.~Corradini, J.~M. Dávila, and C.~Schubert,
\newblock Phys. Lett. {\bf B716}, 345 (2012), arXiv:1202.4502.
%%CITATION = ARXIV:1202.4502;%%

\bibitem{Bastianelli:2012es}
F.~Bastianelli, O.~Corradini, J.~M. Davila, and C.~Schubert,
\newblock Phys. Part. Nucl. {\bf 43}, 630 (2012), arXiv:1203.1689.
%%CITATION = ARXIV:1203.1689;%%

\bibitem{Bastianelli:2000hi}
F.~Bastianelli, S.~Frolov, and A.~A. Tseytlin,
\newblock JHEP {\bf 0002}, 013 (2000), arXiv:hep-th/0001041.
%%CITATION = HEP-TH/0001041;%%

\bibitem{Bastianelli:1999ab}
F.~Bastianelli, S.~Frolov, and A.~A. Tseytlin,
\newblock Nucl.Phys. {\bf B578}, 139 (2000), arXiv:hep-th/9911135.
%%CITATION = HEP-TH/9911135;%%

\bibitem{Bastianelli:2016nuf}
F.~Bastianelli and R.~Martelli,
\newblock JHEP {\bf 11}, 178 (2016), arXiv:1610.02304.
%%CITATION = ARXIV:1610.02304;%%

\bibitem{Bastianelli:2018osv}
F.~Bastianelli and M.~Broccoli,
\newblock arXiv:hep-th/1808.03489.
%%CITATION = HEP-TH/0007222;%%


\bibitem{Bonora:2017gzz}
L.~Bonora {\em et~al.},
\newblock Eur. Phys. J. {\bf C77}, 511 (2017), arXiv:1703.10473.
%%CITATION = ARXIV:1703.10473;%%

\bibitem{Bonora:2018obr}
L.~Bonora {\em et~al.},
\newblock Eur. Phys. J. {\bf C78}, 652 (2018), arXiv:1807.01249.
%%CITATION = ARXIV:1807.01249;%%

\bibitem{Bonora:2014qla}
L.~Bonora, S.~Giaccari, and B.~Lima~de Souza,
\newblock JHEP {\bf 07}, 117 (2014), arXiv:1403.2606.
%%CITATION = ARXIV:1403.2606;%%

\bibitem{2014JHEP...03..111B}
A.~{Bzowski}, P.~{McFadden}, and K.~{Skenderis},
\newblock Journal of High Energy Physics {\bf 3}, 111 (2014), arXiv:1304.7760.

\bibitem{Coriano:2013jba}
C.~Corian\`o, L.~Delle~Rose, E.~Mottola, and M.~Serino,
\newblock JHEP {\bf 1307}, 011 (2013), arXiv:1304.6944.
%%CITATION = ARXIV:1304.6944;%%

\bibitem{Armillis:2009im}
R.~Armillis, C.~Corian\`o, and L.~Delle~Rose,
\newblock Phys. Lett. {\bf B682}, 322 (2009), arXiv:0909.4522.
%%CITATION = 0909.4522;%%

\bibitem{Armillis:2010qk}
R.~Armillis, C.~Corian\`o, and L.~Delle~Rose,
\newblock Phys. Rev. {\bf D82}, 064023 (2010), arXiv:1005.4173.
%%CITATION = ARXIV:1005.4173;%%

\bibitem{Coriano:2014gja}
C.~Corian\`o, A.~Costantini, L.~Delle~Rose, and M.~Serino,
\newblock JHEP {\bf 06}, 136 (2014), arXiv:1402.6369.
%%CITATION = ARXIV:1402.6369;%%

\bibitem{Coriano:2012dg}
C.~Corian\`o, L.~Delle~Rose, C.~Marzo, and M.~Serino,
\newblock Phys.Lett. {\bf B717}, 182 (2012), arXiv:1207.2930.
%%CITATION = ARXIV:1207.2930;%%

\bibitem{Coriano:2012nm}
C.~Corian\`o, L.~Delle~Rose, A.~Quintavalle, and M.~Serino,
\newblock JHEP {\bf 1306}, 077 (2013), arXiv:1206.0590.
%%CITATION = ARXIV:1206.0590;%%

\bibitem{Bzowski:2013sza}
A.~Bzowski, P.~McFadden, and K.~Skenderis,
\newblock JHEP {\bf 03}, 111 (2014), arXiv:1304.7760.
%%CITATION = ARXIV:1304.7760;%%

\bibitem{Bzowski:2018fql}
A.~Bzowski, P.~McFadden, and K.~Skenderis,
\newblock (2018), arXiv:1805.12100.
%%CITATION = ARXIV:1805.12100;%%

\bibitem{Cappelli:2001pz}
A.~Cappelli, R.~Guida, and N.~Magnoli,
\newblock Nucl.Phys. {\bf B618}, 371 (2001), arXiv:hep-th/0103237.
%%CITATION = HEP-TH/0103237;%%

\bibitem{Coriano:2012wp}
C.~Corian\`o, L.~Delle~Rose, E.~Mottola, and M.~Serino,
\newblock (2012), arXiv:1203.1339.
%%CITATION = ARXIV:1203.1339;%%

\bibitem{Bzowski:2017poo}
A.~Bzowski, P.~McFadden, and K.~Skenderis,
\newblock (2017), arXiv:1711.09105.
%%CITATION = ARXIV:1711.09105;%%

\bibitem{Bzowski:2015yxv}
A.~Bzowski, P.~McFadden, and K.~Skenderis,
\newblock JHEP {\bf 02}, 068 (2016), arXiv:1511.02357.
%%CITATION = ARXIV:1511.02357;%%

\bibitem{APPELL}
P.~Appell and K.~Kamp\`e~de Feri\`et,
\newblock Paris : Gauthier-Villars , 434 p. (1926).
%%CITATION = ARXIV:1402.6369;%%

\bibitem{Usyukina:1993ch}
N.~I. Usyukina and A.~I. Davydychev,
\newblock Phys. Lett. {\bf B305}, 136 (1993).
%%CITATION = PHLTA,B305,136;%%

\bibitem{Coriano:2017mux}
C.~Corian\`o, M.~M. Maglio, and E.~Mottola,
\newblock (2017), arXiv:1703.08860.
%%CITATION = ARXIV:1703.08860;%%

\bibitem{Isono:2018rrb}
H.~Isono, T.~Noumi, and G.~Shiu,
\newblock JHEP {\bf 07}, 136 (2018), arXiv:1805.11107.
%%CITATION = ARXIV:1805.11107;%%

\bibitem{Gillioz:2018mto}
M.~Gillioz,
\newblock (2018), arXiv:1807.07003.
%%CITATION = ARXIV:1807.07003;%%

\bibitem{Bastianelli:2000rs}
F.~Bastianelli, G.~Cuoghi, and L.~Nocetti,
\newblock Class. Quant. Grav. {\bf 18}, 793 (2001), arXiv:hep-th/0007222.
%%CITATION = HEP-TH/0007222;%%



\end{thebibliography}

\end{document}